\documentclass[12pt]{nmsuth01}
\usepackage{latexsym}
\usepackage{amsmath,amsthm}
\usepackage{mathtools}
\usepackage{braket}
\usepackage[all]{xy}
\usepackage{graphicx}
\usepackage{caption}
\usepackage{subcaption}

\usepackage[colorlinks=true,linktocpage=true,linkcolor=blue,citecolor=blue]{hyperref}

\usepackage[margin=1in]{geometry}

\usepackage{multirow}

\theoremstyle{plain}

\theoremstyle{definition}
\theoremstyle{remark}

\newtheorem*{convention*}{Convention}
\newlength{\singlespace}
\newlength{\doublespace}
\usepackage[usenames,dvipsnames]{color}
\newcommand{\matt}[1]{ {\color{red} #1 }}

\usepackage[usenames,dvipsnames]{color}

\usepackage[usenames,dvipsnames]{color}

\extrafloats{100}

\usepackage{gensymb}

\begin{document}
\setlength{\baselineskip}{\doublespace}
%
%
\pagenumbering{roman}
\pagestyle{empty}
\renewcommand{\baselinestretch}{2}
\begin{center}
CALCULATING THE TOTAL CHERENKOV RADIATION EMITTED BY LOW ENERGY PROTONS IN LIQUID ARGON AND COMPARING WITH ARGON SCINTILLATION LIGHT AT 128 NM \\ 
\vspace{0.1in}
BY\\
\vspace{0.1in}
HASAN REJOANUR RAHMAN, B.S., M.S.
\end{center}
\vspace{1.0in}
\begin{center}
A dissertation submitted to the Graduate School\\
\vspace{0.1in}
in partial fulfillment of the requirements\\
\vspace{0.1in}
for the degree \\
\vspace{0.1in}
Master of Science
\end{center}
\vfill
\begin{center}
Major Subject: Physics
\end{center}
\vspace{1.0in}
\begin{center}
New Mexico State University\\
\vspace{0.1in}
Las Cruces New Mexico\\
\vspace{0.1in}
March 2024
\end{center}

\pagestyle{plain}









\setlength{\baselineskip}{\singlespace}
\vspace{0.1in}
\begin{flushleft}

Hasan Rahman    \\
\hrulefill
\newline
\emph{{\large Candidate}}
\vspace{0.5in}

Physics \\
\hrulefill
\newline
\emph{{\large Major}}
\vspace{0.5in}

This Thesis is approved on behalf of the faculty of New Mexico State University, and it is acceptable in quality and form for publication:
\vspace{0.2in}

\emph{{\large Approved by the Thesis Committee:}}
\vspace{0.4in}

Dr. Matthew Sievert \\
\hrulefill
\newline
\emph{{\large Chairperson}}
\vspace{0.5in}

Dr. Vassili Papavassiliou  \\
\hrulefill
\newline
\emph{{\large Committee Member}}
\vspace{0.5in}

Dr. Moire Prescott (Dean's Rep) \\
\hrulefill
\newline
\emph{{\large Committee Member}}
\vspace{0.5in}
\end{flushleft}
%
\begin{center}
DEDICATION
\end{center}

I dedicate this work to my father late Md. Khalilur Rahman Miah, and my mother Hosne Ara Bhuiya.


%
\begin{center}
ACKNOWLEDGEMENTS
\end{center}

I thank my advisors, Dr. Matthew Sievert and Dr. Vassili Papavassiliou, for their encouragement, motivation, and patience with this project. Dr. Matthew Sievert has extensively helped me throughout all phases of this project and added speed to the completion of the thesis. 

\hspace{\parindent}

I would like to show my gratitude to all of my family members, including my late father Mohammad Khalilur Rahman Miah, mother Hosne Ara Bhuiya, wife Mehjabeen Sultana, brother Hasan Shahriar Rahman, father-in-law Mohammad Mahbubur Rahman, mother-in-law Syeda Sharmin Sultana, sister-in-law Sara Ahmed, my nephew Shayer and nieces Shayna and Shanaya for their endless support, love, and understanding. I am also grateful to my friends including Mir Mehedi Faruk and Fatema Farjana for their encouragement and support. 

\hspace{\parindent}

Moreover, my best friend Manuel Ca\~nas deserves special thanks for providing initial help and tutorials on programming in Python which was instrumental in this work. Also, my friends from the astronomy department of NMSU, Farhanul Hasan, Bryson Stemock and my research groupmate Joseph Bahder have helped me develop important features of the existing codes related to the visualization of this work.




%
\begin{center}
            VITA
\end{center}
\begin{flushleft}
\begin{tabular}{ll}
June 6, 1987 &  Born at Dhaka, Bangladesh
\\
& \\
2005-2010        &  B.S., University of Dhaka, Dhaka
\\
& \\
2016-2018        &  M.S., \& Graduate Teaching Assistant, Dept. of Physics
\\
                 &  The University of Texas at El Paso, El Paso, Texas.  
\\

& \\
2018-2021        &  Ph.D. \& Graduate Teaching Assistant, Department of Physics, 
\\               &  New Mexico State University, Las Cruces, New Mexico.

\& \\
2021-2022        &  Ph.D. Candidate \& Graduate Research Assistant, Department of Physics, 
\\               &  New Mexico State University, Las Cruces, New Mexico. 
\\
\end{tabular}
\end{flushleft}
\vspace{0.1in}
\begin{center}
PROFESSIONAL  AND HONORARY SOCIETIES
\end{center}
\begin{flushleft}

\begin{itemize}
\item   American Physical Society (APS)
\item   Physics Graduate Students Organization (PGSO), New Mexico State University
\item   Bangladesh Astronomical Society
\end{itemize}

\end{flushleft}
\vspace{0.1in}
\begin{center}PUBLICATIONS [or Papers Presented]
\end{center}

\begin{itemize}

\item   Antiporda, L., Bahder, J., Rahman, H., \& Sievert, M. D. (2022). Jet drift and collective flow in heavy-ion collisions. Physical Review D, 105(5), 054025.


\item   Sievert, M., Rahman, H., Bahder, J., \& Antiporda, L. (2021). Asymmetric Energy Loss of Jets Due to Collective Flow in Heavy-Ion Collisions. In APS Division of Nuclear Physics Meeting Abstracts (Vol. 2021, pp. EH-002).

\item   Rahman, H., \& Banuelos, J. (2019). Solvent and Concentration Effects Governing the Hierarchical Organization of Asphaltenes: A Small-Angle X-Ray Scattering Study. Bulletin of the American Physical Society, 64.

\item   Attended the 2022 TMD Winter School sponsored by the TMD Collaboration (The DOE Topical Collaboration for the Coordinated Theoretical Approach to Transverse Momentum Dependent Hadron Structure in QCD) on January 20-26, 2022 at Santa Fe, New Mexico.

\item   Attended the 37th annual Hampton University Graduate Studies (HUGS 2022) Program at Jefferson Lab from May 31st to June 17th, 2022, and presented a poster with the title “Asymmetric Energy Loss of Jets Due to Collective Flow in Heavy-Ion Collisions”.

\end{itemize}

\hspace{\parindent}

\begin{center}
FIELD OF STUDY
\end{center}
\begin{flushleft}
Major Field: Physics \newline
(Specializing in Theoretical Nuclear \& High-Energy Physics) \newline

\end{flushleft}

Biography: After completing his physics bachelor’s degree from the University of Dhaka in Bangladesh, Hasan started his M.S. in physics at The University of Texas at El Paso (UTEP) in Fall 2016. During this time, he conducted research on the hierarchical structure of Asphaltenes in solution to understand the factors governing the onset of aggregation and their morphology as part of his master’s thesis. He started doctoral studies in Physics at New Mexico State University (NMSU) in the Fall 2018 focusing on High Energy Physics to satisfy his curiosity in this area. He, therefore, joined the Experimental High Energy Physics group led by Dr. Robert Cooper where they collaborated with the CAPTAIN Mills experiment at Los Alamos National Lab which consists of a 10-ton liquid argon scintillation detector to study sterile neutrino phenomenology, search for sub-GeV dark matter and explore physics beyond the Standard Model in the neutrino sector. Since Dr. Cooper left NMSU in 2020, he started working on a project with Dr. Vassili Papavassiliou and his current advisor Dr. Matthew Sievert on calculating the Cherenkov radiation emitted by low-energy Proton (~500 MeV or less) in Liquid Argon (LAr) near resonance (~107 nm) and comparing with argon scintillation light at 128 nm which will put more light on the study of neutrinos. Parallel to this project, he has been doing research in theoretical Nuclear Physics with Dr. Matthew Sievert since Spring 2021 and focusing on the theoretical and phenomenological study of "Quark-Gluon Plasma" and the possible tomography using jets of high energy particles created in ultra-relativistic collisions in large particle detectors like CERN, RHIC, etc. to probe the initial nuclear matter. This research project is funded by a “U.S. Department of Energy (DOE)” grant (DE-SC0024560: Precision Jet Drift and Energy Loss) with close collaboration with the Los Alamos National Lab. He would like to continue making progress in his scientific endeavors in the field of Particle and Nuclear Physics and contribute to the array of collective human knowledge in the search for the laws governing the universe we live in.

%
\begin{center}
ABSTRACT
\end{center}
\vspace{0.3in}
\begin{center}

CALCULATING THE CHERENKOV RADIATION EMITTED BY LOW-ENERGY PROTON IN LIQUID ARGON (LAR) NEAR RESONANCE ($\sim$ 106.6 NM) AND COMPARING WITH ARGON SCINTILLATION LIGHT AT 128 NM.
\\
BY
\\
HASAN REJOANUR RAHMAN, B.S., M.S.
\end{center}
\vspace{0.3in}
\begin{center}
Master of Science

New Mexico State University

Las Cruces, New Mexico, 2024

Dr. Matthew D. Sievert, Chair
\end{center}
\vspace{0.3in}
\hspace{\parindent}
%


Neutrino experiments using liquid argon (LAr) detectors estimate the amount of light produced by different types of particles, but only consider scintillation light, at 128 nm, ignoring Cherenkov light contributions. This research aims to theoretically compare these two contributions to the total amount of light produced between ~ 128 – 500 nm for a proton travelling in LAr and explores how to leverage these under-utilized observables for future detector applications. 

A new theoretical fit of the refractive index of LAr was performed using recent experimental data, which incorporates the physics of anomalous dispersion in the UV resonance for the first time.  Using this fit, we integrate the Frank-Tamm (FT) formula to calculate the instantaneous Cherenkov angular distribution and yield of a proton with a given kinetic energy, as well as the integrated distribution and yield over its trajectory.  We compare our results with those obtained using two other non-absorptive refractive index fits available in the literature.  Because those fits diverge at the resonance, they significantly overestimate the yield.  

Additionally, Cherenkov radiation is highly collimated and emitted instantaneously, whereas scintillation light is isotropic and emitted after a delay. This fundamental difference in their angular distribution (AD) sets the ground for differentiating the Cherenkov signal over the scintillation background, resulting in a substantial observational effect $> 5 \sigma$ for $\sim$ 500 MeV protons in LAr.  This suggests that measurements of the collimated Cherenkov radiation could be used to measure the energy and direction of the incident protons more precisely in detector applications.

\tableofcontents
\newpage
\listoftables
\addcontentsline{toc}{section}{LIST OF TABLES}
\newpage
\listoffigures
\addcontentsline{toc}{section}{LIST OF FIGURES}
\newpage
\pagenumbering{arabic}
\section{INTRODUCTION} \label{intro}
\hspace{\parindent}

\subsection{Neutrinos: A particle entangled with mysteries of our universe} \label{neutrino}

\hspace{\parindent}

Neutrinos are fundamental particles that are abundantly produced in the core of young stars, supernovae, and from cosmic rays that bombard every inch of the Earth’s surface at nearly the speed of light. They can also be created artificially in nuclear reactors and particle accelerators. Despite being abundant in the universe, neutrinos can penetrate through large amounts of matter without interacting, which makes them difficult to detect. Nevertheless, neutrinos hold clues to some of the biggest questions in physics: the origin of matter, unification of forces, black hole formation, and more. The majority of the neutrinos detected on Earth originate from nuclear reactions taking place in our sun. \cite{Bahcall_2005}

Wolfgang Pauli first postulated/hypothesized the neutrinos in 1930 to explain how beta decay could conserve energy, momentum, and angular momentum (spin) \cite{1978PhT....31i..23B}. He considered neutrinos to be emitted from the nucleus together with the electron (or beta particle) in the beta decay process. Later, in Fermi's theory of beta decay, a neutron could decay to a proton, electron, and the smaller neutral particle (now called an electron antineutrino): \cite{1934ZPhy...88..161F}

\begin{figure}[h!]
\begin{centering}
\includegraphics[width=0.75\textwidth]
{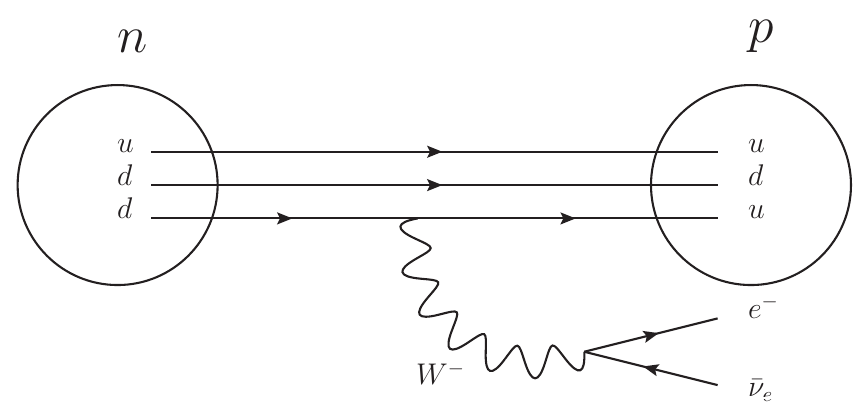}
\caption{The leading-order Feynman diagram for $\beta^-$ decay of a neutron into a proton, electron, and electron antineutrino via an intermediate $W^-$ boson.  The down quark in the neutron decays into an up quark to make a proton.
\label{f:ibd}
}
\end{centering}
\end{figure}
\begin{align}   \label{e:betadecay}
    n^0 \longrightarrow p^+ + e^- + \bar{\nu}_e \: .
\end{align}
The process is shown in Fig.~\ref{f:ibd}.

Neutrinos are fermions: a spin $\frac{1}{2}$ elementary particle predicted by Standard Model of Particle Physics denoted by $\nu$.  
Unlike their charged leptonic counterparts (electron, muon, tau), neutrinos do not have any electric charge or color charge, which restricts them to interact via the weak interaction and gravity only.  
For this reason, the elusive neutrinos are a major focus of research in nuclear and particle physics.

For each neutrino, there also exists a corresponding antiparticle, called an antineutrino, which also has spin of $\frac{1}{2}$ and no electric charge. Antineutrinos are distinguished from neutrinos by having opposite-signed lepton number and weak isospin, and right-handed instead of left-handed chirality. To conserve total lepton number in nuclear beta decay, electron neutrinos only appear together with positrons (anti-electrons), whereas electron antineutrinos only appear with electrons in beta decay. \cite{FClose:2012, RJayawardhana:2015}


Weak interactions create neutrinos in one of three leptonic flavors where each flavor is associated with the corresponding charged lepton: the electron neutrino $\nu_e$, the muon neutrino $\nu_\mu$, and the tau neutrino $\nu_\tau$.  
In nuclear beta decay, electron neutrinos can only occur together with positrons (anti-electrons) in order to conserve the overall lepton number. 

\begin{figure}[t]
\begin{centering}
\includegraphics[width=0.4\textwidth]{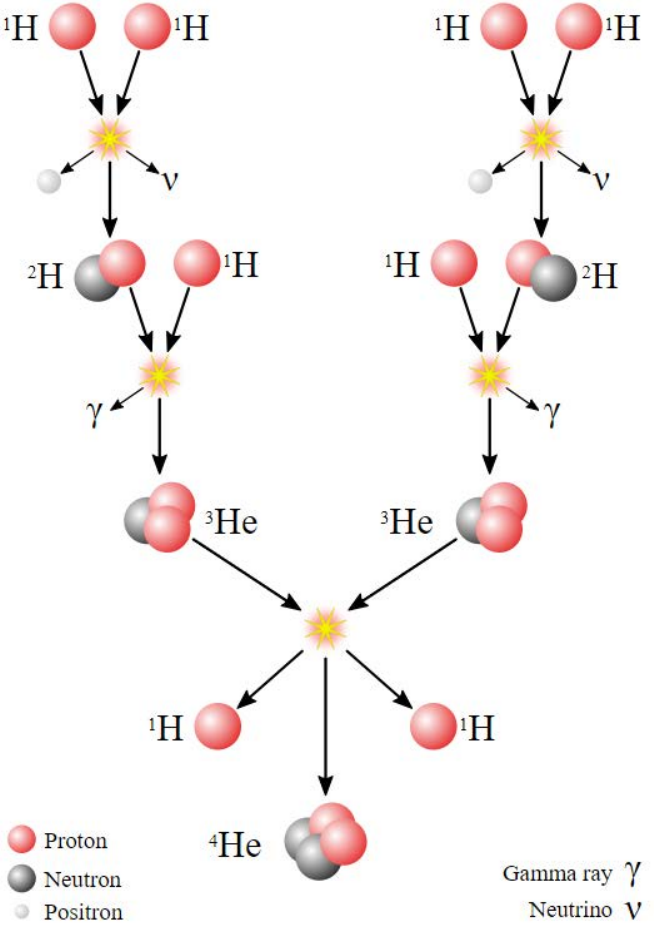}
\caption{Scheme of the proton-proton branch I reaction which can produce neutrinos and gamma rays. 
\label{f:ppchain}
}
\end{centering}
\end{figure}

%
Neutrinos are produced abundantly during stellar nucleosynthesis (Fig.~\ref{f:ppchain}), which fuels stellar evolution.  The fusion reaction of two hydrogen nuclei into a deuterium nucleus, which was studied theoretically by Hans Bethe
%
\cite{1939PhRv...55..434B}, 
\begin{align}   \label{e:ppchain}
    p + p \longrightarrow  {}^{2}_{1}D + e^+ + \nu_e + 0.42 \: \mathrm{MeV} \: ,
\end{align}
produces $0.42$ MeV of charged-particle kinetic energy which is trapped as heat, along with neutrinos as a byproduct.  As a result, the star produces a large flux of neutrinos, such that our sun may be thought of as a huge nuclear reactor which is the major source of neutrinos produced that pass through the earth as cosmic rays.
%
%
%
%
Variable energy fractions go to neutrinos and other products of the process. \cite{Adelberger_2011}



On the other hand, neutrinos also play a crucial role in the collapse of massive stars when they undergo a supernova explosion. When a massive star reaches the end of its life cycle and collapses its core into a neutron star or a black hole, it radiates almost all of its gravitational binding energy in the form of neutrinos and antineutrinos of all flavors \cite{Janka_2017, Cusinato_2022}.  The resulting conversion of protons to neutrons plays an important role in the subsequent formation of a neutron star.  Supernovae are thus one of the strongest and most frequent sources of cosmic neutrinos in the MeV energy range.  




\subsection{Interaction of neutrinos with matter}



The weak interaction is the main method of neutrino detection, consisting of both neutral-current interactions (mediated by the $Z$ boson) and charged-current interactions (mediated by the $W^\pm$ bosons).  

In a neutral-current interaction, the neutrino enters and then leaves the detector after having transferred some of its energy and momentum to a `target' particle.  If the interaction takes place in a scintillation detector, then the recoil of a scattered charged particle can be detected through the subsequent emission of scintillation light.  Moreover, if the target charged particle sufficiently lightweight (e.g. an electron), it might be accelerated to a relativistic speed and consequently emit Cherenkov radiation as well. All three neutrino flavors (electronic, muonic, and tauonic) can participate, regardless of the neutrino energy.  Neutral-current interactions occur equally for all neutrino flavors, while charged-current interactions produce different charged leptons depending on the flavor of the neutrino. \cite{Weinberg:1995mt}

\begin{figure}[t]
\centering
\includegraphics[width=0.6\textwidth]{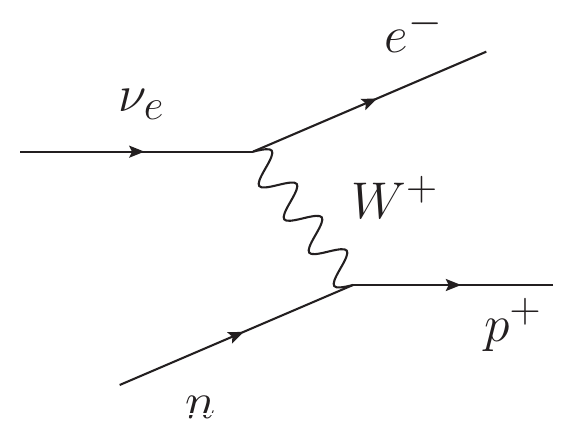}
\caption{Neutrino-initiated inverse beta decay $\nu_e + n \rightarrow e^- + p^+$.}
\label{f:InvBeta}
\end{figure}

In a charged current interaction, a high-energy neutrino transforms into its partner lepton (electron, muon, or tauon). However, if the neutrino does not have sufficient energy to create its heavier partner's mass, the charged current interaction is effectively unavailable to it. Neutrinos from the Sun and from nuclear reactors have enough energy to create electrons. Most accelerator-created neutrino beams can also create muons, and a very few can create tauons. A detector which can distinguish among these leptons can reveal the flavor of the neutrino incident to a charged current interaction; because the interaction involves the exchange of a W boson as shown in Fig.~\ref{f:InvBeta}, the `target' particle also changes (e.g., neutron $\rightarrow$ proton). \cite{Weinberg:1995mt}

\subsubsection{Coherent Elastic Neutrino-Nucleus Scattering (CE$\nu$NS)}

\hspace{\parindent}

\begin{figure}{}
\begin{centering}
\includegraphics[width=0.9\textwidth]{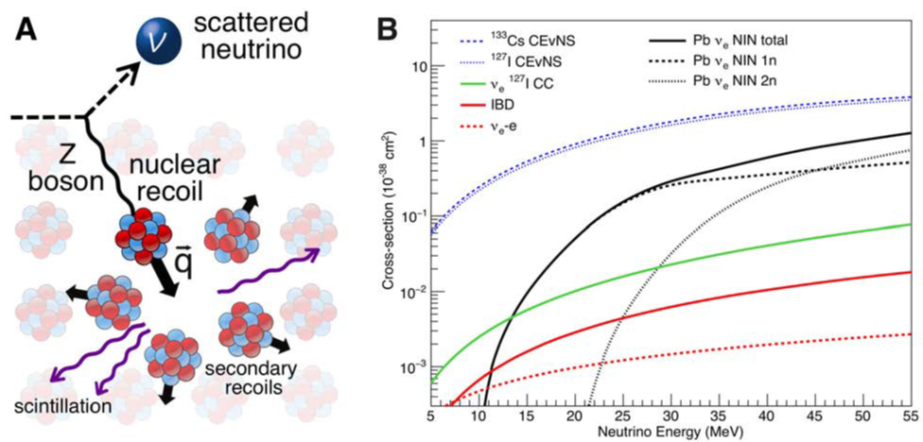}
\caption{(A) Coherent Elastic Neutrino-Nucleus Scattering. (B) Total cross-sections from
CE$\nu$NS process and some known neutrino couplings.  Note the dominance of the CE$\nu$NS cross section due to the $N^2$ coherent enhancement from the nuclear target.
\label{f:cevnsxsec} \cite{Akimov_2017}
}
\end{centering}
\end{figure}
%


One particularly important neutral-current interaction of neutrinos with matter is known as coherent elastic neutrino-nucleus scattering (CE$\nu$NS), first proposed in 1977 by D. Z. Freedman 
\cite{PhysRevD.9.1389}. In a CE$\nu$NS process, a neutrino hits a nucleus via exchange of a neutral Z boson as shown in Fig.~\ref{f:cevnsxsec}, the nucleus recoils as a whole. The interaction occurs coherently on the whole target nucleus at lower neutrino energies, while at higher energies the interaction will occur on a constituent of the nucleus.  For liquid argon, coherent neutrino scattering occurs up to $E_{\nu} \sim 50$ MeV \cite{coherentcollaboration2016coherent}.  In an elastic scattering process like CE$\nu$NS, the kinetic energy of the reaction is conserved, but the particles' directions are modified. The CE$\nu$NS process can be expressed as

\begin{align}   \label{e:cevnseq}
    \nu + A \longrightarrow \nu + A  \: .
\end{align}
%




The coherence of the CE$\nu$NS process results in an enhanced neutrino-nucleus cross section that is approximately proportional to $N^2$, the square of the number of neutrons in the nucleus, due to the small weak charge of the proton \cite{PhysRevD.9.1389}. The coherence condition, in which the neutrino scatters off all nucleons in a nucleus are in phase with each other which requires that the wavelength of the momentum transfer be larger than the size of the target nucleus. In this neutral current elastic process, the initial and final states are the same. 

Because the CE$\nu$NS cross section is cleanly predicted in the SM, deviations can indicate new physics beyond the Standard Model. Studies of beyond-the-standard-model (BSM) physics can be conducted on the CE$\nu$NS cross section by including possible non-standard-interaction (NSI) neutral currents taken from various BSM models. \cite{Akimov_2017, coherentcollaboration2016coherent}

\newpage

\subsubsection{Quasi-Elastic Neutrino Nucleus Scattering (QE)}

\hspace{\parindent}



\begin{figure}[t]
\centering
\begin{subfigure}{.47\textwidth}
    \centering
    \includegraphics[width=1\textwidth]{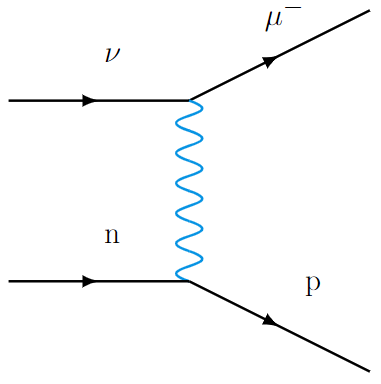}
    \caption{NC Quasi-Elastic scattering
    \label{f:ncqe}
    }
\end{subfigure}
\begin{subfigure}{.47\textwidth}
    \centering
    \includegraphics[width=1\textwidth]{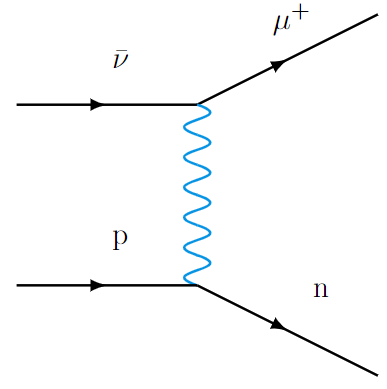}
    \caption{CC Quasi-Elastic scattering
    \label{f:ccqe}
    }
\end{subfigure}
\caption{Neutral current (NC) and charged current (CC) quasi-elastic neutrino-nucleus scattering processes that can create a proton and a neutron respectively, in the final state. 
\label{f:qe}
}    
\end{figure}

In a quasi-elastic (QE) process \cite{Strumia_2003, Sobczyk_2011}, a neutrino can interact with the neutron from a nucleus via neutral current to produce proton and a muon ($\mu^-$) (Fig.~\ref{f:ncqe}) given by: 

\begin{align}   \label{e:ncqe}
    \nu + n^0 \longrightarrow  p^+ + \mu^- \: .
\end{align}

Also, in a similar process, an antineutrino can interact with the nuclear proton quasi-elastically via charged-current (CC) interaction to produce a neutron and muon ($\mu^+$) in the final state (Fig.~\ref{f:ccqe}):

\begin{align}   \label{e:ccqe}
    \Bar{\nu} + p^+ \longrightarrow  n^0 + \mu^+ \: .
\end{align}

\subsubsection{Inverse Beta Decay (IBD)}

\hspace{\parindent}

For energetic neutrinos with higher energies (such as $E_\nu > 50$ MeV in liquid argon), the interaction with matter may not be elastic (like the CE$\nu$NS process described above) anymore and can instead knock out particles from the nuclear target.  One such process would be the conversion of a neutron from the target nucleus into a proton via W exchange.  This process is known as  ``Inverse Beta Decay (IBD),'' (Fig.~\ref{f:InvBeta}) which can result in the ejection of charged protons from the target nucleus if the neutrino energy is sufficient.  One may distinguish both neutrino-induced ($\nu_e + n  \longrightarrow  e^- + p^+$) and antineutrino-induced IBD ($\bar\nu_e + p^+  \longrightarrow e^+ + n$).
%
%
%
%
%
%
Both of these processes can be expressed for the other neutrino flavors as well.  For this thesis, quasi-elastic(QE) processes including neutrino-induced IBD is particularly relevant as a source of low-energy protons in neutrino experiments.  This process can produce protons correlated to the original incoming neutrinos.  


\subsubsection{Resonance Production (RES)}

\hspace{\parindent}

In a resonance production (RES) process \cite{athar2022neutrinosinteractionsmatter}, a neutrino can interact with a proton from a nucleus via charged current ($W^+$ exchange) to produce a lepton and an intermediate resonance particle, $\Delta^{++}$ which subsequently decay into a proton and a pion ($\pi^+$) given by: 

\begin{align}   \label{e:res}
    \nu_l + p^+ \longrightarrow l^- + \Delta^{++} \longrightarrow  l^- + \pi^+ + p^+ \: .
\end{align}
%


\subsubsection{Background Suppression}

To identify elusive neutrino interactions, neutrino experiments must control the background of cosmic ray flux, which dominates over the signal of genuine neutrino reactions by orders of magnitude. 
Theoretically, there are two main sources of astrophysical neutrinos: 

\begin{enumerate}
    \item Cosmic neutrino background (Big Bang originated): Around 1 second after the Big Bang, neutrinos decoupled, giving rise to a background level of neutrinos known as the cosmic neutrino background (CNB).
    \item Diffuse supernova neutrino background (Supernova originated): Many stars have gone supernova in the universe across its history, leaving a  diffuse supernova neutrino background on top of the CNB. \cite{2009APh....31..163P}
\end{enumerate}



To cut down on the cosmic ray background, the higher-energy ($>50 $ MeV or so) neutrino experiments often cover or surround the primary detector with an additional ``veto'' detector, which reveals when a cosmic ray passes into the primary detector.  This allows the corresponding activity in the primary detector to be rejected (``vetoed'').  

For lower-energy experiments (e.g. $E_\nu \leq 50 $ MeV for liquid argon), the cosmic rays themselves are not directly the problem. Instead, the spallation neutrons and radioisotopes produced by the cosmic rays may mimic the desired signals. For these experiments, the solution is to place the detector deep underground so that the earth above can reduce the cosmic ray rate to acceptable levels. \cite{Fusco:2023rvn}



\subsection{Experimental Detection of Neutrinos}


Numerous neutrino experiments are operating worldwide to detect different kinds of neutrinos with different energy ranges. The detectors can be classified by the material used to detect the neutrino interactions.  From this point of view, most of them are either scintillation-based detectors (CCM, JUNO, MiNER$\nu$A, NO$\nu$A, $SNO^+$ etc) or Cherenkov-based detectors (IceCube, MiniBooNE, Hyper-Kamiokande, T2K, SNO etc) but there are also detectors based on Time Projection Chambers (TPCs) (DUNE, ICARUS, MicroBooNE, etc), semiconductors, radiochemicals, and nuclear emulsions.  The various scintillation-based detectors use different kinds of scintillating materials, including argon, xenon, and mineral oil, in different states (solid, liquid, or gaseous). 
%
\begin{figure}[t]
\begin{centering}
\includegraphics[width=0.6\textwidth]{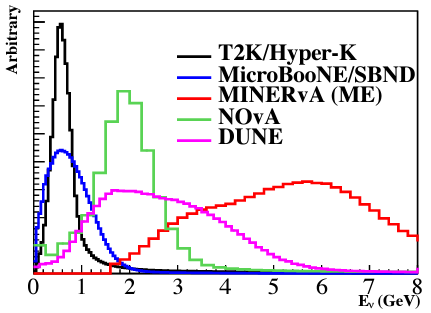}
\caption{Muon neutrino and muon anti-neutrino flux predictions from current and future accelerator based neutrino experiments (Hyper-Kamiokande, MicroBooNE, NOvA, DUNE, MINERvA with medium energy NuMI beamline) \cite{Katori:2016yel, Formaggio_2012}
\label{f:neutrinoexpeflux}
}
\end{centering}
\end{figure}
For example, scintillation-based detectors are often used to detect the nuclear recoil of argon atoms in a CE$\nu$NS process, while Cherenkov detectors are intended to measure the lighter charged particles (like muons) produced directly in charged-current interactions like IBD. The energy distribution of neutrino fluxes of different ongoing and future neutrino experiment programs is shown in Fig.~\ref{f:neutrinoexpeflux}.  

Two examples of important modern neutrino experiments are the Coherent CAPTAIN-Mills (CCM) and the Deep Underground Neutrino Experiment (DUNE). 

\hspace{\parindent}

\newpage

\subsubsection{Deep Underground Neutrino Experiment (DUNE)}

\hspace{\parindent}

\begin{figure}[t]
\begin{centering}
\includegraphics[width=0.95\textwidth]{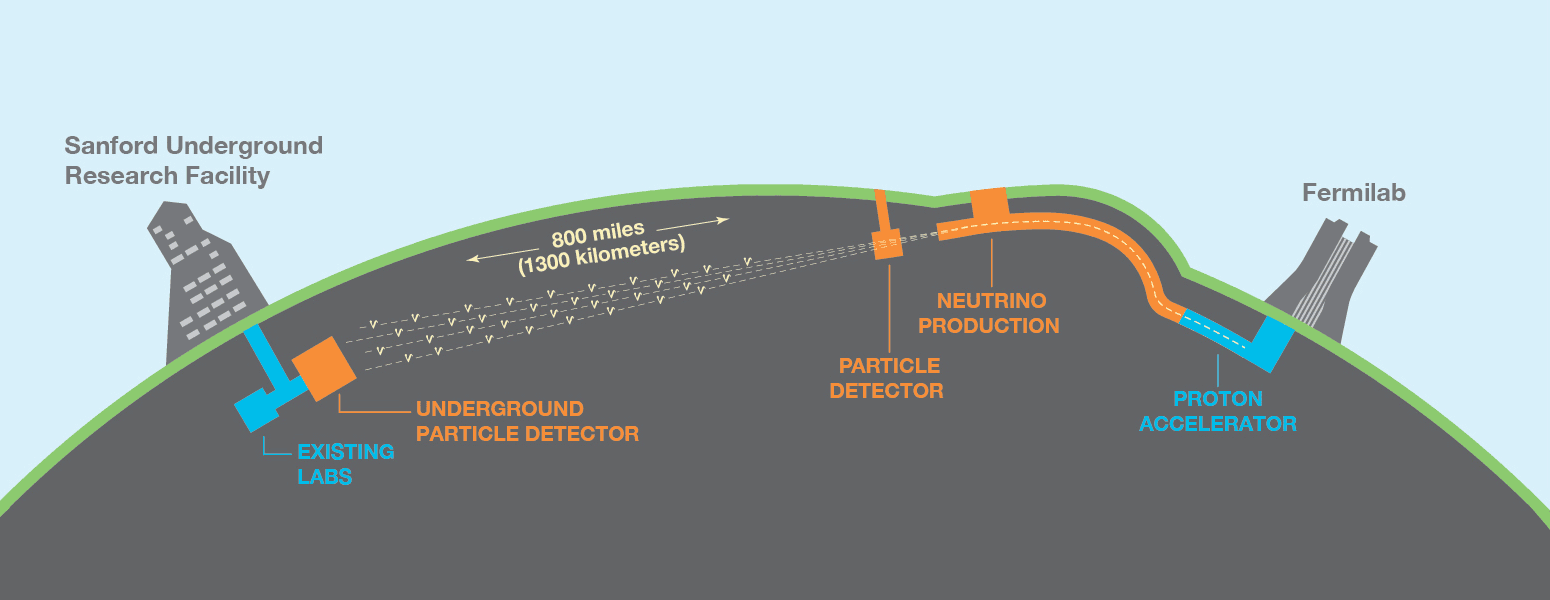}
\caption{Cartoon illustrating the configuration of the LBNF beamline at Fermilab, in Illinois, and the
DUNE detectors in Illinois and South Dakota, separated by 1300 km. \cite{abi2020deep}
\label{f:dune-schematic}
}
\end{centering}
\end{figure}
%

The international DUNE experiment (Fig.~\ref{f:dune-schematic}), hosted by the U.S. Department of Energy’s Fermi National Accelerator Laboratory (Fermilab) in Illinois, comprises three central components: 
\begin{enumerate}
    \item A new, high-intensity neutrino source generated from a megawatt-class proton accelerator at Fermilab, 
    \item A massive far detector (FD) situated 1.5 km underground at the Sanford Underground Research Facility (SURF) in South Dakota, and 
    \item A near detector (ND) installed just downstream of the neutrino source. 
\end{enumerate}
%

\begin{figure}[h!]
\begin{centering}
\includegraphics[width=0.65\textwidth]{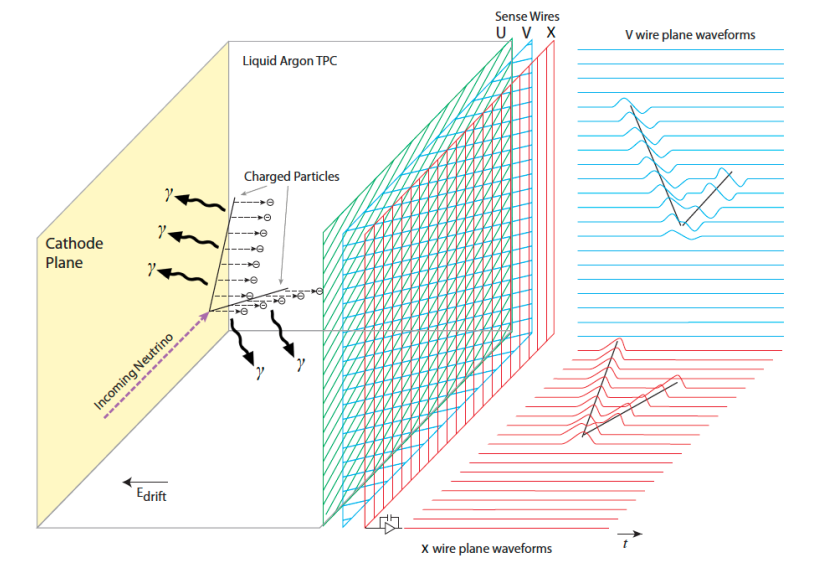}
\caption{The general operating principle of the SP Liquid Argon TPC. Negatively charged ionization electrons from the neutrino interaction drift horizontally opposite to the E field in the LAr and are collected on the anode, which is made up of the U, V and X sense wires. The right-hand side represents the time projections in two dimensions as the event occurs. \cite{abi2020deep}
\label{f:LArTPC}
}
\end{centering}
\end{figure}

The far detector is placed 1300 kilometers away from the neutrino production point to study the oscillations in neutrino flavors during their flight. The far detector will be a modular liquid argon time-projection chamber (LArTPC) used to reconstruct the neutrino interactions \cite{LBNE:2013dhi}. In an LArTPC (Fig.~\ref{f:LArTPC}), the charged particles from neutrino interactions produce ionization electrons which drift in an electric field towards a series of collection wires. The signal on the wires and the drift time are used to reconstruct the interaction \cite{Lowe:2021nkz}. The MicroBooNE detector currently collecting data at Fermilab has 8000 wires, and planned future experiments like DUNE will have 100 times more. \cite{abi2020deep}



\subsubsection{Coherent CAPTAIN-Mills (CCM) Neutrino Experiment}

\begin{figure}[h!]
\centering
\begin{subfigure}{.49\textwidth}
    \centering
    \includegraphics[width=1\textwidth]{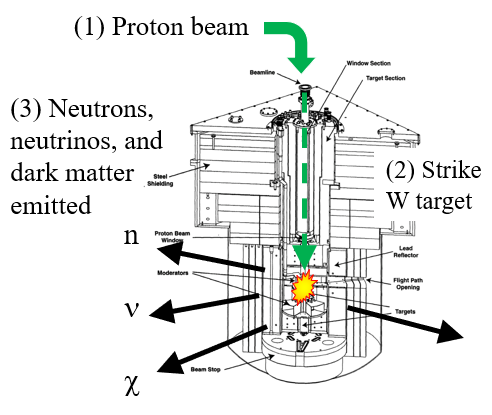}
    \label{f:ccm-onsite}
\end{subfigure}
\begin{subfigure}{.49\textwidth}
    \centering
    \includegraphics[width=1\textwidth]{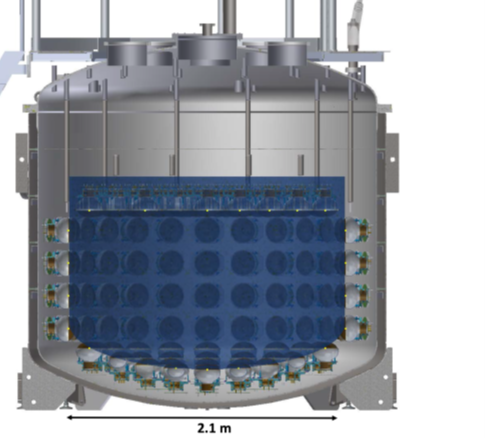}
    \label{f:ccm-schematic}
\end{subfigure}
\caption{The 10-ton CAPTAIN-Mills instrumented with 220 photomultiplier tubes, yielding a 7-ton total fiducial volume of liquid argon. The blue shows the inner neutrino target liquid argon volume and outside active veto region which is also filled with liquid argon \cite{snowmass, Dunton:2022dez, shoemaker2021sailing}. 
\label{f:ccm}
}    
\end{figure}

The Coherent CAPTAIN (Cryogenic Apparatus for Precision Tests of Argon Interactions with Neutrinos)-Mills (CCM) experiment (Fig.~\ref{f:ccm}) consists of an 800 MeV proton beam, a tungsten target, and a 10-ton liquid argon scintillation detector located at the Los Alamos Neutron Science Center at Los Alamos National Laboratory \cite{snowmass, Dunton:2022dez, shoemaker2021sailing}. CCM is located downstream from a neutrino beam produced by pion decay at rest. The LAr detector utilizes 200 photomultiplier tubes, 100 of which are coated in wavelength-shifting foils, to detect both scintillation and Cherenkov light produced by interactions in the liquid argon. 
\cite{Plischke:2001ii}



%


\subsubsection{Scintillation of Liquid Argon}

\hspace{\parindent}

As discussed above, liquid argon is a prominent choice of detector material, especially for scintillation.  See Ref.~\cite{Segreto_2021} for a review.  Scintillation light is produced when neutrino interactions with nuclei produce deliver sufficient kinetic energy to the recoiling nucleus to stimulate scintillation.  Argon is an exceptionally good scintillator; as with most other liquefied noble gases, argon has a high scintillation light yield given by, 
\begin{align} \label{e:scint1}
    N_{scint} &\approx T \times \left(\frac{40,000 \, \gamma}{\mathrm{MeV}}\right) \times \Big( 27.5\% \, \mathrm{prompt} \Big) 
    \notag \\ &=
    \left(\frac{T}{\mathrm{MeV}}\right) \times 11,000   \: ,
\end{align}
where charged particles in liquid argon release approximately $40,000$ photons per MeV of energy lost.  The scintillation photons are emitted from one of two channels: a ``fast component'' carrying 27.5\% of the total scintillation light with a time constant of 8 ns, and a ``slow component'' with a much longer time constant of 1.6 $\mu$s.  The fast component of the scintillation light is used as the primary signal for charged particle detection in the scintillator. Argon is transparent to its own scintillation light and is relatively easy to purify. Compared to xenon, argon is cheaper and has a distinct scintillation time profile, which allows the separation of electronic recoils from nuclear recoils.



\subsection{The Subject of This Work}

\subsubsection{Cherenkov Radiation}


\begin{figure}[t!]
\begin{centering}
\includegraphics[width=0.3\textwidth]{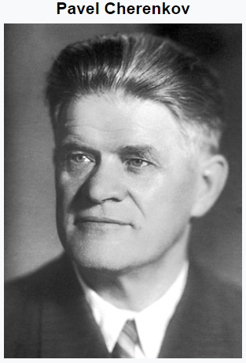}
\hspace{1cm}
\includegraphics[width=0.62\textwidth]{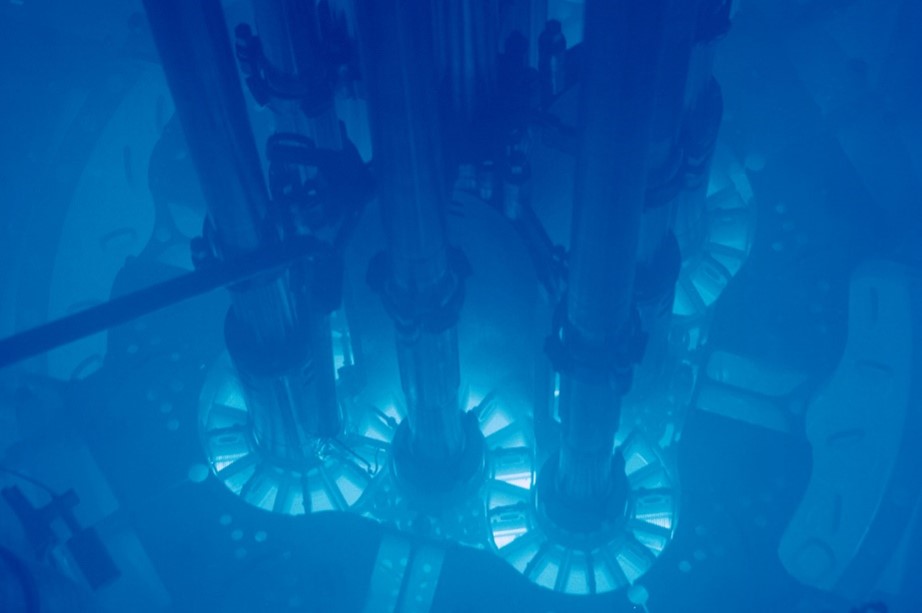}
\caption{Left:  Pavel Cherenkov. Right:  Cherenkov Radiation captured inside the Advanced Test Reactor at Idaho National Laboratory.
\label{f:cherenkovexample}
}
\end{centering}
\end{figure}

When a charged particle passes through a dielectric medium at a speed greater than the phase velocity of light in that medium, an electromagnetic radiation is emitted transverse to the direction of that incoming particle \cite{PhysRev.52.378}. This radiation is named ``Cherenkov radiation'' after the Soviet Nobel laureate Pavel Cherenkov (Fig.~\ref{f:cherenkovexample} (Left)).  Cherenkov photons are emitted primarily at UV and near-UV wavelengths, contributing to their famous blue color. A classic example is the characteristic blue glow of an underwater nuclear reactor shown in Fig.~\ref{f:cherenkovexample} (Right). 


Whereas scintillation photons are emitted isotropically, Cherenkov radiation, on the other hand, has a distinctive angular pattern.  Cherenkov radiation is famously emitted along a Cherenkov cone, with the cone angle being dictated by the charged particle velocity and the refractive index of the medium.  This collimated property of the Cherenkov radiation helps to distinguish it from the isotropic scintillation light, as well as to distinguish different charged particles from each other.  For this reason, Cherenkov detectors are commonly used for particle identification. 




    





\subsubsection{Potential of Cherenkov Radiation to Enhance Neutrino Searches}

Liquid argon scintillators are commonplace in the search for neutrinos in state-of-the-art experiments like CCM and DUNE.  While the main goal of using liquid argon is to take advantage of its extraordinary scintillation properties, these experiments at present neglect the possible contributions of Cherenkov radiation from the secondary particles travelling in that medium.  The amount of additional light produced by Cherenkov radiation from, say, low-energy protons produced by IBD, can thus contribute to the measured signal that has not been investigated in detail before.  

While experimental data on the refractive index $n$ of liquid argon is scarce at short wavelengths, a new data point near the scintillation wavelength of 128 nm has recently been collected, providing a unique opportunity to realistically calculate the Cherenkov radiation emitted by low-energy charged secondary particles in LAr for the first time.  If the additional Cherenkov light produced by secondary particles in liquid argon scintillators can be reliably measured and integrated into the experimental analyses, it could provide important additional information about the energy, direction, and species of charged secondary particles produced.

In this study, we calculate the Cherenkov photon yield and angular distribution emitted by low-energy protons travelling in liquid argon. We exploit the availability of a new data point of the refractive index of liquid argon at the scintillation wavelength (128 nm) to perform our own fit of the liquid argon refractive index, motivated directly from first principles.  Our fit directly accounts for the phenomenon of absorption near the UV resonance $\sim 106.6$ nm, which was neglected in all the previous fits of the refractive index.  

We can better quantify the amount of Cherenkov radiation produced through the ratio of Cherenkov yield to the scintillation yield for a given secondary charged particle.  Remarkably, we find that even including using the most conservative fit to the refractive index and including the effects of absorption, the Cherenkov yield is significant and well above the background of scintillation light alone in certain angular regions.  These results strongly suggest that Cherenkov radiation from low-energy secondary charged particles in neutrino experiments could be used as an untapped experimental lever to improve the worldwide neutrino physics program.

\hspace{\parindent}

%



The rest of this thesis is organized as follows.  In Chapter.~\ref{mathform}, we lay out the mathematical formulation of the problem and the methodology to calculate the Cherenkov yield and angular distribution.  We illustrate the methods using the non-absorptive fits available in the literature, showing that these lead to unrealistically large predictions for the yield due to the unphysical treatment of the refractive index near the UV resonance. In Chapter.~\ref{cerenkovwabs}, we show the solution to this problem by introducing the correct form of the refractive index derived from first-principles theory which includes the physics of absorption and anomalous dispersion.  We study the resulting reduction of the total yield and the modification of the angular distribution, which exhibits qualitative differences between the resonant and absorptive fits.  Despite the much more conservative fit to the refractive index, we find that the Cherenkov yield is still substantial, and that in some angular bins the density of Cherenkov photons far exceeds the background due to scintillation alone. Finally, in Chapter.~\ref{sum}, we compare and summarize the results from the various fits and energies, and we further visualize the angular distribution by projecting it onto a distant screen.  

\hspace{\parindent}

%

\newpage
\newpage

\section{MATHEMATICAL FORMULATION OF THE PROBLEM}\label{mathform}
\hspace{\parindent}
\subsection{Outline of the Problem}
\label{sec:Outline}

A proton travelling in liquid argon (LAr) loses energy during its passage, primarily through inelastic collisions with atomic electrons in the medium. The energy loss $- \frac{dE}{dx}$ per unit length is given by the Bethe-Bloch equation \cite{Workman:2022ynf} as a function of the proton velocity $\beta = v / c$ in units of the speed of light $c$:
\begin{align}   \label{e:BetheBloch1}
-\frac{dE}{dx} = K\frac{\rho Z}{A} \frac{z^2}{\beta^2} \left[ \ln \left(\frac{2m_{e}c^2\gamma^2\beta^2}{I} \right) - \beta^2  \right]   \: .
\end{align}
Here, $z=1$ is the charge of the proton in units of the fundamental charge $e$; $\rho \approx 1.38 \, \mathrm{g/cm}^3$ is the mass density of LAr per unit volume at $T = 89 \, \mathrm{K}$ \cite{BNL:2023,Sinnock:1969zz}, $Z=18$ is its atomic number, and $A=39.948$ is its atomic mass. For LAr, the mean ionization potential $I$ (the mean energy needed to ionize an electron out of the atom) is $23.6 \, \mathrm{eV}$ \cite{Workman:2022ynf}.  This means that, when a charged particle goes through liquid argon (LAr), it takes about 23.6 eV energy to ionize an electron on average. For minimally ionizing particles, the corresponding mean energy loss is about 2.1 MeV/cm, leading to about 100 nm average separation between two adjacent ions.
%
%
Finally, the constant $K$ is given by
\begin{align}   \label{e:constk}
K = \frac{4\pi \alpha_{EM}^2 (\hbar c)^2 N_{A} (10^3 \, \mathrm{kg}^{-1})}{m_{e} c^2} = 30.7 \: \mathrm{keV \, m^2/kg} = 0.0307 \: \mathrm{MeV \, m^2 / kg }     \: ,
\end{align}
where $\alpha_{EM} = 1/137$ is the fine structure constant, $\hbar$ is the reduced Planck constant, $N_A$ is Avogadro's number, and $m_e$ is the mass of the electron \cite{Workman:2022ynf}. 

\begin{figure}[ht]
\centering
\begin{subfigure}{.47\textwidth}
\centering
\includegraphics[width=1\textwidth]{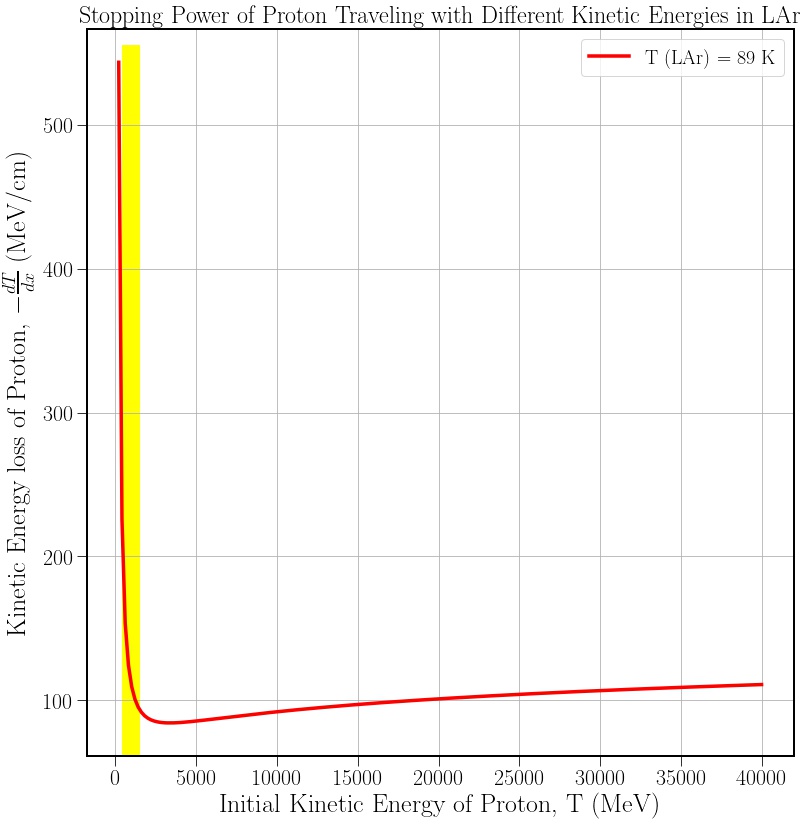}
\caption{Normal Plot  
\label{f:dTbydxvsKEnormal}
}
\end{subfigure}
%
\begin{subfigure}{.5\textwidth}
\centering
\includegraphics[width=1\textwidth]{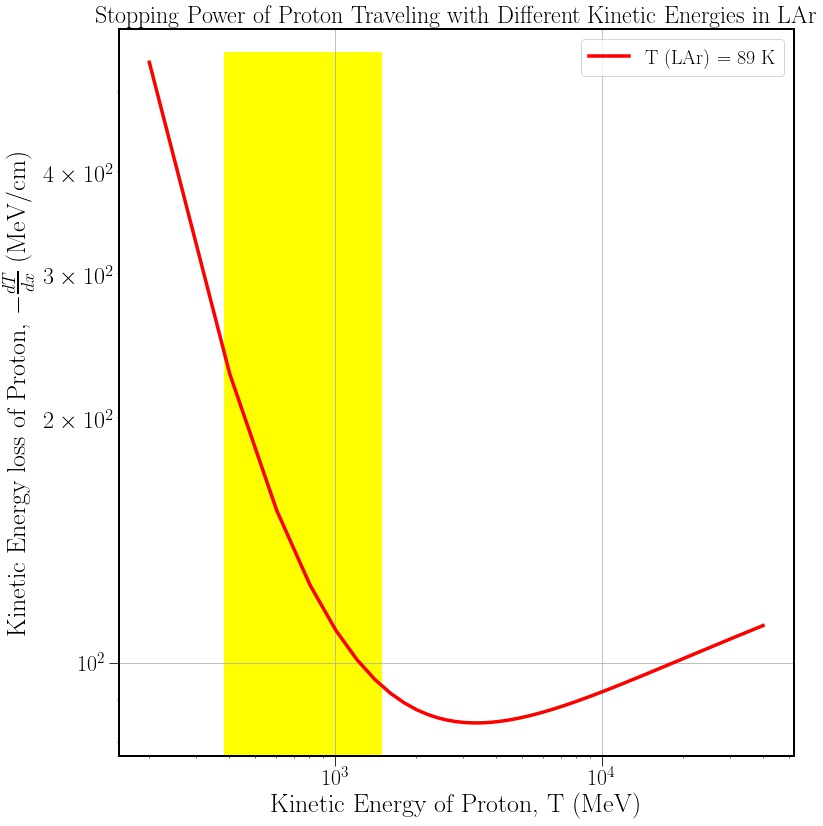}
\caption{Log/Log Plot
\label{f:dTbydxvsKElog}
}
\end{subfigure}

\caption{Plot of $-\frac{dT}{dx}$ of a Proton Traveling in LAr as a Function of its K.E. The range of interest for this work (383 - 1500 MeV) is highlighted in yellow.  
\label{f:dTbydxvsKE}
}
\end{figure}

After fixing the various parameters, the Bethe-Bloch equation \eqref{e:BetheBloch1} describes the energy loss (or ``stopping power'') of protons as a function of their instantaneous velocity $\beta = v/c$.  The Bethe-Bloch stopping power for LAr using the constants described above is shown in Fig.~\ref{f:dTbydxvsKE} as a function of the proton energy.  This non-uniform stopping power leads to a non-uniform rate of loss of the proton kinetic energy.  In particular, we note that in the low-energy range of interest to us (383 - 1500 MeV, shaded in yellow in Fig.~\ref{f:dTbydxvsKE}), the stopping power is a \textit{decreasing} function of the kinetic energy.


The decrease in the kinetic energy of the proton corresponds to a decrease in $\beta$, such that the Bethe-Bloch equation constitutes an implicit differential equation for the velocity $\beta(x)$ as a function of distance $x$ propagated in LAr.  By solving the Bethe-Bloch equation with fully relativistic kinematics, we determine the velocity $\beta(x) = v(x) / c$ of the proton in units of the speed of light $c$ in vacuum. 
The energy lost by the proton is deposited in the LAr medium, and much of this deposited energy appears in the form of isotropic scintillation light: photons emitted during the relaxation of excited atomic states.  LAr is an exceptional scintillator, emitting approximately 40 photons per keV of energy deposited at a wavelength of 128 nm for minimally ionizing particles \cite{Gastler_2012}.  Of these scintillation photons, approximately $27.5\%$ of them are prompt \cite{Segreto_2021}, while the rest experience a significant delay before emission and are not relevant for this study.  For a 500 MeV proton, this leads to the emission of $5.5 \times 10^6$ prompt scintillation photons; since each such photon carries away an energy $\frac{h c}{\lambda} = \frac{1239.8 \, \mathrm{eV}\,\mathrm{nm}}{128 \, \mathrm{nm}} = 9.69 \, \mathrm{eV}$, this amounts to approximately 10.7\%  of the proton energy converted to prompt scintillation photons. These isotropic scintillation photons are considered the background to the effect which is the main subject of this work: Cherenkov radiation.

\begin{figure}[t!]
\centering
\begin{subfigure}{.35\textwidth}
\centering
\includegraphics[width=1\textwidth]{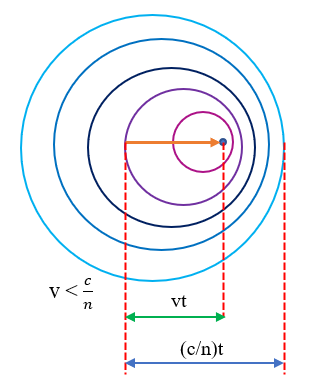}
\caption{$v < c/n$
\label{f:coherentwaveform2}
}
\end{subfigure}
\centering
\begin{subfigure}{.64\textwidth}
\centering
\includegraphics[width=1\textwidth]{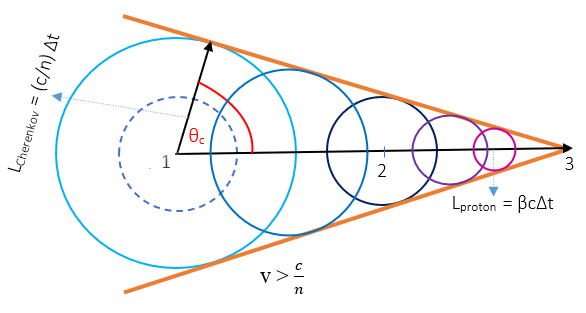}
\caption{$v > c/n$
\label{f:coherentwaveform1}
}
\end{subfigure}
\caption{Spherical wavelets of fields of a particle travelling less than (left fig) and greater than (right fig) the velocity of light in the medium. For $v > c/n$, an electromagnetic "shock" wave appears known as Cherenkov radiation, moving in the direction given by the Cherenkov angle. Deriving the Cherenkov condition from Coherent Electromagnetic wave construction by expanding the spherical wavefront at three instances of time denoted by 1,2, and 3 (right fig).
\label{f:Cherenkovsphericalwavelets}
}
\end{figure}

In addition to the scintillation photons generated by the relaxation of LAr, the proton will also emit Cherenkov photons \cite{PhysRev.52.378} (an instantaneous process \cite{Jackson:1998nia}) at any wavelength $\lambda$ for which the speed $v = \beta c$ of the proton exceeds the speed of light $c_{LAr} (\lambda) = \frac{1}{n(\lambda)} c$ in medium (see Fig.~\ref{f:Cherenkovsphericalwavelets}).  Here $n(\lambda)$ is the index of refraction of liquid argon at wavelength $\lambda$ and $c$ is the speed of light in vacuum.  This Cherenkov condition $v > c_{LAr}$ may be expressed in terms of the index of refraction $n(\lambda)$ as
\begin{align}   \label{e:CherenCond1}
    \frac{v}{c_{LAr} (\lambda)}  
    = \frac{v}{c} \, \frac{c}{c_{LAr} (\lambda)} = \beta \, n(\lambda)
    > 1 \: .
\end{align}
This shows that the speed of the particle must be larger than the phase velocity of the electromagnetic fields at frequency $\omega$ (or wavelength $\lambda$) in order to have an emission of Cherenkov radiation of that frequency.

    %

\begin{figure}[t!]
\begin{centering}
\includegraphics[width=0.42\textwidth]{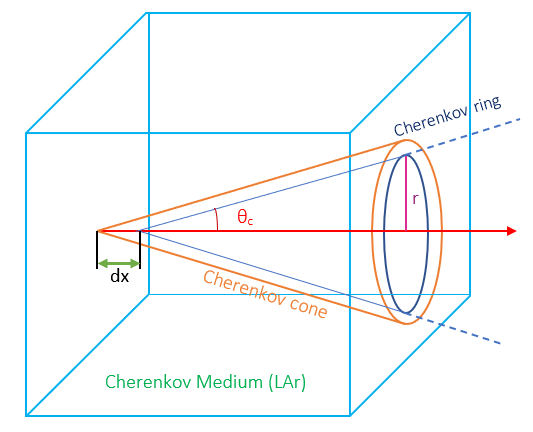}
\hspace{1cm}
\includegraphics[width=0.49\textwidth]{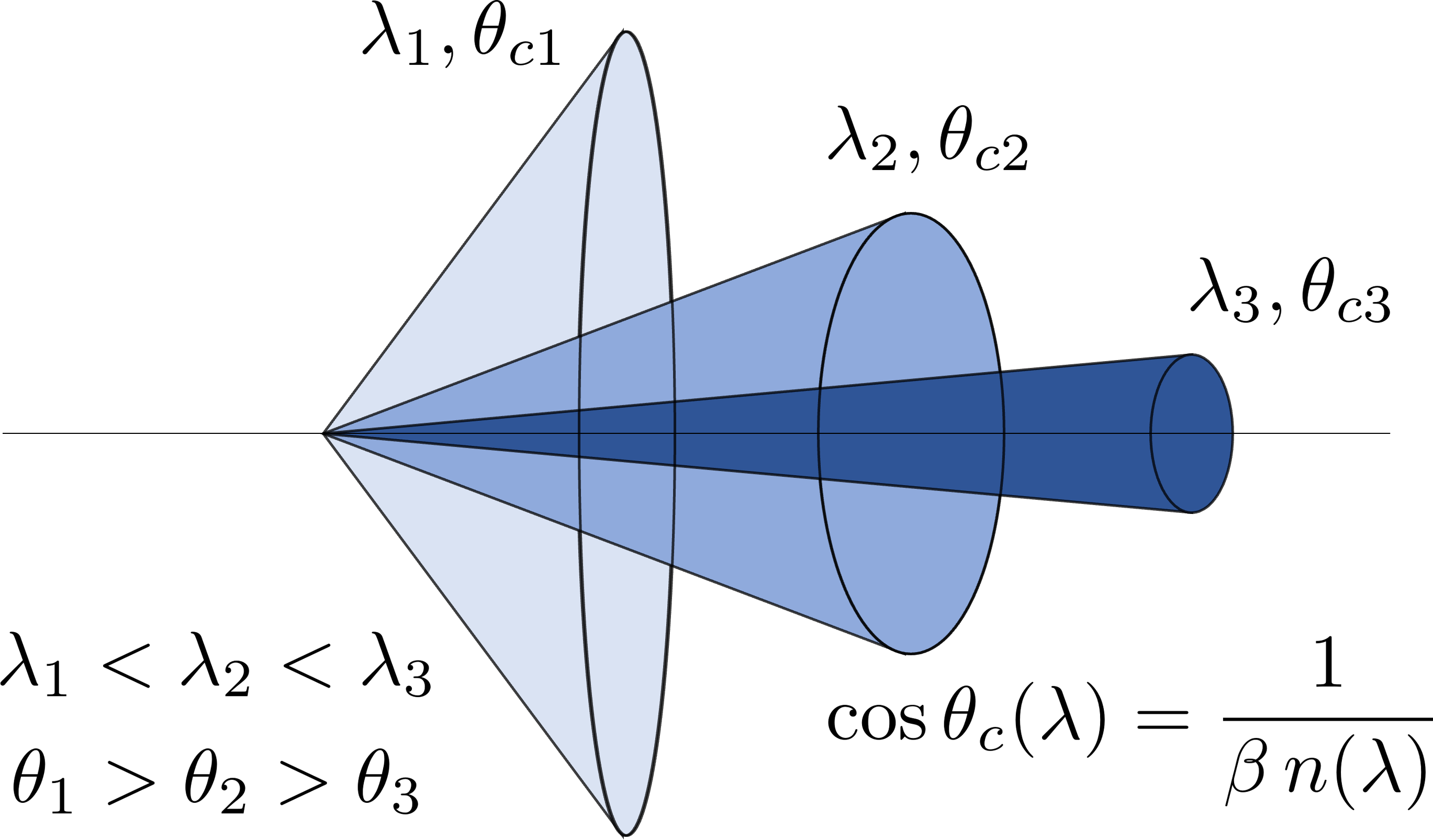}
\caption{Characteristic cone-shaped Cherenkov radiation of a charged particle exceeding the speed of light in a medium.  Left:  A single Cherenkov cone at fixed $\theta_c$; this occurs when the index of refraction $n(\lambda) \approx \mathrm{const}$(i.e., far away from a resonance). Right:  A superposition of many Cherenkov cones emitted at different angles $\theta_c$ corresponding to different wavelengths $\lambda$ with different indices of refraction $n(\lambda)$, shown here for the case of normal dispersion $dn/d\lambda < 0$.
\label{f:cherenkovcone}
}
\end{centering}
\end{figure}

A characteristic feature of Cherenkov radiation is its angle of emission. At large distances from the path, the fields become transverse radiation fields. Fig~\ref{f:Cherenkovsphericalwavelets} shows that coherent electromagnetic waves, known as Cherenkov radiation, can be produced from the superposition of expanding spherical waves if the speed of the particle is greater than the speed of light in the medium.  Such a coherent waveform can only be constructed while the charged particle (in this case, a proton) travels through a medium, where the speed of light is less than its vacuum value $c$.  In Fig.~\ref{f:coherentwaveform1}, an example of such a waveform is depicted produced by expanding the spherical wavefront at three instances of time denoted by 1, 2, and 3. Constructive interference (net emission of Cherenkov photons) occurs only at a specific angle $\theta_c$ relative to the trajectory given by the Cherenkov condition,
\begin{align}   \label{e:CherenAngle1}
    \cos\theta_c = \frac{L_{Cherenkov}}{L_{Proton}} = \frac{(c/n)\Delta t}{\beta c \Delta t}= \frac{1}{\beta n(\lambda)}= \frac{1}{n(\lambda) \, \sqrt{1 - \frac{1}{(1 + \frac{T}{mc^2})^2}}}   \: ,
\end{align}
unlike scintillation photons, which are emitted isotropically. Note that the Cherenkov angle $\theta_c$ describes the direction of \textit{propagation} of the wavefront, which is normal to the wavefront itself. Thus $\theta_c$ is \textit{not} the opening angle of the wavefront cone shown in Fig.~\ref{f:coherentwaveform1}, but rather the opening angle of the cone defining the photon \textit{momenta} seen in Fig.~\ref{f:cherenkovcone}.  

In contrast, if the proton speed is less than $c/n$, then one has the situation shown in Fig.~\ref{f:coherentwaveform2}, in which the proton advances too slowly to produce overlapping wave fronts and no Cherenkov radiation is emitted.
%
%
%
The emission angle $\theta_c$ can be interpreted qualitatively in terms of a \textit{shock} wavefront akin to the familiar shock wave (sonic boom) produced by an aircraft in supersonic flight.

The standard picture of Cherenkov radiation as drawn in the left panel of Fig.~\ref{f:cherenkovcone} shows a single Cherenkov cone at a fixed angle given by \eqref{e:CherenAngle1}.  This is the case if the index of refraction $n$ is a constant, as occurs at wavelengths $\lambda$ far away from any resonances.  But near a resonance (and in general), the index of refraction is a continuous function $n(\lambda)$ of wavelength, so that there can be a continuous range of frequencies $\lambda$ which radiate.  For the same proton velocity $\beta$, the various frequencies will have different indices of refraction $n(\lambda)$, resulting in the emission of each wavelength at its own Cherenkov angle \eqref{e:CherenAngle1}.  Thus, for a proton with instantaneous velocity $\beta$ moving through LAr, there will not be a single Cherenkov cone as in Fig.~\ref{f:cherenkovcone}, but rather a continuous distribution of Cherenkov photons of different wavelengths $\lambda$, all emitted at different angles $\theta_c (\lambda)$ (right panel of Fig.~\ref{f:cherenkovcone}).

\begin{figure}[t] 
\begin{centering}
\includegraphics[width=0.55\textwidth]{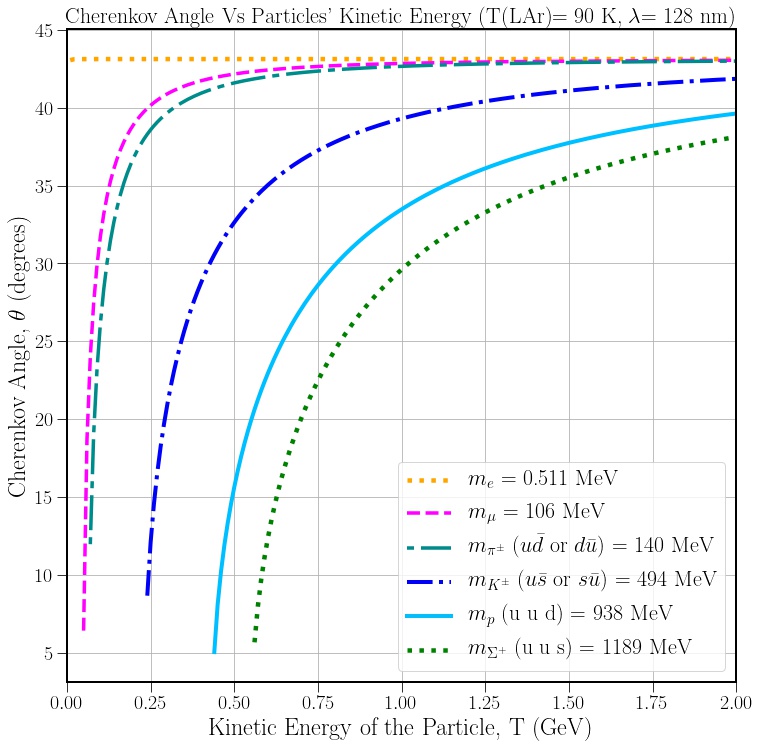}
\caption{Cherenkov Angle ($\theta_{c}$) as a function of kinetic energy $T$ for the proton (cyan curve) and other particles. 
\label{f:thetavstforall}
}
\end{centering}
\end{figure}

Both energy loss \cite{Bichsel:2006cs} and Cherenkov radiation \cite{Lippmann_2012} are useful methods for performing particle identification (PID), which refers to an experimental technique to identify and discriminate between different particles based on their physical properties like mass, charge, spin, etc. Overall, particle identification is a crucial part of subatomic particle research, allowing scientists to understand the properties and behavior of particles that make up the universe.
Various methods exist for PID, but Cherenkov radiation is a powerful one, since the Cherenkov angle \eqref{e:CherenAngle1} depends on a particle's velocity at a given kinetic energy -- and therefore its mass.  As seen in Fig.~\ref{f:thetavstforall}, the shape of the curve relating the Cherenkov angle $\theta_c$ and the kinetic energy $T$ depends on the mass of the particle.  The Cherenkov angle \eqref{e:CherenAngle1} depends only on the velocity $\beta$ and the index of refraction $n(\lambda)$, so the mass enters only through the relation between kinetic energy $T$ and velocity $\beta$ \eqref{e:beta1}.  Since heavier particles have a lower $\beta$ at the same kinetic energy, this corresponds to a Cherenkov angle $\theta_c$ closer to zero.  Thus for light particles like electrons, the Cherenkov angle is nearly flat as a function of kinetic energy, while for heavier particles like protons the angle changes significantly as a function of $T$.  These differences can then be used to distinguish between different species of particles charged particles through the separation of the curves in Fig.~\ref{f:thetavstforall}.  This illustrates that Cherenkov radiation is a valuable method for PID, but it can only be applied to charged particles which emit Cherenkov photons. The Cherenkov radiation technique is most useful for identifying particles with high velocities, such as muons, pions, and electrons, and for distinguishing them from other particles that have similar energies but different masses and charges.

%
%


As seen from Eq.~\eqref{e:CherenCond1}, the refractive index $n(\lambda)$ of LAr is particularly important to determine the wavelength range over which Cherenkov emission will occur.  This further determines the angular distribution of the Cherenkov photons through the condition \eqref{e:CherenAngle1}.  The number distribution $d^2 N / dx d\lambda$ of Cherenkov photons per unit distance $dx$ and per unit wavelength $d\lambda$ is given by the Frank-Tamm formula \cite{Workman:2022ynf},
\begin{align}   \label{e:FrankTamm1}
\frac{d^2N}{dx d\lambda} = \frac{2 \pi \alpha z^2}{\lambda^2} \left( 1 - \frac{1}{\beta^2 n^2(\lambda)} \right)     \: .
\end{align}
Here $\alpha = 1/137$ is the fine structure constant and $z = 1$ is the charge of the proton in units of $e$.  The Frank-Tamm formula was first derived in 1937 and published to explain the radiation measured in 1934 by Cherenkov. 

\begin{figure}[t]  
\begin{centering}
\includegraphics[width=0.55\textwidth]{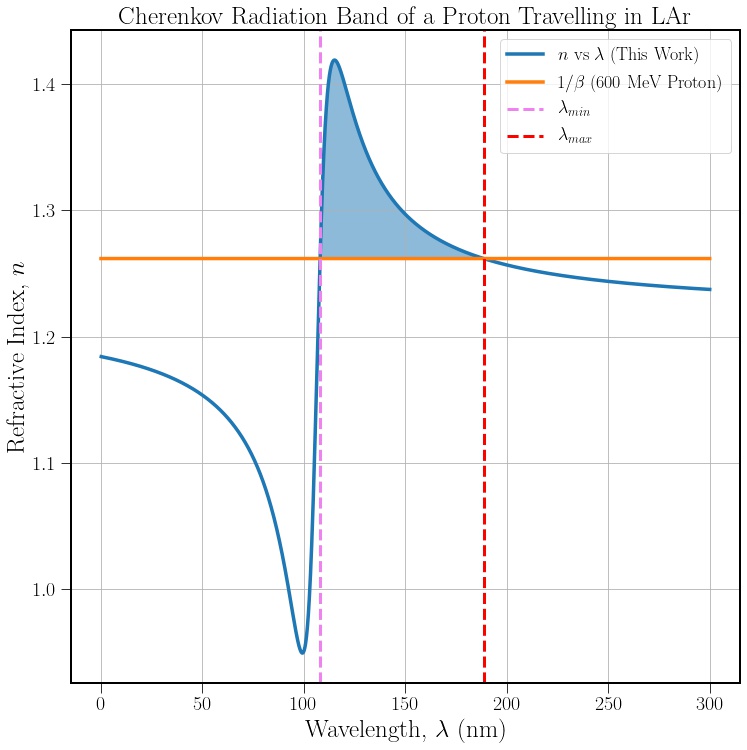}
\caption{Cherenkov band: Radiation is emitted 
only in the shaded wavelength range, where the Cherenkov condition $n > \beta^{-1}$ is satisfied.} 
\label{f:Cherenkovband}
\end{centering}
\end{figure}

The number of Cherenkov photons $dN$ emitted per distance $dx$ travelled by the proton in LAr is found by integrating \eqref{e:FrankTamm1} over the wavelengths $\lambda_{min} \leq \lambda \leq \lambda_{max}$ satisfying the Cherenkov condition \eqref{e:CherenCond1}; that is,
\begin{align}   \label{e:ftintegrand}
 \frac{dN}{dx} = \int_{\lambda_{min}}^{\lambda_{max}} \frac{2 \pi \alpha z^2}{\lambda^2} \left( 1 - \frac{1}{\beta^2 n^2(\lambda)} \right) d\lambda  = I(\beta, x)  \: .
\end{align}    
We refer to this quantity as the ``instantaneous Cherenkov yield.''  This procedure is illustrated in Fig.~\ref{f:Cherenkovband} for the particular absorptive model (Ch.~\ref{cerenkovwabs}) where the index of refraction $n(\lambda)$ is computed using the damped harmonic oscillator (see Appendix~\ref{sec:HarmonicOscillator}).  Here, the wavelength band satisfying the Cherenkov condition $n(\lambda) > \beta^{-1}$ is shown as a shaded region.  For this model, the boundaries $\lambda_{min}$ and $\lambda_{max}$ are found by the intersection of the line $1/\beta$ with the index $n(\lambda)$; models including an undamped resonance at 128 nm will have $\lambda_{min}$ set by limits of validity of the fit.  Finally, we note that the Frank-Tamm integrand \eqref{e:FrankTamm1} contains additional factors, such as the prefactor $1/\lambda^2$, which influence the amount of Cherenkov radiation emitted at a given wavelength.  While the index $n(\lambda)$ may be the most important, it is not the only factor in determining the instantaneous Cherenkov yield.




%
%
 

We can further generalize Eq.~\eqref{e:FrankTamm1} to include the angular condition \eqref{e:CherenAngle1} through the use of the Dirac delta function:
\begin{align}   \label{e:FrankTamm2}
\frac{d^3N}{dx \: d\lambda \: d\cos\theta} = \frac{2 \pi \alpha z^2}{\lambda^2} \left( 1 - \frac{1}{\beta^2 (x)n^2(\lambda)} \right) \, \delta\left( \cos\theta - \frac{1}{\beta(x) n(\lambda)} \right)     \: .
\end{align}
Our generalization \eqref{e:FrankTamm2} is one of the most important equations in this work, because it allows us to simultaneously incorporate both the Cherenkov cone condition \eqref{e:CherenAngle1} and the intensity given by the Frank-Tamm formula \eqref{e:FrankTamm1} on an equal footing.  As we will show, the use of the Dirac delta function in \eqref{e:FrankTamm2} is powerful because it allows us to integrate over any of the variables $dx, d\lambda, d\cos\theta$ in a precise, exact way.  We will take advantage later of this to integrate over all emitting wavelengths $\lambda$ and obtain $\frac{d^2 N}{dx \, d\cos\theta}$ -- the instantaneous angular distribution of emitted Cherenkov photons at a particular point in the medium.  Further integration over the range $dx$ of the proton will yield the total angular distribution $\frac{dN}{d\cos\theta}$, and a final integration over the angle yields the total number $N$ of Cherenkov photons.  Therefore, Eq.~\eqref{e:FrankTamm2} will serve as the master equation for our studies.  The three necessary inputs which underlie our analysis are the master equation \eqref{e:FrankTamm2} for the Cherenkov spectrum, the solution $\beta(x)$ to the Bethe-Bloch equation \eqref{e:BetheBloch1} and experimental data determining the index of refraction $n(\lambda)$.

The first major goal of this work is to compute the total number $N$ of emitted Cherenkov photons (fully integrated over wavelength, position, and angle) and compare it with the total number of scintillation (background) photons for protons of the same kinetic energy.  The total number $N$ of Cherenkov photons can be derived by integrating the Frank-Tamm formula \eqref{e:FrankTamm1} twice: first over the appropriate wavelength range $\lambda_{min} \leq \lambda \leq \lambda_{max}$ determined from the Cherenkov condition \eqref{e:CherenCond1} and then again integrating over the range $0 \leq x \leq R$ of the proton.  Given an input for the index of refraction $n(\lambda)$ and the solution to the Bethe-Bloch equation for $\beta (x)$, this determines the total number $N$ of Cherenkov photons.  

Beyond just the total number of photons, exploring the unique directional properties of Cherenkov radiation is the second major goal of this work.  By integrating the angular-differential version \eqref{e:FrankTamm2} of the Frank-Tamm formula over the wavelength $\lambda$ and range $x$, we will determine the total angular distribution $dN/d\cos\theta$ of Cherenkov photons.  As we will show, the structure present in this angular distribution reveals new physics sensitive to anomalous dispersion in the index of refraction $n(\lambda)$.  We further validate this angular distribution through convergence testing and comparison with the integrated number of photons.

The third and final major goal of this work is to study how the physics of absorption modifies the emitted Cherenkov spectrum.  As we will discuss, the physics of absorption is encoded in the behavior of the index of refraction $n(\lambda)$ near the resonance, leading to a region of anomalous dispersion.  This both leads to additional emission of Cherenkov photons from the anomalous dispersion region as well as the reabsorption of those photons before they reach the detector.  We study the effects of absorption in two ways: by fitting the index of refraction $n(\lambda)$ to a functional form which includes the absorptive / anomalous behavior near the resonance, and by using a ``black disk model'' for reabsorption, in which all Cherenkov photons emitted close to the resonance are considered to be reabsorbed.  This comparison, while simple, will allow us to estimate the effects of reabsorption on the Cherenkov distribution.  As we will show, the suppression of Cherenkov photons due to reabsorption can be significant.



\hspace{\parindent}


    
    
    
    
    
    


%
\subsection{Bethe-Bloch Formula:  Stopping Power of Proton in LAr}
\label{BB}
%

\begin{figure}[t!] 
\begin{centering}
\includegraphics[width=0.45\textwidth]{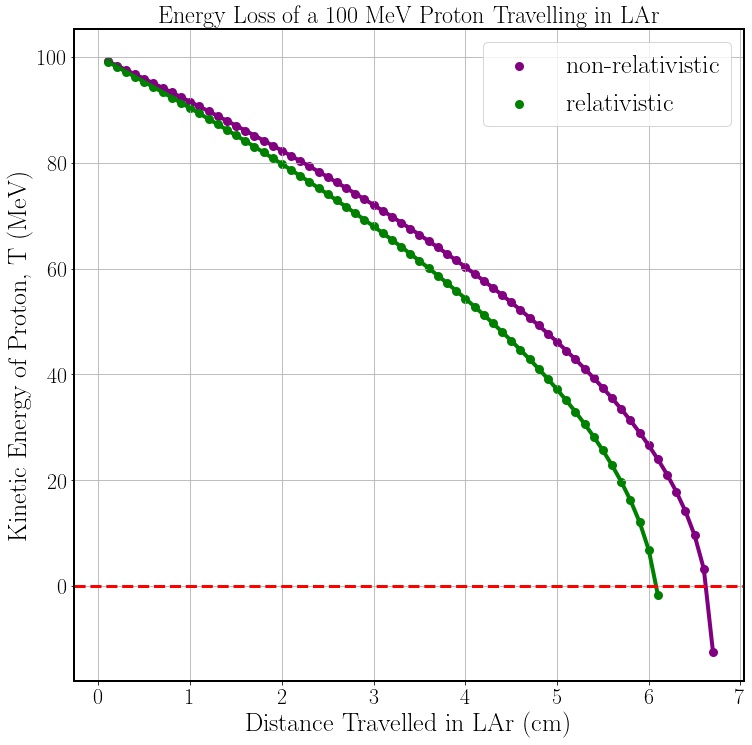}
\hspace{1cm}
\includegraphics[width=0.45\textwidth]{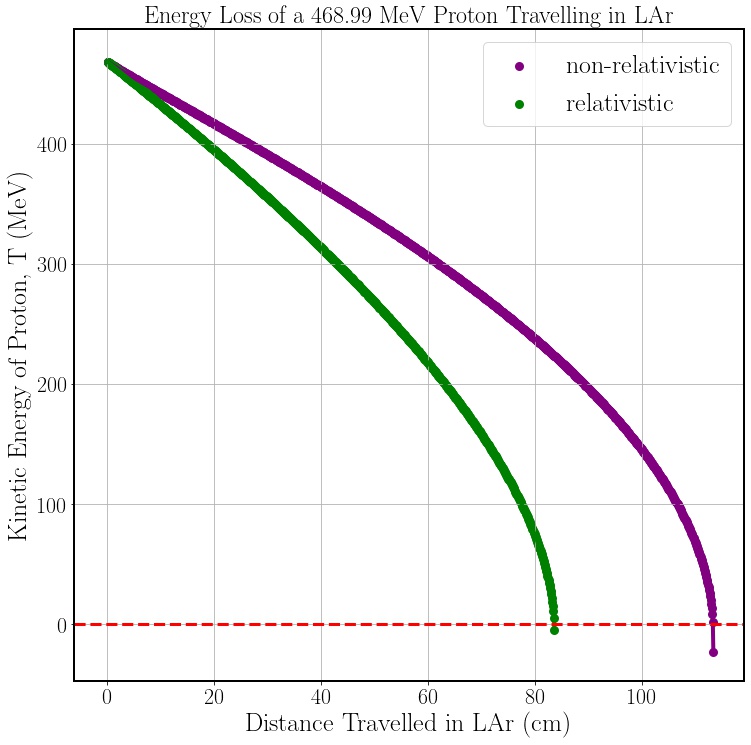}
\caption{Comparison of Energy loss of a 100 MeV (Left) and 468.99 MeV (Right) Proton in LAr Calculated using Classical Vs Relativistic Method. 
\label{f:KEloss_proton_relvsnonrel}
}
\end{centering}
\end{figure}

The first key ingredient we will need is a solution to the Bethe-Bloch equation \eqref{e:BetheBloch1} for the velocity $\beta (x)$ as a function of distance $x$ travelled by the proton.  Since the Bethe-Bloch equation \eqref{e:BetheBloch1} determines the trajectory in terms of the $E$ energy of the proton, we will need to convert from the energy $E(x)$ to the corresponding velocity $\beta(x)$.  For fully relativistic kinematics, this is done by writing the kinetic energy $T$ as the difference between the total energy $E = \gamma m c^2$ and the rest mass energy $m c^2$:
\begin{align}   \label{e:KE1}
T = \gamma mc^2 - mc^2 = (\gamma - 1) mc^2 =
\Big( (1-\beta^2)^{-1/2} \Big) m c^2 \: ,
\end{align}
with the relativistic gamma factor $\gamma = (1-\beta^2)^{-1/2}$.  Solving \eqref{e:KE1} for $\beta$ in terms of the kinetic energy $T$ gives
\begin{align}   \label{e:beta1}
\beta = \sqrt{1 - \frac{1}{(1 + \frac{T}{mc^2})^2}} \: .
\end{align}
We note that in the limit of small kinetic energies $T \ll m c^2$, this reduces to
\begin{align}   \label{e:beta2}
\beta \approx \sqrt{\tfrac{2T}{mc^2}} \: ,
\end{align}
which just reproduces the nonrelativistic kinetic energy $T \approx \frac{1}{2} m v^2 = \frac{1}{2} m \beta^2 c^2$. From \eqref{e:beta2}, it is clear that for $2T \geq mc^2$, this non-relativistic formula will yield unphysical $\beta \geq 1$.  Therefore it is useless for high energy particles, for example, for protons with $T \geq$ 469 MeV (half of the Proton's mass). But even for comparatively lower energy particles i.e. a 100 MeV Proton in LAr, we can see that the relativistic and non-relativistic approach produces significantly different results  in Fig~\ref{f:KEloss_proton_relvsnonrel}. So, in this work, reasonably we have adopted the relativistic approach \eqref{e:beta1} for all of our calculations.      

Using \eqref{e:KE1}, we can convert from a solution $T(x)$ to the Bethe-Bloch equation to the velocity $\beta(x)$. Since the rest mass energy $m c^2$ is constant, we can equally well write the Bethe-Bloch equation \eqref{e:BetheBloch1} in terms of the kinetic energy $T$ rather than the total energy $E = T + m c^2$:
\begin{align}   \label{e:BetheBloch2}
-\frac{dT}{dx} = K\frac{\rho Z}{A} \frac{z^2}{\beta^2} \left[ \ln \left(\frac{2m_{e}c^2\gamma^2\beta^2}{I} \right) - \beta^2 \right]   \: .
\end{align}
A finite-differencing approximation to \eqref{e:BetheBloch2} makes it suitable for numerical implementation:

\begin{align}   \label{e:BetheBloch3}
    T(x_{i+1}) =  T(x_i) - 
    \Big( K\frac{\rho z^2 Z}{A} \Delta x \Big) 
    \frac{1}{\beta^2(x_i)} \left[ \ln \left(\frac{2m_{e}c^2 \beta^2(x_i)}{I} (1 - \beta^2 (x_i))\right) - \beta^2(x_i)  \right]   \: .
\end{align}
Given the initial velocity $T(x_0)$ or initial velocity $\beta(x_0)$, we can iterate \eqref{e:BetheBloch3} to construct all later values of $T(x_i)$ or $\beta(x_i)$.  The kinetic energy as a function of distance is shown in Fig.~\ref{f:KEloss_proton_relvsnonrel} using both the relativistic \eqref{e:beta1} and nonrelativistic \eqref{e:beta2} kinetic energies. Perhaps surprisingly, the relativistic formula leads to a \textit{shorter} prediction for range.  The reason for this is that the nonrelativistic formula overestimates the velocity $\beta$ for a given kinetic energy, while the Bethe-Bloch formula \eqref{e:BetheBloch1} is a \textit{decreasing} function of $\beta$ in this energy range, as shown in Fig.~\ref{f:dTbydxvsKElog}. Consequently, the nonrelativistic formula \textit{underestimates} the energy loss, leading to a \textit{larger} range prediction than for the true relativistic formula.  Surprisingly, the relativistic corrections are significant (about an $8\%$ correction) even for a 100 MeV proton.

\begin{figure}[t!] 
\begin{centering}
\includegraphics[width=0.6\textwidth]{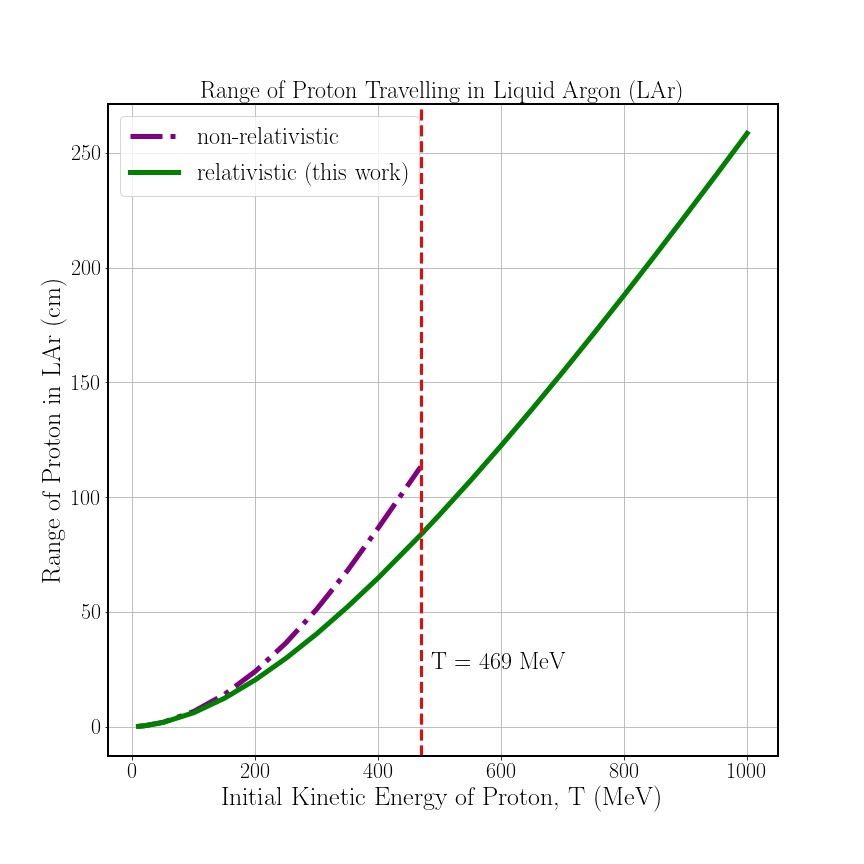}
\caption{Range of Proton in LAr from different methods (Relativistic Vs Non-Relativistic). The vertical red line indicates the energy $T = \frac{1}{2} m \beta^2 c^2 =$ 469 MeV at which the non-relativistic formula yields $\beta = 1$ and becomes unusable.   
\label{f:rangenonrelvsrel}
}
\end{centering}
\end{figure}

The energy loss process should stop when the kinetic energy goes to zero.  Because of the finite-differencing approximation \eqref{e:BetheBloch3}, it is unlikely that we will encounter $T(x_i) = 0$ exactly.  Instead, the recursion terminates when $T(x_i)$ first becomes negative, and then we use a linear interpolation strategy to determine the range $R$ for which $T(R) = 0$.  In this method, shown in Fig.~\ref{f:KEloss_proton_relvsnonrel}, we use the last two data points in $x$ and $T$ (one positive and one negative value) to interpolate the $x$ for which $T = 0$.  Let's suppose these two datapoints (the two points on the lower right: one above and one below the T = 0 axis) are labelled as $(x_{n-1}, T_{n-1})$ and $(x_{n}, T_{n})$. Then the straight line joining these two points is given in point-slope form by
\begin{align}   \label{e:lininterpol}
T - T_{n-1} = \left(
    \frac {T_{n} - T_{n-1}}{x_{n} - x_{n-1}}
    \right) (x - x_{n-1}) \: .
\end{align}   
Now setting $T = 0$ in \eqref{e:lininterpol} will give us the range $x = R$ of the proton: 
\begin{align}   \label{e:rangeformula}
    R = x_{n-1} +  
    \left(  \frac{T_{n-1}} {T_{n-1} - T_{n}} \right)   \Delta x
\end{align}
In this way, we can determine the range $R$ as a function of the initial kinetic energy $T$ of the proton, given different assumptions about the Bethe-Bloch stopping power.  Those results are shown in Fig.~\ref{f:rangenonrelvsrel} for two cases: using the parameters specified in Sec.~\ref{sec:Outline} using the Relativistic Vs Non-relativistic formula. 

We observe that the range increases slowly at low kinetic energies, but grows more linearly at higher kinetic energies.  This reflects the fact that the Bethe-Bloch stopping power is highest at low $T$, as seen in Fig.~\ref{f:dTbydxvsKE}.  The important outputs from solving the Bethe-Bloch equation in this way are the determination of the range $R$ and the discrete positions $x_i$ and velocities $\beta(x_i)$ describing the trajectory of the decelerating protons.  These are then used as inputs for determining the spectrum of Cherenkov radiation using the generalized Frank-Tamm formula \eqref{e:FrankTamm2}.

%
\hspace{\parindent}
%
\subsection{Fitting the Refractive Index Curve}


The index of refraction $n(\lambda)$, which describes the optical properties of a medium, is obtained from the real part of the wavelength-dependent permittivity $\epsilon(\lambda)$ and permeability $\mu(\lambda)$:
\begin{align}   \label{e:ndef}
    n(\lambda) = \frac{c}{c_{LAr}(\lambda)}
    = c \: \mathrm{Re} \Big[ \mu(\lambda) \epsilon(\lambda) \Big]^{1/2}
    = \mathrm{Re} \left[ \frac{\epsilon(\lambda)}{\epsilon_0} \right]^{1/2}
    \: ,
\end{align}
where the last equality assumes a non-magnetic material $\mu(\lambda) = \mu_0$, with $\epsilon_0 (\mu_0)$ the permittivity (permeability) of free space.  
%
We emphasize that the wavelength-dependent functions $\mu (\lambda) , \epsilon (\lambda)$ can be complex, such that $n(\lambda)$ is just the real part of a more general function.  For this reason, we write the complex root $[ \epsilon(\lambda) / \epsilon_0 ]^{1/2}$ rather than the ordinary square root $\sqrt{\epsilon(\lambda) / \epsilon_0}$.  The imaginary part determines the closely-related function $\alpha(\lambda)$ called the absorption coefficient:
\begin{align}   \label{alphadef}
    \alpha(\lambda) = \frac{4\pi \, c}{\lambda} \, \mathrm{Im} \Big[ \mu(\lambda) \epsilon(\lambda) \epsilon_0 \Big]^{1/2} 
    =
    \frac{4\pi}{\lambda} \, \mathrm{Im} \left[ \frac{\epsilon(\lambda)}{\epsilon_0} \right]^{1/2} 
    \: .
\end{align}
The imaginary part $\mathrm{Im} \: \epsilon(\lambda)$ is related to the physics of absorption because the imaginary part of the wave number $k = [ \mu(\lambda) \epsilon(\lambda)]^{1/2} \, \omega = \mathrm{Re} \: k + i \: \mathrm{Im} \: k$ produces a decaying wave,
\begin{align}
    e^{i k z} = e^{i (\mathrm{Re} \: k) z} \, e^{- (\mathrm{Im} \: k) z}  \: ,
\end{align}
with the imaginary part proportional to the inverse attenuation length.

\begin{figure}[t!] 
\begin{centering}
\includegraphics[width=0.49\textwidth]{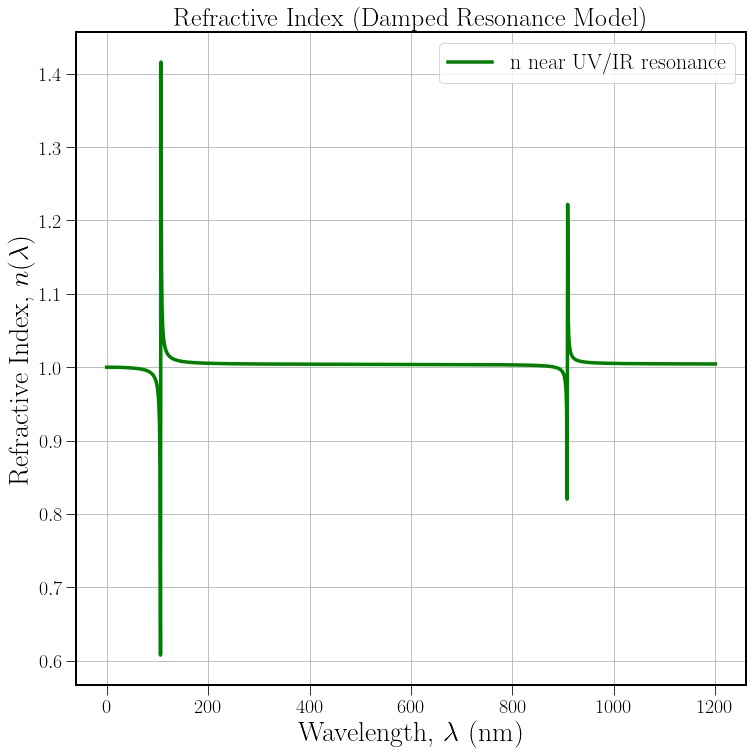}
\includegraphics[width=0.49\textwidth]{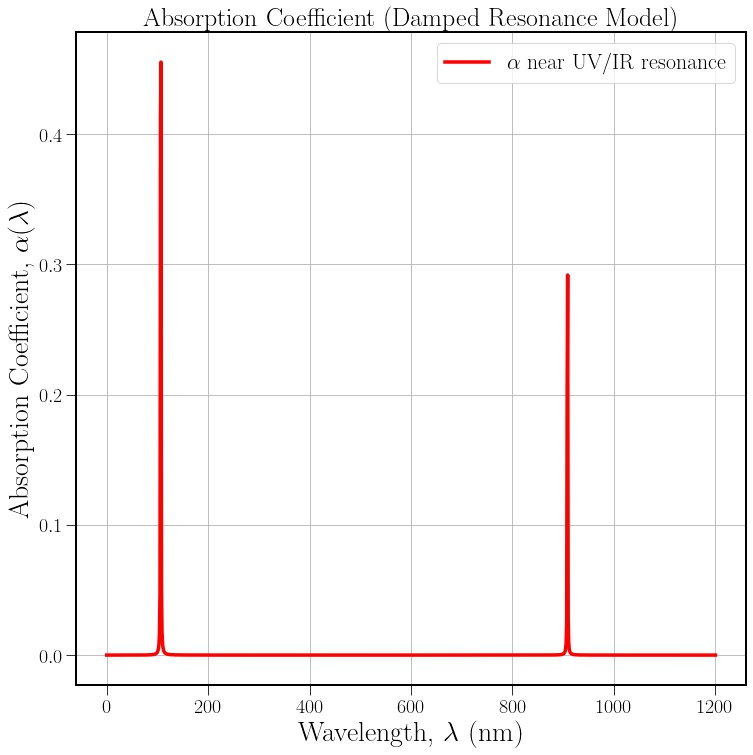}
\caption{Refractive index (left) and absorption coefficient (right) near UV/IR resonances in the Damped Resonance Model.  We illustrate the form of $n(\lambda), \alpha(\lambda)$ for the case of two resonances at $\lambda_{UV} = 106.6 \: \mathrm{nm}$ and $\lambda_{IR} = 908.3 \, \mathrm{nm}$.  For the strength and width of the resonances, we take $f_{UV} = 1250 , f_{IR} = 1, \gamma_{UV} = 1500000, \gamma_{IR} = 200000, N = 6.023\times 10^{23}$ as an example.  
\label{f:Index_Real_Imag_Theorist}
}
\end{centering}
\end{figure}

\begin{figure}[ht] 
\begin{centering}
\includegraphics[width=0.55\textwidth]{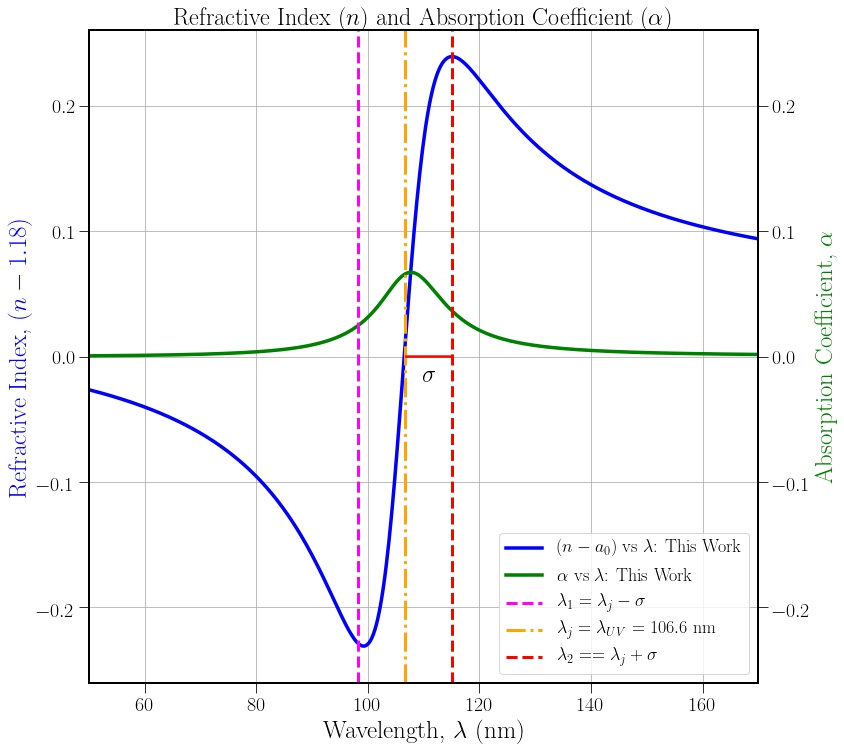}
\caption{Refractive Index and Absorption Coefficients near UV resonanace wavelength $\lambda_{UV} =$ 106.6 nm.
\label{f:absorption_coeff_theoryplot}
}
\end{centering}
\end{figure}

Being real and imaginary parts of the same analytic function, the refractive index $n(\lambda)$ and absorption coefficient $\alpha(\lambda)$ are not independent, but are related by the Kramers-Kronig relation \cite{Jackson:1998nia}. As sketched in Fig.~\ref{f:Index_Real_Imag_Theorist}, the real part specifying the refractive index $n(\lambda)$ oscillates when it goes through a resonance, rapidly changing between large positive and negative values.  The imaginary part specifying the absorption coefficient $\alpha(\lambda)$ is large very close to the resonance, but quickly decays to zero away from it.  The interrelated shapes of the two functions $n(\lambda), \alpha(\lambda)$ can be understood in a simple damped-driven oscillator model of electrons bound to an Ar nucleus in an oscillating external electric field \cite{griffiths_2017} which is discussed in detail in Appendix~\ref{sec:HarmonicOscillator}.  The damping arises from the imaginary part of $\epsilon(\lambda)$, leading to nonzero absorption absorption $\alpha > 0$; the corresponding impact on the real part of $\epsilon(\lambda)$ also modifies the index of refraction $n(\lambda)$ and prevents it from becoming infinite at the resonance
(See Fig.~\ref{f:absorption_coeff_theoryplot}).

The index of refraction and the absorption coefficient are given explicitly by the formulas \eqref{e:nlambdauvir} and \eqref{e:alphalambdauvir} derived in Appendix~\ref{sec:HarmonicOscillator}.  We plot those equations in the vicinity of two resonances ($\lambda_{UV} = $ 106.6 nm, $\lambda_{IR} = $ 908.3 nm ) in Fig. \ref{f:Index_Real_Imag_Theorist} to show that absorption coincides with the location of the resonances.  We further detail the shape of the two curves by zooming in on the UV resonance ($\lambda_{UV} = $ 106.6 nm) in Fig.~\ref{f:absorption_coeff_theoryplot}.  For our purposes, what is important is that the real part $\mathrm{Re} \: n(\lambda)$ determines not only the refraction of light, but also the intensity of Cherenkov radiation from the Frank-Tamm formula \eqref{e:FrankTamm1}.  The imaginary part is only appreciable near a resonance; in our case, the relevant UV resonance of LAr is $\lambda = 106.6 \mathrm{nm}$.  

For visible light, refraction indices $n$ of most transparent materials (e.g, air, glasses) decrease with increasing wavelength $\lambda$: $ 1 < n (\lambda_{red}) < n (\lambda_{yellow}) < n (\lambda_{blue})$.  More generally, when ${\frac {dn}{d\lambda }}<0$, the medium is said to have \textit{normal dispersion}.  This is the case far away from the resonance, for $\lambda \gg \lambda_{UV}$ in  Fig.~\ref{f:absorption_coeff_theoryplot}.  On the other hand, if the index increases with increasing wavelength, the medium is said to have \textit{anomalous} dispersion.  This is the case very close to the resonance.  Until recently, experimental data on $n(\lambda)$ for LAr only existed for $\lambda > 350 \: \mathrm{nm}$, far away from the UV resonance, so the imaginary part of $n(\lambda)$ was not previously considered.


\subsubsection{Theory of the Dispersive Refractive Index}

In optics and in wave propagation in general, dispersion is the phenomenon in which the phase velocity of a wave depends on its frequency. The spatial separation of a white light into components with different wavelengths is called dispersion, and the most well-known example of this phenomenon is undoubtedly a rainbow (different colors). The splitting of white light into a color spectrum by a prism is the most frequently observed effect of dispersion in optics. The phase velocity $v$ of a wave in a given uniform medium is given by $v={\frac {c}{n}}$, where $c$ is the speed of light in vacuum, and $n$ is the refractive index of the medium.

In general, the refractive index is some function of the frequency $f$ of the light, thus $n = n(f)$, or alternatively, with respect to the wave's wavelength $n = n(\lambda)$.  Because of the Kramers–Kronig relations \cite{deL.Kronig:26}, the wavelength dependence of the real part of the refractive index is related to the material absorption, described by the imaginary part of the refractive index (also called the extinction coefficient). In particular, for non-magnetic materials ($\mu = \mu_0$), the susceptibility $\chi$ that appears in the Kramers–Kronig relations is the electric susceptibility $\chi_e = n^2 - 1$.


The dependence of the refractive index on the wavelength was described  in 1836 by A.L. Cauchy’s semi-empirical dispersion formula 
\begin{align}   \label{e:Cauchysn}
    n (\lambda) = a_0 + \frac{a_1}{\lambda^2} + \frac{a_2}{\lambda^4} + \frac{a_3}{\lambda^6} + \cdots \: , 
\end{align}
where the coefficients $a_i$ are constants to be determined experimentally (see Ref.~\cite{Kragh2018TheLF} for a historical review.)  The coefficient $a_0$ thus denotes the refractive index evaluated far from the resonances (at infinite wavelength or zero frequency): $n (\lambda) \rightarrow a_0$ for $\lambda \rightarrow \infty$. If only the first two terms on the right hand of \eqref{e:Cauchysn} are used, then $a_0$ can be calculated from measurements of two values of n corresponding to two wavelengths $\lambda_1$ and $\lambda_2$ with the result that
\begin{align}   \label{e:a_0red}
    a_0 = \frac{\lambda_1^2 n_1 - \lambda_2^2 n_2}{\lambda_1^2 - \lambda_2^2}   \: .
\end{align}
In the early 1860s, Ludvig V. Lorenz established a general, phenomenological theory of light from which he claimed that all optical phenomena could be deduced. \matt{[Refs]}  Parameterizations such as \eqref{e:Cauchysn} have been the basis of prior experimental fits to the index of refraction.  While these fits provide a reasonable starting point, we note that fits of the form \eqref{e:Cauchysn} have only normal dispersion and miss the physics of absorption near the resonance.

\subsubsection{Experimental Data of Refractive Index of LAr}

\begin{figure}[t!]  \label{f:expsetup_Sinnock} 
\begin{centering}
\includegraphics[width=0.35\textwidth]{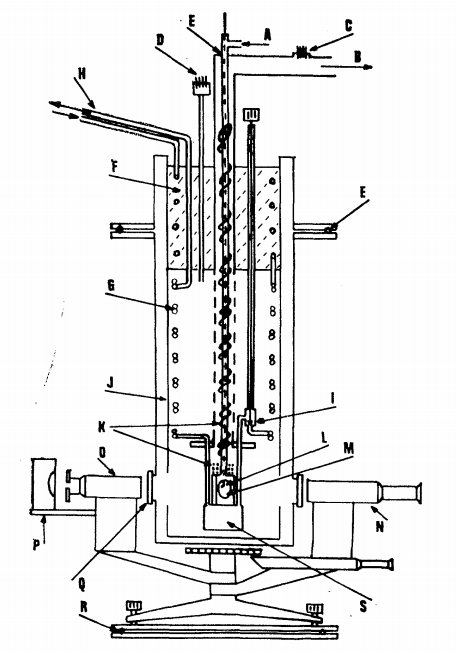}
\caption{Experimental Setup for determining Refractive Index of LAr by Sinnock and Smith. Apparatus used in the investigation. (A) Gas sample inlet, (B) outlet to vacuum system, (C) heater lead in, (D) thermocouple lead in, (E) o-ring seals, (F) liquid-nitrogen reservoir, (G) heat exchangers, (H) hydrogen inlet and outlet tubes,(I) throttling needle valve, (I) radiation shield at 77 K, (K) stem and prism heaters, (L) thermocouple mounted inside prism cell, (M) prism cell with sapphire windows, (N) telescope and cross wires, (0) collimator and slit, (P) spectral lamp, (Q) vacuum jacket windows, (R) co-axial turntable, and (S) liquid-hydrogen can.
\label{expsetup_Sinnock}
}
\end{centering}
\end{figure}

The refractive index of LAr was first experimentally measured by Sinnock and Smith \cite{Sinnock:1969zz} for higher wavelengths far from the UV resonance. They measured the refractive index for wavelengths between 361.2 and 643.9 nm, by the spectroscopic method of minimum deviation. In this method (see Fig.~\ref{expsetup_Sinnock}), a prism-shaped specimen was confined in an optical cell contained in a cryostat and located at the axis of rotation of a spectrometer mounted outside the cryostat on a coaxial turntable. The refractive index n was determined from the angle of minimum deviation D and the angle of the prism A given by the relation
\begin{align}   \label{e:nSinnock}
    n(\lambda) = \frac{\sin (\frac{A + D}{2})}{\sin\frac{A}{2}}   \: .
\end{align}

Using spectral lines from a mercury-cadmium-zinc lamp, measurements were done at various wavelengths in the range of 435.8-643.9 nm, fewer observations were made than those for the 546.1 nm line.  
Additionally, the spectrum was photographed, making it possible to determine angles of minimum deviation that correspond to wavelengths between 361.2 and 435.8 nm.



\begin{table}[h!]
  \begin{center}
    \caption{Experimental Refractive Index Data of LAr by Sinnock and Smith}
    \label{tab:ndataSinnock}
    \begin{tabular}{|c|c|c|c|c|c|c|c|c|c|c|} 
      \hline
      \multicolumn{10}{ |c| }{Wavelength (nm)} \\
      \hline
      \textbf{$T(K)$} & \textbf{643.9} & \textbf{578.0} & \textbf{546.1} & \textbf{508.6} & \textbf{475.3} & \textbf{435.8} & \textbf{406.3} & \textbf{365.0} & \textbf{361.2} \\
      \hline
      90  & 1.2256 & 1.2264 & 1.2269 & 1.2277 & 1.2285 & 1.2297 & 1.2308 & 1.2331 & 1.2326  \\ 
      \hline
    \end{tabular}
  \end{center}
\end{table}
Now, we start with the two refractive index fit function available in the literature i.e. Grace, Nikkel (2017) \& Babicz et al (2020).

\subsubsection{Grace's Refractive Index Fit (2017)}

Grace's refractive fit \cite{Grace:2015yta} formula used existing data (by Sinnock and Smith) in liquid argon to extrapolate the optical properties at the scintillation wavelengths using the Sellmeier dispersion relationship.  The functional form of the refractive index $n(\lambda)$ is given by,
\begin{align}   \label{e:ngrace}
n^2 = a_{0} + \frac {a_{UV}\lambda^{2}} {\lambda^{2} - \lambda_{UV}^{2}} + \frac {a_{IR}\lambda^{2}} {\lambda^{2} - \lambda_{IR}^{2}}       \: .
\end{align}
The UV and IR resonance wavelengths ($\lambda_{UV}$, $\lambda_{IR}$)  are 106.6 nm and 908.3 nm respectively. Here, $a_{0}$, $a_{UV}$, $a_{IR}$ are Sellmeier coefficients, and the values extracted for at $T = 90 \, \mathrm{K}$ are listed in Table~\ref{tab:Sellmeiergrace}.


\begin{table}[h!]
  \begin{center}
    \caption{Sellmeier Coefficients of Grace Fit for LAr}
    \label{tab:Sellmeiergrace}
    \begin{tabular}{|c|c|c|} 
      \hline
      \textbf{$a_0$} & \textbf{$a_{UV}$} & \textbf{$a_{IR}$}\\
      \hline
      1.26 $\pm$ 0.09 & 0.23 $\pm$ 0.09 & 0.0023 $\pm$ 0.007 \\ 
      \hline
    \end{tabular}
  \end{center}
\end{table}

The formula has been derived by simply considering UV and IR resonances in The Sellmeier dispersion relation for liquids and solids at constant temperature and density \cite{born_wolf:1999}, which relates wavelength $\lambda$ to the index of refraction $n$ given by, 

\begin{align}   \label{e:dispersionreln}
n^2 = a_{0} + \sum \frac {a_{i}\lambda^{2}} {\lambda^{2} - \lambda_{i}^{2}}    
\end{align}

In this instance, $a_0$ is a Sellmeier coefficient accounting for the effect of UV resonances not included in the sum whereas $a_i$s are the Sellmeier coefficients corresponding resonances, occurring at wavelength $\lambda_i$. The Sellmeier dispersion equation was derived from the Lorentz-Lorenz equation \cite{Lorentz:1880} and \cite{landau2013electrodynamics} but the coefficients $(a_0, a_i)$ are experimentally determined for a given medium. \cite{Grace:2015yta}

The Sellemeier coefficients obtained by fitting the LAr data from Sinnock and Smith \cite{Sinnock:1969zz} were then used to extrapolate the index of refraction of LAr plotted in Fig.~\ref{f:n_Grace_vs_Babicz}. 


\subsubsection{Babicz's Refractive Index Fit (2020)}

Babicz et al. \cite{Babicz:2020den} measured the propagation velocity of scintillation light in liquid argon $v_g$ at $\lambda \sim 128$ nm wavelength for the first time in a dedicated experimental setup at CERN. The obtained result $v_g^{-1}= 7.46 \pm 0.03$ ns/m, was then used to derive the value of the refractive index (n) and the Rayleigh scattering length (L) for LAr in the VUV region. For the scintillation wavelength, $\lambda = 128$ nm, they found $n = 1.357 \pm 0.001$. The measured values are of interest for a variety of experiments searching for rare events like neutrino and dark matter interactions. The derived quantities also represent key information for theoretical models describing the propagation of scintillation light in liquid argon. 

\begin{figure}[ht]   
\begin{centering}
\includegraphics[width=0.6\textwidth]{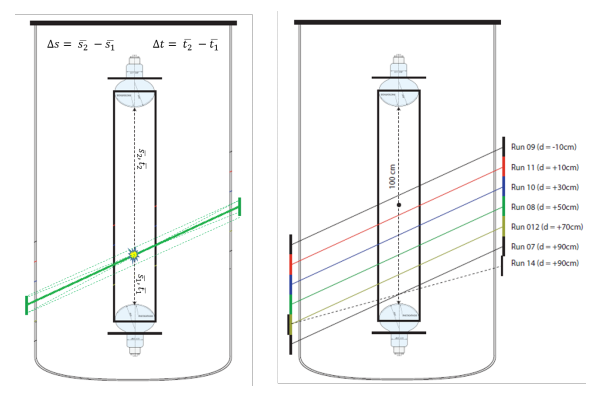}
\caption{Schematic drawing of the experimental setup of Babicz et al. together with the movable trigger system. (Right) The two internal PMTs are located at the extremities of a box and immersed in liquid argon. Outside the cryostat, located symmetrically with respect to its diameter, two scintillator bars (colored lines) play the role of cosmic hodoscope and allow selecting cosmic ray tracks crossing the box at given distances with respect to the internal PMTs. For a given distance, different track slopes can also be selected (e.g. Run 07 and Run 14 in Figure).
(Left) Sketch of how $\delta s$ and $\delta t$ are obtained in the measurement for a single position. 
\label{f:expsetup_Babicz} 
}
\end{centering}
\end{figure}

Details of the Experiment: The measurement of the velocity of LAr scintillation light presented in Babicz et al. is based on a simple setup.  Two photomultiplier tubes (PMT) were placed in LAr at 1 m apart, facing each other. The cryostat was surrounded symmetrically by an external, adjustable cosmic hodoscope that allowed muons to cross the dewar at different distances from the PMTs. The scintillation light velocity was determined using a straightforward linear fit by measuring the difference in path lengths and of the light arrival time at the PMTs for a range of positions of the external hodoscope. Fig.~\ref{f:expsetup_Babicz} displays a schematic diagram of the experimental setup.

Babicz et al. calculated the propagation speed according to the sampled photon wavelength. As a starting point, the parametrization of the refractive index given in \cite{BIDEAUMEHU1981395} has been used, and the group velocity is related to the refractive index through its derivative:


\begin{align}   \label{e:groupveln}
    v_g = \frac{c}{n - \lambda_0 \frac{dn}{d\lambda_0}}
\end{align}

The refractive index given by Babicz et al. \cite{Babicz:2020den} is,  

\begin{align}   \label{e:nBabicz}
  n = \sqrt{1 + \frac{3x}{3-x}}  
\end{align}

where,

\begin{align}   \label{e:Babiczx}
    x &= 4\pi N \alpha = \sum_k \frac{\rho_k}{v_k^2 - v^2} 
\end{align}

Since for LAr, the closest resonances to its scintillation wavelength are only two, one in the UV region (106.6 nm) and one in the IR region (908.3 nm), Equation \eqref{e:Babiczx} can thus be re-written as, 

\begin{align}   \label{e:Babiczxre}
    x &= a \sum_k \frac{b_k}{\lambda^2 - \lambda_k^2} 
    = a_{0} + \frac {a_{UV}\lambda^{2}} {\lambda^{2} - \lambda_{UV}^{2}} + \frac {a_{IR}\lambda^{2}} {\lambda^{2} - \lambda_{IR}^{2}}   
\end{align}

Here, $a_{0}$, $a_{UV}$, $a_{IR}$ are Sellmeier coefficients for Babicz fit and at T = 90 K, the values are listed here:  

\begin{table}[h!]
  \begin{center}
    \caption{Sellmeier Coefficients of Babicz Fit}
    \label{tab:SellmeierBabicz}
    \begin{tabular}{|c|c|c|} 
      \hline
      \textbf{$a_0$} & \textbf{$a_{UV}$} & \textbf{$a_{IR}$}\\
      \hline
      0.335 $\pm$ 0.003 & 0.099 $\pm$ 0.003  & 0.008 $\pm$ 0.003 \\
      \hline
    \end{tabular}
  \end{center}
\end{table}

\begin{figure}[h!]
    \centering
    \begin{subfigure}{.47\textwidth}
      \centering
      \includegraphics[width=1\linewidth]{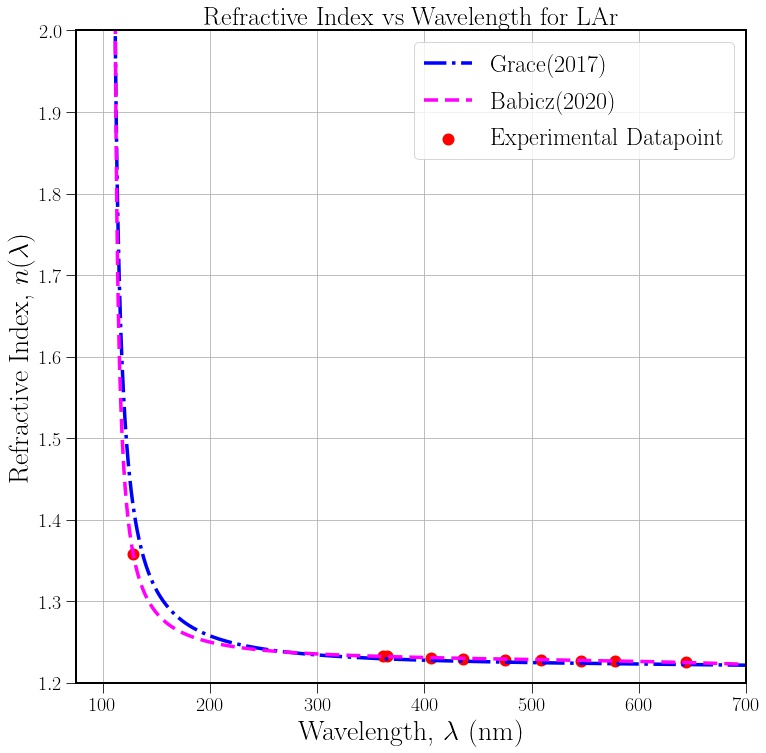}
      \caption{$\lambda = $ 100-700 nm}
      \label{f:n_Grace_vs_Babicz1}
    \end{subfigure}
    \begin{subfigure}{.53\textwidth}
      \centering
      \includegraphics[width=1\linewidth]{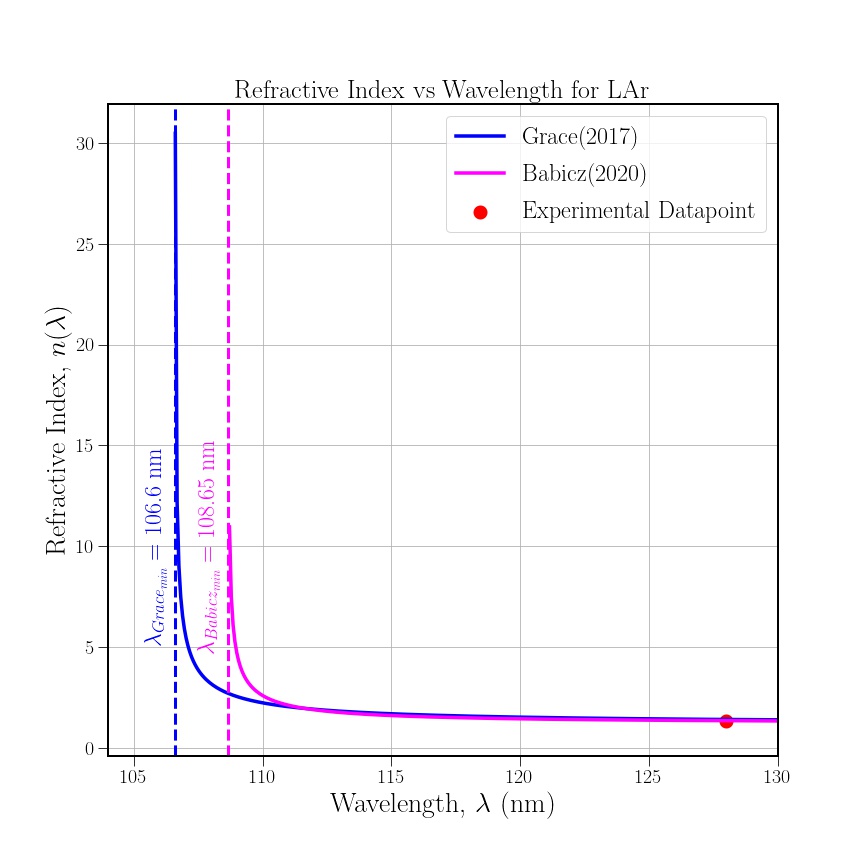}
      \caption{$\lambda = $ 105-130 nm}
      \label{f:n_Grace_vs_Babicz2}
    \end{subfigure}
    \caption{Refractive Index Curve: Grace vs Babicz for $\lambda = $ 100-700 nm (left) and 105-130 nm (right)}
    \label{f:n_Grace_vs_Babicz} 
    \end{figure}

From Fig.~\ref{f:n_Grace_vs_Babicz2}, it is visible that the Grace fit diverges at the resonance
$\lambda_{UV} =$ 106.6 nm, but the Babicz fit blows up to infinity earlier, at $\lambda =$ 108.65 nm.  The reason is that, while the Sellmeier form \eqref{e:Babiczxre} of $x$ diverges properly at $\lambda_{UV}$, its translation \eqref{e:nBabicz} to the index $n(\lambda)$ introduces an additional divergence at $x = 3$.  As a result, the Babicz fit is inapplicable for $106.6 \, \mathrm{nm} \leq \lambda \leq 108.65 \, \mathrm{nm}$, and we lose some valuable optical information of the medium near the UV resonance with Babicz' fit. 


\subsection{Calculating the Cherenkov Yield} \label{sec:noabs}


In this work, we use various fits to the index of refraction $n(\lambda)$ to compute the Cherenkov radiation in liquid argon.  Here, we demonstrate the methodology of our analysis using the pre-existing fits \eqref{e:ngrace} of Grace and \eqref{e:nBabicz} of Babicz.  Note that neither of these fits incorporates the physics of anomalous dispersion / absorption shown in Fig.~\ref{f:absorption_coeff_theoryplot}.  Later on, in Chap. \ref{cerenkovwabs}, we will repeat this procedure with our own fit, which goes beyond the previous fits to include the effects of absorption.

\subsubsection{Yield Methodology: Grace Fit} 
\label{sec:YieldMethod}


\hspace{\parindent}

\begin{figure}[t!]  
    \label{f:cerenkov_lamdmax_all}
    \centering
    \begin{subfigure}{.45\textwidth}
      \centering
      \includegraphics[width=1\linewidth]{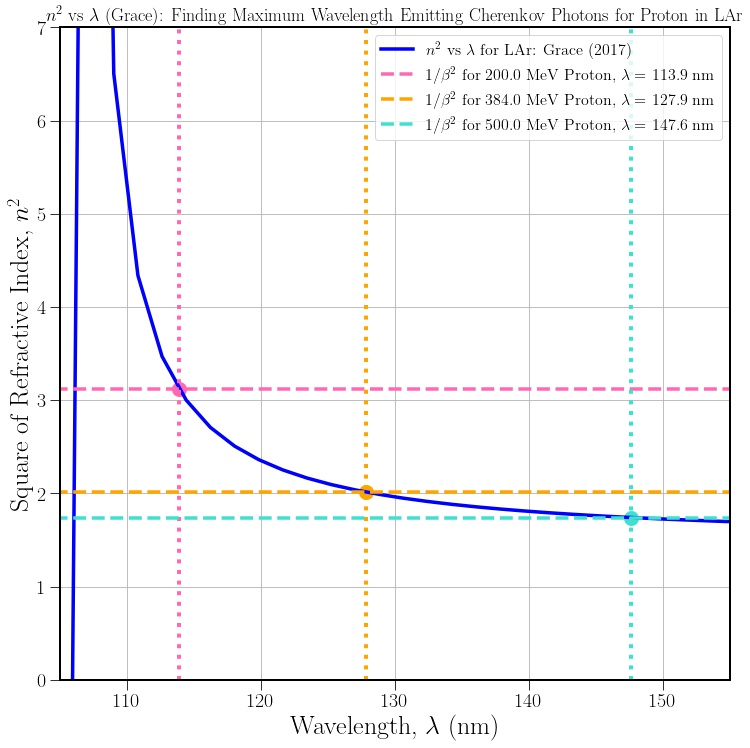}
      \caption{200-500 MeV Protons}
      \label{f:cerenkov_lamdmax_200to500MeV}
    \end{subfigure}\hfill %
    \begin{subfigure}{.45\textwidth}
      \centering
      \includegraphics[width=1\linewidth]{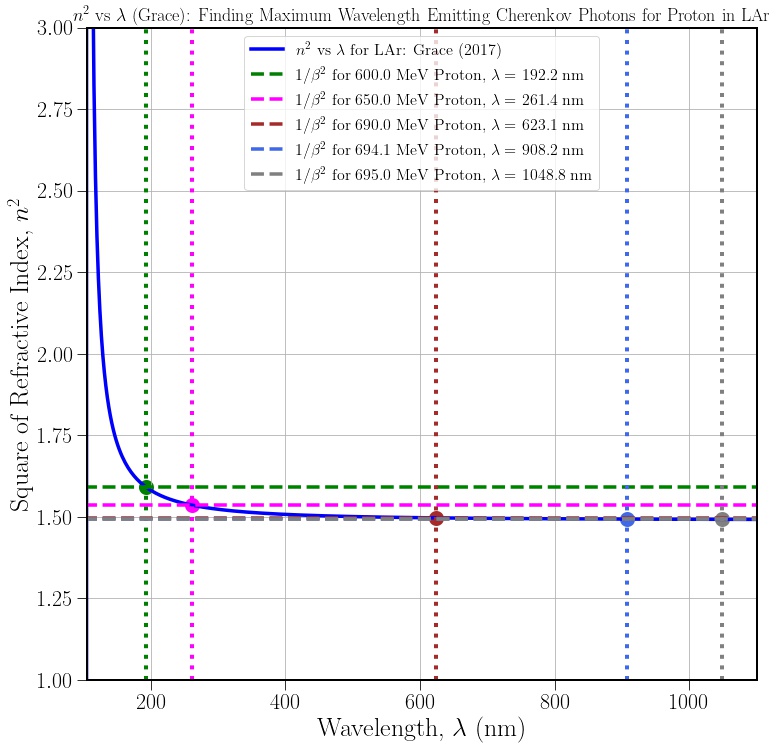}
      \caption{600-695 MeV Proton}
      \label{f:cerenkov_lamdmax_600to695MeV}
    \end{subfigure}
    \caption{Intersection Method applied to Grace's Refractive Index Fit.  Dashed horizontal lines indicate the value of $1/\beta^2$ at a given kinetic energy, and dotted vertical lines with the same color schemes indicate the value of $\lambda_{max}$ given by \eqref{e:lamdmaxgrace1}. }
    \label{f:cerenkov_lamdmax_grace} 
\end{figure}

To compute the instantaneous yield \eqref{e:ftintegrand} of Cherenkov photons, we must integrate the Frank-Tamm formula \eqref{e:FrankTamm1} over all wavelengths $\lambda$ which satisfy the Cherenkov condition \eqref{e:CherenCond1}:
\begin{align}   \label{e:ftintegrand2}
 \frac{dN}{dx} = \int_{\lambda_{min}}^{\lambda_{max}} \frac{2 \pi \alpha z^2}{\lambda^2} \left( 1 - \frac{1}{\beta^2 n^2(\lambda)} \right) d\lambda  = I(\beta, x)  \: ,
\end{align}    
where we write $dN/dx$ as $I(\beta, x)$ to emphasize its dependence on the velocity $\beta(x)$ that decreases with distance.
%
%
%
We determine the wavelength range for integrating Eq.~\eqref{e:ftintegrand2} using the ``intersection method.''  This method imposes the square of the Cherenkov condition $n^2(\lambda) \geq \frac{1}{\beta^2}$ by solving for the intersection of the squared refractive index fit $n^2(\lambda)$ with this the constant $\frac{1}{\beta^2(x)}$ at a given distance $x$ (see Fig.~\ref{f:cerenkov_lamdmax_grace}).  Working with the square of the Cherenkov condition is just for convenience, given the form of the Grace and Babicz fits.  Since both the Grace and Babicz fits are divergent in the UV limit, those divergences (at $\lambda_{min} = \lambda_{UV} = 106.6 \, \mathrm{nm}$ and $\lambda_{min} = 108.65 \, \mathrm{nm}$, respectively; see Fig.~\ref{f:n_Grace_vs_Babicz2}) determine the minimum radiating wavelength.  The intersection method is used to determine the maximum wavelength $\lambda_{max}$.

Both Grace's and Babicz's fits include the effects of two resonances -- one in the UV at $\lambda_{UV} = 106.6 \, \mathrm{nm}$ and the other in the IR at $\lambda_{IR} = 908.3 \, \mathrm{nm}$.  The impact of the IR resonance on the Cherenkov radiation is expected to be small for two reasons.  The first is that the strength of the IR resonance (coefficient $a_{IR}$) in the fit is 1-2 orders of magnitude less than the strength of the UV resonance (coefficient $a_{UV}$) so that away from the resonance, this contribution is comparatively small.  The second is that the Cherenkov spectrum is less sensitive to the index of refraction at large $\lambda$ due to the $1/\lambda^2$ weighting of the F-T integrand \eqref{e:ftintegrand2}.  For these reasons, we will neglect the IR resonance in both the Grace and Babicz fits for the purposes of computing the Cherenkov emission. 


The comparison of the total yield, $N_{Grace}$ for different choices of $\lambda_{max}$ will be presented in Tab.~\ref{tab:NGracelambmax}. 

For Grace's fit \eqref{e:ngrace}, we neglect the IR resonance to write the Cherenkov condition as
\begin{align}   \label{e:ngracecherenkov1}
n^2 (\lambda) = a_{0} + \frac {a_{UV}\lambda^{2}} {\lambda^{2} - \lambda_{UV}^{2}} 
\geq \frac{1}{\beta^2}  \:.
\end{align}
Solving this inequality for $\lambda^2$, we obtain
\begin{align}   \label{e:lambdasolngrace1}
   \lambda^2 \leq
   \left(
    \frac{\beta^{-2} - a_0}
    {\beta^{-2} - a_0 - a_{UV}} \right)
    \lambda_{UV}^2   \: , 
\end{align}
which explicitly determines the maximum radiating wavelength
\begin{align}   \label{e:lamdmaxgrace1}
    \lambda_{max} &= 
    \sqrt{\frac{\beta^{-2} - a_0}
    {\beta^{-2} - a_0 - a_{UV}}} 
    \lambda_{UV}
\end{align}
from the intersection of $n(\lambda)$ with $1/\beta$.  

The intersection method fails when \eqref{e:ngracecherenkov1} has no solution; that requires special care because the index of refraction (neglecting the IR resonance) has an asymptote at large $\lambda$.  For $\beta$ large enough, $1/\beta$ lies below the asymptote $\lim_{\lambda \rightarrow \infty} n^2 (\lambda) = a_0 + a_{UV}$, and there is no intersection.  This means that all wavelengths above $\lambda_{UV}$ pass the Cherenkov condition, meaning $\lambda_{max} = \infty$.  To solve for where \eqref{e:lamdmaxgrace1} fails and $\lambda_{max} \rightarrow \infty$, we set the denominator of \eqref{e:lamdmaxgrace1} to zero, giving
\begin{align}   \label{e:betaltgrace1}
    \beta = \frac{1}{\sqrt{(a_0 + a_{UV})}}
\end{align}
Using the values of the Sellmeier coefficients for Grace's fit from Table~\ref{tab:Sellmeiergrace} in \eqref{e:betaltgrace1}, we obtain $\beta = 0.819$ corresponding to $T = 698$ MeV using \eqref{e:KE1}.  In summary, the wavelength limits of integration for Grace's refractive index fit are
\begin{align}   \label{e:wlltgrace}
    \lambda_{min} &= 106.6 \, \mathrm{nm} \: ,
    \notag 
    \\
    \lambda_{max} &= 
    \begin{cases}
        \sqrt{\frac{a_0 - \beta^{-2}}{a_0 - \beta^{-2} + a_{UV}}} \lambda_{UV}
        & \mathrm{if} \qquad \beta 
         < \frac{1}{\sqrt{(a_0 + a_{UV})}}   \\
        \infty & \mathrm{else} 
    \end{cases} \: .
\end{align}

The result of the intersection method is shown in Fig.~\ref{f:cerenkov_lamdmax_200to500MeV} and Fig.~\ref{f:cerenkov_lamdmax_600to695MeV} applied to Grace's refractive index fit.  As the K.E. of the proton increases, so does $\beta$, leading to a smaller $\frac{1}{\beta^2}$ as shown in the dashed line in \ref{f:cerenkov_lamdmax_200to500MeV} (from 200 MeV (pink), 384 MeV (orange) to 500 MeV (turquoise)).  This results in a higher $\lambda_{max}$ with increasing kinetic energy, growing slowly at low kinetic energies (Fig.~\ref{f:cerenkov_lamdmax_200to500MeV}) and quickly as we approach the asymptote at $T = 698 \, \mathrm{MeV}$ (Fig ~\ref{f:cerenkov_lamdmax_600to695MeV}).  Above this threshold, $T > 698$  MeV, there is no intersection point denoting $\lambda_{max}$ in the finite wavelength range and $\lambda_{max} = \infty$, indicating that all IR wavelengths satisfy the Cherenkov condition and are lead to Cherenkov radiation.


\newpage

\subsubsection{Instantaneous \& Total Cherenkov Yield}
\hspace{\parindent}

\begin{figure}[h!] 
\begin{subfigure}{.48\textwidth}
\centering
\includegraphics[width=1\textwidth]{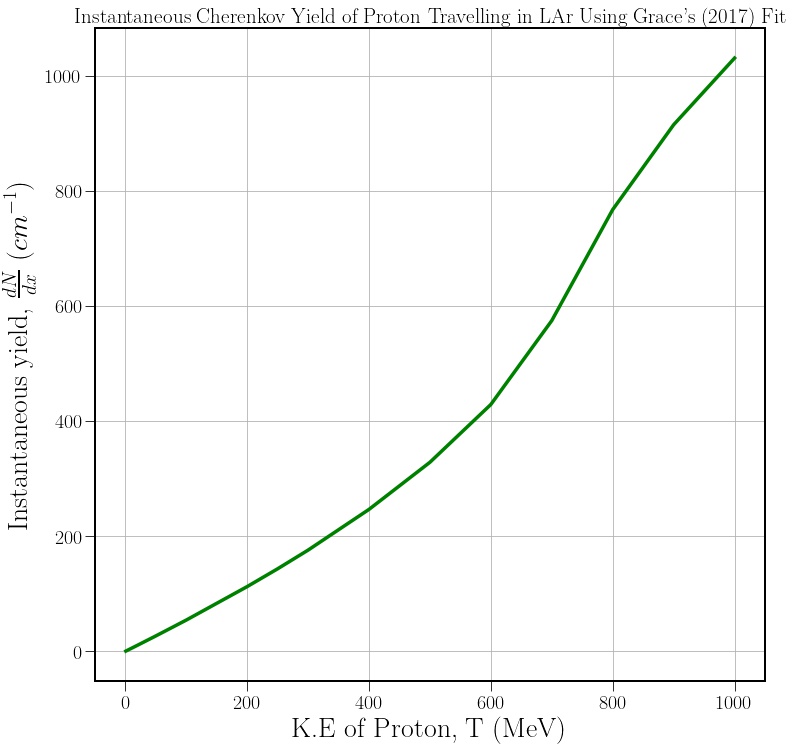}
\caption{Instantaneous yield ($\frac{dN}{dx}$)   
\label{f:cerenkov_length_grace}
}
\end{subfigure}
\begin{subfigure}{.48\textwidth}
\centering
\includegraphics[width=1\textwidth]
{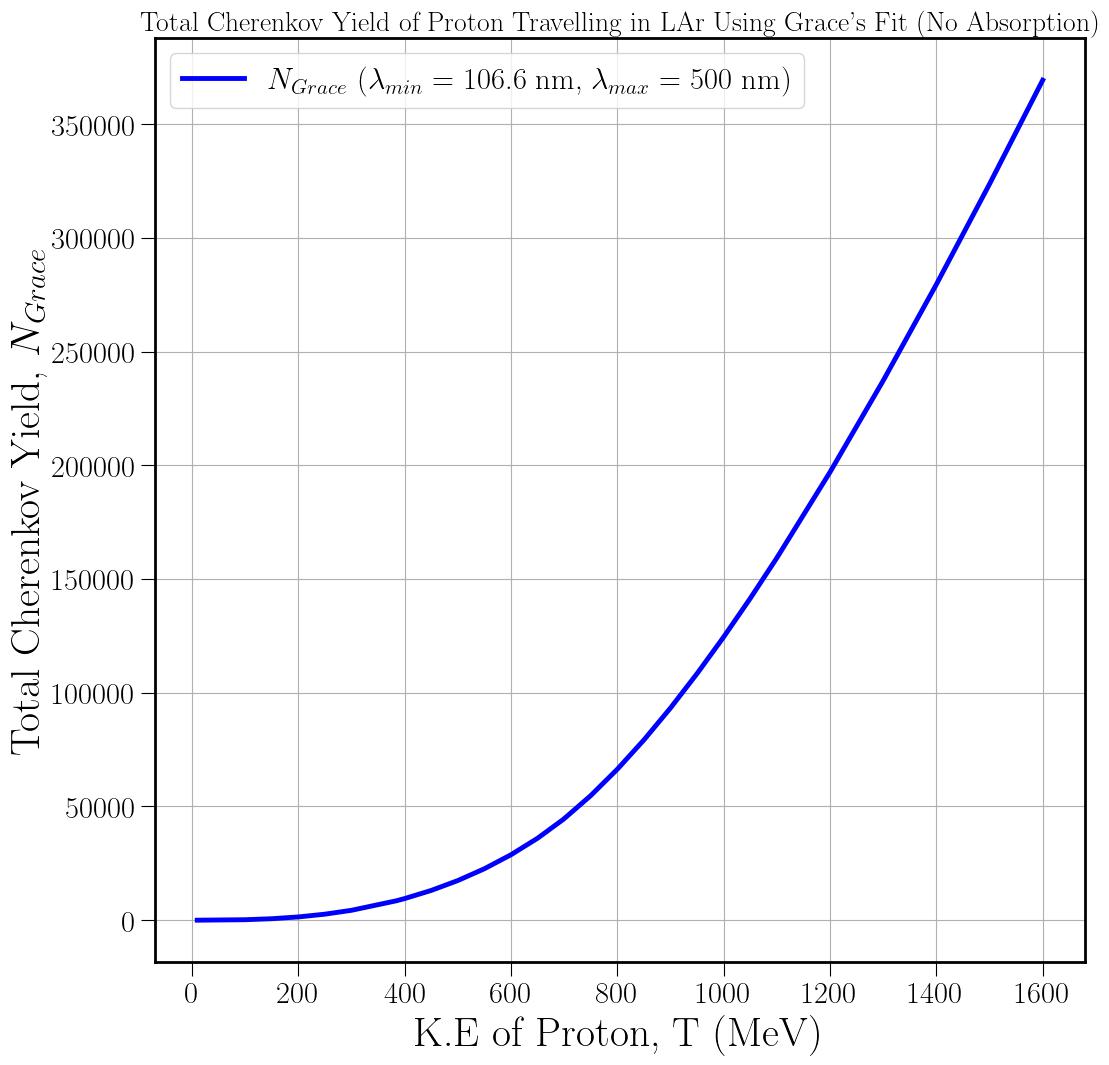}
\caption{Total/Integrated Yield ($N_{Grace}$)   
\label{f:cerenkov_total_Grace}
}
\end{subfigure}
\caption{Cherenkov Yield of Proton in LAr Calculated Using Grace's Refractive Index Fit.
\label{f:cerenkov_Grace}
}
\end{figure}

The instantaneous Cherenkov yield \eqref{e:ftintegrand2} gives us the number of Cherenkov photons emitted by the proton per unit length in LAr. The wavelength limits of integration \eqref{e:wlltgrace} are uniquely determined from the intersection method for a given K.E. of the proton. The instantaneous yield calculated for protons with different K.E.s from this method using Grace's refractive index fit is plotted in Fig.~\ref{f:cerenkov_length_grace}. 

We note that the instantaneous Cherenkov yield is an increasing function of the proton kinetic energy, as expected.  But interestingly, there is a change in the curvature of the Cherenkov yield seen in Fig.~\ref{f:cerenkov_length_grace}.  This inflection point occurs exactly at the threshold value $T = 698 \, \mathrm{MeV}$ at which $\lambda_{max} \rightarrow \infty$.  This allows us to identify the reason for the change in curvature.  At low energies, the yield grows rapidly, both due to the increasing number of Cherenkov photons with a given wavelenth, and to the growth of the wavelength range itself.  But above $T = 698 \, \mathrm{MeV}$, the entire wavelength range above $\lambda_{UV}$ is already radiating, so further increases in $T$ do not result in as great an increase to the yield $\frac{dN}{dx}$.  Above $T = 698 \, \mathrm{MeV}$, the growth in  yield is much slower and is due only to the $\beta$ dependence of the Frank-Tamm integrand \eqref{e:ftintegrand2}.




\begin{table}[h!]
  \begin{center}
    \caption{Comparison (\% difference) of Total Cherenkov yield (N) using Grace's Fit Using Frank-Tamm Integral (FT) with $\lambda_{min} =$ 106.6 nm but different choices of $\lambda_{max}$ (500 nm, 908.3 nm ($\lambda_{IR}$), and $\infty$). \% diff1 = $ (2 (N_{908.3nm} - N_{500nm})/(N_{908.3nm} + N_{500nm})) \times 100$ and \% diff2 = $(2 (N_{\infty} - N_{500nm})/(N_{\infty} + N_{500nm})) \times 100$
    (Imp: need to include average of two N's  in the denominator for \% diff)
}
    \label{tab:NGracelambmax}
    \begin{tabular}{|c|c|c|c|c|c|} 
      \hline
      \textbf{$T$}(MeV) & \textbf{$N_{\lambda_{max}=500nm}$} & \textbf{$N_{\lambda_{max}=908.3nm}$} &
      \textbf{$N_{\lambda_{max}=\infty}$} &
      \textbf{\% diff1} &
      \textbf{\% diff2} \\
      \hline
      200	& 1434 & 1434 & 1434 & 0.00 & 0.00 \\
      \hline
      500	& 17425	& 17425	& 17425	& 0.00	& 0.00 \\
      \hline
      800	& 66364	& 66825	& 67323	& 0.69	& 1.43 \\
      \hline
      1000 & 124517 & 127754 & 131531	& 2.56 & 5.47 \\
      \hline
      1200 & 196997 & 204841 & 214141	& 3.90 & 8.34 \\
      \hline
      1400 & 279538 & 293287 & 309702	& 4.80 & 10.24 \\
      \hline
      1600 & 369320 & 389920 & 414609 & 5.42 & 11.55 \\
      \hline
      
    \end{tabular}
  \end{center}
\end{table}

\begin{figure}[t!] 
\begin{centering}
\includegraphics[width=0.65\textwidth]{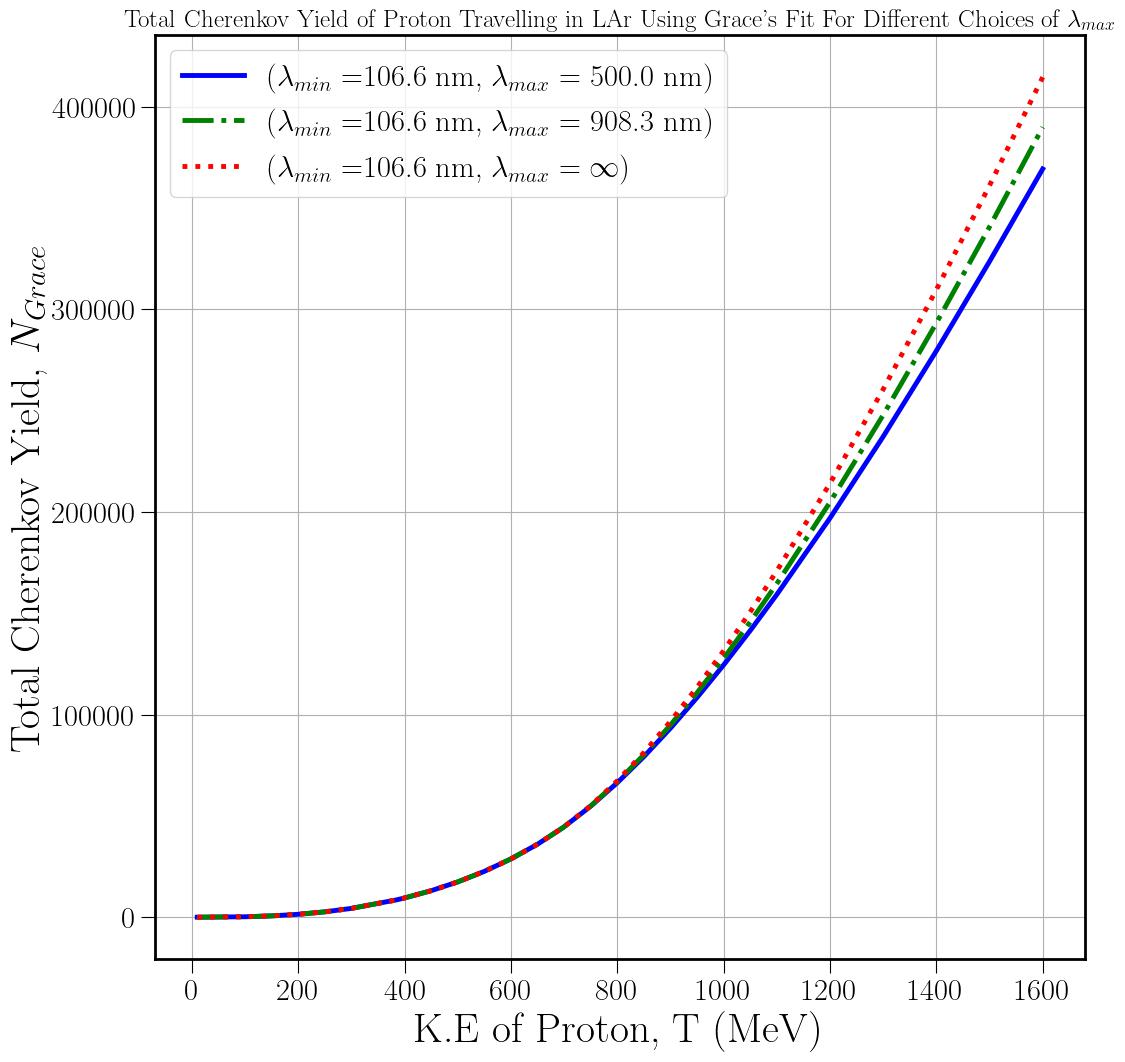}
\caption{Total Yield Calculated Using Grace's Refractive Index Fit with $\lambda_{min} = \lambda_{UV} = $ 106.6 nm for Different Choices of $\lambda_{max}$.   
\label{f:cerenkov_total_Grace_difflamdmax}
}
\end{centering}
\end{figure}

The choice of $\lambda_{max}$ have an insignificant effect on the total calculated yield using Grace's refractive index fit for protons with lower T which is shown in Fig.~\ref{f:cerenkov_total_Grace_difflamdmax}. For $T > 800$ MeV, $N_{Grace}$ calculated for $\lambda_{max} =$ 908.3 nm (green dashdotted curve) and $\lambda_{max} = \infty$ (red dotted curve) gets bigger compared to $\lambda_{max} =$ 500 nm (blue solid curve)
for increasing T.

%




Finally, the total (or ``integrated'') number of Cherenkov photons emitted can be determined by integrating the instantaneous yield $\frac{dN}{dx}$ once again over the total distance trajectory of the proton in LAr:
\begin{align}   \label{e:totalcherenkov}
N = \int dN = \int_{0}^{Range} I(\beta, x) dx   \: .
\end{align}
The total Cherenkov yield calculated numerically using the Gaussian quadrature method is plotted in Fig.~\ref{f:cerenkov_total_Grace}.  Again, we note a distinct change in curvature at the threshold $T = 698 \, \mathrm{MeV}$, although the effect is not as pronounced as in the instantaneous yield shown in Fig.~\ref{f:cerenkov_length_grace}.  This can be understood because the integrated yield probes the Cherenkov emission over a range of energies.  For protons with initial kinetic energies less than $698 \, \mathrm{Mev}$, the active wavelength range is changing over the proton's trajectory, shrinking to a smaller range as the proton slows down.  This results in a large positive curvature of the integrated yield with increasing energy at low $T$ seen in Fig.~\ref{f:cerenkov_length_grace}.  But for protons at high initial kinetic energies greater than $698 \, \mathrm{MeV}$, there is initially Cherenkov radiation at all wavelengths above the resonance. Very high energy protons spend longer amounts of their trajectory in this regime where the wavelength range is static, leading to a curve that is nearly linear at high energies.  Both the instantaneous (Fig.~\ref{f:cerenkov_length_grace}) and integrated Cherenkov yields (Fig.~\ref{f:cerenkov_total_Grace}) reflect the interplay of the active wavelength range and the number of photons emitted at a given wavelength.  Both factors are affected by the form of the refractive index $n(\lambda)$.

\subsubsection{Comparison: Grace vs. Babicz Yields}
%

Here we will similarly apply the methodology described in the previous section to find the total Cherenkov yield using Babicz's fit given in Eqs.~\eqref{e:nBabicz} and \eqref{e:Babiczxre}.  As with Grace's fit, we neglect the IR resonance to write 
\begin{align}   \label{e:BabiczTruncated}
    n^2(\lambda) = 1 + \frac{
        3\left(
        a_0 + \frac{a_{UV} \lambda^2}{\lambda^2 - \lambda_{UV}^2}
        \right)
    }{
        3 - \left(a_0 + \frac{a_{UV} \lambda^2}{\lambda^2 - \lambda_{UV}^2} \right)
    }   \: ,
\end{align}
with the Sellmeier coefficients given in Table~\ref{tab:SellmeierBabicz}.  As noted previously, Babicz's refractive index fit has different functional form than Grace's, which diverges not at the UV resonance $106.6 \, \mathrm{nm}$ but rather at the larger value $\lambda_{min} =$ 108.7 nm.  The same intersection method shown in Fig.~\ref{f:cerenkov_lamdmax_grace} has been used to find the maximum wavelength $\lambda_{max}$ satisfying the Cherenkov condition \eqref{e:CherenCond1} for Babicz's fit.   

\begin{figure}[t!] 
\centering
\begin{subfigure}{.45\textwidth}
\centering
\includegraphics[width=1\textwidth]{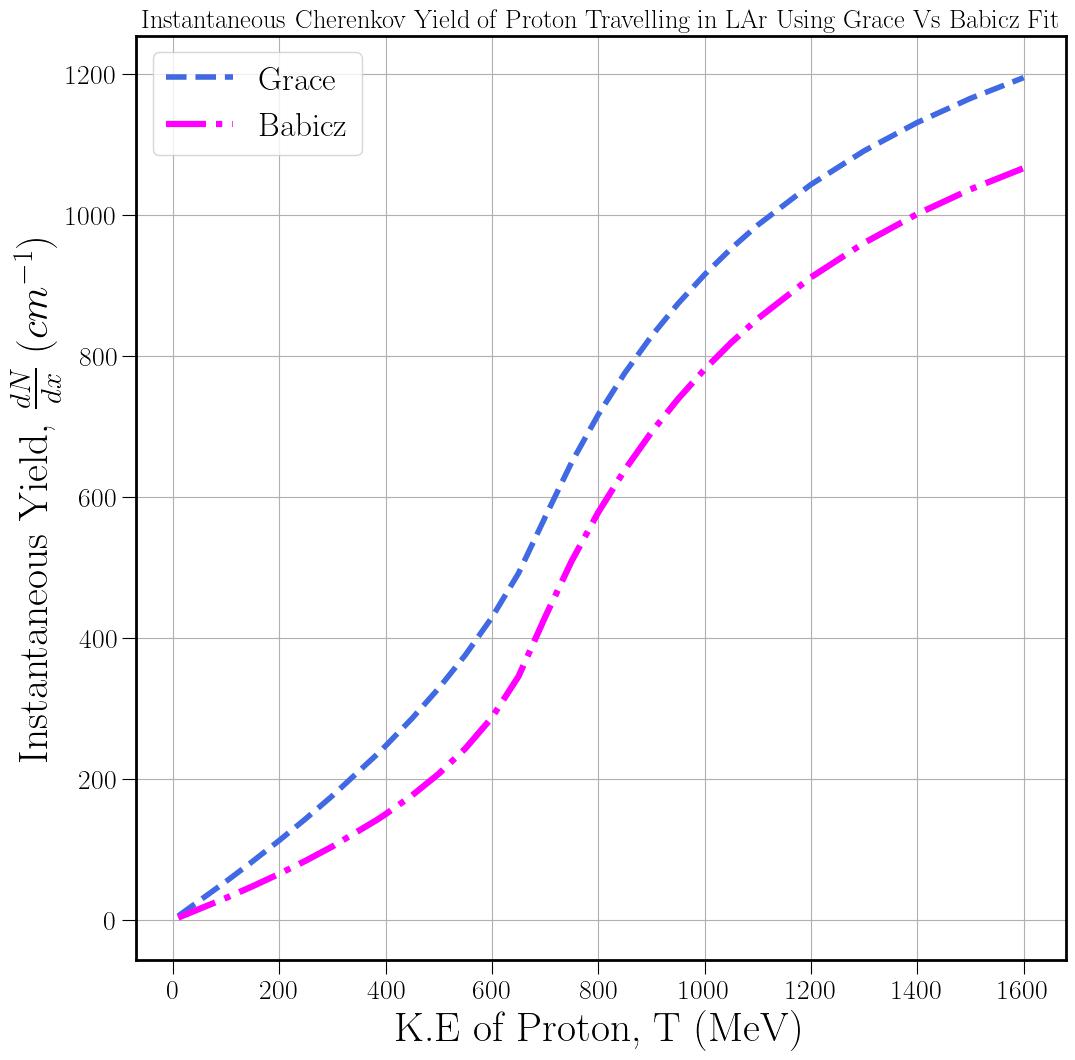}
\caption{ Instantaneous Cherekov Yield 
\label{f:nfitslog}
}
\end{subfigure}
\begin{subfigure}{.46\textwidth}
\centering
\includegraphics[width=1\textwidth]{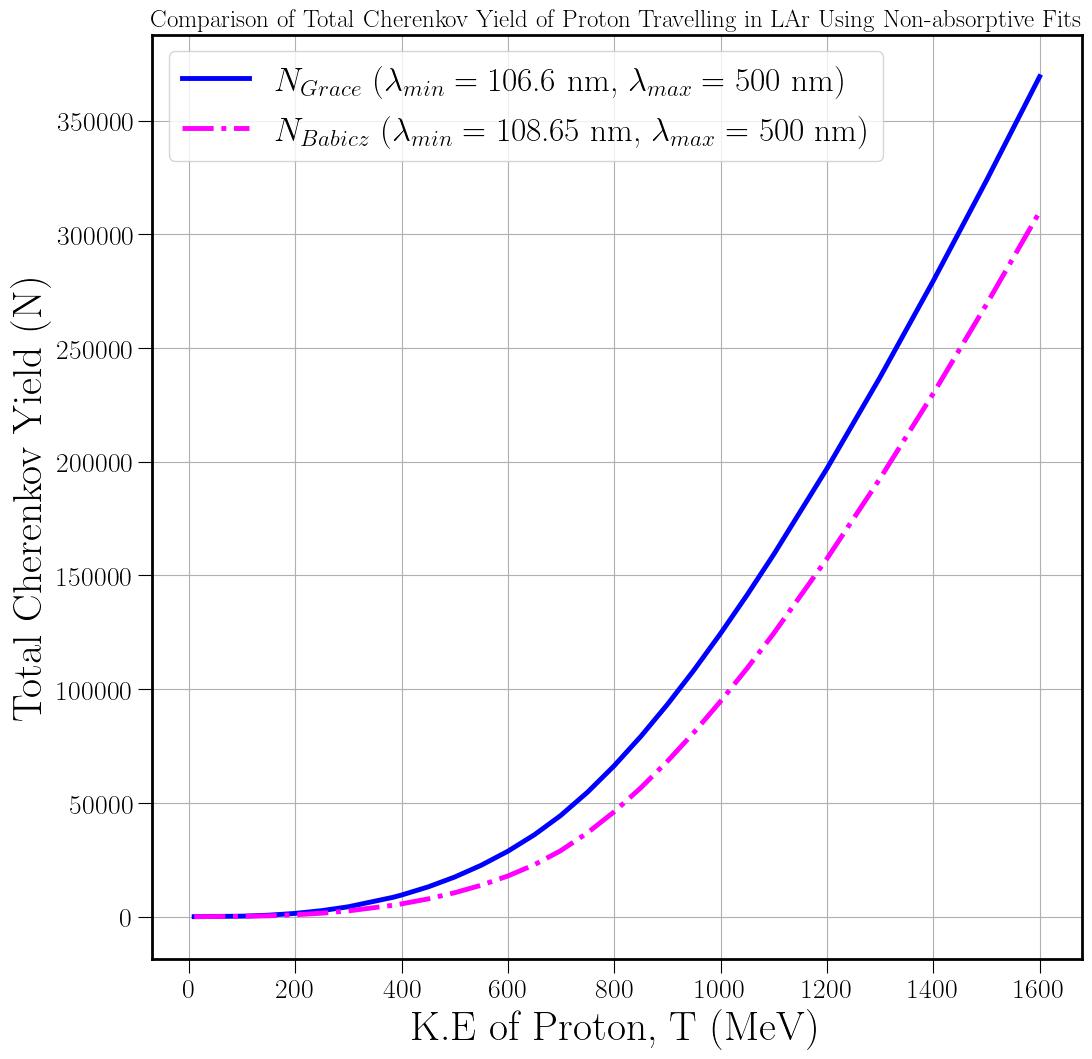}
\caption{Total Cherenkov Yield  
\label{f:cerenkovtotalcomparisonmidt}
}
\end{subfigure} 
\caption{Comparison of the instantaneous (left) and total (right) number of Cherenkov photons calculated using Babicz' vs. Grace's fits
\label{f:cerenkovtotalcomparisongracevsBabicz}
}
\end{figure}

The instantaneous yield $\frac{dN}{dx}$ and integrated yield $N_{tot}$ of Cherenkov photons computed from Babicz' fit are shown in Fig.~\ref{f:cerenkovtotalcomparisongracevsBabicz}.From Fig.~\ref{f:nfitslog} we see that the instantaneous yield for Babicz' fit is significantly reduced compared to Grace's fit.  More than any other feature of the fit, this reduction can be attributed to the divergence of Babicz' fit early, at $\lambda_{min} =$ 108.65 nm, compared to using Grace's fit ($\lambda_{min} =$ 106.6 nm).  When integrated over the proton range to obtain the total yield shown in Fig.~\ref{f:cerenkovtotalcomparisonmidt}, we see that reduction persists.  The reduction in yield of Babicz relative to Grace becomes nearly constant at high kinetic energies.

\begin{figure}[t!] 
\begin{centering}
\includegraphics[width=0.65\textwidth]{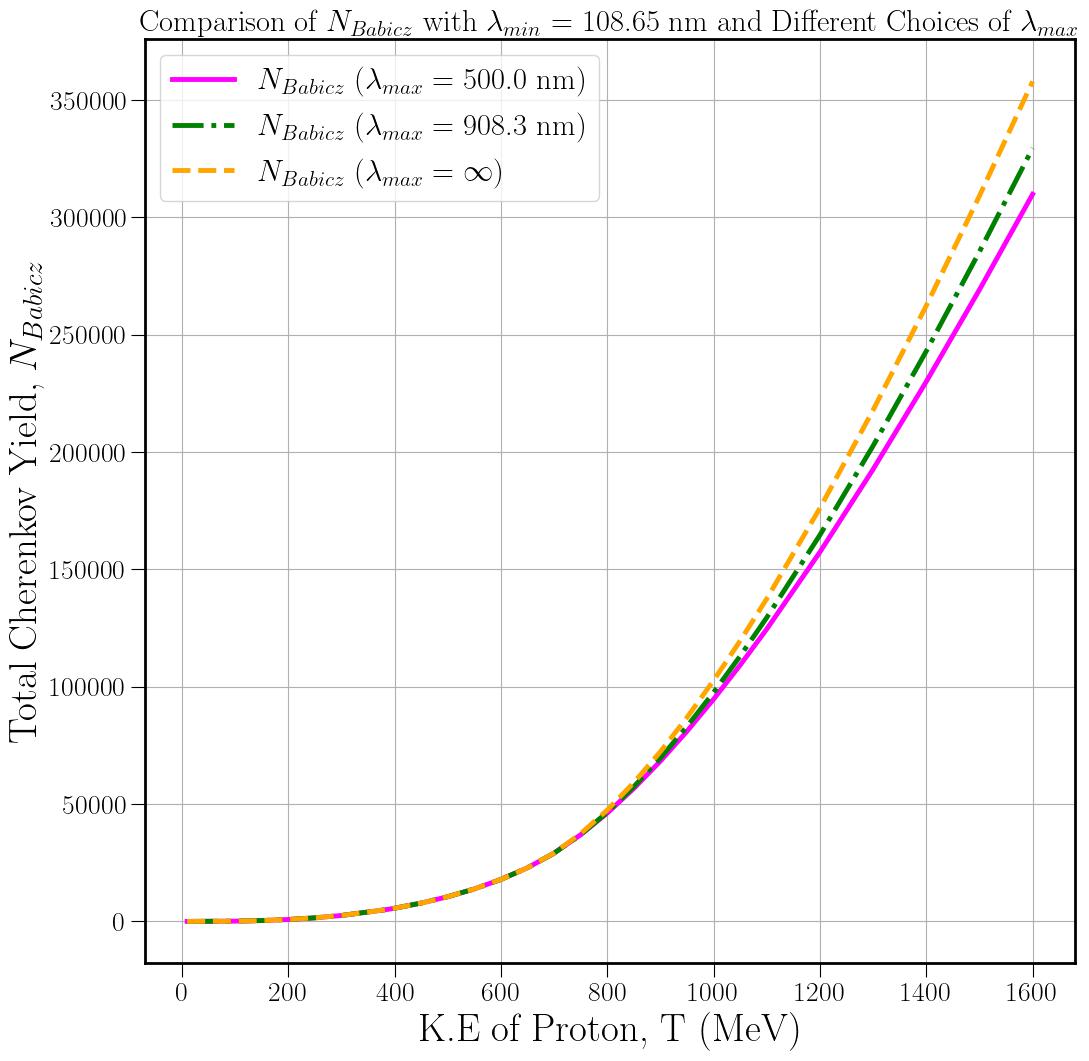}
\caption{Total Yield Calculated Using Babicz's Refractive Index Fit with $\lambda_{min} = \lambda_{UV} = $ 108.65 nm for Different Choices of $\lambda_{max}$ (500 nm, 908.3 nm($\lambda_{UV}$), and $\infty$).   
\label{f:cerenkov_total_Babicz_difflamdmax}}
\end{centering}
\end{figure}

The changes between the Grace and Babicz fits is due primarily to their different treatment of the lower cutoff $\lambda_{min} = 106.6 \, \mathrm{nm}$ vs $\lambda_{min} = 108.7 \, \mathrm{nm}$, resulting in the decrease in yield seen in Fig.~\ref{f:cerenkovtotalcomparisongracevsBabicz}.  On the other hand, the choice of $\lambda_{max}$ shown in  Fig.~\ref{f:cerenkov_total_Babicz_difflamdmax} also makes a similar reduction in yield in going from $\lambda_{max} = \infty$ to $\lambda_{max} = 500 \, \mathrm{nm}$, but for very different reasons.  The $\lambda_{max}$ sensitivity is fairly weak and only affects protons at high energies; for $T < 800 \, \mathrm{MeV}$, the value of $\lambda_{max}$ is determined by the intersection method and is the same for all three conventions.  Only when the energy is large enough that the intersection method gives $\lambda_{max} \rightarrow \infty$ does it need to be cut off.  In contrast, there is much greater sensitivity to $\lambda_{min}$, especially at low kinetic energies.  As seen in Fig.~\ref{f:cerenkovtotalcomparisonmidt}, the large change in yield comes from changing $\lambda_{min}$ only by $2 \, \mathrm{nm}$, and it immediately impacts the Cherenkov yield of low-energy protons because of the resonant form to the index of refraction.

. 


%


\subsubsection{Comparison: Cherenkov vs. Scintillation Yields}

Now that we have the ``signal'' -- the total Cherenkov yield for protons traveling in LAR computed from two different fits of $n(\lambda)$ -- we can compare it against the ``background -- the fast component of scintillation photons which are normally utilized for particle detection.  As discussed in Sec.~\ref{sec:Outline}, the prompt scintillation yield is approximately \cite{Segreto_2021},
\begin{align}   \label{scintcount}
    N_{scint} &\approx \left(\frac{T}{\mathrm{MeV}}\right) \times \Big( 40,000 \, \gamma \Big) \times \Big( 27.5\% \, \mathrm{prompt} \Big) 
    \notag \\ &=
    \left(\frac{T}{\mathrm{MeV}}\right) \times 11,000   \: .
\end{align}
The scintillation yield is much larger than the Cherenkov yield at low kinetic energies, but increases only linearly with $T$.  In contrast, the Cherenkov yield for either fit of the refractive index starts small but grows rapidly with increasing $T$.  Assuming that the distribution of scintillation photons is Poissonian, the uncertainty $\delta N_{scint}$ in the number of detected scintillation photons is given by the square root:
\begin{align}   \label{e:ScintError1}
    \delta N_{scint} \equiv \sqrt{N_{scint}} = \sqrt{\left(\frac{T}{\mathrm{MeV}}\right) \times 11,000} \: .
\end{align}
When the Cherenkov yield is less than the uncertainty $\delta N_{scint}$ of the background measurement, it is not possible to discern the Cherenkov radiation above background.  But if $N > \delta N_{scint}$, then the Cherenkov signal becomes measurable and increases in statistical significance.

\begin{figure}[t!] 
\centering
\begin{subfigure}{.46\textwidth}
    \centering
    \includegraphics[width=1\textwidth]{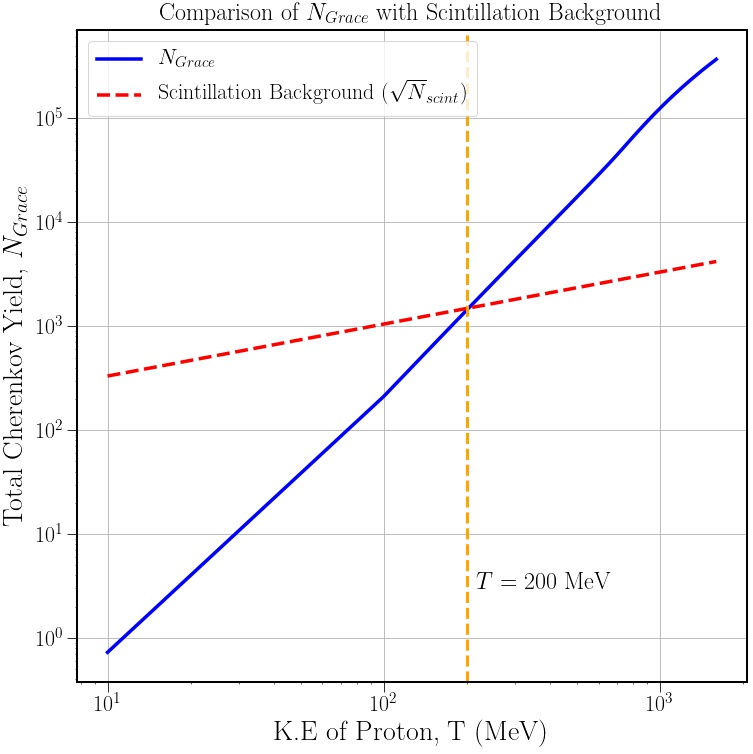}
    \caption{Log Plot  
    \label{f:CerenkovGracevsScintillationback_log}
    }
\end{subfigure} 
\begin{subfigure}{.45\textwidth}
    \includegraphics[width=1\textwidth]{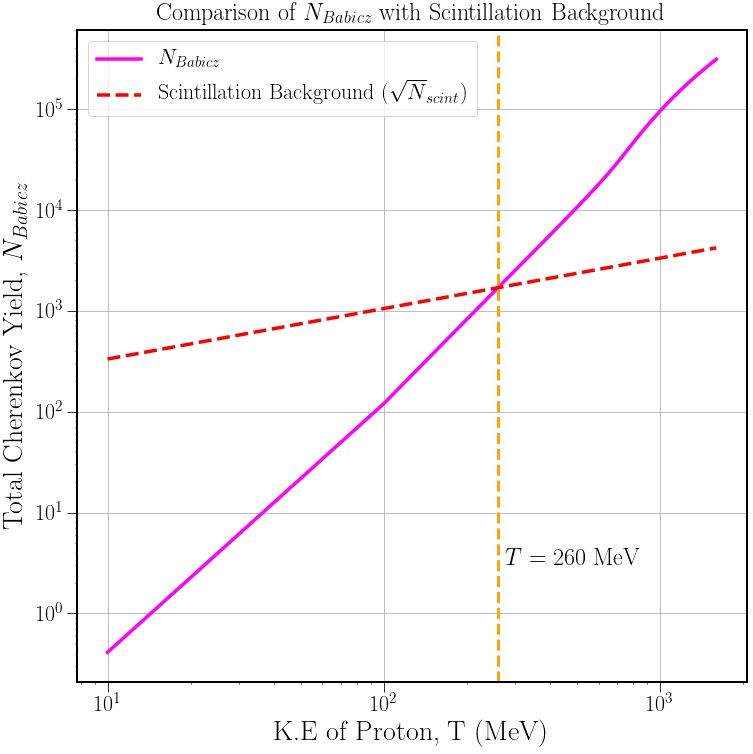}
    \caption{Babicz FT Log Plot  
    \label{f:BabiczFTvsscintback_log}
    }
\end{subfigure} 
%
\caption{Comparison of total Cherenkov yield (solid lines) and uncertainty in the scintillation yield (dashed line) for both Grace (left) and Babicz (right) fits.  The kinetic energy $T$ at which the Cherenkov yield first exceeds the scintillation uncertainty is noted in the figures.
\label{f:CerenkovGracevsScintillationback}
}
\end{figure}

In Fig.~\ref{f:CerenkovGracevsScintillationback}, a comparison between Cherenkov yield and scintillation uncertainty \eqref{e:ScintError1} is plotted as a function of kinetic energy for both the Grace (left) and Babicz (right) refractive index fits. As expected, for high enough kinetic energies, the Cherenkov yield overtakes the scintillation background, with the transition occurring for $T > 200 \, \mathrm{MeV}$ for Grace's fit and $T > 260 \, \mathrm{MeV}$ for Babicz'.  While the existence of such a crossover in which the Cherenkov signal starts to become significant over the scintillation background is perhaps expected, the rather low values of $T$ at which this occurs may well be surprising.  As we will show in Chap.~\ref{cerenkovwabs}, both the Grace and Babicz fits severely overestimate the number of Cherenkov photons due to a fully resonant fit for $n(\lambda)$.  The existence of a complete resonance with $n \rightarrow \infty$ allows even a proton with infinitesimal kinetic energy to radiate Cherenkov photons, and (as we will show next) it even allows these photons to be radiated far from the axis.  These unrealistic features strongly motivate the development of a new, absorptive fit that remedies these deficiencies.


%

%

%
\subsection{Extracting the Angular Distribution}
\hspace{\parindent}

As Cherenkov radiation is instantaneous and directional, unlike the isotropic scintillation background, deriving the angular distribution of the Cherenkov radiation can be extremely useful to distinguish the two contributions.  This angular distribution can reveal new physics, such as anomalous dispersion, and can increase the statistical significance of the Cherenkov signal.  Exploiting this fundamental difference in the angular properties of these two types of radiation, we can analyze the role of signal (Cherenkov) and background (scintillation) in a different way.  For convenience, we recall that the instantaneous angular distribution of Cherenkov radiation is given by integrating the master formula \eqref{e:FrankTamm2} over the wavelength $d\lambda$, together with a delta function enforcing the Cherenkov condition:
\begin{align}   \label{e:angdistformulagen}
    \frac{dN}{d\Omega \, dx} &= \alpha_{EM} \,
    \int_{\lambda_{min}}^{\lambda_{max}} \frac{d\lambda}{\lambda^2}
    \left( 1 - \frac{1}{\beta^2 n^2(\lambda)} \right)
     \delta\left( \cos\theta - \frac{1}{\beta n(\lambda)} \right)   \: .
\end{align}

\subsubsection{Angular Distribution Methodology: Grace Fit}
\hspace{\parindent}

\begin{figure}[h!]
\begin{centering}
\includegraphics[width=0.55\textwidth]{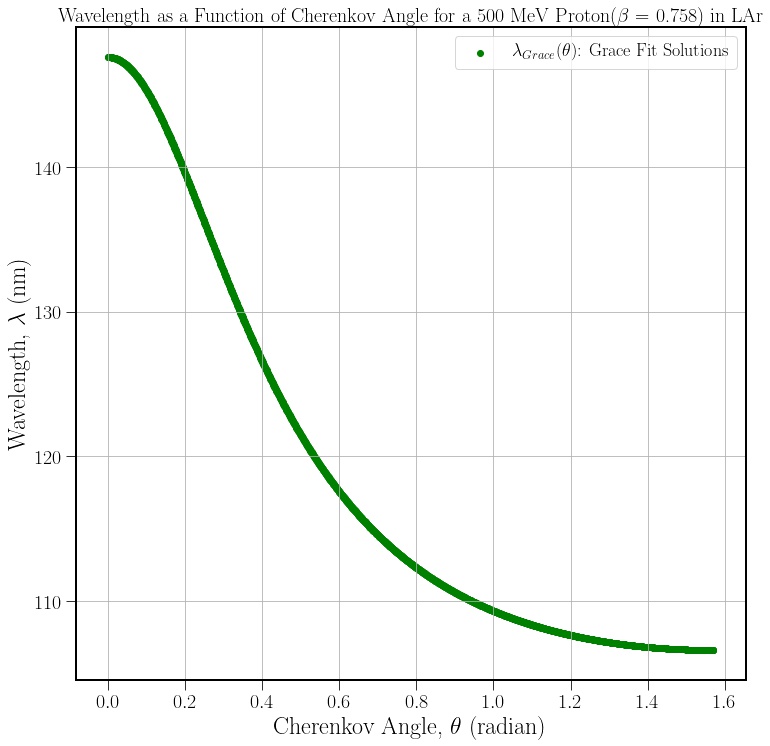}
\caption{Wavelength Solutions Derived from Grace's Refractive Index Fit
\label{f:lambda_Grace}
}
\end{centering}
\end{figure}

As in Sec.~\ref{sec:YieldMethod} for the Cherenkov yield, here we illustrate the methodology for computing the angular distribution first for the case of Grace's refractive index fit \eqref{e:ngracecherenkov1}.  The strategy is to simplify \eqref{e:angdistformulagen} by performing the $d\lambda$ integral using the delta function, which sets $\lambda = \lambda_\theta$ to be the particular wavelength(s) capable of radiating at a given angle $\theta$.  To do so, we will first need to solve for when the delta function is satisfied:
\begin{align}   \label{e:cherenkovcondntheta}
    \cos\theta \equiv \frac{1}{\beta n(\lambda_\theta)} \: .
\end{align}
Taking the square of the Cherenkov condition \eqref{e:cherenkovcondntheta} and substituting the square of the Grace fit \eqref{e:ngracecherenkov1}, we find
\begin{align}   \label{e:cherenkovcondcosthetagrace}
    \cos^2\theta = \frac{(\lambda_{\theta}^2 - \lambda_{UV}^2)} {\beta^2 [a_0 (\lambda_{\theta}^2 - \lambda_{UV}^2) + a_{UV} \lambda_{\theta}^2]}   \: ,
\end{align}
and solving for $\lambda_\theta$ gives the solution
\begin{align}   \label{e:lambdathetagrace}
    \lambda_{\theta} = \sqrt{\frac{
        1 - a_0 \beta^2 \cos^2\theta
    }{
        1 - (a_0 + a_{UV}) \beta^2 \cos^2\theta
    } } \: \lambda_{UV} \: .
\end{align}
The wavelength solutions \eqref{e:lambdathetagrace} derived from the Grace fit are shown in Fig.~\ref{f:lambda_Grace}.  Note first that, since the form of Grace's $n(\lambda)$ decreases monotonically with increasing $\lambda$, the wavelength solution $\lambda_\theta$ for a given angle $\theta$ is unique (see Fig.~\ref{f:n_Grace_vs_Babicz}).  Only a single wavelength is radiated at a given angle at any given moment, and the emitting wavelength is a monotonically decreasing function of the Cherenkov angle $\theta$.  This property is also true of Babicz' fit for $n(\lambda)$, but it is \textit{not} true in general, and we will show explicitly that it fails when the physics of anomalous dispersion is included (see Fig.~\ref{f:absorption_coeff_theoryplot}).  

For a given fit to $n(\lambda)$ such as Grace's fit, imposing the Cherenkov condition \eqref{e:cherenkovcondntheta} allows us to solve for the angular wavelength solutions $\lambda_\theta$.  Rather than immediately substituting the particular form of $n(\lambda)$ and $\lambda_\theta$ for Grace's fit, it is useful to use the properties of the delta function to perform the $d\lambda$ integral for an arbitrary $n(\lambda)$.  This will make it easier to substitute in specific models later on.

The essential properties of the Dirac delta function we need are the change-of-variables formula
\begin{align}   \label{e:deltaseqsimplify}
    \delta\left( \cos\theta - \frac{1}{\beta n(\lambda)} \right) = 
        \frac{1}{\left| \frac{d}{d\lambda} \left(
            \cos\theta - \frac{1}{\beta n(\lambda)} \right) \right|_{\lambda = \lambda_\theta}} \:
    \delta (\lambda - \lambda_{\theta}) 
\end{align}
and the integral formula
\begin{align}   \label{e:deltaseq}
   \int d\lambda \: \delta (\lambda - \lambda_{\theta}) \: f(\lambda)  = f(\lambda_{\theta}) 
\end{align}
for any $f(\lambda)$, as long as $\lambda_\theta$ falls within the integration range.  Together, these properties allow us to perform the $d\lambda$ integral as follows:
\begin{align}   \label{e:angdistdergrace0}
    \frac{dN}{d\Omega \, dx} &= \alpha_{EM} \,
    \frac{1}{\lambda_\theta^2}
    \left( 1 - \frac{1}{\beta^2 n^2(\lambda_\theta)} \right)
    \frac{1}{\left| \frac{d}{d\lambda} (\cos\theta - \frac{1}{\beta n(\lambda)}) \right|_{\lambda = \lambda_\theta}}
    \notag \\ &=
    \alpha_{EM} \,
    \frac{1}{\lambda_\theta^2}
    \left( 1 - \frac{1}{\beta^2 n^2(\lambda_\theta)} \right)
    \frac{{\beta n^2(\lambda_\theta)}}{\left| \frac{dn}{d\lambda}
    \right|_{\lambda= \lambda_{\theta}}}
    \notag \\ &=
    \frac{2 \alpha_{EM}}{\beta^2} \frac{1}{\lambda_\theta^2} \left(\frac{\sin^2 \theta}{\cos^3 \theta}\right)
                    \frac{1}{2 n(\lambda_\theta)\left| \frac{dn}{d\lambda}
                    \right|_{\lambda= \lambda_{\theta}}}    \: ,
\end{align}
where we have routinely substituted $\frac{1}{\beta n(\lambda_\theta)} = \cos\theta$.  Finally, rewriting the derivative in terms of $n^2$ using  $\frac{d}{d\lambda} n^2 = 2 n \, \frac{dn}{d\lambda}$, we obtain the general formula for the instantaneous angular distribution of Cherenkov radiation,
\begin{align}   \label{e:angdistdergrace}
    \frac{dN}{d\Omega \, dx} &=
    \frac{2 \alpha_{EM}}{\beta^2} \frac{1}{\lambda_\theta^2} \left(\frac{\sin^2 \theta}{\cos^3 \theta}\right)
    \frac{1}{\left| \frac{dn^2}{d\lambda} \right|_{\lambda= \lambda_{\theta}}}
    \: ,
\end{align}
which is valid for any fit $n(\lambda)$ to the index of refraction.

\subsubsection{Instantaneous and Integrated Angular Distributions}
                                              
Now, we simply need to evaluate the wavelength solutions $\lambda_\theta$ and the derivative of $n$ with respect to $\lambda$ for a given fit to the refractive index.  For Grace's fit, we already found the wavelength solutions \eqref{e:lambdathetagrace}, and differentiating the Grace fit \eqref{e:ngracecherenkov1} straightforwardly gives
\begin{align}   \label{e:nderivativegrace}
    \frac{dn}{d\lambda} 
    = 
    - \frac{\lambda a_{UV} \lambda_{UV}^2}{\sqrt{(\lambda^2 - \lambda_{UV}^2)^3 [a_0 (\lambda^2 - \lambda_{UV}^2) + a_{UV} \lambda^2]}}  \: .
\end{align}
Substituting this result back into \eqref{e:angdistdergrace} then gives the instantaneous angular distribution
\begin{align}
    \frac{dN}{d\Omega \, dx} = \alpha_{EM} \left(\frac{a_{UV}}{\lambda_{UV}} \right) \frac{\sin^2\theta}{\cos^3\theta} \:
    \frac{\beta^{-2}}{
        \left( a_0 + a_{UV} - \beta^{-2} \cos^{-2}\theta \right)^{1/2}
        \left( a_0 - \beta^{-2} \cos^{-2}\theta \right)^{3/2}   \: .
    }
\end{align}
%
%

Equivalently, this can be written as the instantaneous angular distribution per unit $\cos\theta$ as
\begin{align}   \label{e:angdistfinalgrace}
    \frac{dN}{d\cos\theta \, dx} = 2\pi \: \alpha_{EM} \left(\frac{a_{UV}}{\lambda_{UV}} \right) \frac{\sin^2\theta}{\cos^3\theta} \:
    \frac{\beta^{-2}}{
        \left( a_0 + a_{UV} - \beta^{-2} \cos^{-2}\theta \right)^{1/2}
        \left( a_0 - \beta^{-2} \cos^{-2}\theta \right)^{3/2}   \: ,
    }
\end{align}
and as usual, the integrated angular distribution $\frac{dN}{d\Omega}$ or $\frac{dN}{d\cos\theta}$ is obtained by further integration over the range $x$ of the proton.





%
\begin{figure}[h!] 
\centering
\begin{subfigure} {.49\textwidth}
    \centering
    \includegraphics[width=1\textwidth]{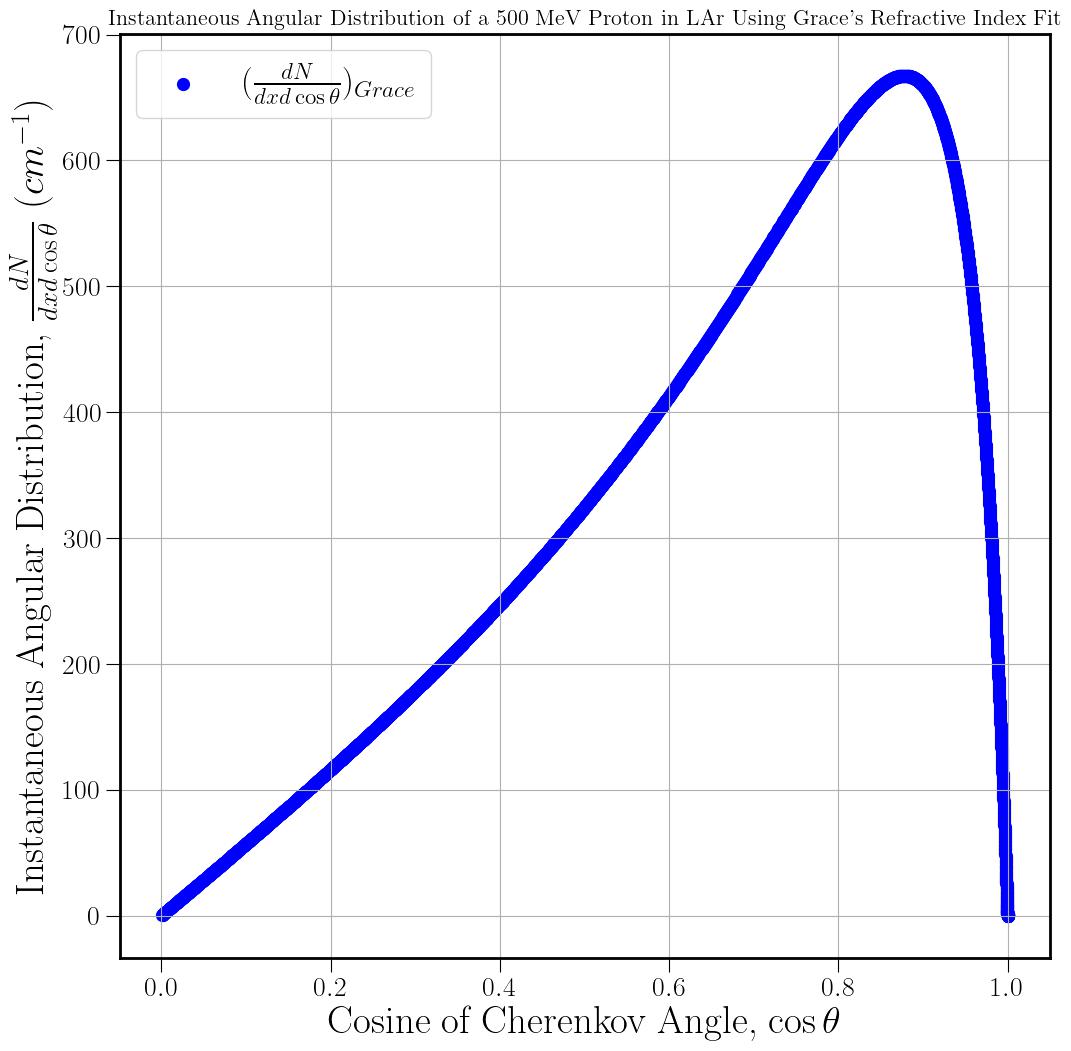}
    \caption{$T =$ 500 MeV
    \label{f:IADGrace500MeV}
    }
\end{subfigure}   
\begin{subfigure} {.49\textwidth}
    \centering
    \includegraphics[width=1\textwidth]{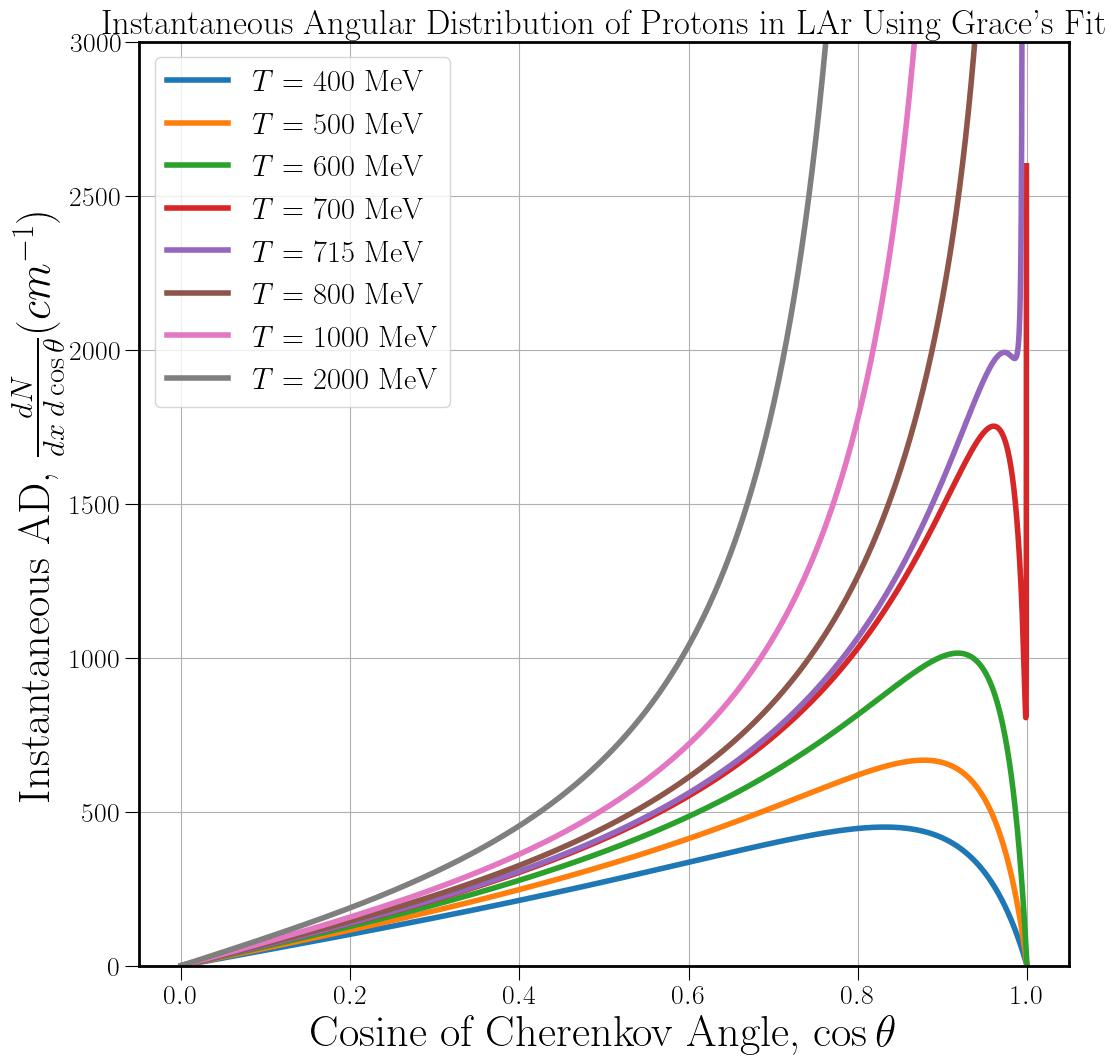}
    \caption{$400 \, \mathrm{MeV} \leq T \leq 2000 \, \mathrm{MeV}$)
    \label{f:IADGraceallts}
    }
\end{subfigure}
\caption{Instantaneous Cherenkov Angular Distribution (IAD) of protons in LAr using Grace's refractive index fit.
\label{f:IADGrace}
}
\end{figure}

The instantaneous angular distribution (IAD) for protons with velocity $\beta=$ 0.758 (i.e $T = 500$ MeV) is plotted in Fig.~\ref{f:IADGrace500MeV}.  First, we note that Grace's fit to $n(\lambda)$ has resulted in significant Cherenkov radiation -- at all angles $0 \leq \theta \leq \tfrac{\pi}{2}$ -- for a 500 MeV proton.  This occurs because the chosen fit $n(\lambda)$ is fully resonant at $\lambda_{UV}$ (see Fig.~\ref{f:n_Grace_vs_Babicz}), so that $n(\lambda)$ can become arbitrarily large near the resonance.  This, in turn, allows the Cherenkov angle $\theta$ to be large, since $\cos\theta = \frac{1}{\beta n}$ can become arbitrarily small.  This results in the paradoxical prediction that even protons with tiny velocities will emit significant large-angle Cherenkov radiation.  This is clearly an unreasonable artifact of the use of a divergent $n(\lambda)$ in the calculation; getting reasonable predictions for the large-angle radiation will require substantially modifying the form of the index fit.  We also note the emergence of a significant peak at small Cherenkov angles near $\cos\theta \approx 0.9$ ($\theta \approx 26^\circ$).  This peak resembles the peak for the emission of Cherenkov light from a single, fixed value of the index $n$, and it appears because deep in the infrared, $n(\lambda) \rightarrow \sqrt{a_0 + a_{UV}} = \mathrm{const}$.  Since the refractive index ``freezes'' at large wavelengths, this results in an accumulation of many IR photons at roughly the same Cherenkov angle, resulting in the sizeable small-angle peak seen in Fig.~\ref{f:IADGrace500MeV}.  In Fig.~\ref{f:IADGraceallts}, the IAD of different kinetic energies from 400 MeV to 2000 MeV is shown. For $T < 700$ MeV, the shape of the IAD is similar to the shape at 500 MeV shown in Fig.~\ref{f:IADGrace500MeV}. But for $T \geq 700$ MeV, the IAD gets distorted in the small-angle region ( $\cos \theta \rightarrow 1$) and sharply increases. For higher energies, the peak of the IAD at small angles continues to increase.

%


\begin{figure} [h!]
\centering
\begin{subfigure} {.49\textwidth}
    \centering
    \includegraphics[width=1\textwidth]{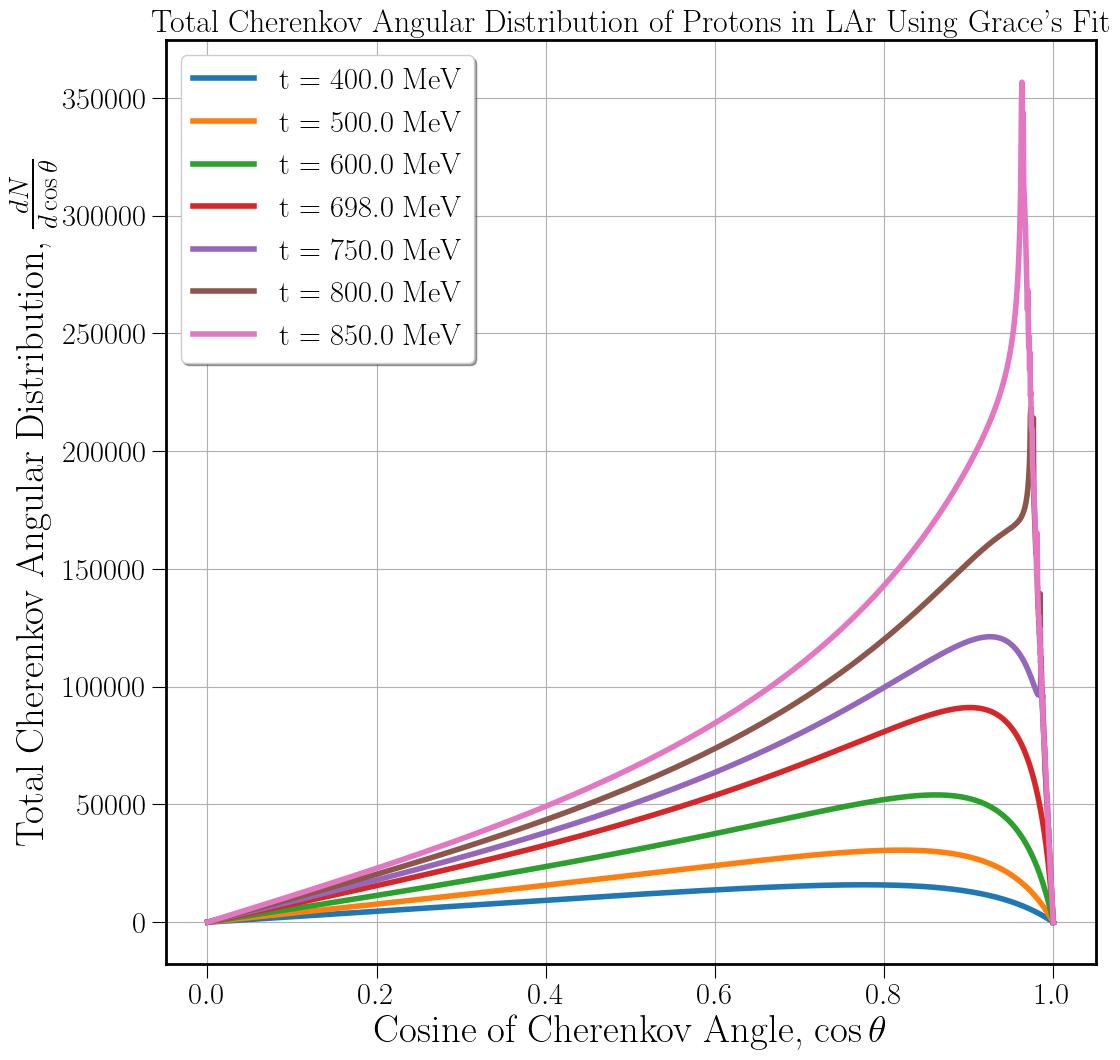}
    \caption{400 to 850 MeV (Normal Plot)
    \label{f:ADGracenormalts}
    }
\end{subfigure}
\begin{subfigure} {.48\textwidth}
    \centering
    \includegraphics[width=1\textwidth]{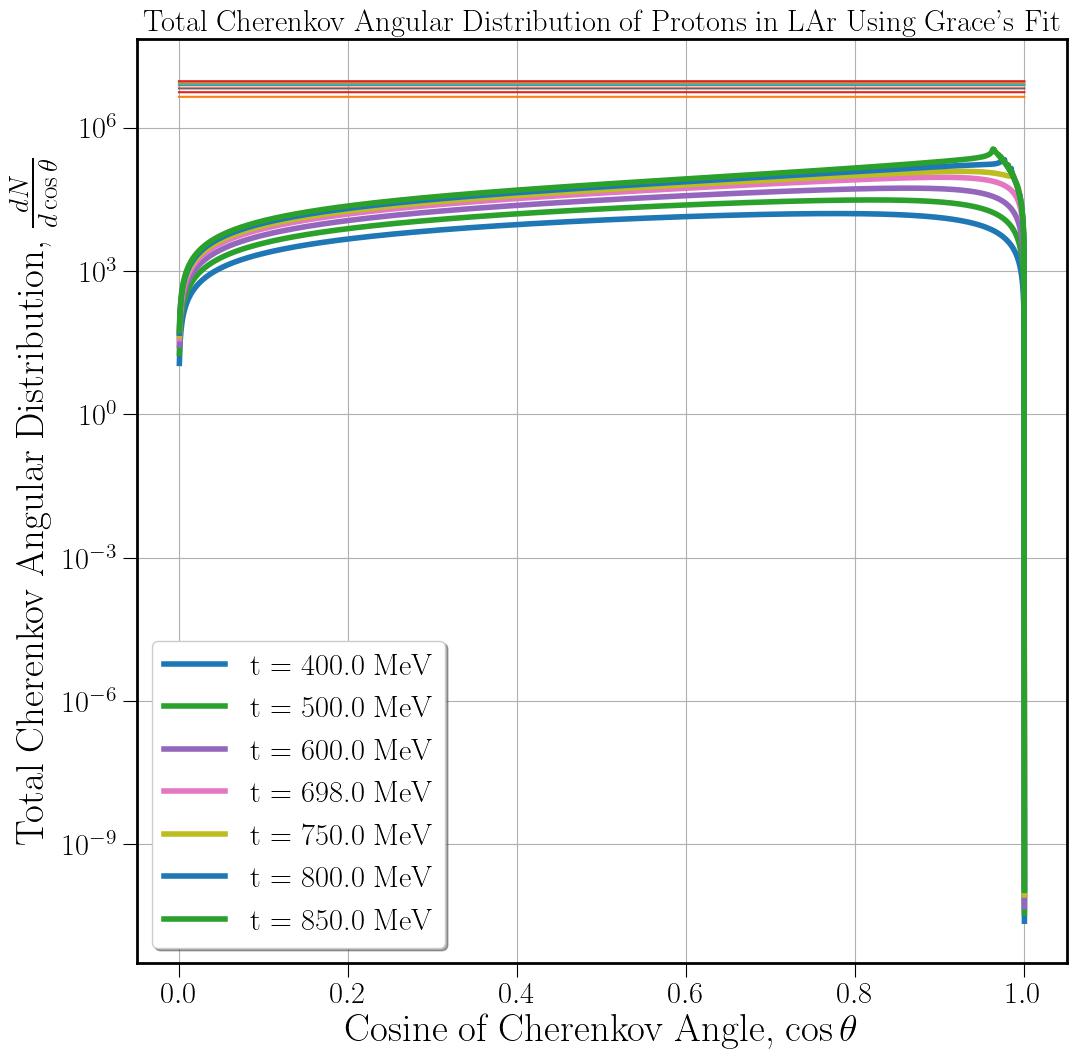}
    \caption{400 to 850 MeV (Log Plot)
    \label{f:ADGracelogts}
    }
\end{subfigure}
\caption{Total Cherenkov AD Derived from Grace's Refractive Index Fit for a Proton with different K.E.s (400 to 850 MeV) travelling in LAr.
\label{f:AD_Gracefit}
}
\end{figure}

The integrated angular distribution $\frac{dN}{d\cos\theta}$, obtained by numerically integrating the instantaneous angular distribution \eqref{e:angdistfinalgrace} over the proton range (see Fig.~\ref{f:KEloss_proton_relvsnonrel}), is plotted in Fig.~\ref{f:AD_Gracefit}.  For the main energies of interest ($T \sim 500 \, \mathrm{MeV}$), the shape of the integrated angular distribution strongly resembles the shape of the initial instantaneous distribution.  The peak location has been shifted to somewhat larger angles (from about $26^\circ$ in the instantaneous distribution to $35^\circ$ in the integrated distribution) due to smearing the initial distribution with the angular profiles of the low-energy protons at the end of their trajectories.  We similarly note a significant accumulation of additional large-angle Cherenkov photons ($\cos\theta \rightarrow 0$) due to integration over the range.  This is another indication that the resonant form of Grace's index fit allows arbitrarily low-energy protons to continue emitting large-angle Cherenkov radiation.

Continuing to higher energies in Fig.~\ref{f:ADGracenormalts}, we see a qualitatively new feature emerge when the kinetic energy exceeds the threshold scale $T_{Grace} = 698$ MeV.  While $n(\lambda)$ has no maximum in Grace's fit, it does have a minimum $n_{min} = \sqrt{a_0 + a_{UV}}$ coming from the asymptote at large wavelengths.  As a result, for sufficiently large velocities $\beta > \beta_{Grace} \approx 0.8$, the Cherenkov condition imposes a finite maximum value on $\cos\theta$,
\begin{align}   \label{e:CherenkovBdy}
    (\cos\theta)_{max} = \min\left(
        \frac{1}{\beta n_{min}} \: , \: 1
    \right) \: .
\end{align}
This gives an instantaneous angular distribution which is highly peaked at forward angles $\cos\theta \rightarrow 1$, but is cut off by the angular boundary \eqref{e:CherenkovBdy} of the Cherenkov condition.  Integrating these very sharp features over the proton range results in cutting off the integrated angular distribution near the axis.  For higher kinetic energies, the peak disappears completely, and the curve takes an upward turn near $\cos\theta = 1$, with the Cherenkov boundary \eqref{e:CherenkovBdy} gradually pushing the angular distribution to larger angles for higher energy protons.

In Fig~\ref{f:ADGracelogts}, we show a logarithmic plot of the total Cherenkov angular distribution energies between 400 and 850 MeV and compare with the corresponding distribution of scintillation photons.  The scintillation photons are distributed isotropically, appearing as a flat line in the angular distribution, and are far above the Cherenkov angular distribution.  This clearly shows the scale difference between these two sources of radiation for our problem, with the scintillation background greatly exceeding the yield of Cherenkov photons.  While the scintillation curves are always above the Cherenkov curves for a given $T$, the gap between them narrows down with increasing $T$. The reason is that the scintillation yield scales linearly with $T$, whereas the Cherenkov yield is anisotropic, with a peak that grows rapidly (non-linearly) as $T$ increases.

As a consistency check between our results for the yield and the angular distribution, we recalculate the total Cherenkov yield by numerically integrating the angular distributions and compare with the previously calculated total yields\footnote{
We note that the total yield depends on our numerical resolution through the step sizes $dx$ and $d\theta$.  Smaller step sizes, especially in $dx$, can significantly affect the total yield by better resolving the time evolution, but at the cost of higher computational time.  We perform a detailed convergence test studying the effects of varying our step sizes in App.~\ref{App:ConvergenceTest}.
}.  Recall that the total yields were directly obtained from integrating the Frank-Tamm formula \eqref{e:ftintegrand} without any reference to the emission angle.  Thus this can serve as a valuable cross-check of the validity of the methods applied to the angular distribution.  The results are summarized in Table~\ref{tab:NGrace}.  We see that there is excellent ($<0.1 \%$) agreement these two methods, confirming that the numerical error is small and remains controlled across the entire energy range of interest.  This serves as a validation of our computed angular distributions.

\begin{table}[t!]
  \begin{center}
    \caption{Comparison (\% difference) of Total Cherenkov yield (N) using Grace's Fit from Two Different Methods: Frank-Tamm Integral (FT) and Angular Distribution (AD).}
    \label{tab:NGrace}
    \begin{tabular}{|c|c|c|c|} 
      \hline
      \textbf{$T$}(MeV) & \textbf{$N(FT)_{\lambda_{min}=106.6nm}$ } & \textbf{$N(AD)_{Grace}$} & \textbf{\% difference} \\
      \hline
      200	& 1434.20 & 1433.48	& 0.05 \\
      \hline
      300	& 4349.27 & 4347.09	& 0.05 \\
      \hline
      500	& 17425.13 & 17416.40 & 0.05 \\
      \hline
      800	& 67322.79 & 67281.96 & 0.06 \\
      \hline
      1000 & 131530.96 & 131372.37 & 0.12 \\
      \hline
      2000 & 643325.31 & 642693.65 & 0.10 \\
      \hline
      3000 & 1270896.95 & 1270062.97 & 0.06 \\
      \hline
      
    \end{tabular}
  \end{center}
\end{table}

\subsubsection{Comparison: Grace vs. Babicz Angular Distributions}

Having demonstrated our methodology for obtaining the instantaneous and integrated angular distributions for Grace's fit to $n(\lambda)$, we now repeat the process for Babicz's fit \eqref{e:BabiczTruncated}.  Substituting the square of the Babicz's fit \eqref{e:BabiczTruncated} into the square of the Cherenkov condition \eqref{e:cherenkovcondntheta} gives
\textbf{\begin{align}   \label{e:cherenkovcondcossqthetaBabicz}
    \cos^2\theta &= 
    \frac{1}{\beta^2} \: 
    \frac{
        (3 - a_0 - a_{UV}) \lambda_{\theta}^2 + (a_0 - 3) \lambda_{UV}^2
        }{
            (3 + 2 a_0 + 2 a_{UV})\lambda_{\theta}^2 - (2 a_0 + 3) \lambda_{UV}^2
        }  \: .
\end{align}}
As before, solving this formula gives us the wavelength solutions $\lambda_\theta$ as a function of the Cherenkov angle $\theta$ for Babicz's fit:
\begin{align}   \label{e:lambdathetaBabicz}
    \lambda_{\theta} = \sqrt{ \frac{
        (a_0 - 3) + (3 + 2 a_0) \, \beta^2 \cos^2\theta 
    }{
        (3 + 2 a_0 + 2 a_{UV}) \, \beta^2 \cos^2\theta - (3 - a_0 - a_{UV})
    }} \: \lambda_{UV}    \: .
\end{align}
%

\begin{figure}[h]
\begin{centering}
\includegraphics[width=0.55\textwidth]{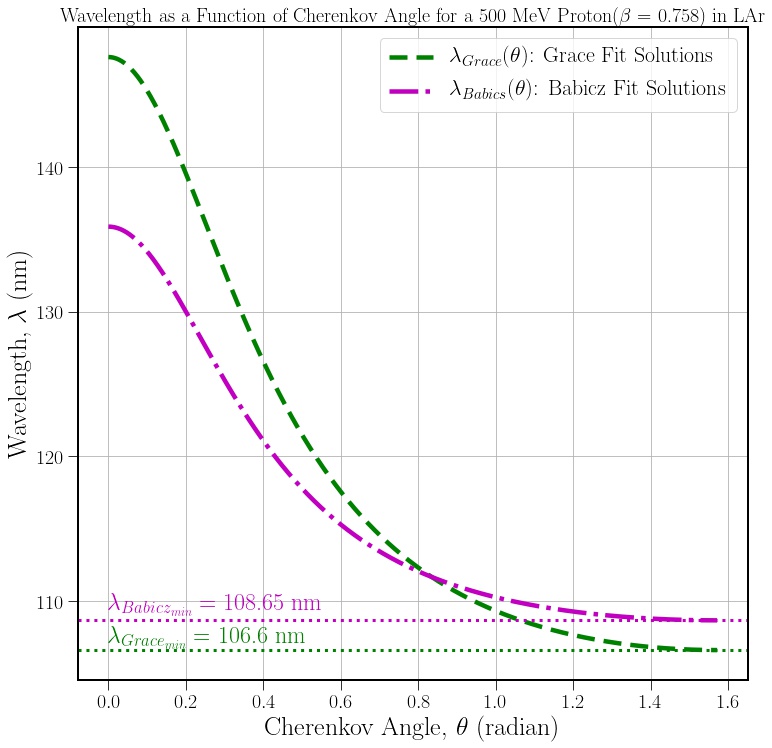}
\caption{Comparison of Wavelength Solutions Derived from Babicz Vs Grace Refractive Index Fit
\label{f:lambdaGracevsBabicz}
}
\end{centering}
\end{figure}

The wavelength solutions for Cherenkov photons emitted at a given angle $\theta$ for both Grace's (green curve) and Babicz's (magenta curve) refractive index fits are plotted in Fig.~\ref{f:lambdaGracevsBabicz}.  We note that, while the qualitative shape of the two curves is the same, there are important quantitative differences.  For small-angle radiation ($\theta \rightarrow 0$, or $\cos\theta \rightarrow 1$), Babicz' fit radiates at a shorter wavelength than Grace's fit.  Equivalently, the large-wavelength tail of Babicz' fit is downshifted to smaller angles compared to Grace's.  The two curves cross, however, at around $0.83$ radians ($48^\circ$), with Babicz' fit emitting at much shorter wavelengths than Grace's.  This is a consequence of the different singularities $\lambda_{Grace} = 106.6 \, \mathrm{nm}$ vs. $\lambda_{Babicz} = 108.7 \, \mathrm{nm}$ of the two index fits, which are reflected in the indicated horizontal asymptotes in Fig.~\ref{f:lambdaGracevsBabicz}.  As noted previously in Fig.~\ref{f:cerenkovtotalcomparisongracevsBabicz}, this difference in the UV cutoff produces significantly lower total yield for Babicz' fit than for Grace's.  This is also the case for the angular distribution, as we discuss in the next section. 

To compute the angular distribution for Babicz' fit, we simply evaluate the derivative of $n$ with respect to $\lambda$ for the Babicz fit function,   
\begin{align}   \label{e:nderivativeBabicz}
    \frac{dn}{d\lambda} &= \frac{1}{
        \left(1 + \frac{3 \left(a_{0} + \frac {a_{UV}\lambda^{2}} {\lambda^{2} - \lambda_{UV}^{2}}\right)}{\left(3 - a_0 - \frac {a_{UV}\lambda^{2}} {\lambda^{2} - \lambda_{UV}^{2}}\right)}\right)
    } 
    \frac{d}{d\lambda} \left( \frac{3 \left(a_{0} + \frac {a_{UV}\lambda^{2}} {\lambda^{2} - \lambda_{UV}^{2}}\right)}{3 - a_0 - \frac {a_{UV}\lambda^{2}} {\lambda^{2} - \lambda_{UV}^{2}}}\right) 
    \notag \\
    %
    &= \frac{- 9 a_{UV} \lambda \lambda_{UV}^2}{
        \sqrt{[(3 + 2 a_0 + 2 a_{UV}) \lambda^2 - (2 a_0 + 3) \lambda_{UV}^2] [(3 - a_0 - a_{UV}) \lambda^2 + (a_0 - 3) \lambda_{UV}^2]^3
        } }
\end{align}
and substitute the derivative \eqref{e:nderivativeBabicz} and the wavelength solutions \eqref{e:lambdathetaBabicz} into the angular distribution formula \eqref{e:angdistdergrace} or \eqref{e:angdistdergrace0}.  The explicit result can be written
\begin{align} \label{e:adpolarBabicz}
    \frac{dN}{d\cos \theta \, dx} &= 
    \frac{2 \pi\alpha_{EM}}{9 \beta a_{UV} \, \lambda_\theta^3 \lambda_{UV}^2} \,
    \sqrt{(3 + 2 a_0 + 2 a_{UV}) \lambda_\theta^2 - (2 a_0 + 3) \lambda_{UV}^2}
    \notag \\ &\times 
    \sqrt{(3 - a_0 - a_{UV}) \lambda_\theta^2 + (a_0 - 3) \lambda_{UV}^2}    
    \bigg\{\beta^2 \left[{(3 + 2 a_0 + 2 a_{UV}) \lambda_\theta^2 - (2 a_0 + 3) \lambda_{UV}^2}\right]
    \notag \\ & \hspace{1cm}
    - [(3 - a_0 - a_{UV}) \lambda_\theta^2 + (a_0 - 3) \lambda_{UV}^2] \bigg\}
\end{align} 
where the Sellmeier coefficients for Babicz' refractive index fit are given in Tab.~\ref{tab:SellmeierBabicz}.

\begin{figure}[h!] 
\begin{centering}
\includegraphics[width=0.7\textwidth]{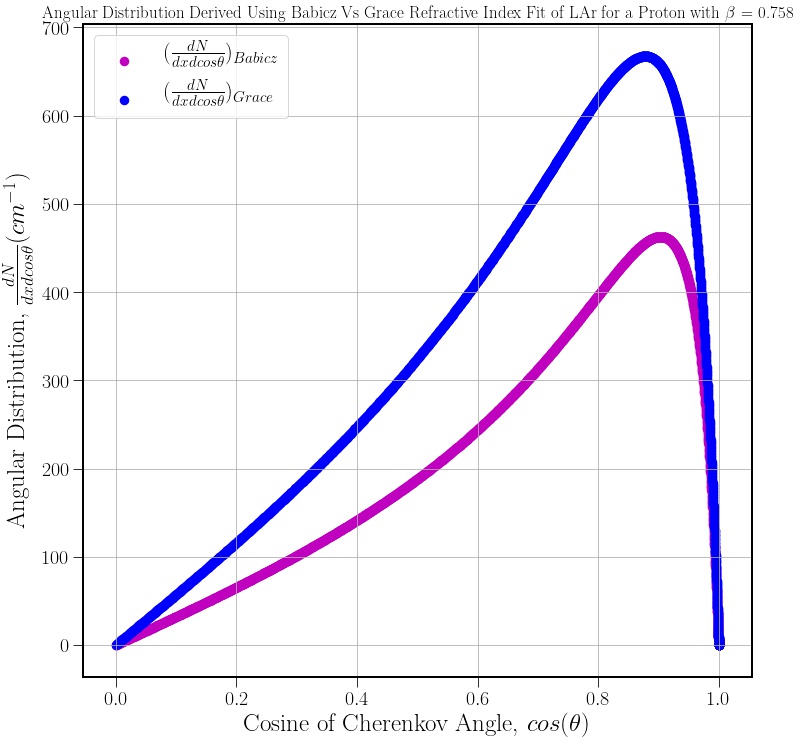}
\caption{Cherenkov AD Derived from Babicz's vs Grace's Refractive Index Fit for a Proton with constant $\beta = 0.758$ travelling in LAr.
\label{f:ADBabiczvsGracebp758}
}
\end{centering}
\end{figure}

The instantaneous angular distribution \eqref{e:adpolarBabicz} for Babicz' fit (magenta) is plotted in Fig.~\ref{f:ADBabiczvsGracebp758} and compared against the angular distribution for Grace's fit.  We see that Babicz' fit produces an angular profile with $\sim 30 - 40\%$ fewer photons but similar in shape compared to Grace's fit.  This result is expected due to the different cutoffs $\lambda_{min} = $108.657 nm for Babicz vs. $\lambda_{min} = $106.6 nm for Grace noted previously.  We also note that the two angular distributions mostly converge at small angles (arising from the long-wavelength IR regime) but differ significantly at large angles due to the different treatment of the resonance in both fits.

\begin{figure} [h!]
\centering
\begin{subfigure} {.50\textwidth}
\centering
\includegraphics[width=1\textwidth]{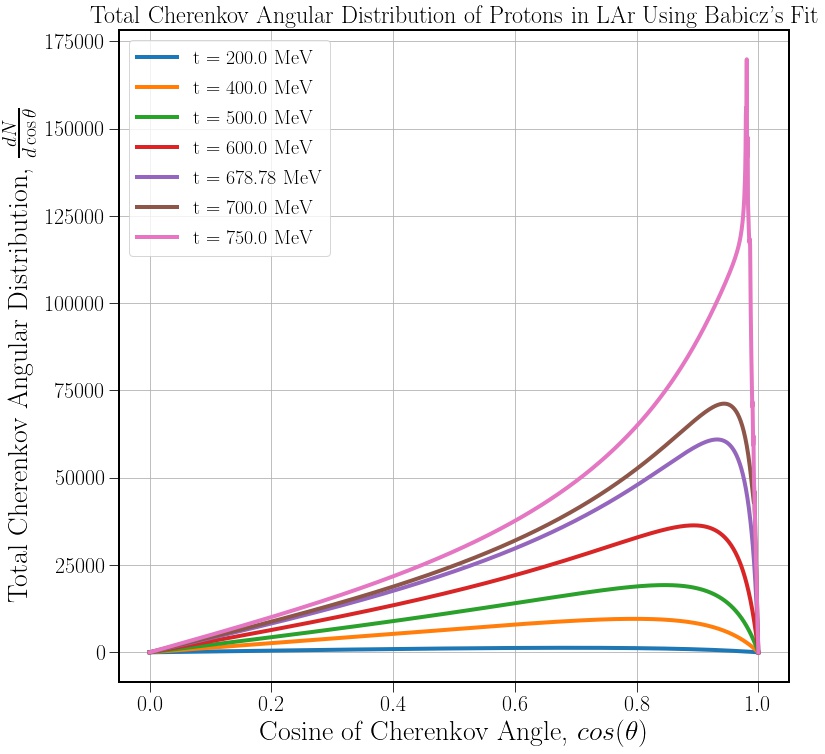}
\caption{200 to 750 MeV 
\label{f:ADBabicz200to750MeV}
}
\end{subfigure}
\begin{subfigure} {.47\textwidth}
\centering
\includegraphics[width=1\textwidth]{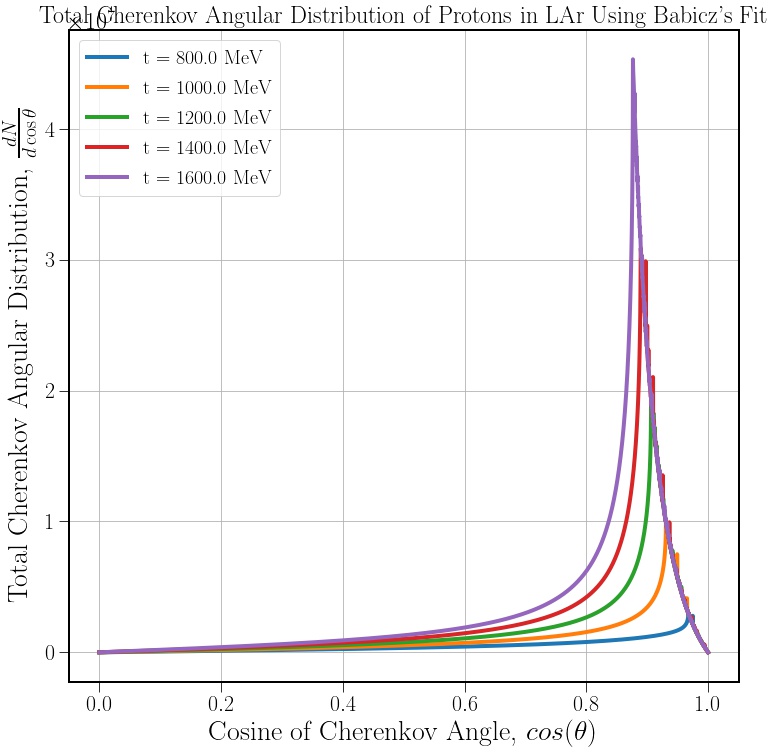}
\caption{800 to 1600 MeV
\label{f:AD_Babicz_800to1600MeV}
}
\end{subfigure}
\caption{Cherenkov AD Derived from Babicz's Refractive Index Fit for a Proton with different K.E.s traveling in LAr.
\label{f:AD_Babiczfit}
}
\end{figure}

Similarly, the integrated angular distributions produced by Babicz' fit for different kinetic energies are shown in Fig.~\ref{f:AD_Babiczfit}.  For low kinetic energies $T < T_{Babicz} = 679 \, \mathrm{MeV}$, the shape of the angular profile strongly resembles the initial instantaneous angular distribution (Fig.~\ref{f:ADBabiczvsGracebp758}), with just a smearing of the overall shape.  This pattern is consistent with what was seen in Fig.~\ref{f:ADGracenormalts} for Grace's fit at low energies $T < T_{Grace}$.  Also being consistent with the results from Grace's fit for $T > T_{Grace}$, we note in Fig.~\ref{f:AD_Babicz_800to1600MeV} the appearance of a new qualitative feature at high energies $T > T_{Babicz}$.  Above this characteristic scale, there is a nonzero minimum  emission angle (maximum $\cos\theta$) for the Cherenkov photons, which grows with increasing energy.  This effect pushes the peak of the integrated angular distribution toward higher angles and sharply cuts off the Cherenkov spectrum at small angles.  Because of the very similar shapes of Grace's and Babicz' fits, the two angular distributions have the same qualitative features.

\begin{table}[ht!]
  \begin{center}
    \caption{Comparison (\% error) of Total Cherenkov yield (N) using Babicz's Fit from Two Different Methods: Frank-Tamm Integral (FT) and Angular Distribution (AD).}
    \label{tab:NBabicz}
    \begin{tabular}{|c|c|c|c|} 
      \hline
      \textbf{$T$}(MeV) & \textbf{$N(FT)_{\lambda_{min}=108.65nm}$} & \textbf{$N(AD)_{Babicz}$} & \textbf{\% error} \\
      \hline
      200	 &	822.75	&  822.27 & 0.06 \\
      \hline
      300  &  2529.18	 & 2527.98 & 0.05 \\
      \hline
      500	 &  10520.88 & 10517.59 & 0.03 \\
      \hline
      800	 &  47460.83 & 47541.72 & 0.17 \\
      \hline
      1000 & 102686.12 & 102871.27 & 0.18 \\
      \hline
      2000 & 568046.46 & 568018.31 & 0.01 \\
      \hline
      3000 & 1149944.19 & 1147871.67 & 0.18 \\
      \hline
    \end{tabular}
  \end{center}
\end{table}

As with Grace's fit, we can cross-check the angular distributions by benchmarking them against the total Cherenkov yield found directly from the Frank-Tamm formula.  The total yield $N(AD)$ computed by numerically integrating the angular distribution is compared with the known yield $N(FT)$ computed from the Frank-Tamm method in Tab.~\ref{tab:NBabicz}.  We again see that there is an excellent agreement ($<1 \%$)  these two methods, which demonstrates the consistency of the methods applied in this study.





\begin{figure}[p!] 
\centering
\begin{subfigure}{.45\textwidth}
\includegraphics[width=1\textwidth]{images/total_cerenkov_photon_vs_scintback_Grace_hight_log.jpg}
\caption{Grace FT Log Plot  
\label{f:GraceFTvsscintbacklog}
}
\end{subfigure} 
\centering
\begin{subfigure}{.50\textwidth}
\centering
\includegraphics[width=1\linewidth]{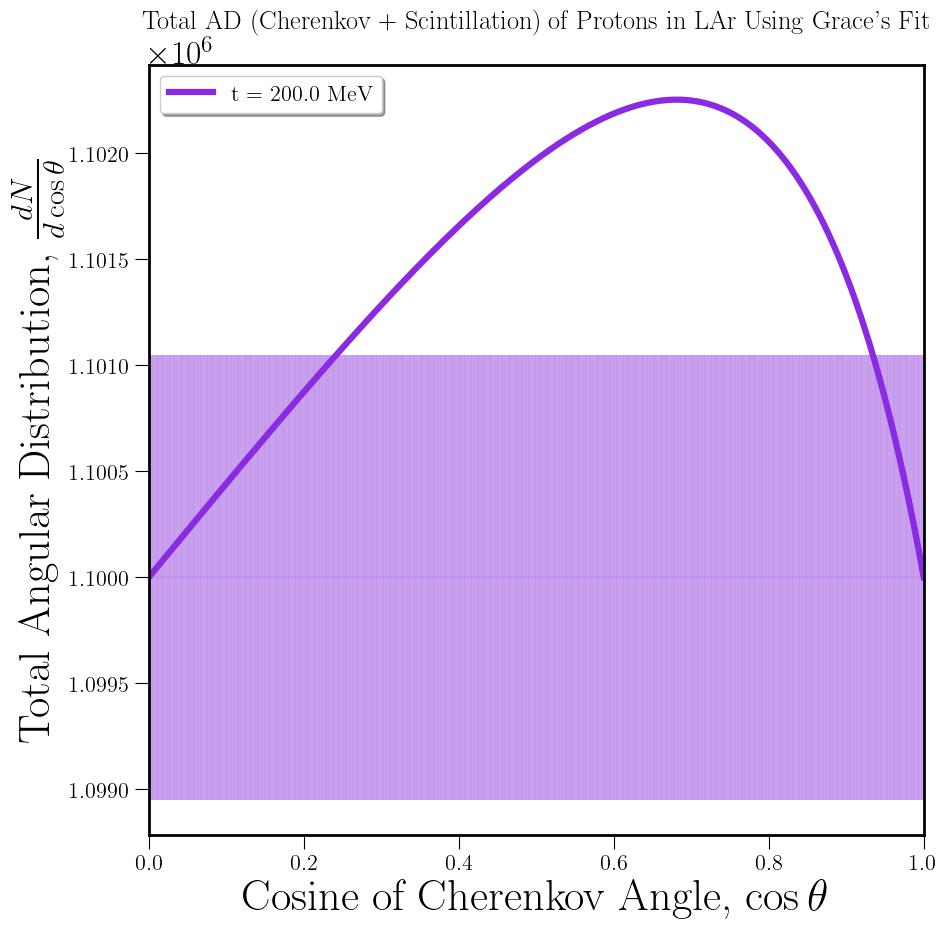}
\caption{Grace AD Normal Plot
\label{f:GraceADvsscintback_normal}
}
\end{subfigure}

\medskip

\begin{subfigure}{.45\textwidth}
\includegraphics[width=1\textwidth]{images/total_cerenkov_photon_vs_scintback_Babicz_hight_log.jpg}
\caption{Babicz FT Log Plot  
\label{f:BabiczFTvsscintbacklog}
}
\end{subfigure} 
\centering
\begin{subfigure}{.50\textwidth}
\centering
\includegraphics[width=1\linewidth]{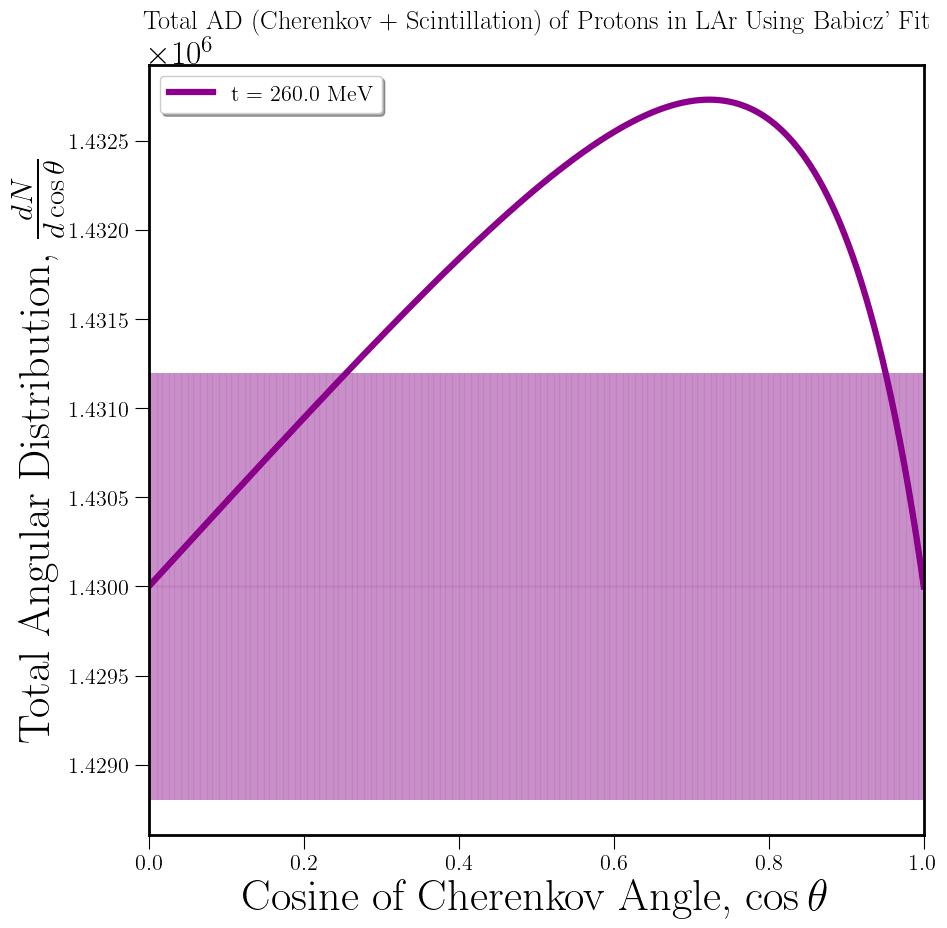}
\caption{Babicz AD Normal Plot
\label{f:BabiczADvsscintbacknormal}
}
\end{subfigure}

\caption{Comparison of Cherenkov yield and angular distribution with scintillation background.  
\label{f:GraceBabiczScint}
}

\end{figure}

\subsubsection{Comparison: Cherenkov vs. Scintillation Angular Distributions}

As with the total Cherenkov yields we computed for the Grace and Babicz fits, we can now compare the angular distributions of Cherenkov photons with the expected background from scintillation photons.  The distribution of Cherenkov photons (signal) is to be compared with the uncertainty $\delta N_{scint} = \sqrt{N_{scint}}$ from \eqref{e:ScintError1} in the number of scintillation photons (background).  That comparison, both for the integrated yield $N$ and for the integrated angular distribution $dN/d\cos\theta$, is shown in Fig.~\ref{f:GraceBabiczScint}.

The left panels of Fig.~\ref{f:GraceBabiczScint}, showing the integrated yields for the Grace and Babicz fits, are the same as the plots shown in Fig.~\ref{f:CerenkovGracevsScintillationback}.  Both the Cherenkov yield $N$ and the background uncertainty $\delta N_{scint}$ increase with the proton energy, but the Cherenkov yield grows faster.  The intersection between the Cherenkov excess and the uncertainty $\delta N_{scint}$ in the total yield is marked by a vertical line; this occurs at $T = 200 \, \mathrm{MeV}$ for Grace's fit and $T = 260 \, \mathrm{MeV}$ for Babicz' fit.  In both cases, however, at sufficient energy for the integrated Cherenkov yield to just begin to compete with the background, the angular distributions are already much more statistically significant at the same energy.  This is clear from the right panels of Fig.~\ref{f:GraceBabiczScint}, which show the integrated angular distributions for Grace's and Babicz' fits at the same energies marked on the integrated yields.  In both cases, a sizeable peak develops at $\cos\theta \approx 0.7$ ($\theta \approx 45^\circ$) with an excess above background spanning $0.4 \leq \cos\theta \leq 0.9$  ($66^\circ \geq \theta \geq 26^\circ$).





\begin{figure}[p]
\centering
\begin{subfigure}{0.35\columnwidth}
\centering
\includegraphics[width=\textwidth]{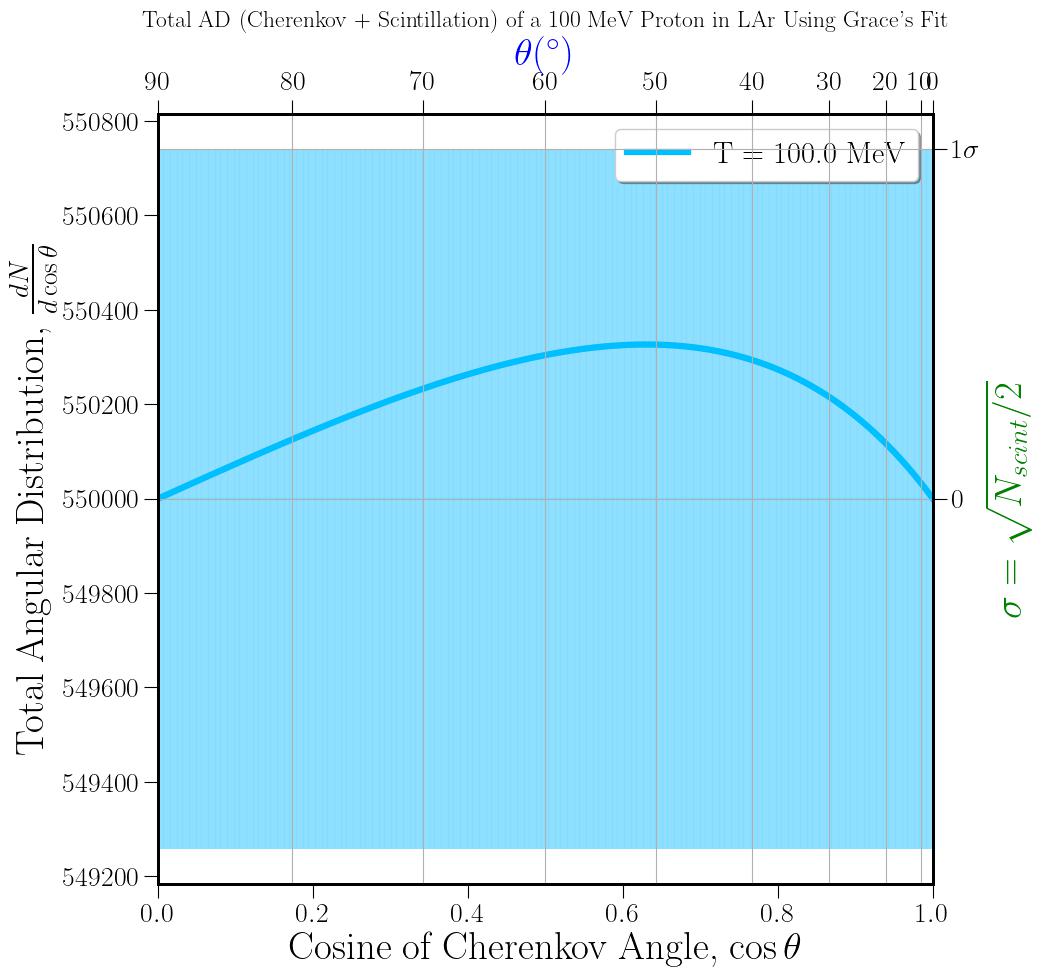}
\caption{100 MeV Proton}
\label{fig:Grace100 MeV}
\end{subfigure}\hfill
\begin{subfigure}{0.35\columnwidth}
\centering
\includegraphics[width=\textwidth]{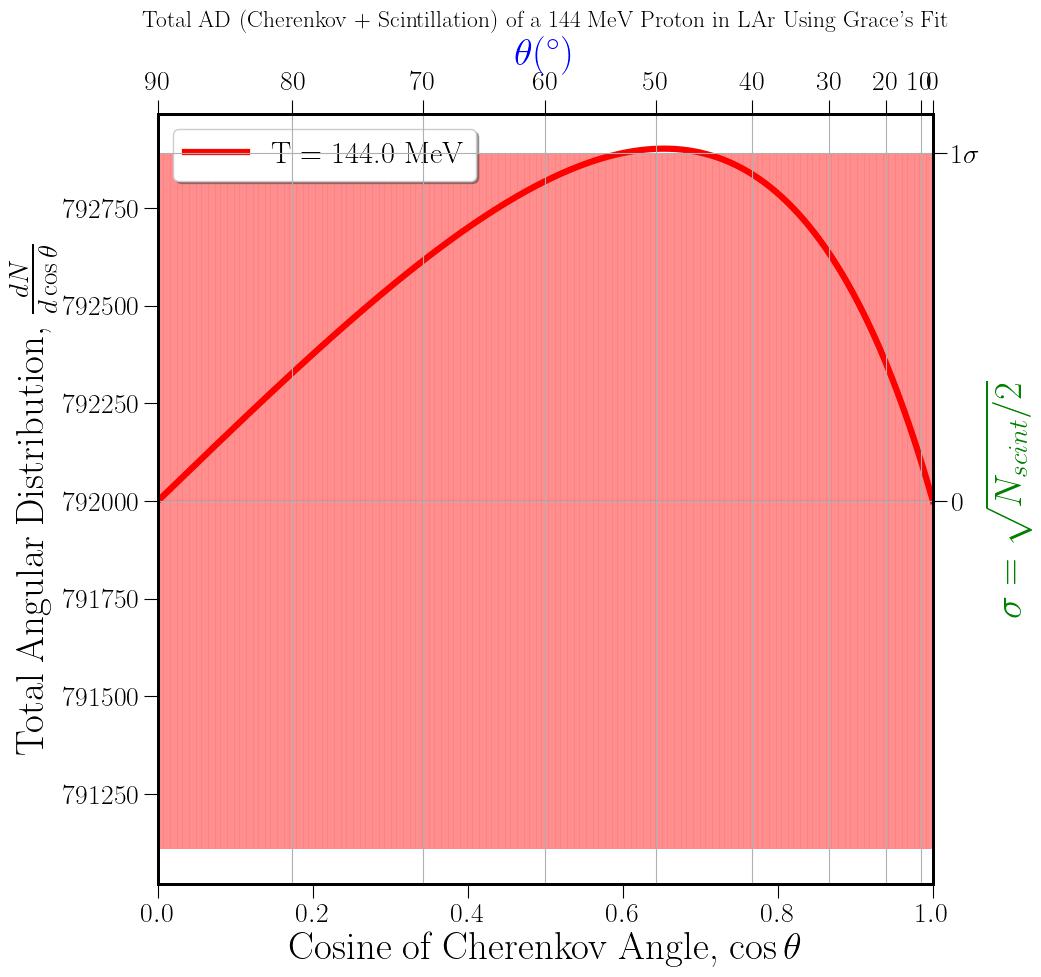}
\caption{144 MeV Proton}
\label{fig:Grace144MeV}
\end{subfigure}

\medskip

\begin{subfigure}{0.35\columnwidth}
\centering
\includegraphics[width=\textwidth]{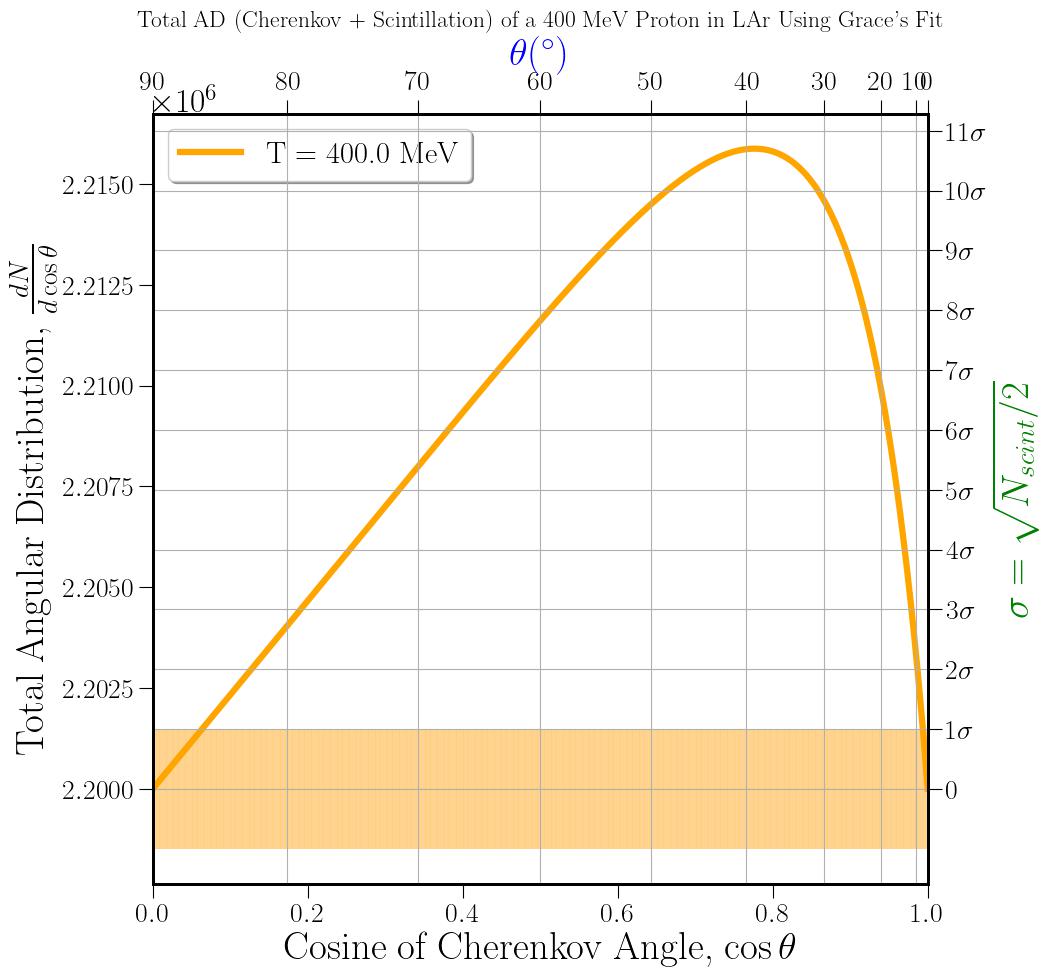}
\caption{400 MeV Proton}
\label{fig:Grace400MeV}
\end{subfigure}\hfill
\begin{subfigure}{0.35\columnwidth}
\centering
\includegraphics[width=\textwidth]{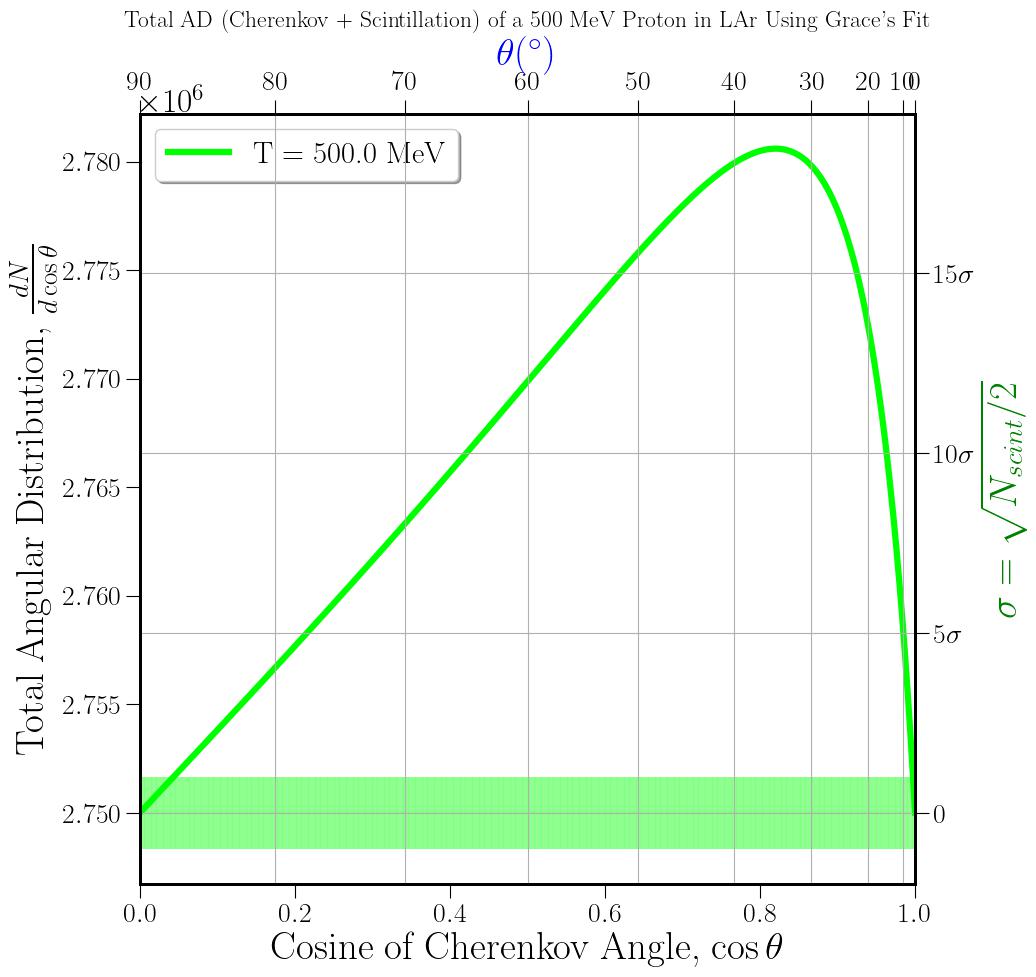}
\caption{500 MeV Proton}
\label{fig:Grace500MeV}
\end{subfigure}

\medskip

\begin{subfigure}{0.35\columnwidth}
\centering
\includegraphics[width=\textwidth]{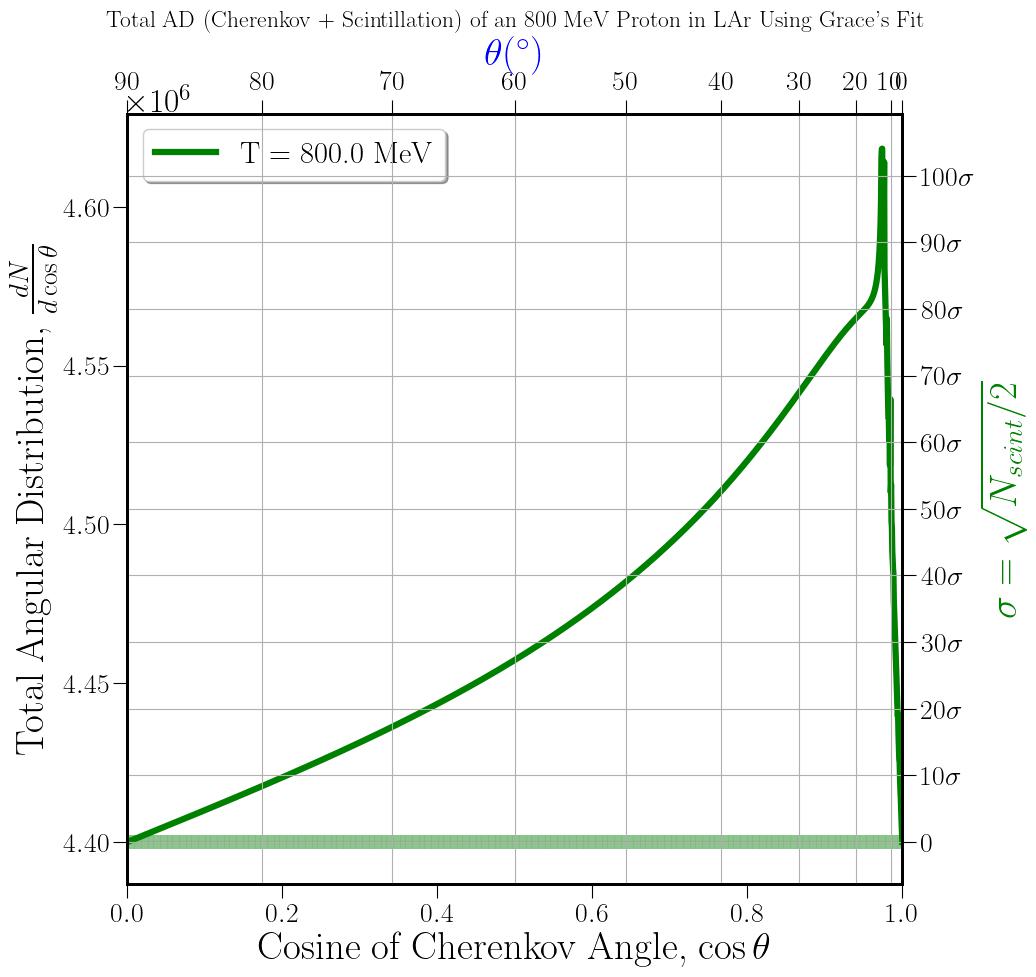}
\caption{800 MeV Proton}
\label{fig:Grace800MeV}
\end{subfigure}\hfill
\begin{subfigure}{0.35\columnwidth}
\centering
\includegraphics[width=\textwidth]{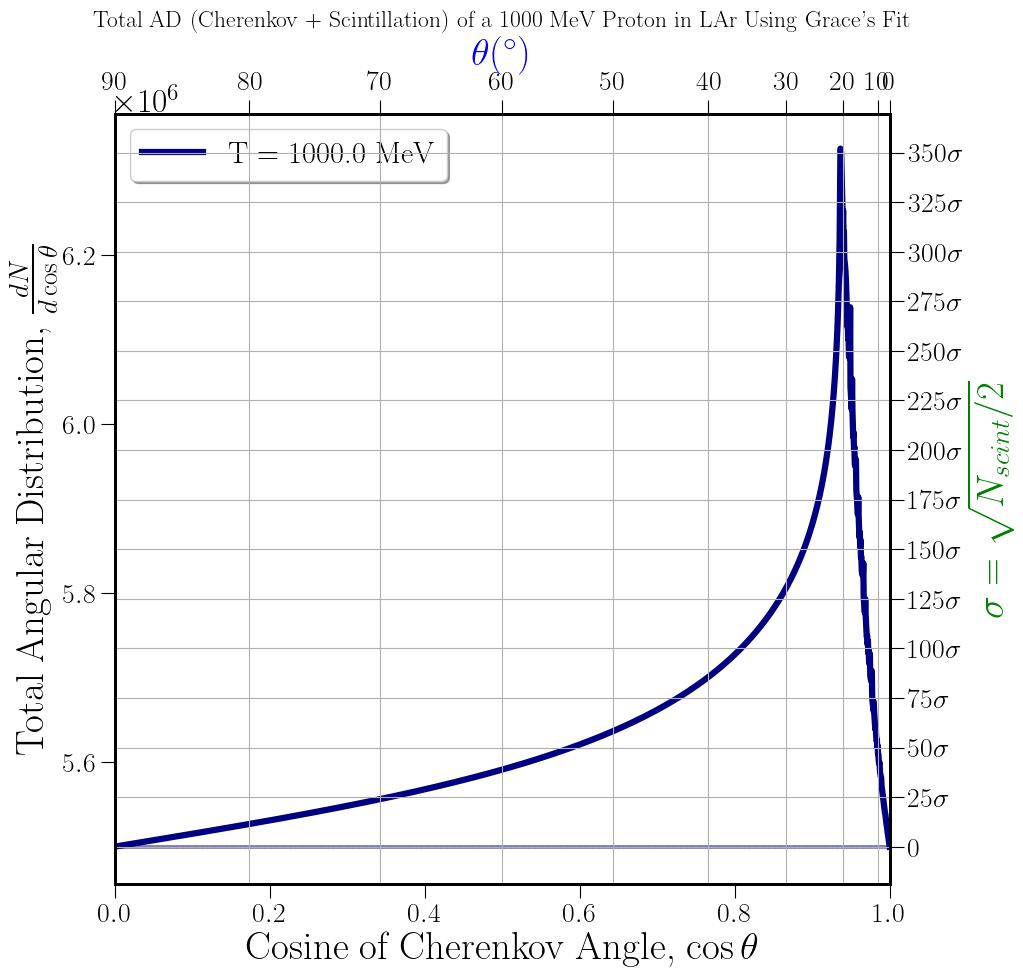}
\caption{1000 MeV Proton}
\label{fig:Grace1000MeV}
\end{subfigure}
\caption{Comparison of Cherenkov Angular Distribution with Scintillation Background Using Grace's Fit}
\label{fig:cherenkovoverscintbackGrace}
\end{figure}

With increasing energy, the statistical significance of the angular distribution only grows.  The same comparison of the integrated Cherenkov angular distribution with the background $\delta N_{scint}$ is shown in Fig.~\ref{fig:cherenkovoverscintbackGrace} for a variety of kinetic energies.  Only Grace's fit is shown for simplicity; Babicz' fit is similar but with a lower magnitude.  As we saw from Fig.~\ref{f:GraceBabiczScint}, the integrated yield for Grace's fit just crosses the background at $T = 200 \, \mathrm{MeV}$.  The angular distribution, on the other hand, begins to cross the scintillation background at a lower energy of $167 \, \mathrm{MeV}$, well before the integrated yield becomes statistically significant.  The significance increases quickly with energy, growing to \textit{greater than $10\sigma$} already by $T = 500 \, \mathrm{MeV}$.  We also note the appearance of the spiky feature at small angles for $T > 700 \, \mathrm{MeV}$ as noted previously.

The picture of the rates of Cherenkov emission of low-energy protons in LAr painted by Grace and Babicz fits is very optimistic.  Both fit predict that the total Cherenkov yield grows rapidly, exceeding the scintillation background already by $\sim 250 \, \mathrm{MeV}$ protons.  For $500 \, \mathrm{MeV}$ protons, the yield is statistically significant on the order of tens of $\sigma$'s. For the angular distributions, the significance is even better, since the highly collimated Cherenkov radiation is concentrated in certain angular ranges.

Clearly, this rosy picture is overly optimistic.  As we have pointed out, both the Grace and Babicz fits approximate the form of $n(\lambda)$ near the UV resonance with an unshielded singularity.  This result, which follows if $n(\lambda)$ were to be purely real, crucially neglects the physics of absorption.  Rather than diverging to infinity at a finite wavelength (see Fig.~\ref{f:n_Grace_vs_Babicz}), absorptive effects arising from the imaginary part of $n(\lambda)$ must stop its growth at a \textit{finite maximum value} (see Fig.~\ref{f:absorption_coeff_theoryplot}).  This shielding of the UV divergence due to absorptive effects will radically change the Cherenkov emission, both in total yield and in angular distribution.  In particular, we expect to see major modifications of the UV region of the spectrum, which for the Grace and Babicz fits resulted in large-angle Cherenkov radiation even for very low-energy protons.  The modification of the Cherenkov yield due to absorption is studied in detail next, in Chap.~\ref{cerenkovwabs}.









\hspace{\parindent}
\pagebreak




\newpage
\newpage

\pagebreak
\newpage
\section{CHERENKOV RADIATION WITH ABSORPTION}\label{cerenkovwabs}

\hspace{\parindent}

In this Chapter, we will focus on applying the same methodology used in Chap.~\ref{mathform} 
to a new fit to the refractive index $n(\lambda)$ which includes absorptive effects.  We perform this new fit of the available experimental data ourselves, similar to the procedure of Grace and Babicz, but using a different functional form for the fit.  We use the shape of $n(\lambda)$ predicted from first principles, including the vital anomalous behavior near the resonance where the imaginary part of $n(\lambda)$ grows (reflecting absorption) and the real part of $n$ decreases instead of diverging  (see Fig.~\ref{f:absorption_coeff_theoryplot}).  Thus, unlike the existing fits of Grace and Babicz, our fit to $n(\lambda)$ rises to a \textit{finite maximum} $n_{peak} = 1.36$, rather than diverging at the resonance (see Fig.~\ref{f:absorption_coeff_theoryplot}).  As anticipated in Chap.~\ref{mathform}, regulating the singular behavior at the resonance due to absorptive effects leads to a huge impact on the emission of Cherenkov light, both in terms of the cut in the number of Cherenkov photons and the modification of their angular distribution. We thus argue that the inclusion of absorptive effects in our fit constitutes a major improvement over the previously-available fits on the market.








\subsection{Absorptive Fits of the Refractive Index}
\hspace{\parindent}

The relationship between the refractive index $n(\lambda)$ and the absorption coefficient $\alpha(\lambda)$ arises from the fact that they are the real and imaginary parts, respectively, of the same analytic function.  Both functions can be computed explicitly, together, for a given model of the medium.

\subsubsection{Harmonic Oscillator (HO) Model Fit}

We use the standard textbook treatment (\cite{Jackson:1998nia}, \cite{griffiths_2017}) which models the atoms as simple harmonic oscillators, as summarized in Appendix~\ref{sec:HarmonicOscillator}.  Applying this model to LAr, the frequency-dependent index of refraction is derived to have the functional form
\begin{align}   \label{e:nfinalho1}
     n = 1 + \frac{Nq^2}{2m_e \epsilon_0} \sum_{j}\frac{f_j (\omega_{j}^2 - \omega^2)}{(\omega_{j}^2 - \omega^2)^2 + \gamma_j^2 \omega^2} \: , 
\end{align}
where $N$ is the number of electrons per unit volume, $q$ and $m_e$ are the charge and mass of the electron, respectively, and $\epsilon_0$ is the permittivity of free space.  In the general case \eqref{e:nfinalho1} there is a summation over many possible resonances $j$ with respective strengths $f_j$.  From the derivation, Eq.~\eqref{e:nfinalho1} is expressed in terms of the angular frequency $\omega = 2\pi c_{LAr} / \lambda$ and the corresponding resonance position  $\omega_{UV} = 2\pi c_{LAr} / \lambda_{UV}$.  See the full derivation of Eq.~\eqref{e:nfinal} in  Appendix~\ref{sec:HarmonicOscillator}.   






To employ the theoretical form \eqref{e:nfinalho1} for LAr, we consider only a single resonance $j=1$, allow for flexible coefficients (including an overall constant shift mimicking the presence of distant resonances), and change variables to the wavelengths $\lambda$ rather than frequencies $\omega$.  This gives the parameterization we use for our new absorptive fit: 
\begin{align}   \label{e:nlambdauvhofinal}
     n_{HO} = a_{0(HO)} + a_{UV(HO)} \left(\frac{\lambda_{UV}^{-2} - \lambda^{-2}}{(\lambda_{UV}^{-2} - \lambda^{-2})^2 + \gamma_{UV}^2 \lambda^{-2}}\right) \: , 
\end{align}
with the corresponding absorption coefficient given by 
\begin{align}   \label{e:alphalambdauv}
    \alpha \cong \frac{Nq^2}{m \epsilon_0 c} \left( \frac{f_{UV} \gamma_{UV} \lambda^2 \lambda_{UV}^4}{(2 \pi c)^2 (\lambda^2 - \lambda_{UV}^2)^2 + \gamma_{UV}^2 \lambda^2 \lambda_{UV}^4} \right). 
\end{align}
where the best-fit parameters for $a_{0(HO)}$, $a_{UV(HO)}$, and $\gamma_{UV}$ are listed in Table~\ref{tab:bestfitho}.
%
\begin{table}[h!]
  \begin{center}
    \caption{Best Fit Parameters of Our Refractive Index Fit} 
    \label{tab:bestfitho}
    \begin{tabular}{|c|c|c|} 
      \hline
      \textbf{$a_{0(HO)}$} & \textbf{$a_{UV(HO)}$} ($nm^{-2}$) & \textbf{$\gamma_{UV} (nm^{-1}$)}\\
      \hline
      1.10232 & 0.00001058 & 0.002524\\  
      \hline
    \end{tabular}
  \end{center}
\end{table}
%
%
\begin{figure}[h!]
\centering
\begin{subfigure}{.48\textwidth}
\centering
\includegraphics[width=1\textwidth]{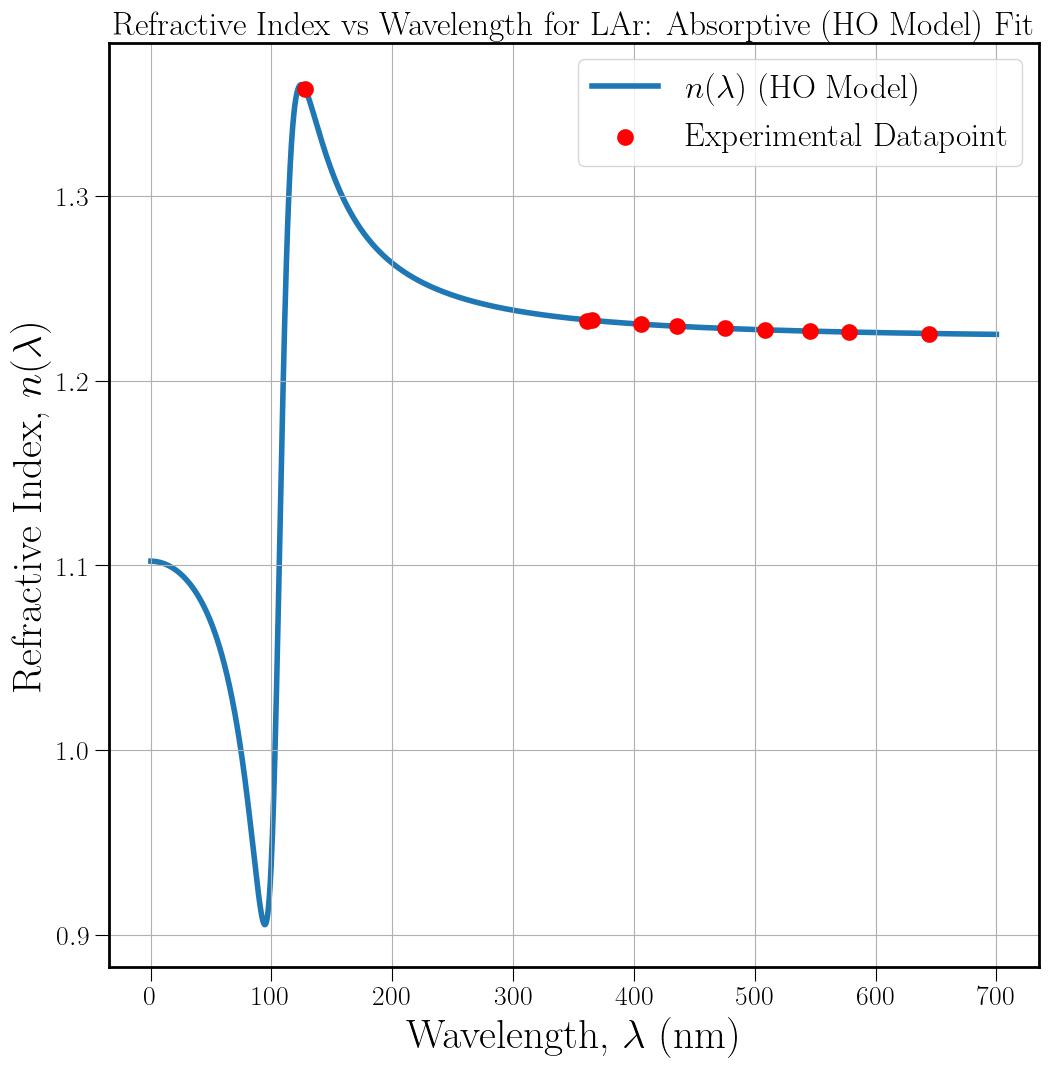}
\caption{HO Model Refractive Fit
\label{f:nvslambdahofit}
}
\end{subfigure}
\begin{subfigure}{.49\textwidth}
\centering
\includegraphics[width=1\textwidth]
{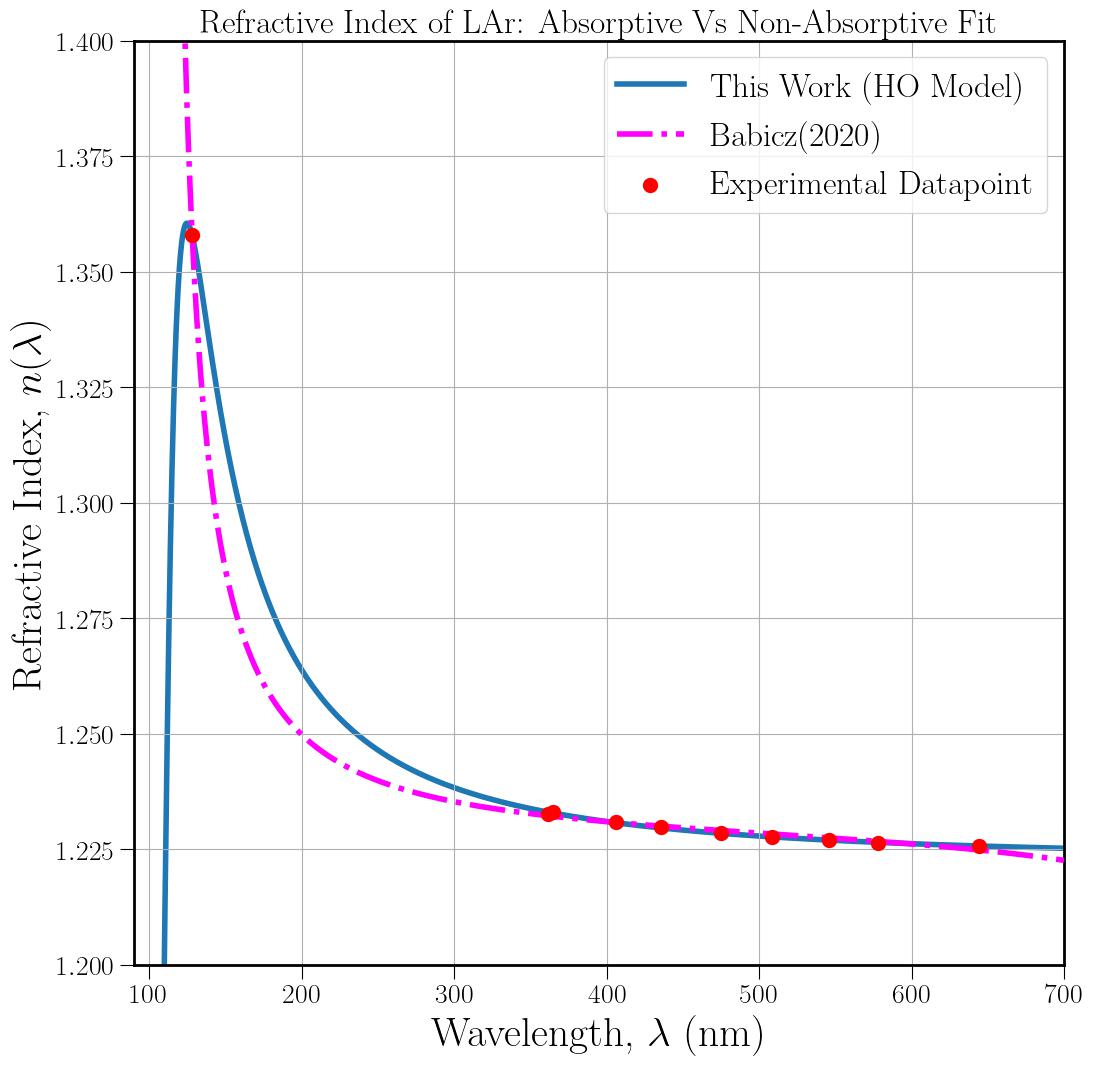}
\caption{HO Fit Vs Babicz. 
\label{f:nvslambdahofitvsBabicz}
}
\end{subfigure}
\caption{Comparison of Our (HO Model) Refractive Fit Vs Babicz. It is very clear that, when the wavelength approaches UV resonance, Babicz's refractive index fit blows to infinity, but our fit does not. Also, This work has a clear peak ($n_{peak} = 1.36$) at $\lambda = 124.68$ nm.
\label{f:nvslambdahoandBabiczfit}
}
\end{figure}
The absorptive fit \eqref{e:nfinalho1} with these best-fit parameters is plotted in Fig.~\ref{f:nvslambdahofit} and compared against Babicz' non-absorptive fit in Fig.~\ref{f:nvslambdahofitvsBabicz}.  Clearly, when the wavelength approaches UV resonance, non-absorptive fits to the refractive index like Babicz's and Grace's blow up to infinity, while absorptive fits like the HO model do not. Instead, our harmonic oscillator fit has a clear peak ($n_{peak} = 1.36$) at $\lambda = 124.68$ nm.

%
\begin{figure}[t!]
\begin{centering}
\includegraphics[width=0.6\textwidth]{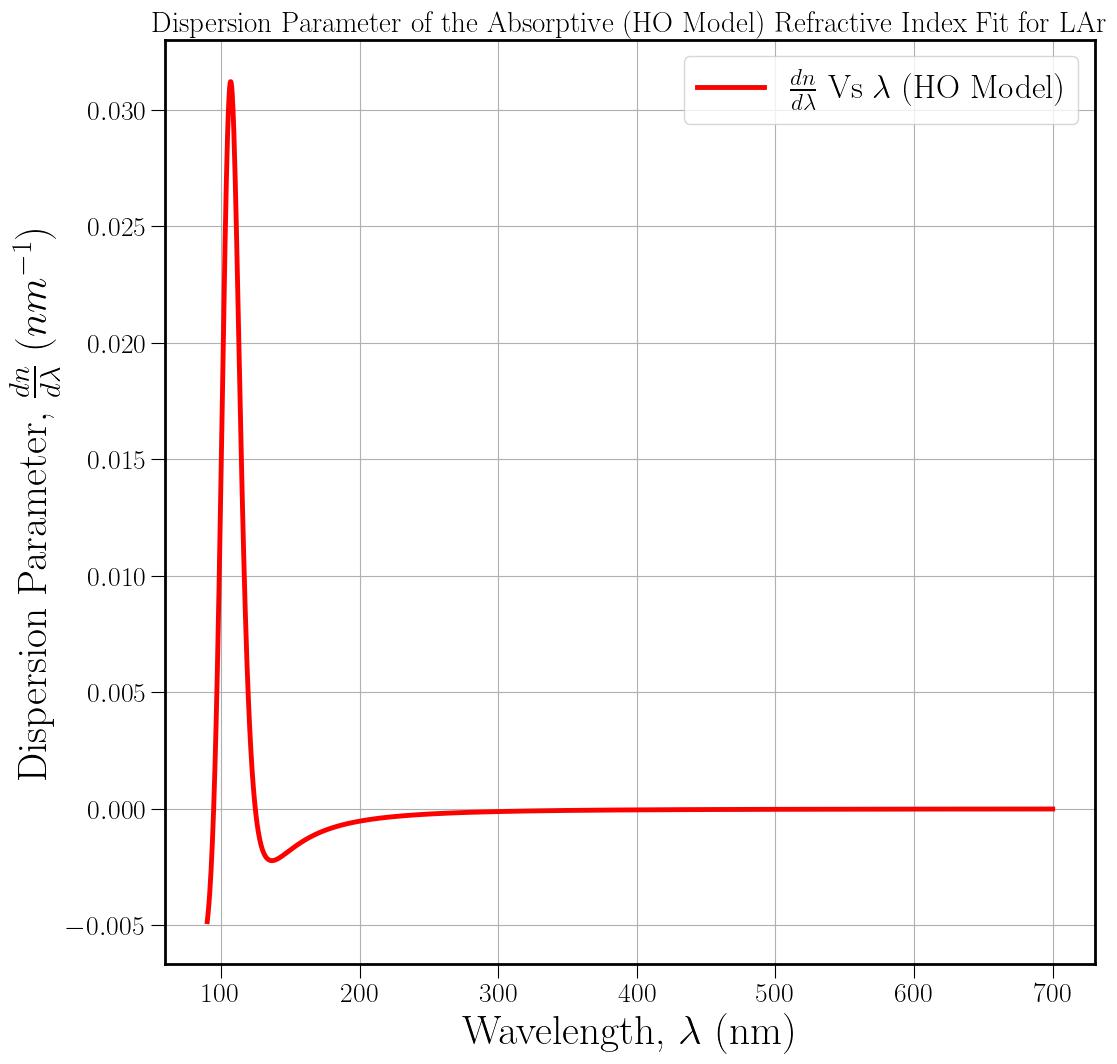}
\caption{Determining Peak of our Refractive Index Fit from the Derivative ($\frac{dn}{d\lambda}$) known as the Dispersion Parameter, a measure of the dispersion phenomenon occurring at different wavelengths ($\lambda$). 
\label{f:derivativenjackson}
}
\end{centering}
\end{figure}
This change from a total divergence $n \rightarrow \infty$ to a finite maximum $n_{peak}$ leads to major changes in the emitted Cherenkov spectrum.  Instead of $n(\lambda)$ continuing to grow as the wavelength approaches the UV resonance $\lambda_{UV}$, for absorptive fits $n(\lambda)$ instead \textit{decreases} between the peak and the resonance.  As introduced in Ch.~\ref{mathform}, an index of refraction for which ${\frac {dn}{d\lambda }}>0$ is said to characterize \textit{anomalous} dispersion.  Absorptive fits with a maximum $n_{peak}$ therefore automatically introduce regions of anomalous dispersion to the refractive index, adding a \textit{second pattern} of emitted Cherenkov radiation beyond that which would have been produced by resonant fits like Babicz.  The dispersion parameter $\frac{dn}{d\lambda}$ for our absorptive fit is plotted in Fig.~\ref{f:derivativenjackson}, from which one can clearly identify the peak (zero of the dispersion parameter) at $\lambda_{peak} = 124.68$ nm, separating the regions of normal ($\lambda > 124.68$ nm) and anomalous ($106.6 \: \mathrm{nm} < \lambda < 124.68 \: \mathrm{nm}$) dispersion.  We also note from Fig.~\ref{f:derivativenjackson} that the dispersion parameter has a maximum value at $\lambda = \lambda_{UV} = 106.6$ nm, corresponding to the resonance itself, as well as a second zero crossing at $\lambda = 94.58$ nm.  The second zero of the dispersion parameter corresponds to the \textit{minimum} of $n(\lambda)$ seen in Fig.~\ref{f:nvslambdahofit}.  Because of the sharp rise in absorption near the resonance (see Fig.~\ref{f:absorption_coeff_theoryplot}), Cherenkov photons emitted with $\lambda < \lambda_{UV}$ will be absorbed in the medium; therefore, we do not consider Cherenkov emission at wavelengths below the UV resonance.

%





\subsection{Calculating the Cherenkov Yield} \label{sec:abs}

As mentioned in section \ref{sec:noabs} in Ch.~\ref{mathform}, we will calculate the total Cherenkov yield using absorptive fits discussed in the previous section to include the important physics of absorption for the first time. I have calculated the Cherenkov yield of Protons in LAr using our own "Absorptive (HO model)" refractive index fit.

\subsubsection{Yield Methodology: Intersection Method}

Imposing the Cherenkov condition \eqref{e:CherenCond1} on the parameterization \eqref{e:nlambdauvhofinal}, we find that Cherenkov photons are emitted for any wavelengths $\lambda$ satisfying the inequality for our harmonic oscillator (HO) model fit, 

\begin{align}   \label{e:cherenkovncond}
n_{HO} = a_{0(HO)} + a_{UV(HO)} \left(\frac{\lambda_{UV}^{-2} - \lambda^{-2}}{(\lambda_{UV}^{-2} - \lambda^{-2})^2 + \gamma_{UV}^2 \lambda^{-2}}\right) \geq \frac{1}{\beta}  \: .  
\end{align}
For simpler fits, the intersection of $n(\lambda)$ from \eqref{e:cherenkovncond} with the constant $1/\beta$ could be solved analytically for the $\lambda$ for a given $\beta$, but due to the quartic denominator in \eqref{e:nlambdauvhofinal}, in this case the intersection must be computed numerically.  Doing so gives the lower and upper bounds of $\lambda$ that will emit Cherenkov radiation (at any angle) for a given instantaneous kinetic energy of the proton.

\begin{figure}[p!]
\centering
    \begin{subfigure}{.49\textwidth}
      \centering
      \includegraphics[width=1\linewidth]{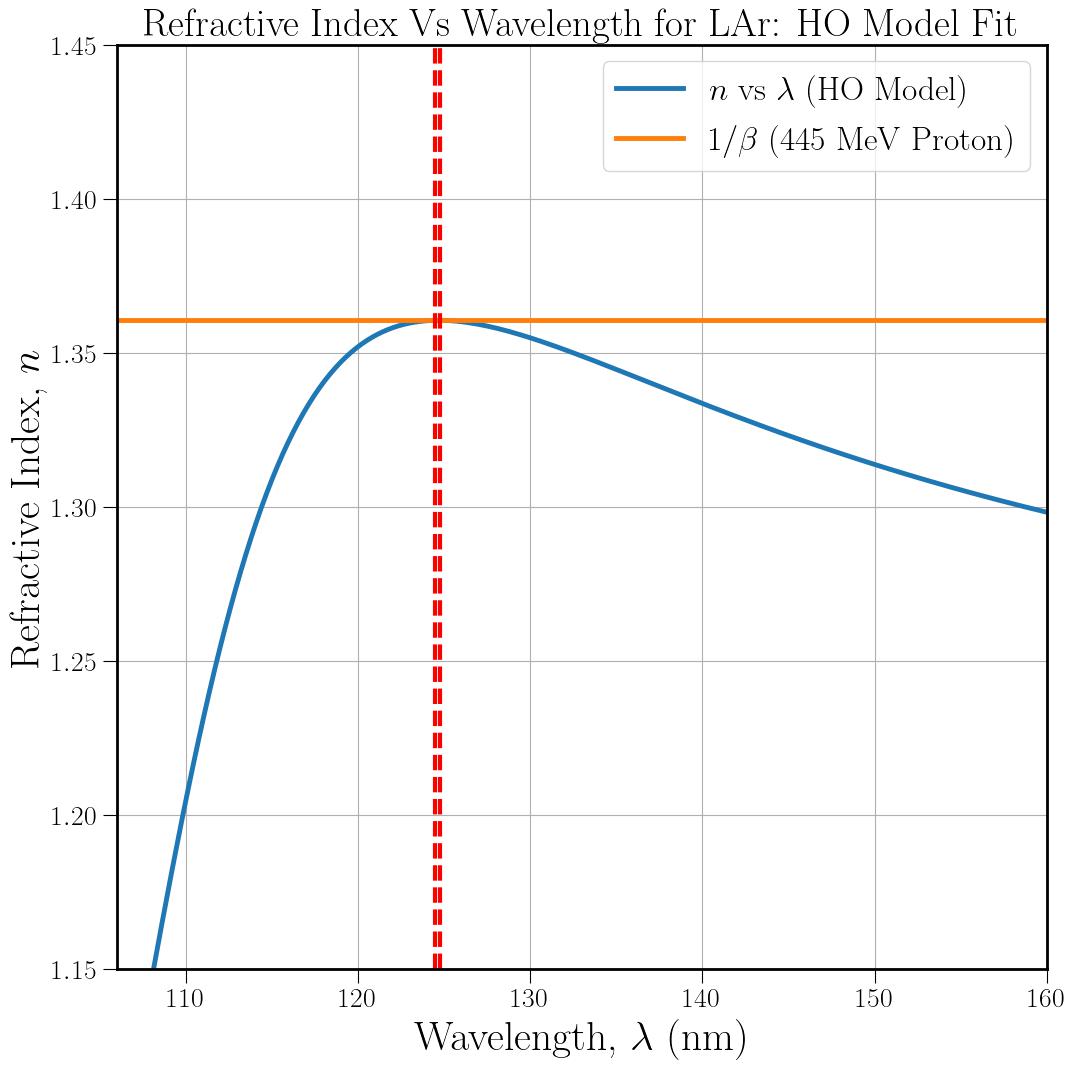}
      \caption{445.4 MeV Proton}
      \label{f:ho445MeV}
    \end{subfigure}
    \begin{subfigure}{.49\textwidth}
      \centering
      \includegraphics[width=1\linewidth]{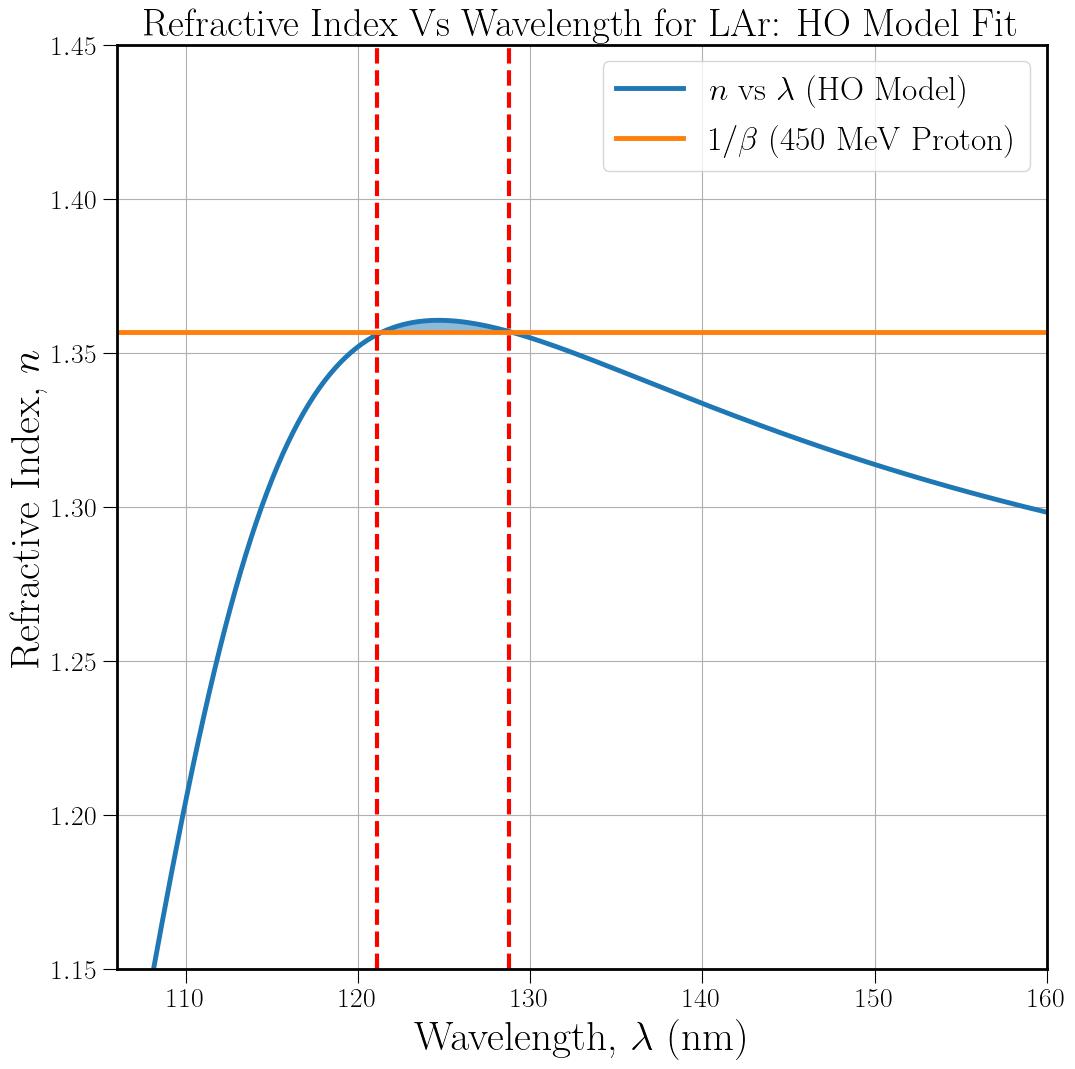}
      \caption{450 MeV Proton}
      \label{f:ho450MeV}
    \end{subfigure}
    \medskip
    \centering
    \begin{subfigure}{.49\textwidth}
      \centering
      \includegraphics[width=1\linewidth]{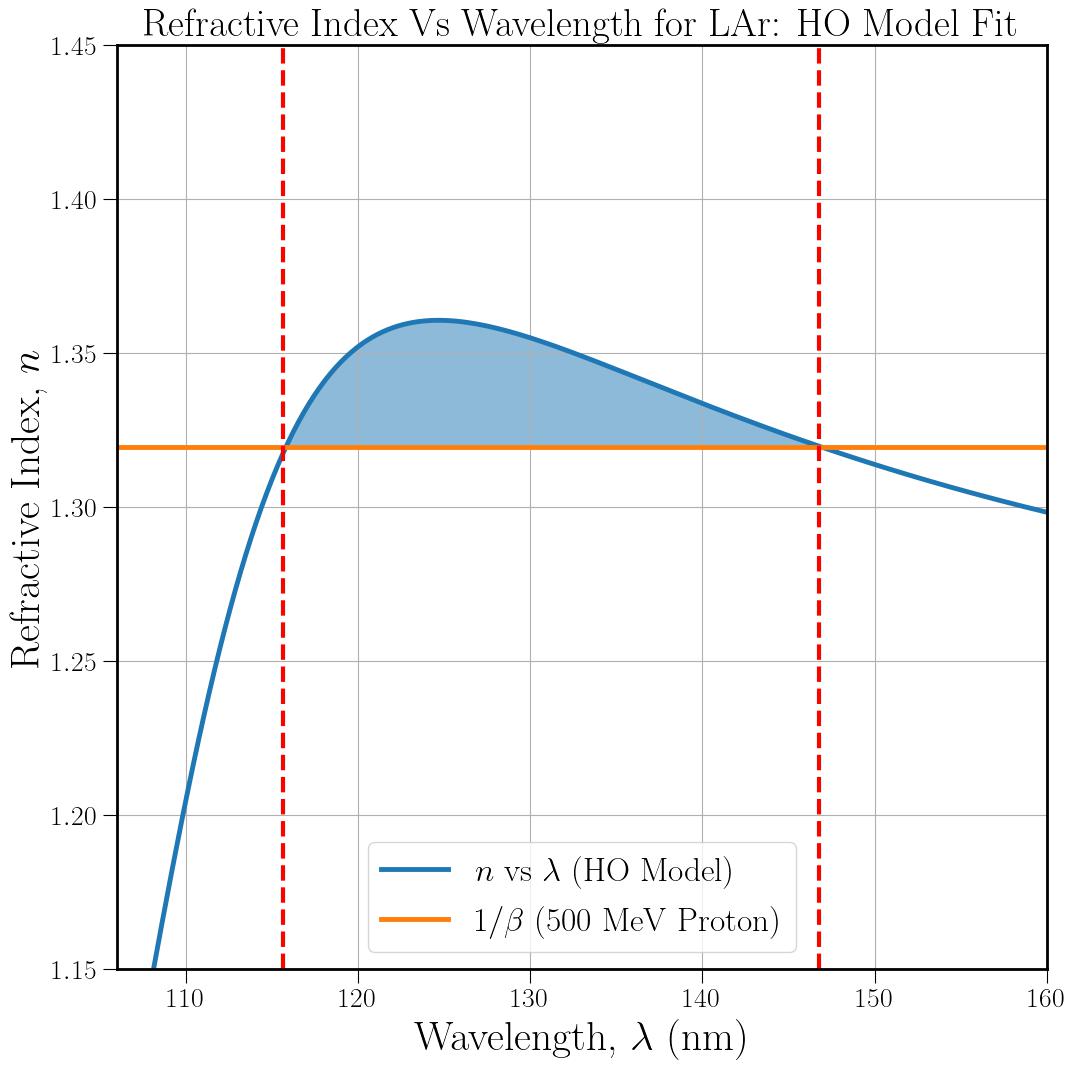}
      \caption{500 MeV Proton}
      \label{f:ho500MeV}
    \end{subfigure}
    \begin{subfigure}{.49\textwidth}
      \centering
      \includegraphics[width=1\linewidth]{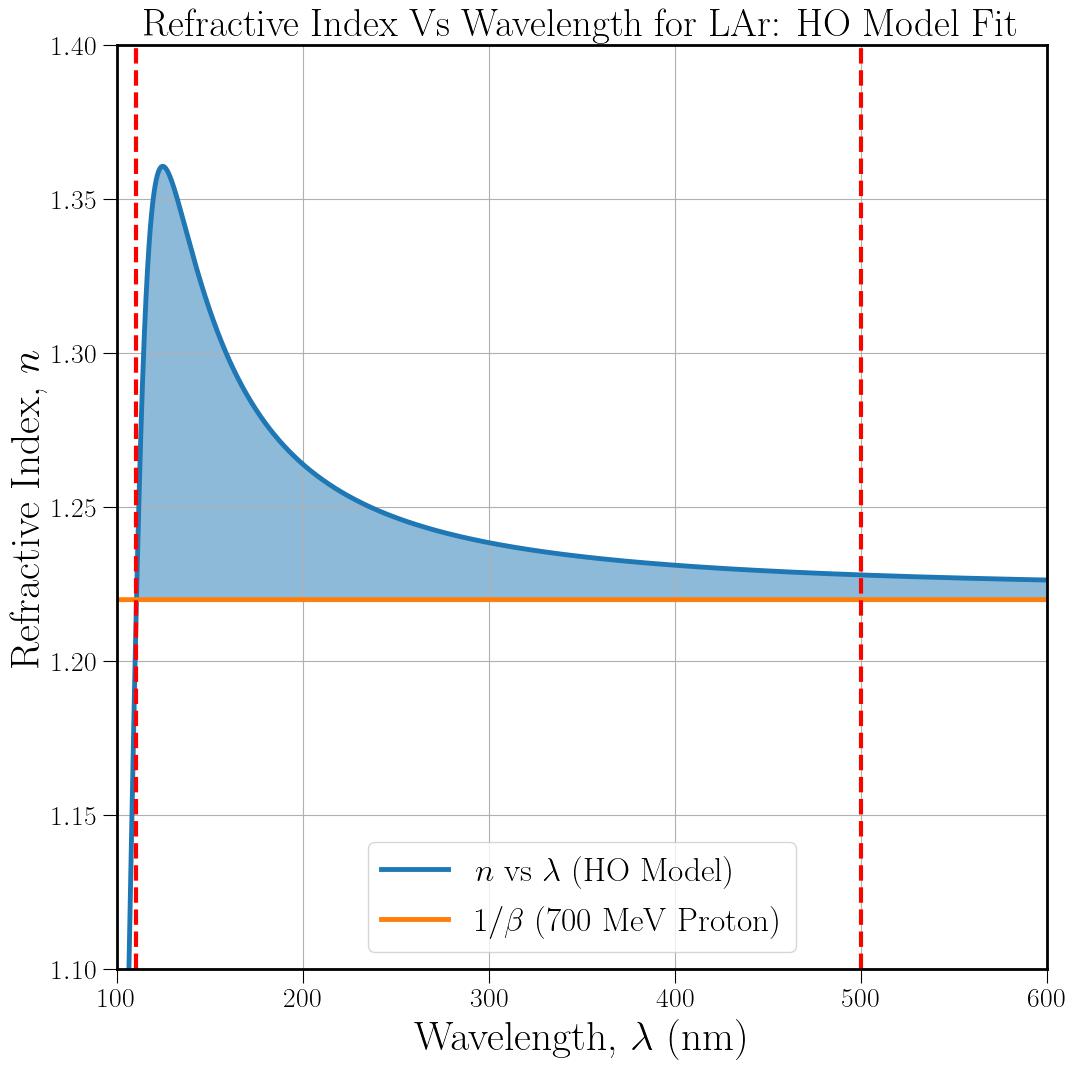}
      \caption{700 MeV Proton}
      \label{f:ho700MeV}
    \end{subfigure}%
\caption{Finding the Wavelength Range of Cerenkov Radiation from Intersection Method}
\label{f:intersectionarrayho} 
\end{figure}

The intersection method and the wavelength range which emits Cherenkov radiation is shown in Fig.~\ref{f:intersectionarrayho} for various proton kinetic energies.  At an energy of $445 \: \mathrm{MeV}$ (Fig.~\ref{f:ho445MeV}), the orange $1/\beta = n_{peak}$ line just barely touches the peak of the refractive index; this is the minimum kinetic energy at which Cherenkov photons begin being emitted.  For higher kinetic energies, the the $1/\beta$ line moves downward and the range of wavelength solutions broadens, resulting in a small band of radiating wavelengths between the vertical dashed lines in Fig.~\ref{f:ho450MeV}.  By $500 \: \mathrm{MeV}$ (Fig.~\ref{f:ho500MeV}), the wavelength range has grown to $115.63 \: \mathrm{nm} < \lambda < 146.76 \: \mathrm{nm}$, and by $700 \: \mathrm{MeV}$ (Fig.~\ref{f:ho700MeV}), the maximum wavelength $\lambda_{max}$ has formally diverged to infinity.  As discussed in Ch.~\ref{mathform}, photon detectors have finite acceptance, and to account for that we cut off $\lambda_{max}$ at 500 nm, not counting any Cherenkov photons emitted at higher wavelengths.





%
\begin{figure}[h!]
\begin{centering}
\includegraphics[width=0.5\textwidth]{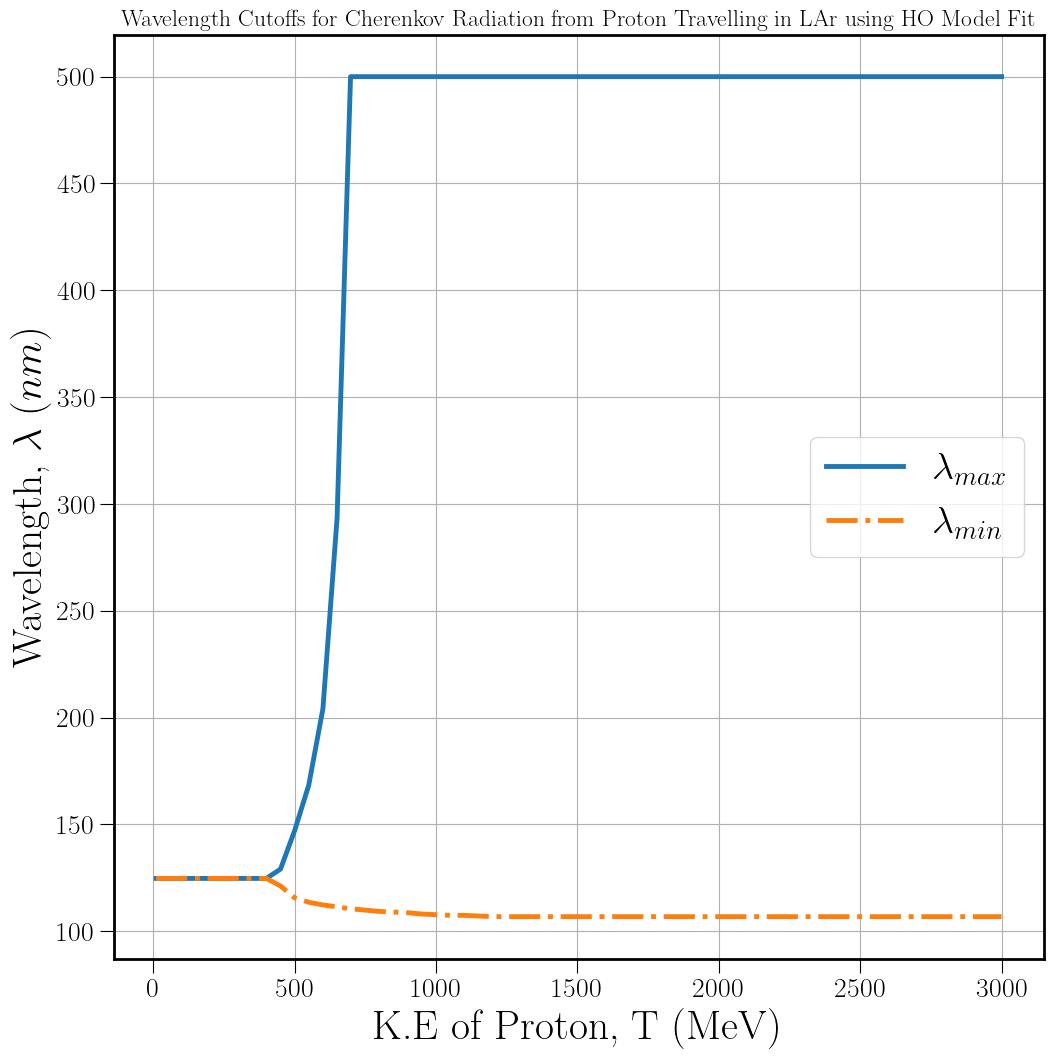}
\caption{Normal and Anomalous Wavelength Solutions Derived Using Our Absorptive (HO Model) Refractive Index Fit for Protons with Different K.Es Travelling in LAr. 
\label{f:lambdaminmaxallTHO}
}
\end{centering}
\end{figure}
As we saw in Ch.~\ref{mathform}, determining the proper wavelength limits $\lambda_{min}, \lambda_{max}$ is crucial for the calculation of the Cherenkov yield.  The wavelength limits determined by the intersection method (see Fig.~\ref{f:intersectionarrayho}) are plotted in Fig.~\ref{f:lambdaminmaxallTHO}.  Below the threshold energy of $445$ MeV, there is no Cherenkov radiation, due to the finite maximum value $n_{peak}$ the refractive index.  We treat this case by setting $\lambda_{min} = \lambda_{max} = \lambda_{peak} = 125$ nm, such that the integration range is zero.  As the kinetic energy increases, a range of wavelengths begins to emit Cherenkov radiation (all at various angles) in between the limit values $\lambda_{min}$ and $\lambda_{max}$.  The upper limit $\lambda_{max}$ is set by the part of $n(\lambda)$ characterized by normal dispersion, such that $\lambda_{max}$ grows as a function of energy.  Conversely, $\lambda_{min}$ is set by the anomalous dispersion part of $n(\lambda)$ and decreases as a function of energy.  Finally, we note that we freeze $\lambda_{max}$ at 500 nm to account for finite detector acceptance.  This results in the wavelength range $\lambda_{max} - \lambda_{min}$ mostly (although not completely) saturating at approximately $678$ MeV.


\begin{figure}
\centering
    \begin{subfigure}{.48\textwidth}
    \includegraphics[width=1\textwidth]{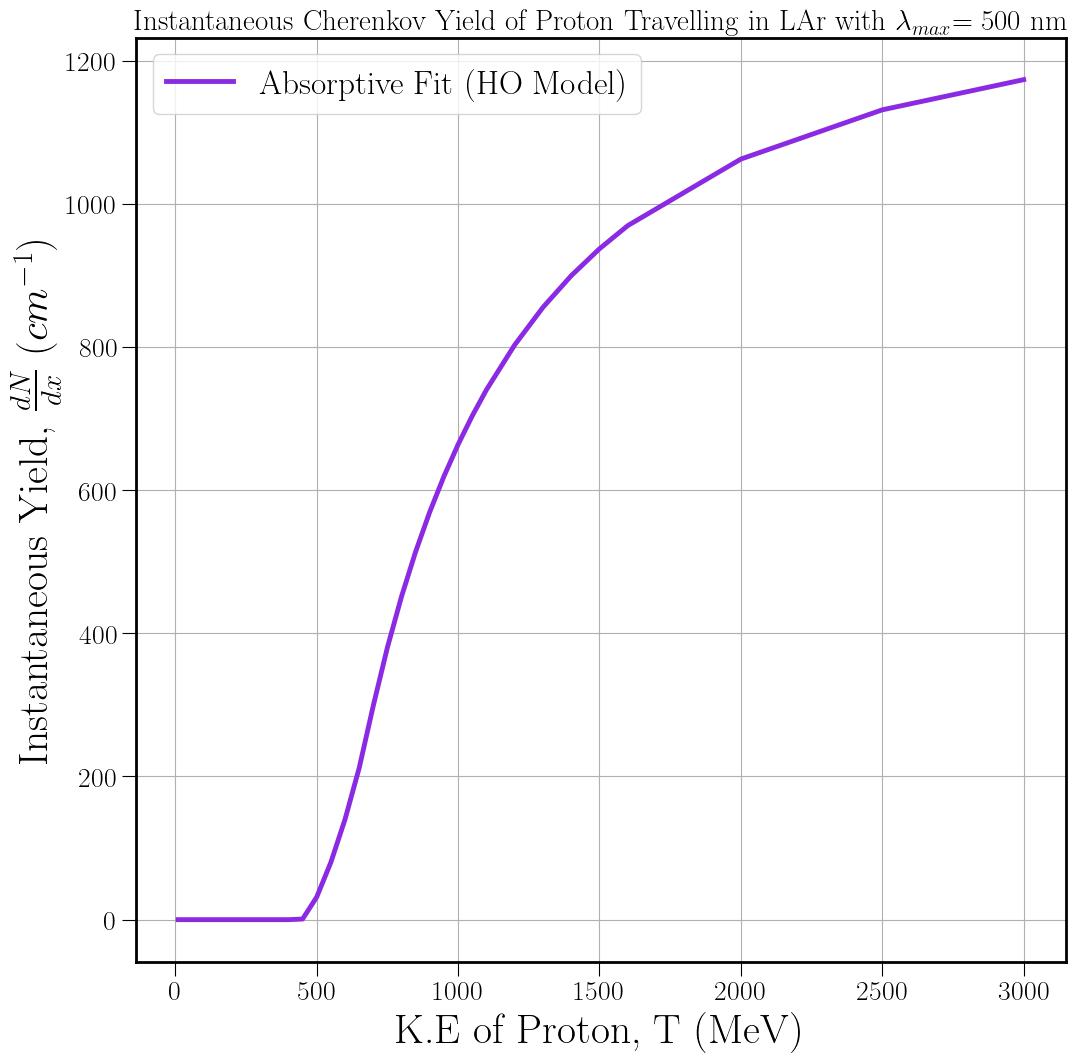}
    \caption{Instantaneous Yield
    \label{f:abshoinstyield}
    }
    \end{subfigure} 
\centering
    \begin{subfigure}{.48\textwidth}
    \centering
    \includegraphics[width=1\linewidth]{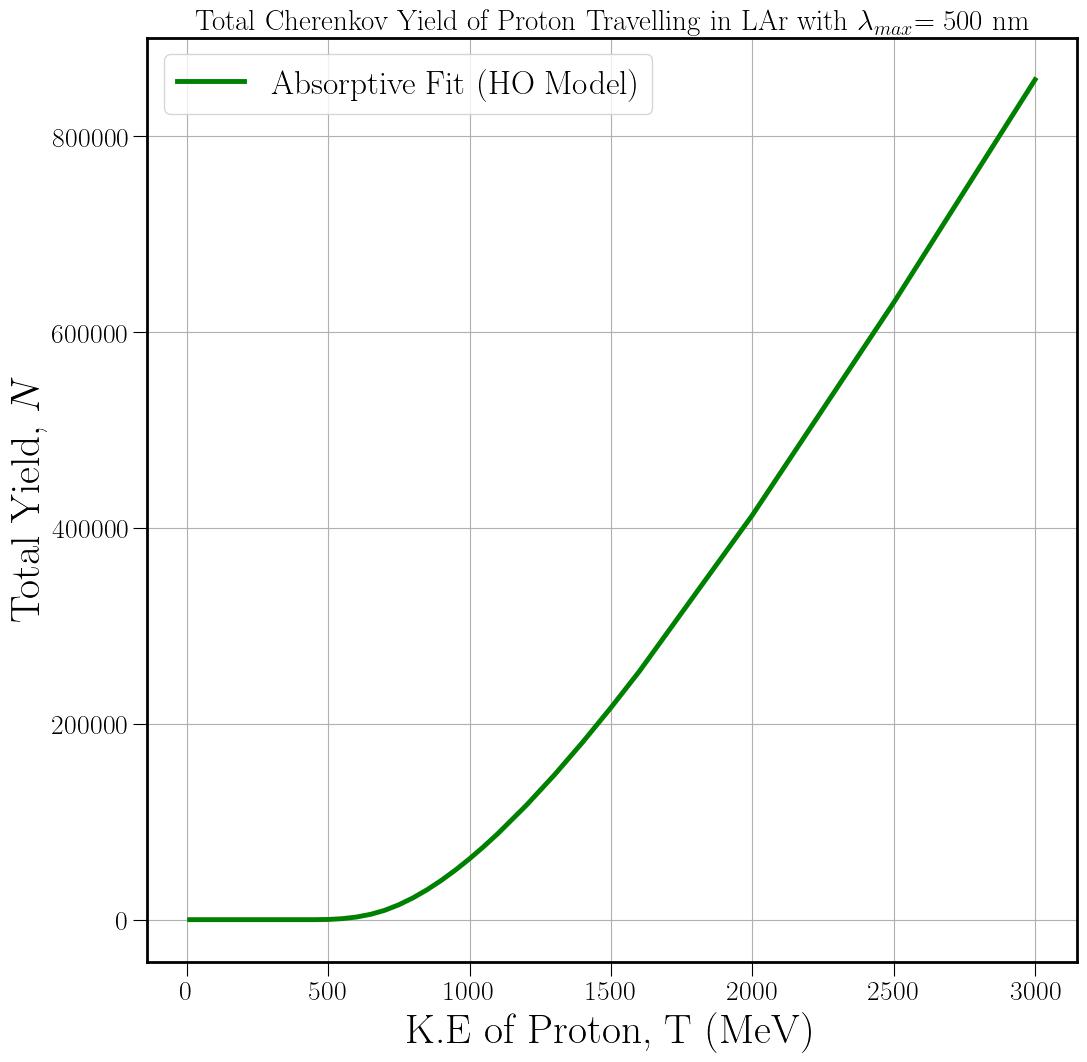}
    \caption{Integrated Yield
    \label{f:abshointegyield}
    }
    \end{subfigure}
\caption{Instantaneous and Total (Integrated) Cherenkov Yield for Protons with Different K.Es Travelling in LAr Using Our Absorptive Fit.
\label{f:Cherenkovnoabsorptivefit}
}
\end{figure}
%
\begin{figure} 
\centering
    \begin{subfigure}{.48\textwidth}
    \includegraphics[width=1\textwidth]{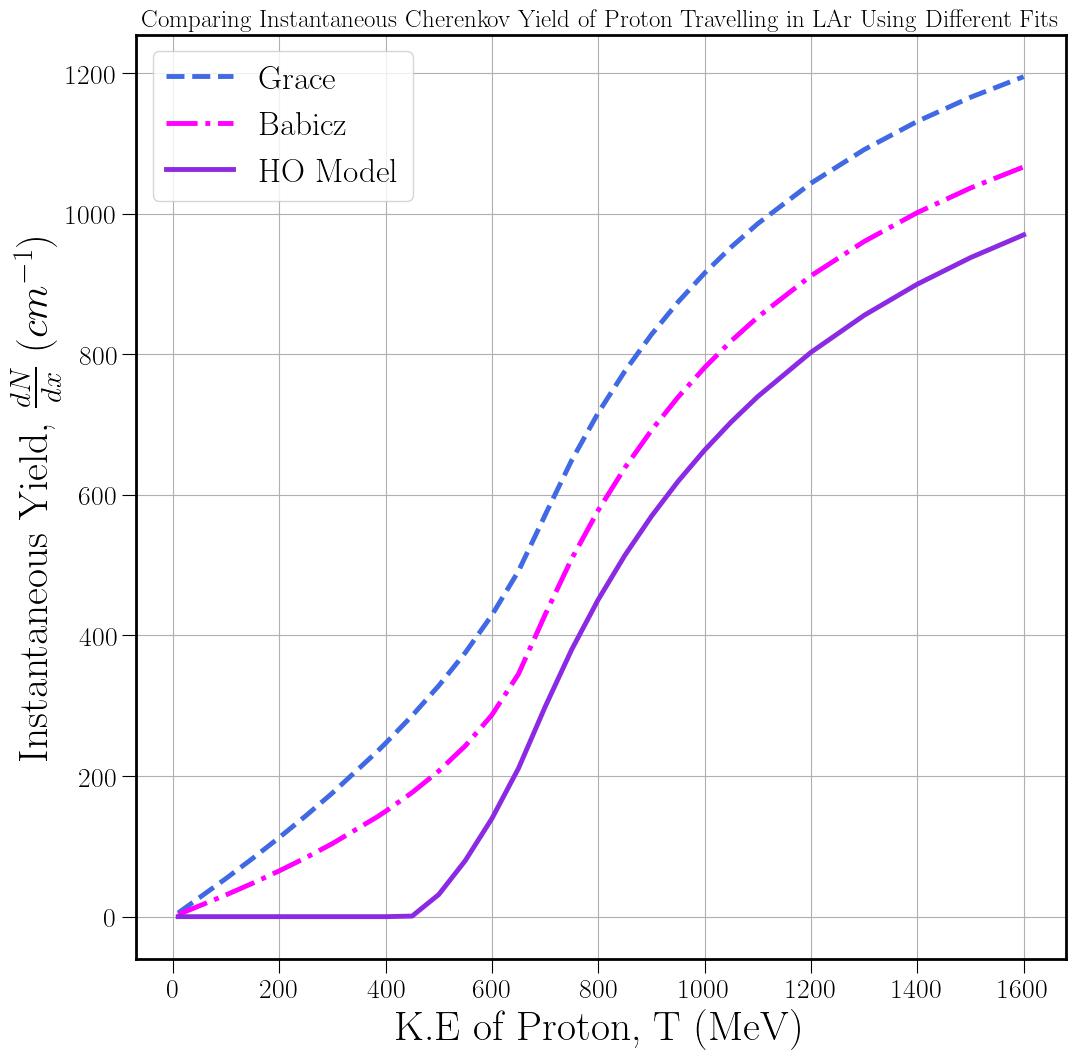}
    \caption{Instantaneous Yield
    \label{f:abshoinstyieldallfits}
    }
    \end{subfigure} 
\centering
    \begin{subfigure}{.48\textwidth}
    \centering
    \includegraphics[width=1\linewidth]{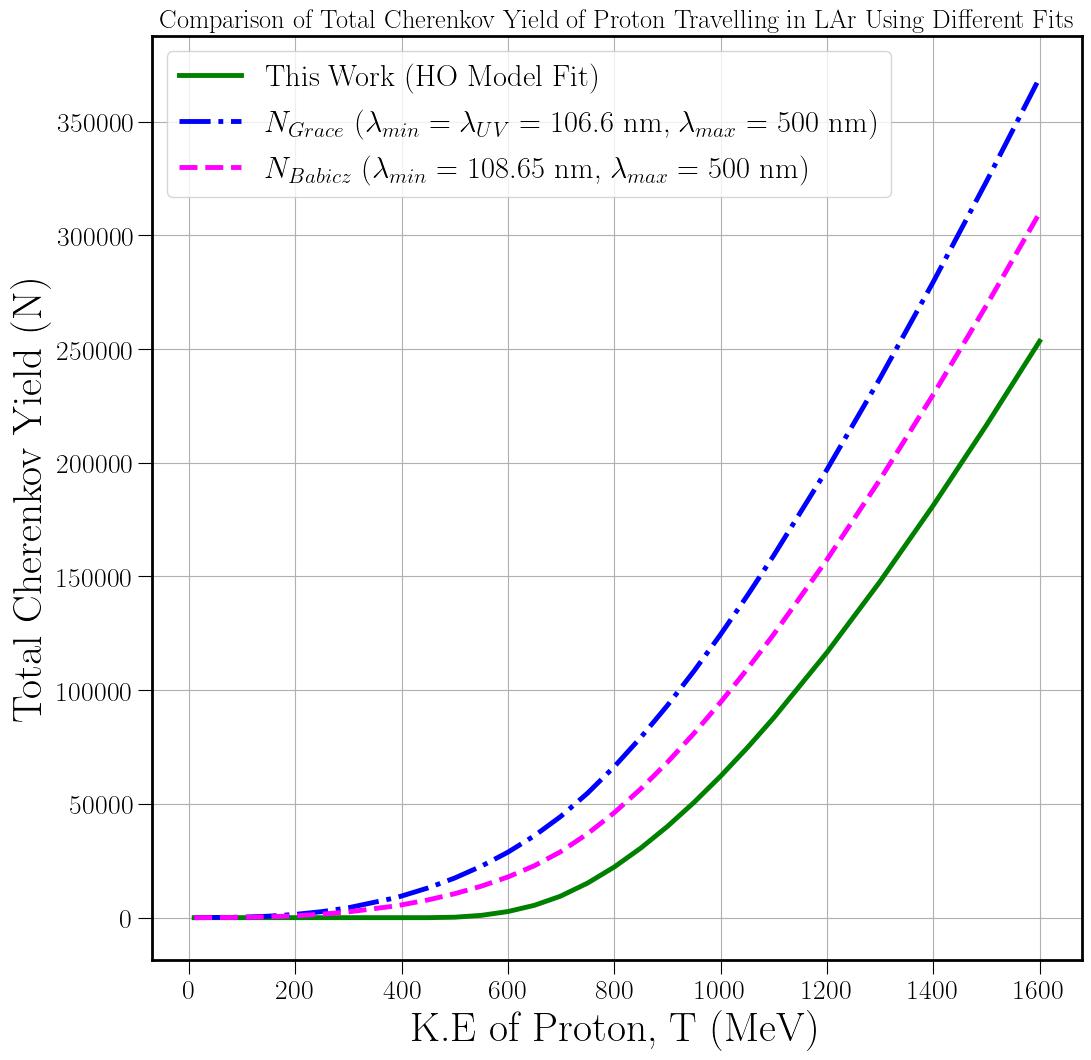}
    \caption{Integrated Yield
    \label{f:abshointegyieldallfits}
    }
    \end{subfigure}
\caption{Comparison of Total Cherenkov Yield of Protons with Different K.Es in LAr Using Absorptive (This work) Vs Non-absorptive Fits (Grace, Babicz). 
\label{f:Cherenkovnoabsvsnonabs}
}
\end{figure}

\subsubsection{Instantaneous and Total Cherenkov Yield}

As in Ch.~\ref{mathform}, we compute the instantaneous yield of Cherenkov photons by  integrating the Frank-Tamm formula \eqref{e:FrankTamm1} over the wavelength range $\lambda_{min} \leq \lambda \leq \lambda_{max}$ determined from the intersection method:
\begin{align}   \label{e:ftintegrand3}
 \frac{dN}{dx} = \int_{\lambda_{min}}^{\lambda_{max}} \frac{2 \pi \alpha z^2}{\lambda^2} \left( 1 - \frac{1}{\beta^2 n^2(\lambda)} \right) d\lambda   \: ,
\end{align}    
and we subsequently determine the integrated yield $N_{tot}$ by numerically integrating \eqref{e:ftintegrand3} over the range of the proton.  Both of these observables are plotted in Fig.~\ref{f:Cherenkovnoabsorptivefit}.

\subsubsection{Yield Comparison: Absorptive Vs Non-absorptive Fits}

In Fig.~\ref{f:Cherenkovnoabsvsnonabs} we further compare the instantaneous and integrated Cherenkov yields for our absorptive fit \eqref{e:cherenkovncond} based on the Harmonic Oscillator Model (HO) with Grace and Babicz' resonant fits described in Ch.~\ref{mathform}.  We can immediately see that, by both measures, the absorptive fit has a significantly reduced total yield.  This reduction in Cherenkov yield was anticipated, because resonant fits to the refractive index (like Grace and Babicz) grossly overestimate $n(\lambda)$ -- and therefore $dN/dx$ -- near the UV resonance.  In fact, for high energies, the loss in yield is perhaps surprisingly small, with the absorptive fit only losing $\sim\mathcal{O}(30\%)$ of the yield compared to Grace's most generous resonant fit.  

The other major difference in yield between the resonant and absorptive fits is that the absorptive fit has a minimum threshold energy before any Cherenkov radiation is emitted.  That existence of threshold energy (approximately $445$ MeV) is nicely illustrated in Fig.~\ref{f:ho445MeV}, which clearly shows that for an absorptive fit with a finite $n_{peak} < \infty$, there is always a threshold velocity $\beta_{min} = 1/n_{peak}$ before any Cherenkov radiation can be emitted.  The presence of a Cherenkov threshold at $445$ MeV for the absorptive fit, as well as the absence of any threshold whatsoever for the resonant fits, is clearly seen from the instantaneous Cherenkov yields in Fig.~\ref{f:abshoinstyieldallfits}.  Obviously, for energies below or close to the Cherenkov threshold of our fit, there is a huge difference between the yield predicted by the resonant and absoprtive fits.  But with increasing kinetic energy -- and especially after the wavelength range $\lambda_{min} - \lambda_{max}$ saturates around $678$ MeV -- the yields qualitatively converge, leading to a net loss of yield of only $\sim\mathcal{O}(30\%)$ at high energies.


\subsubsection{Yield Comparison: Different Absorptive Fits} \label{compabs}

At this point, it is very well understood that the observables we are interested in calculating in this study such as yields, angular distributions, etc. very much depend on the form of the refractive index (e.g., absorptive or non-absorptive) employed in the calculation. Even within the absorptive fits, the yields and angular distributions can change/vary depending on the height of the peak, the asymptotic value of the fit (value of the refractive index at $\lambda \to \infty$), and other physical features of the fit being used. The reason for the difference is clear from the intersection method of enforcing the Cherenkov condition, which determines whether a given wavelength emits Cherenkov photons for a given $\beta$ value from the intersection of the refractive index with line $1 / \beta$.  

In this section, we compare two different absorptive fits which have qualitatively similar shapes and describe the experimental data equally well, but differ quantitatively, such as having positions of the peak.  One of these is the absorptive harmonic oscillator model \eqref{e:nlambdauvhofinal} derived from first principles in Appendix~\ref{sec:HarmonicOscillator} already discussed in this chapter.  The other is a simpler, approximate form sometimes employed in the literature, with a somewhat different shape:
\begin{align}   \label{e:nmodifiedlambda}
    n(\lambda)  &=  a_0 + a_{approx} \:
    \frac{(\lambda_{UV}^{-1} - \lambda^{-1})}{(\lambda_{UV}^{-1} - \lambda^{-1})^2 + \Gamma^2}
\end{align}
with corresponding absorption coefficient, 
\begin{align}   \label{e:alphamodified2}
    \alpha(\lambda) &=
    a_{approx}
    \left(\frac{4\pi}{\lambda} 
    n(\lambda) \right)
    \:
    \frac{\Gamma}{(\lambda^{-1}_{UV} - \lambda^{-1} )^2 + \Gamma^{2}}
\end{align}
where, $a_{0}$, $a_{approx}$, and $\Gamma$ are parameters of the fit.



%
\begin{table}[t!]
  \begin{center}
    \caption{Best fit parameters of the  approximate absorptive refractive Index fit 
     }
    \label{tab:bestfitpvs}
    \begin{tabular}{|c|c|c|} 
      \hline
      \textbf{$a_0$} & \textbf{$a_{approx}$} ($nm$) & \textbf{$\Gamma$ ($nm^{-1}$)}\\
      \hline
      1.18416 & 0.000325985 & 0.000693652\\  
      \hline
    \end{tabular}
  \end{center}
\end{table}
%
%

\begin{figure}[h]
\begin{centering}
\includegraphics[width=0.5\textwidth]{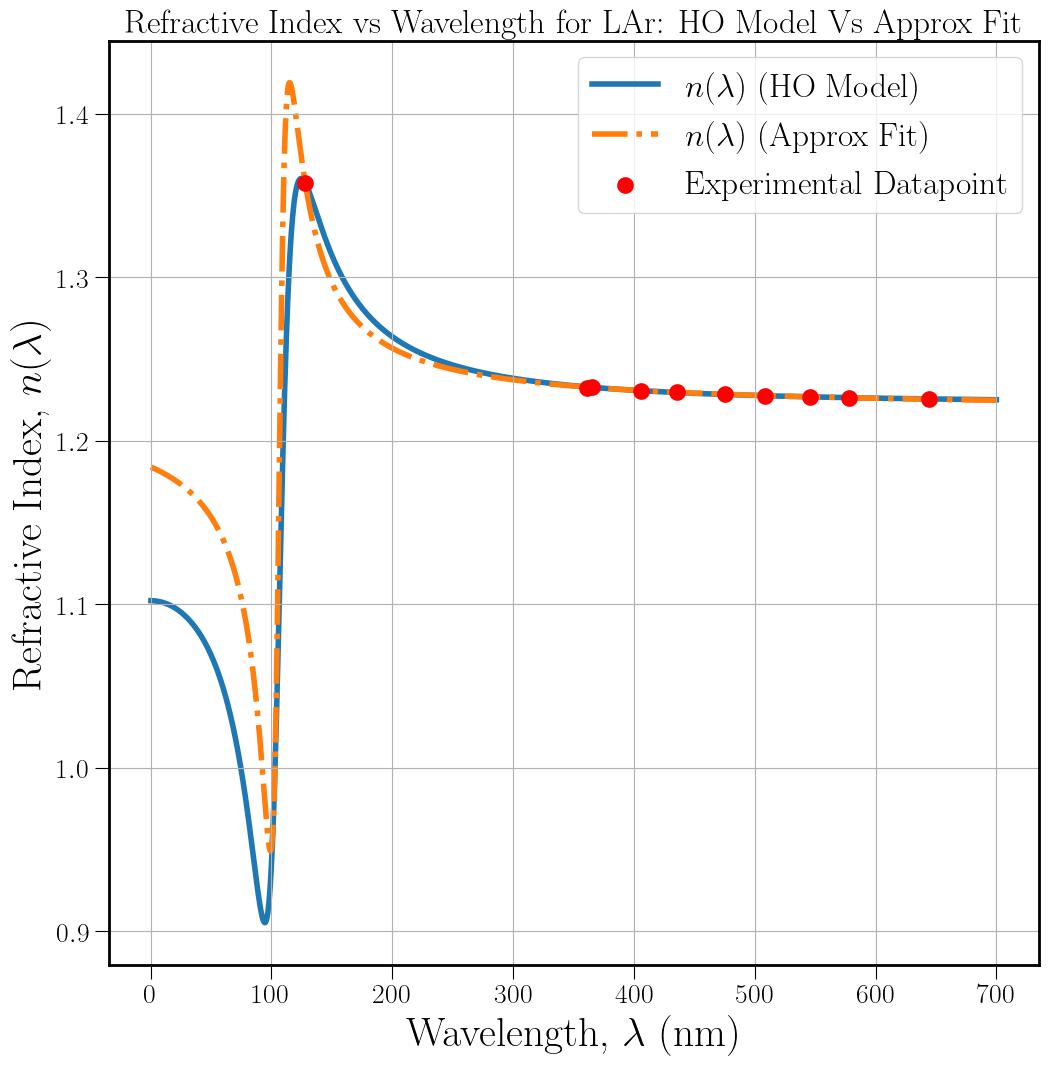}
\caption{Comparison of Absorptive Refractive Fits.
\label{f:n_vs_lambda_absfits}
}
\end{centering}
\end{figure}

The approximate form \eqref{e:nmodifiedlambda} can also describe well the experimental data on the refractive index, as shown in Fig.~\ref{f:n_vs_lambda_absfits}.  The best-fit parameters for the approximate form \eqref{e:nmodifiedlambda} are given in Table~\ref{tab:bestfitpvs}.  While two forms of $n(\lambda)$ are qualitatively similar, they differ in important quantitative measures -- most importantly the location and value of the peak index of refraction.  This peak value controls the onset of Cherenkov radiation at the minimum sped $\beta$, as discussed in Fig.~\ref{f:intersectionarrayho}.  For the harmonic oscillator model \eqref{e:nlambdauvhofinal} the index had a maximum of $n_{peak} = 1.36$ at $\lambda = 125$ nm, whereas the approximate form \eqref{e:nmodifiedlambda} has a maximum of ($n_{peak} = 1.42$) at $\lambda = 115$ nm.  The increased height $n_{peak}$ and the corresponding earlier activation of Cherenkov radiation (at lower energies) are the main differences between the harmonic oscillator and approximate fits to $n(\lambda)$.

\begin{figure}[h!] 
\centering
    \begin{subfigure}{.5\textwidth}
    \includegraphics[width=1\textwidth]{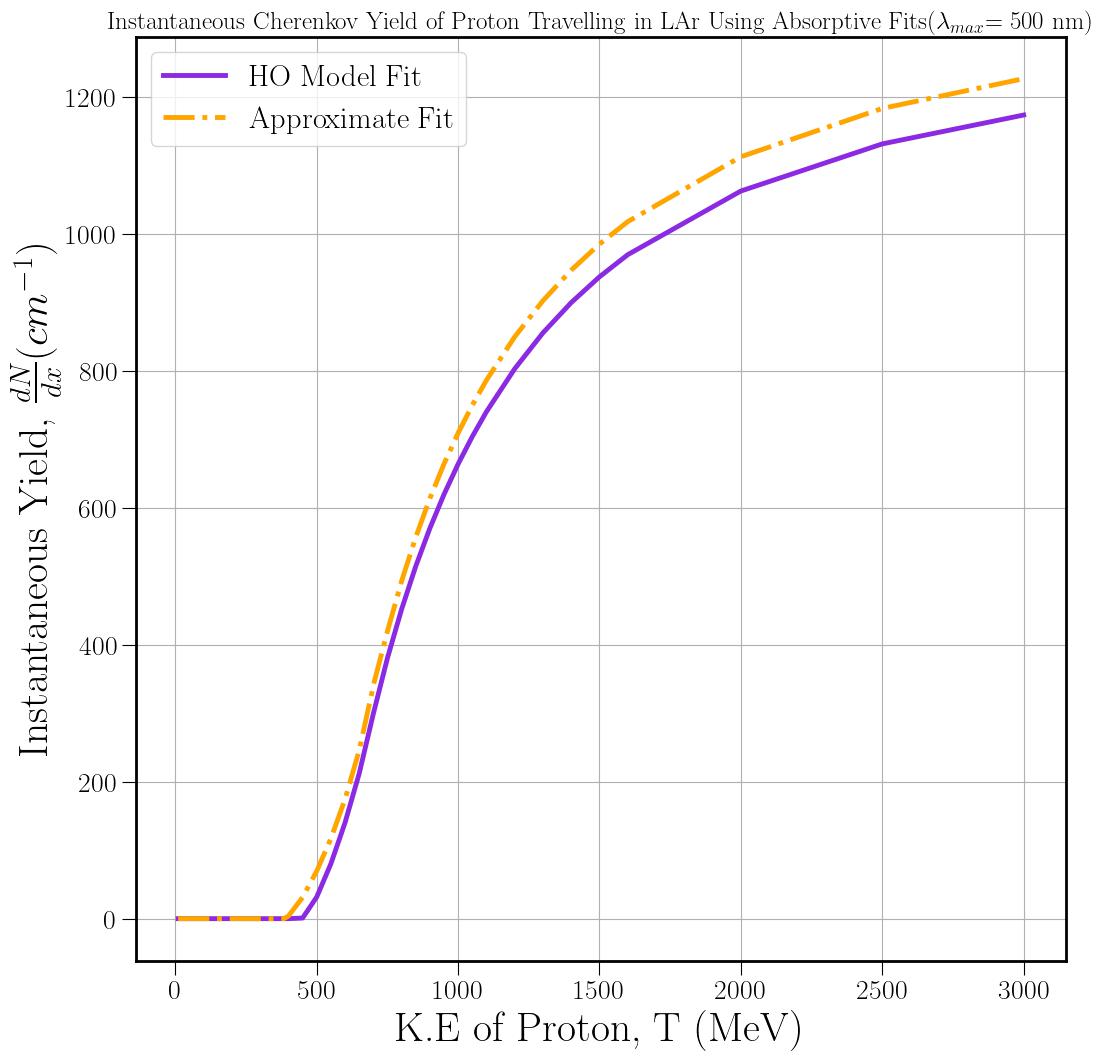}
    \caption{Our Absorptive Fit 
    \label{f:absorptivefits_inst}
    }
    \end{subfigure} 
\centering
    \begin{subfigure}{.48\textwidth}
    \centering
    \includegraphics[width=1\linewidth]{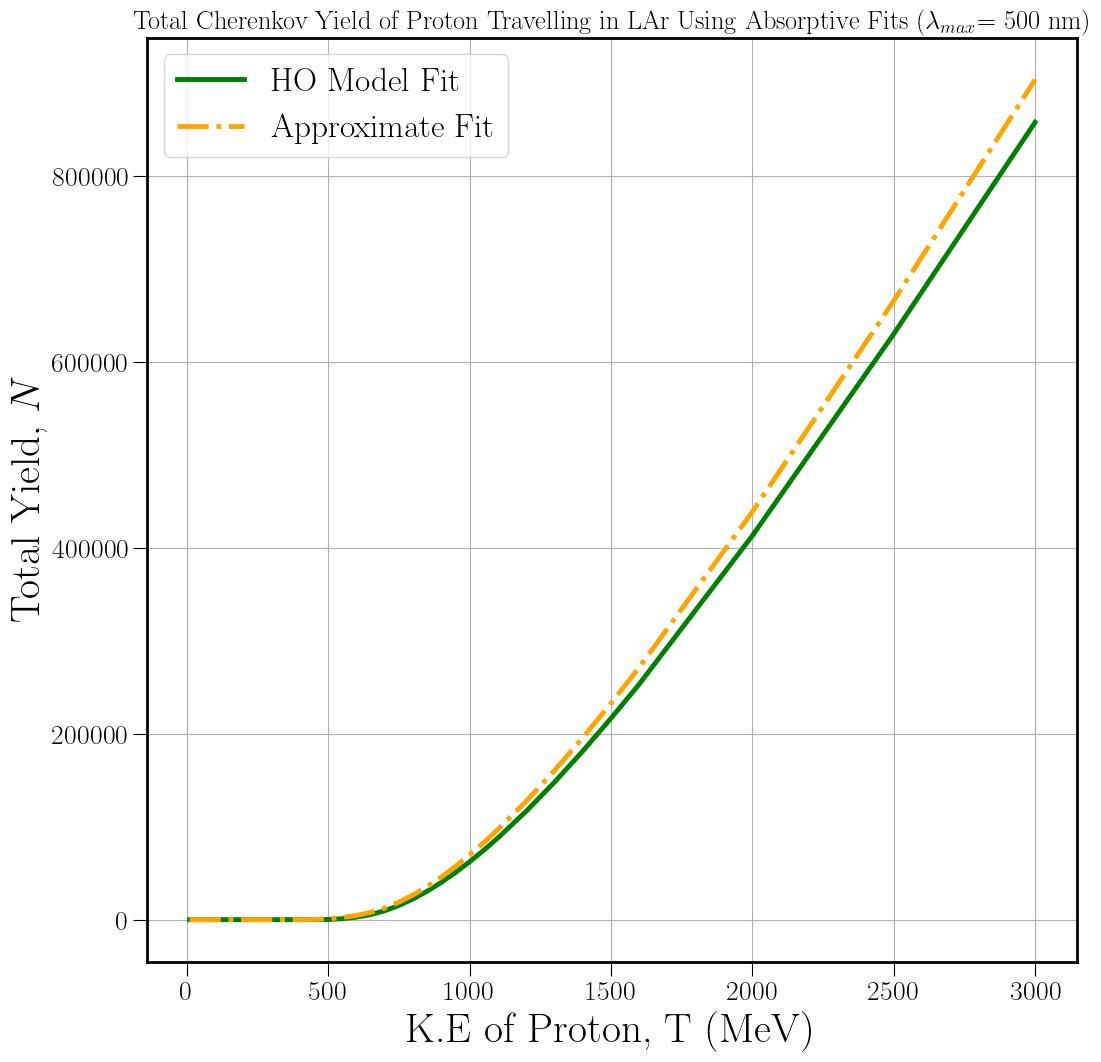}
    \caption{Total Cherenkov Yield
    \label{f:absorptivefits_integrated}
    }
    \end{subfigure}
\caption{Instantaneous and Total (Integrated) Cherenkov Yield for Protons with Different K.Es Travelling in LAr Using Two Different Absorptive Fits.
\label{f:Cherenkovyieldabsorptivefits}
}
\end{figure}
The instantaneous and integrated yields for the harmonic oscillator and approximate absorptive fits are compared in Fig.~\ref{f:Cherenkovyieldabsorptivefits}.  As anticipated, Cherenkov emission begins at a slightly lower kinetic energy ($384$ vs. $445$ MeV) for the approximate fit, and so it has a larger Cherenkov yield at smaller energies.  This excess predicted by the approximate fit becomes less important with increasing kinetic energies, such that the total integrated yields differ only by $\sim 25$\% at kinetic energies above $\sim 700$ MeV, when the wavelength range of both fits is saturated.

\subsubsection{Simple Black Disk Model for Reabsorption}

Because the refractive index $n(\lambda)$ for the absorptive fits discussed in this chapter do not diverge to infinity at the resonance, they are a substantial improvement over the prior resonant fits.  While the peak in $n(\lambda)$ and subsequent turnover as $\lambda \rightarrow \lambda_{UV}$ are a reflection of the anomalous dispersion produced by absorption, they do not fully take into account the actual ``reabsorption'' of Cherenkov photons after they are emitted.  The reabsorption of Cherenkov photons after emission is controlled by the absorption coefficient $\alpha(\lambda)$, given by Eqs.~\eqref{e:alphalambdauvir} or \eqref{e:alphamodified2} for the harmonic oscillator and approximate models, respectively.  

In a proper treatment of reabsorption 
\cite{griffiths_2017}, the number $N(\lambda, x)$ of Cherenkov photons of frequency $\lambda$ will attenuate exponentially with distance $x$, according to the differential equation
\begin{align}   \label{e:absprob}
    \frac{dN(\lambda)}{dx} = - \alpha(\lambda) \: N(\lambda, x)    \:. 
\end{align}
However, solving \eqref{e:absprob} would require evolving the entire spectrum $N(\lambda, x)$ of Cherenkov photons dynamically over the extent $x$ of the LAr environment.  This would be a major increase in the complexity of the calculation, and such a full calculation would be beyond the scope of the present analysis.

However, given the large impact of the absorptive fits compared to the resonant fits, it seems important to at least estimate the potential impacts from reabsorption.  As clearly shown in Figs.~\ref{f:Index_Real_Imag_Theorist} and \ref{f:absorption_coeff_theoryplot}, the absorption coefficient $\alpha(\lambda)$ is only significant close to the vicinity of the resonance.  Moreover, the half-width at half-maximum of the absorption coefficient coincides precisely with the peak position of the refractive index $n(\lambda)$.  Therefore, as a crude estimate for the impact of reabsorption of emitted Cherenkov photons, we can simply consider any Cherenkov photons with $\lambda < \lambda_{peak}$ to be reabsorbed before ultimately reaching the detector.  This simple estimate can be achieved by merely cutting off the wavelength range for Cherenkov photons at the peak index of refraction: $\lambda_{min} = \lambda_{peak}$.  

%
\begin{figure}[t]
\begin{centering}
\includegraphics[width=0.65\textwidth]{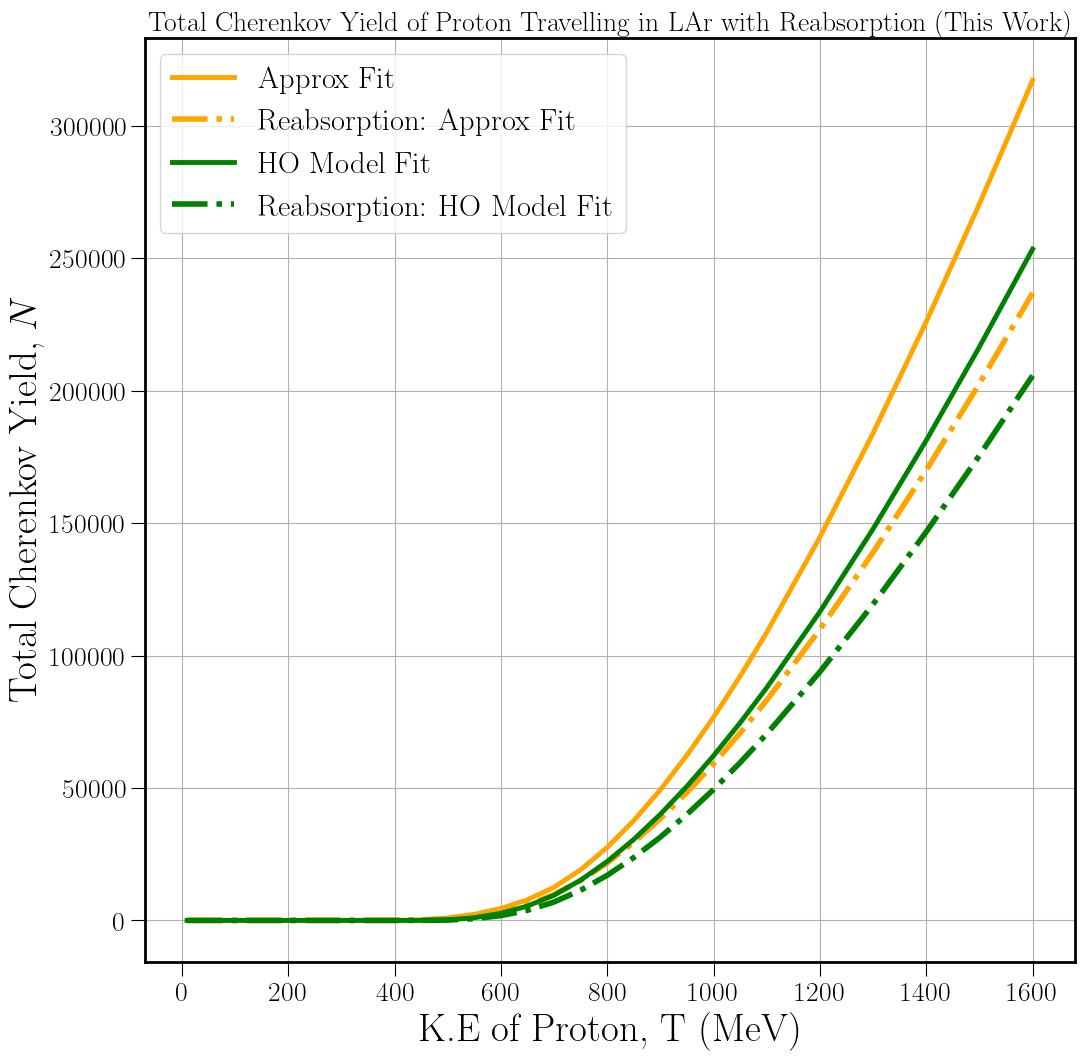}
\caption{Comparison of the total Cherenkov yield for two different absorptive fits (harmonic oscillator vs. approximate) and showing the effect of reabsorption (solid vs. dashed curves).
\label{f:comparisoncherenkovreabs}
}
\end{centering}
\end{figure}

The integrated yields for the harmonic oscillator and approximate forms of the refractive index are shown in Fig.~\ref{f:comparisoncherenkovreabs}, with and without reabsorption included.  For both fits, reabsorption decreases the integrated Cherenkov yield by a modest factor of $\sim 25$\% at high energies.  These integrated yields, after including the effects of reabsorption, can serve as a valuable conservative estimate (lower bound) on the expected Cherenkov yields.  Interestingly, the uncertainty from the effects of reabsorption is comparable to the uncertainty from different choices of absorptive forms for the refractive index (both about $\sim 25\%$).





\subsubsection{Yield Comparison: Cherenkov vs. Scintillation}

\begin{figure}[t] 
\centering
\begin{subfigure}{.49\textwidth}
\includegraphics[width=1\textwidth]{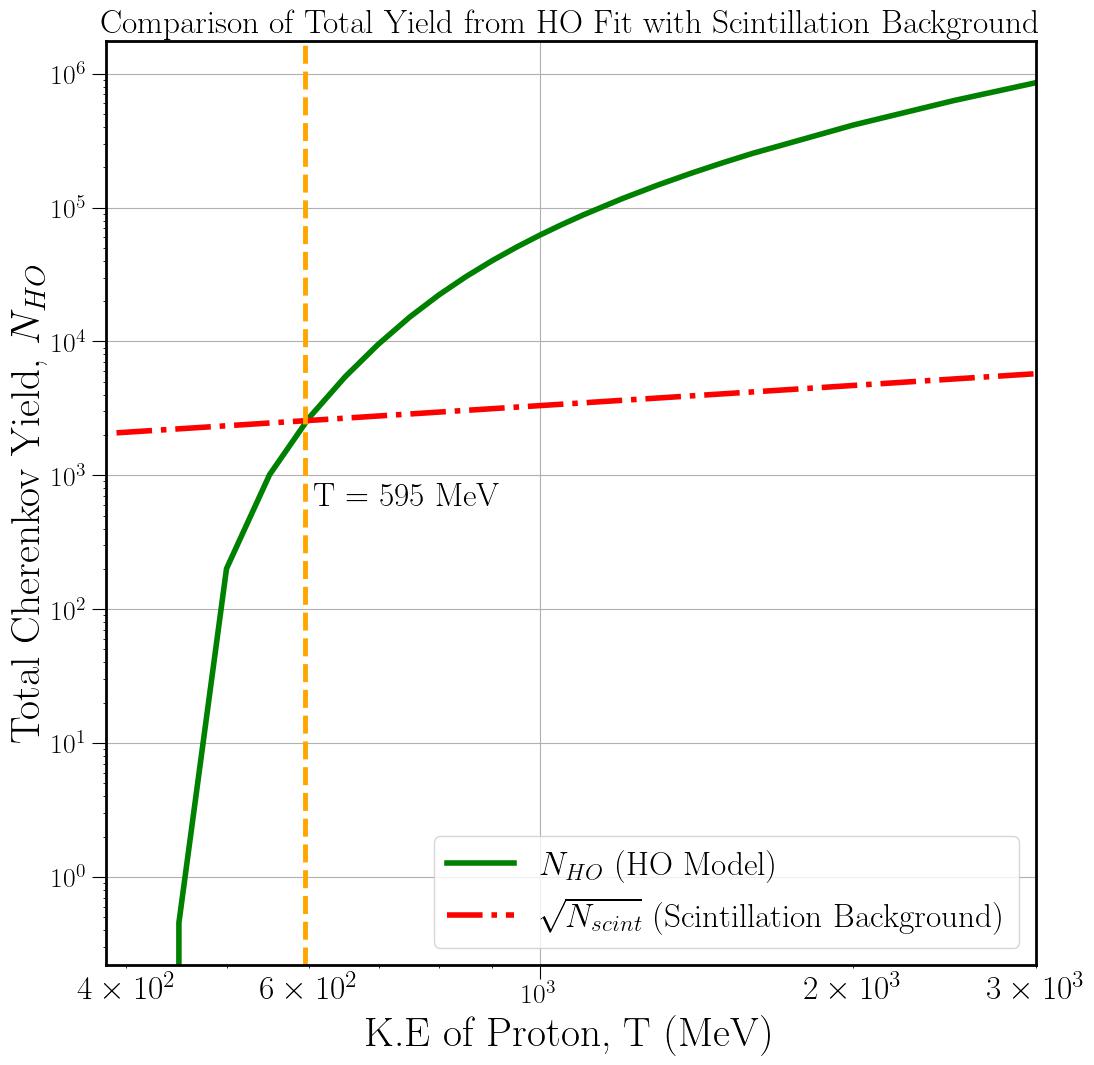}
\caption{Our Absorptive Fit  
\label{f:ourFTvsscintback_log}
}
\end{subfigure} 
\centering
\begin{subfigure}{.48\textwidth}
\centering
\includegraphics[width=1\linewidth]{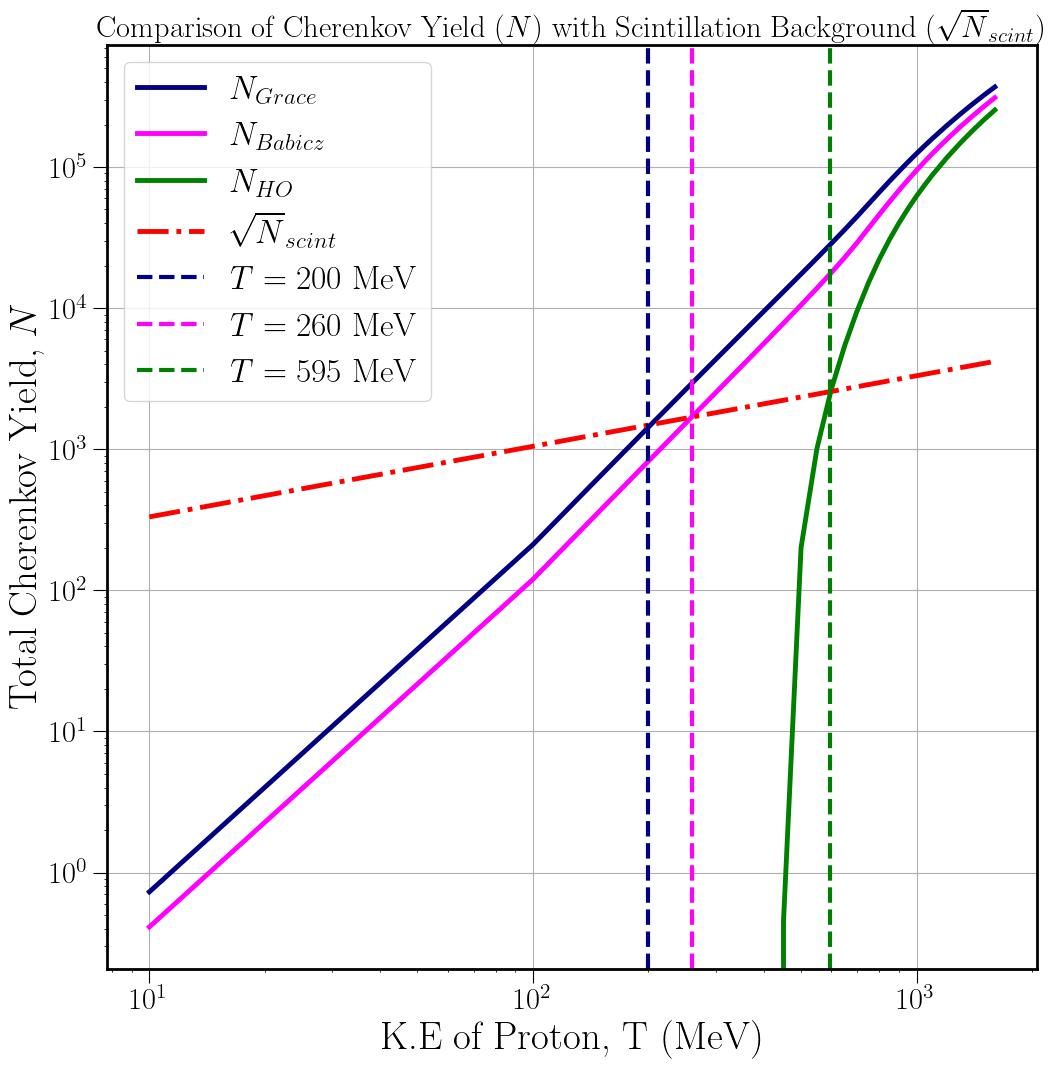}
\caption{Absorptive Vs Resonant Fits
\label{f:allFTvsscintback_log}
}
\end{subfigure}
\caption{Comparison of Total Cherenkov Yield from FT with Scintillation Background ($\sqrt{N_{scint}}$) of Proton in LAr Calculated Using Our Absorptive Refractive Index Fit Vs Resonant Fits.
\label{f:allFTwithscintback}
}
\end{figure}
Finally, in Fig.~\ref{f:allFTwithscintback} we further compare the integrated Cherenkov yield computed from the harmonic oscillator fit \eqref{e:nlambdauvhofinal} with the standard error $\sqrt{N_{scint}}$ in the scintillation background.  Despite the reduction in yield compared to the resonant fits (Grace, Babicz), the absorptive fit \eqref{e:nlambdauvhofinal} still predicts that the total Cherenkov yield will exceed the detection threshold over the background for sufficiently high-energy protons.  Whereas for the resonant fits, this detection threshold was crossed very quickly, at $200 - 300$ MeV (Fig.~\ref{f:allFTvsscintback_log}), for the absorptive fit, the crossing of the detection threshold is delayed to $595$ MeV.  Interestingly, this delayed onset by about $\sim 400$ MeV can be explained almost entirely by the $485$ MeV delayed onset of Cherenkov radiation due to the finite $n_{peak}$ as shown in Figs.~\ref{f:ho445MeV} and \ref{f:abshoinstyieldallfits}.  While we do not explicitly show the yield for the approximate fit \eqref{e:nmodifiedlambda}, it is clear from the discussion around Fig.~\ref{f:comparisoncherenkovreabs} how the approximate fit would compare.  Since the approximate fit has a higher peak value $n_{peak}$, its Cherenkov emission turns on sooner than the harmonic oscillator model, resulting in a yield which is $\sim 25\%$ larger and falls in between the harmonic oscillator and resonant fits.

%



%

%
\subsection{Extracting the Angular Distribution (AD)}
\hspace{\parindent}

As we have seen from the previous Ch.~\ref{mathform}, the angular distribution (AD) of radiation can differentiate the Cherenkov signal from the overall background dominated by scintillation photons even more effectively than the integrated yield alone. In this section, we will calculate the Cherenkov angular distribution of protons in LAr using our absorptive harmonic oscillator (HO) model fit \eqref{e:nfinalho1} and explore the physics that it may bring.

\subsubsection{Instantaneous and Integrated Angular Distributions} 
\label{Sec:AbsorptiveAD}

\begin{figure}[t]
\begin{centering}
\begin{subfigure}{0.49\textwidth}
\centering
\includegraphics[width=1\textwidth]{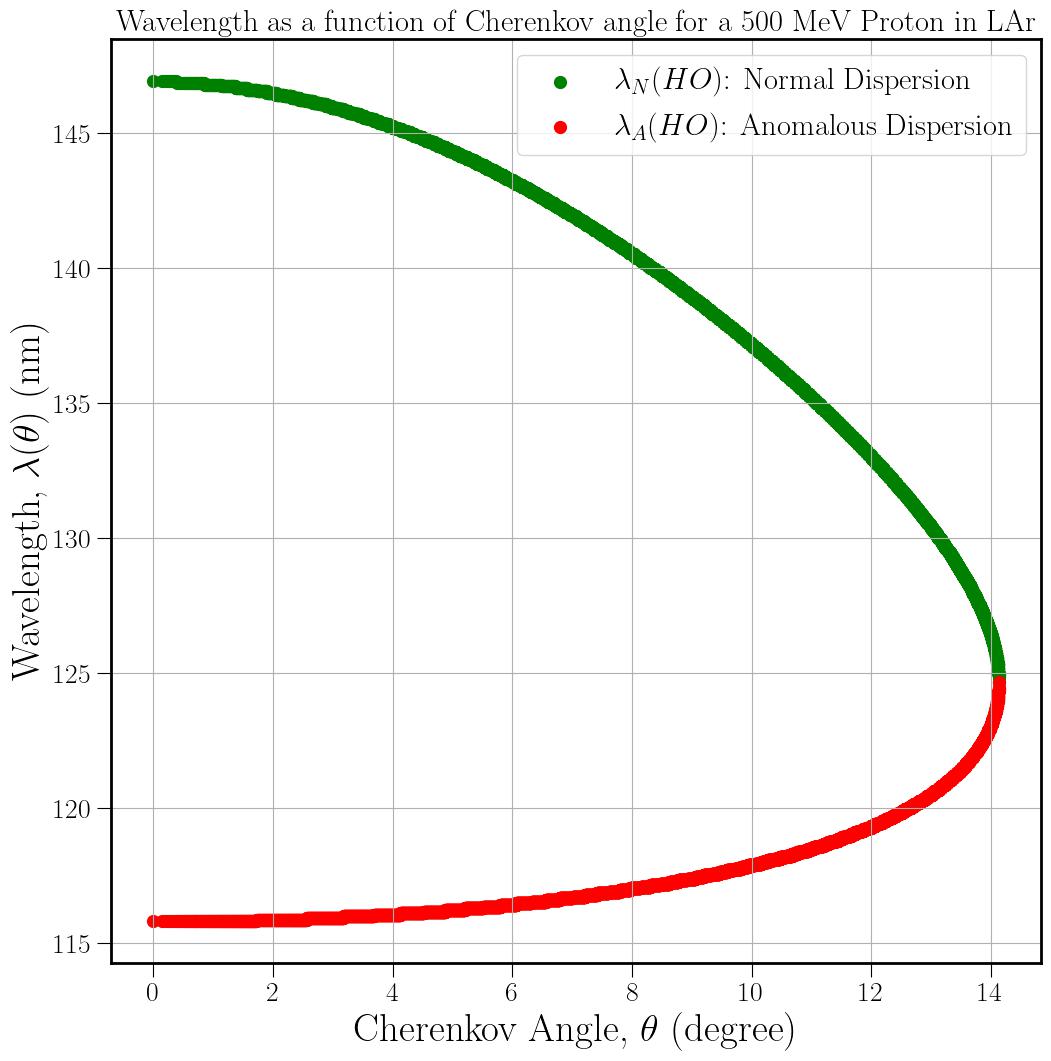}
\caption{500 MeV
\label{f:lambdanaho500MeV}
}
\end{subfigure}
\begin{subfigure}{0.49\textwidth}
\centering
\includegraphics[width=1\linewidth]{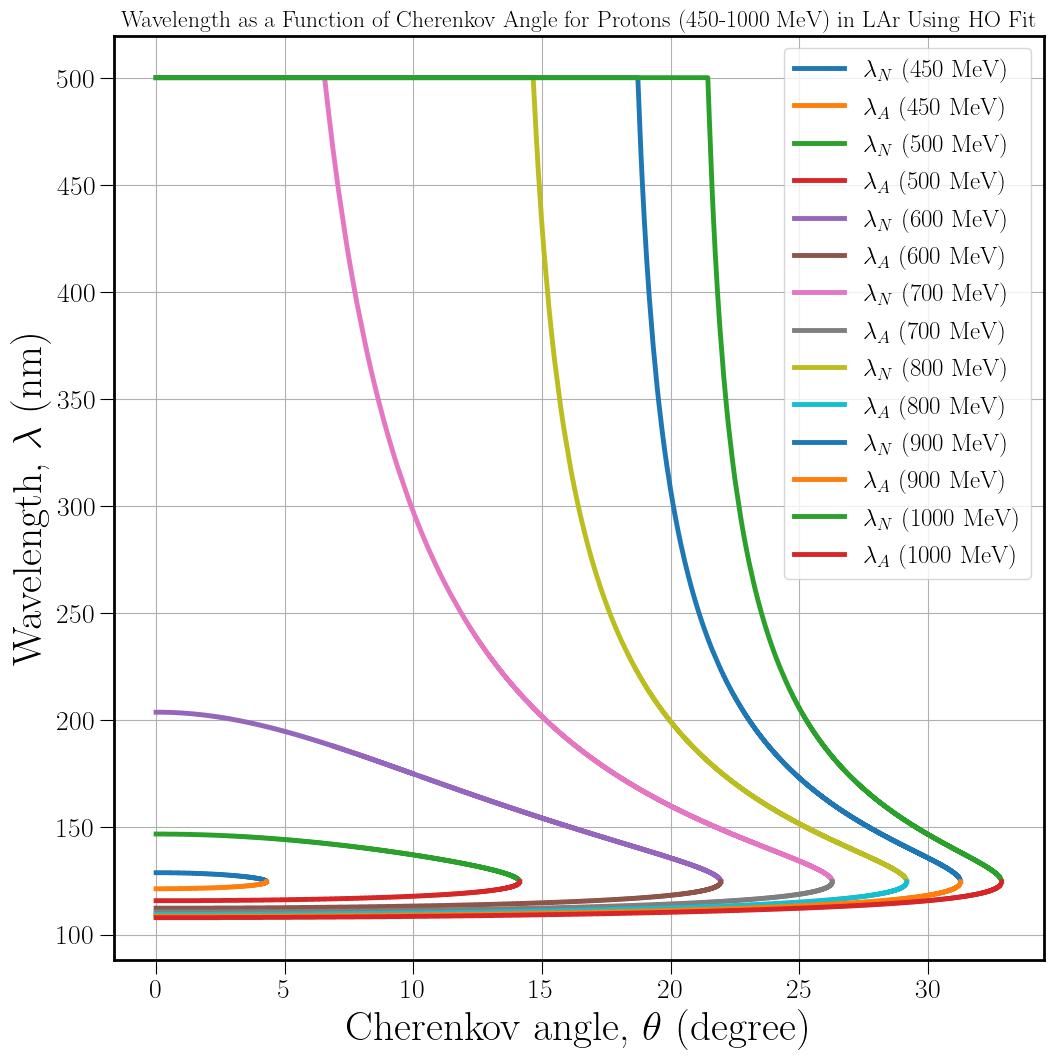}
\caption{450-1000 MeV
\label{f:lambdanahoallt}
}
\end{subfigure}
\caption{Normal and Anomalous Wavelength Solutions Derived Using Our Refractive Index Fit (HO Model) for Protons Travelling in LAr. 
\label{f:lambdanaho}
}
\end{centering}
\end{figure}

Applying the Cherenkov condition to our HO model \eqref{e:nfinalho1} refractive fit gives the wavelength solutions $\lambda(\theta)$ that emit Cherenkov radiation for a given $\beta$ (or equivalently, K.E.) of the proton in LAr,  
\begin{align}   \label{e:nourhofit}
    n_{HO} (\lambda) = a_{0(HO)} + a_{UV(HO)} \left(\frac{\lambda_{UV}^{-2} - \lambda^{-2}}{(\lambda_{UV}^{-2} - \lambda^{-2})^2 + \gamma_{UV}^2 \lambda^{-2}}\right) = \frac{1}{\beta \cos\theta}  \: .
\end{align}
As before, we solve \eqref{e:nourhofit} numerically using the intersection method, exactly as shown in Fig.~\ref{f:intersectionarrayho}.  The resulting wavelength solutions obtained this way are shown in Fig.~\ref{f:lambdanaho}.
                                    

%

Fig.~\ref{f:lambdanaho500MeV} clearly illustrates the appearance of the new, second solution corresponding to anomalous dispersion (red), along with the familiar normal dispersion solution (green) seen previously in Fig.~\ref{f:lambdaGracevsBabicz} for the resonant fits.  Agreeing with the definition of dispersion, both sets of solutions touch at $\lambda_{peak} =$ 124.68 nm where $\lambda_{anomalous} \leq \lambda_{peak}$ and $\lambda_{normal} \geq \lambda_{peak}$. The angular distribution of Cherenkov radiation as seen in Fig.~\ref{f:lambdanaho} can thus reveal the physics of anomalous dispersion which is present in absorptive refractive index fits like \eqref{e:nfinalho1}. 
%
%
Fig.~\ref{f:lambdanahoallt}, on the other hand, shows the evolution of normal and anomalous wavelength solutions for protons with kinetic energies between 450-1000 MeV. We can see that the anomalous wavelength components ($\lambda < \lambda_{peak}$) have similar shapes for different energies, extending to larger angles with increasing energy, although the wavelengths themselves change very little.  In contrast, the wavelength of Cherenkov photons with normal dispersion ($\lambda > \lambda_{peak}$) grows rapidly with increasing energy.  We again cut off the $\lambda_{max}$ at 500 nm to account for finite detector acceptance; this results in the flat horizontal line at $\lambda =$ 500 $nm$ for $T \geq 678 MeV$ shown in Fig.~\ref{f:lambdanahoallt}.         




    

In Ch.~\ref{mathform}, we derived a general formula \eqref{e:angdistdergrace} for the instantaneous angular distribution of Cherenkov radiation, 
%
\begin{align}   \label{e:angdistformulagenho}
    \frac{dN}{d\cos\theta \, dx} &= 2\pi \: \alpha_{EM} \,
    \int_{\lambda_{min}}^{\lambda_{max}} \frac{d\lambda}{\lambda^2}
    \left( 1 - \frac{1}{\beta^2 n^2(\lambda)} \right)
     \delta\left( \cos\theta - \frac{1}{\beta n(\lambda)} \right)   \:,
    \notag \\
    &= \frac{4\pi \alpha_{EM}}{\beta^2} \frac{1}{\lambda_\theta^2} \left(\frac{\sin^2 \theta}{\cos^3 \theta}\right)
    \frac{1}{\left| \frac{dn^2}{d\lambda} \right|_{\lambda= \lambda_{\theta}}}
    \: ,
\end{align} 
where we have multiplied by a factor of $(2\pi)$ compared to Eq.~\eqref{e:angdistdergrace} to write the angular distribution per unit $\cos\theta$.



Eq.~\eqref{e:angdistformulagenho} gives the number density of Cherenkov photons emitted at an angle $\theta$ from a given solution $\lambda_\theta$ of the Cherenkov condition \eqref{e:nourhofit}.  Since there are two such solutions $\lambda_\theta$ -- one $\lambda_N(\theta)$ from normal dispersion, the other $\lambda_A(\theta)$ from anomalous -- we must add the two contributions together:
%


\begin{subequations}
\label{e:angdistformulhonormanom}
\begin{align}   \label{e:angdistformulhonorm}
    \mathrm{Normal:} \qquad
    \frac{dN}{dx \: d\cos\theta} &= \frac{4\pi\alpha_{EM}}{\beta^2}
    \frac{1}{\lambda_{N}^2(\theta)}
    \left( \frac{\sin^2 \theta}{\cos^3 \theta} \right)
    \frac{1}{\left|\frac{dn^2}{d\lambda}\right|}_{\lambda = \lambda_{N}(\theta)}    \: ,
\end{align}
\begin{align}   \label{e:angdistformulhoanom}
    \mathrm{Anomalous:} \qquad
    \frac{dN}{dx \: d\cos\theta} &= \frac{4\pi\alpha_{EM}}{\beta^2}
    \frac{1}{\lambda_{A}^2(\theta)}
    \left( \frac{\sin^2 \theta}{\cos^3 \theta} \right)
    \frac{1}{\left|\frac{dn^2}{d\lambda}\right|}_{\lambda = \lambda_{A}(\theta)}    \: ,
\end{align}
\end{subequations}
where the normal and anomalous wavelength solutions are plotted in Fig.~\ref{f:lambdanaho}.  As usual, the integrated angular distribution $\frac{dN}{d\cos\theta}$ is obtained by further integration over the range $x$ of the proton.


    \begin{figure}[p]
    \centering
    \begin{subfigure}{.5\textwidth}
      \centering
      \includegraphics[width=1\linewidth]{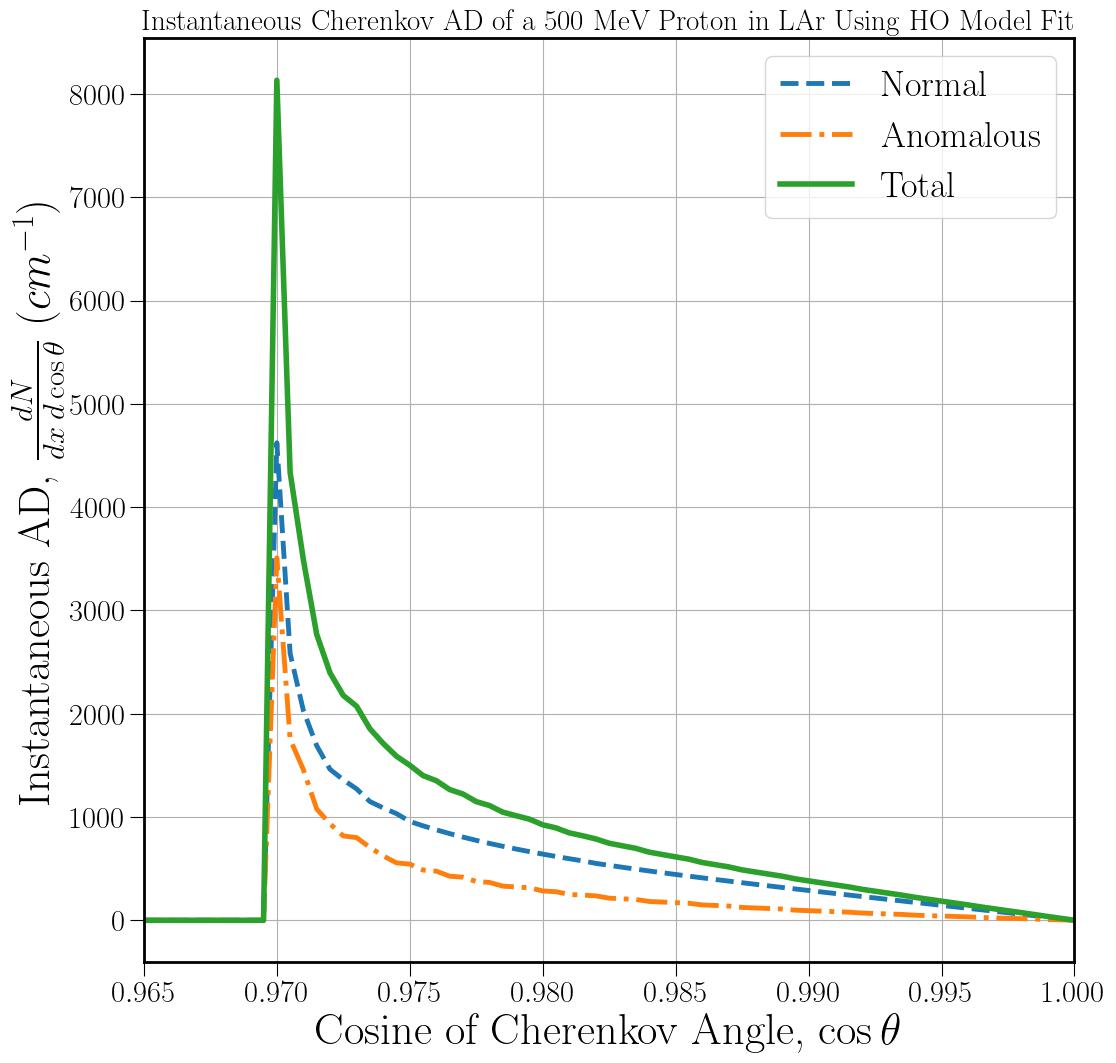}
      \caption{500 MeV}
      \label{f:ADinstHO500MeV}
    \end{subfigure}%
    \begin{subfigure}{.5\textwidth}
      \centering
      \includegraphics[width=1\linewidth]{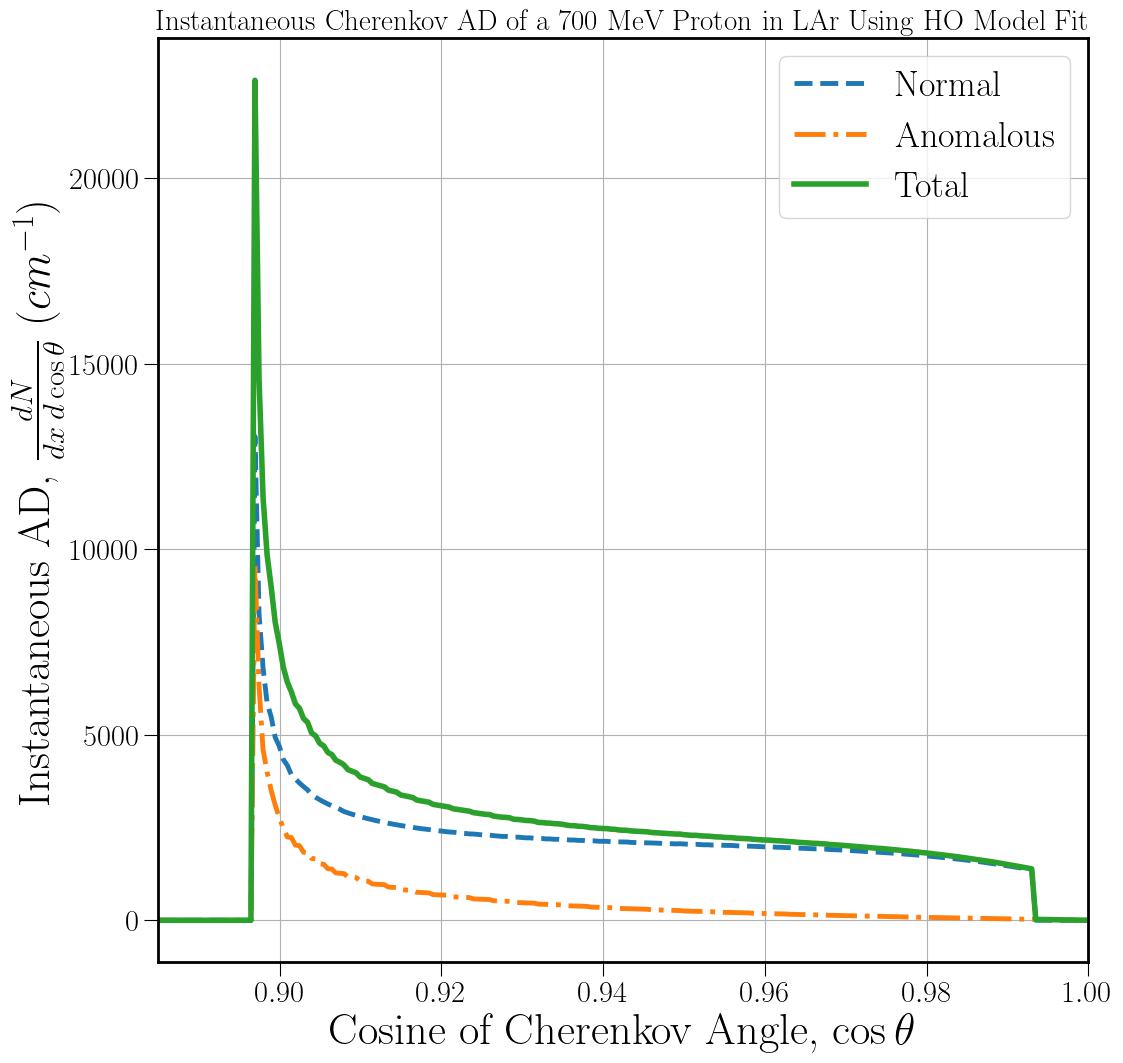}
      \caption{700 MeV}
      \label{f:ADinstHO700MeV}
      
    \end{subfigure}

    \medskip

    \begin{subfigure}{.5\textwidth}
      \centering
      \includegraphics[width=1\linewidth]{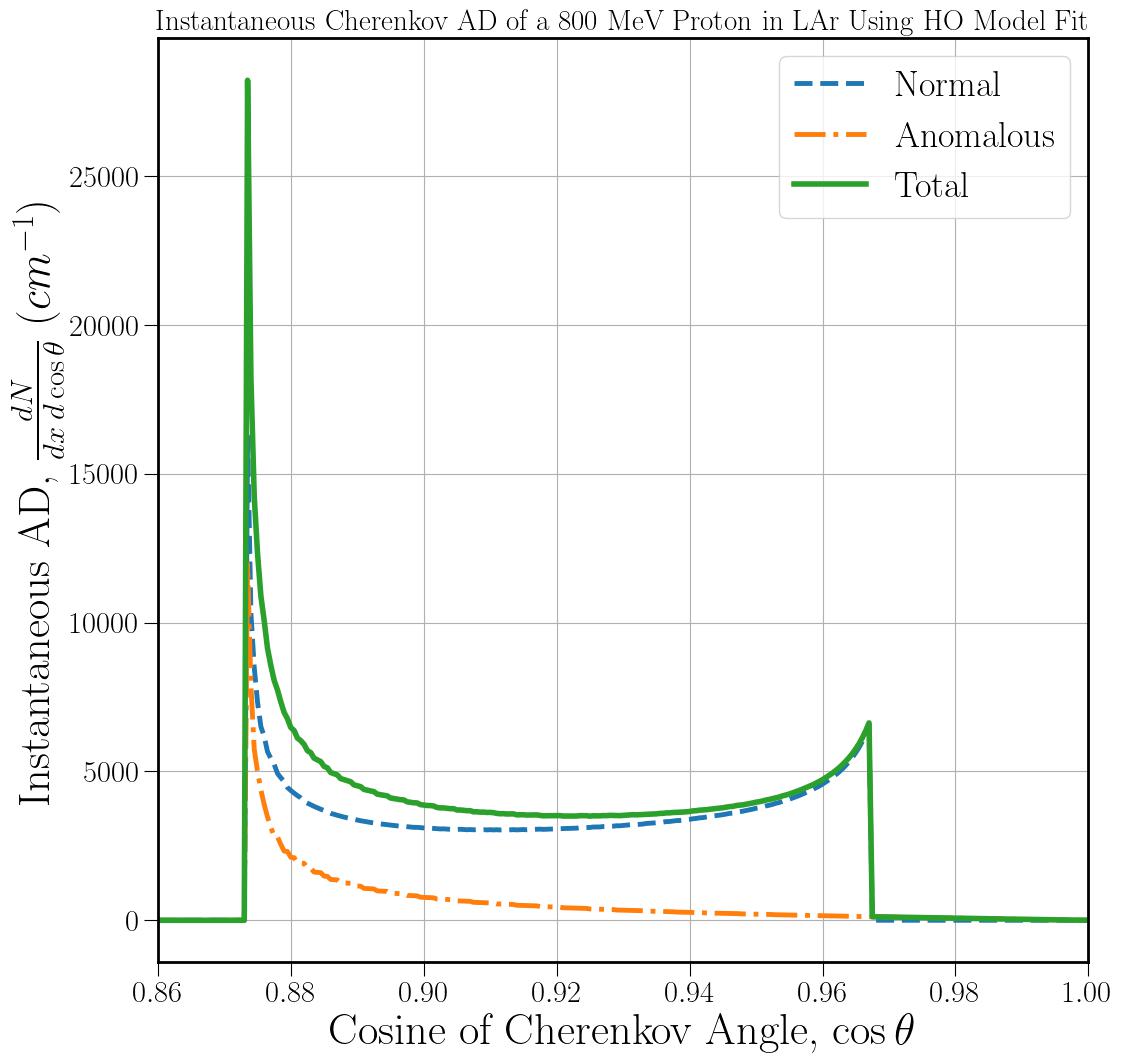}
      \caption{800 MeV}
      \label{f:ADinstHO800MeV}
    \end{subfigure}%
    \begin{subfigure}{.5\textwidth}
      \centering
      \includegraphics[width=1\linewidth]{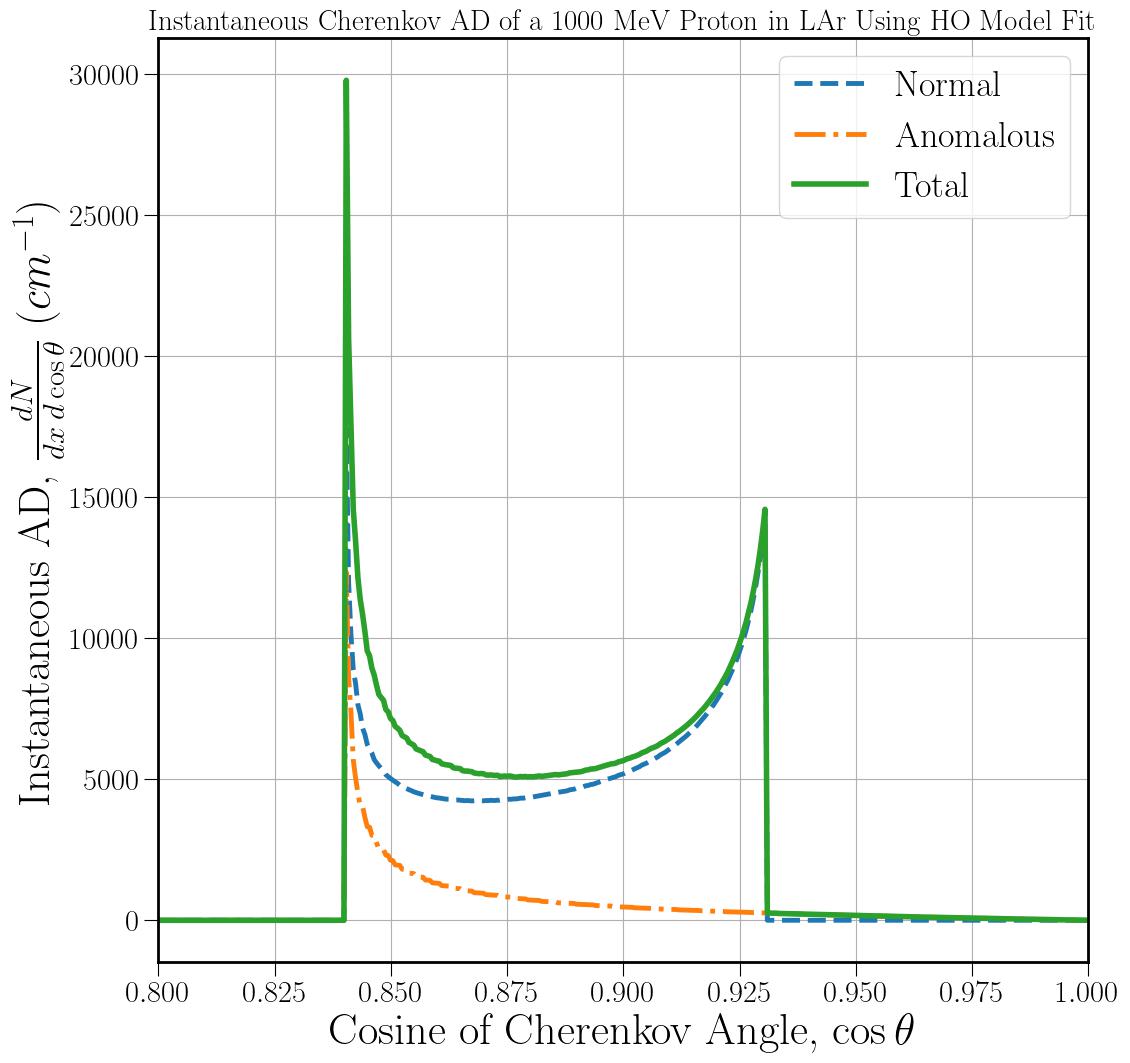}
      \caption{1000 MeV}
      \label{f:ADinstHO1000MeV}
      
    \end{subfigure}
    \caption{Instantaneous Cherenkov AD calculated using Absorptive (HO Model) Fit}
    \label{f:ADHOinst} 
    \end{figure}

The instantaneous angular distributions (IAD) $dN/dx d\cos \theta$, broken down into the normal component, anomalous component, and total, are shown in Fig.~\ref{f:ADHOinst} for various kinetic energies ranging from 500 MeV to 1000 MeV. We instantly notice that the absorptive model for the refractive index generates a significantly different angular pattern compared to the resonant fits studied in Chap.~\ref{mathform}.  Unlike those smooth distributions shown in Fig.~\ref{f:IADGrace500MeV} which cover the whole angular range $0 \leq \theta \leq \tfrac{\pi}{2}$, the IAD for the HO model has a hard cutoff at a maximum theta (i.e. minimum $\cos\theta$) for a given energy $T$.  For example, in Fig.~\ref{f:ADinstHO500MeV}, the IAD of a 500 MeV proton clearly has a hard cutoff near $\cos \theta \geq 0.97$ i.e. for $\theta \leq 14 \degree$, confining the Cherenkov photons to fall between 0 to 14 degrees.  Moreover, unlike the IAD for the resonant fits which peaked at small $\theta$, the IAD from the absorptive HO model possesses a sharp peak precisely at this $\theta_{max}$ (or $(\cos\theta)_{min}$ ).  This divergence can be understood from the formula \eqref{e:angdistformulagenho}: the finite maximum for $\theta$ corresponds to the finite maximum of the refractive index $n(\lambda)$, and at the peak of $n(\lambda)$, the derivative vanishes.  (See the dispersion parameter plotted in Fig.~\ref{f:derivativenjackson}.)

All the plots in Fig.~\ref{f:ADHOinst} indicate the normal component with a blue dashed line and the anomalous component with an orange dash-dotted line. For a 500 MeV proton (Fig.~\ref{f:ADinstHO500MeV}), the normal and anomalous components contribute about equally to the total, and both become singular at $\theta_{max}$ as both approach the same peak in $n(\lambda)$.  But when the proton energy is increased to 700 MeV, the normal AD increases significantly in yield and dominates the interior of the Cherenkov cone $\theta \leq \theta_{max}$).  For energies above $678$ Mev, a new hard cutoff appears at a \textit{maximum} value of $\cos\theta$, corresponding to a \textit{minimum value} $\theta_{min}$ for Cherenkov emission.  As a result, the IAD of 700 MeV proton does not smoothly go to zero, filling the full interior of the cone $\cos \theta \rightarrow 1$ as it did for the 500 MeV proton in Fig.~\ref{f:ADinstHO500MeV}.  This $\theta_{min}$ cutoff is the same as the one that appeared for the instantaneous angular distributions above $\sim 700$ MeV (see Fig.~\ref{f:IADGraceallts}).  So instead of radiating in a filled cone that ranges from $\theta_{min} = 0$ to $\theta_{max}$, a 700 MeV proton will radiate in a conic band that ranges from a non-zero $\theta_{min} \sim 6.5\degree$ to $\theta_{max} \sim 26.29\degree$. As we go higher in T, this effect is even more prominent, with the cutoff at $\theta_{min}$ growing larger and larger.  The intensity at this $\theta_{min}$ cutoff also grows with increasing energy. 

Interestingly, the anomalous part (orange dash-dotted curve) still behaves the same as it did for $T = 500$ MeV, filling in the interior of the Cherenkov cone with no $\theta_{min}$, and it is only the normal component (blue dashed curve) which is sensitive to the $\lambda_{max}$ cutoff.  This occurs because the anomalous Cherenkov photons are emitted in the UV region with $\lambda < \lambda_{peak}$, so they are insensitive to IR cutoff effects at $\lambda_{max} = 500$ nm.  This would suggest that interesting experimental separation could be achieved between the normal and anomalous contributions to the Cherenkov pattern through their concentration at different angular regions.





\begin{figure}[p]
\centering
\begin{subfigure}{.5\textwidth}
\centering
\includegraphics[width=1\linewidth]{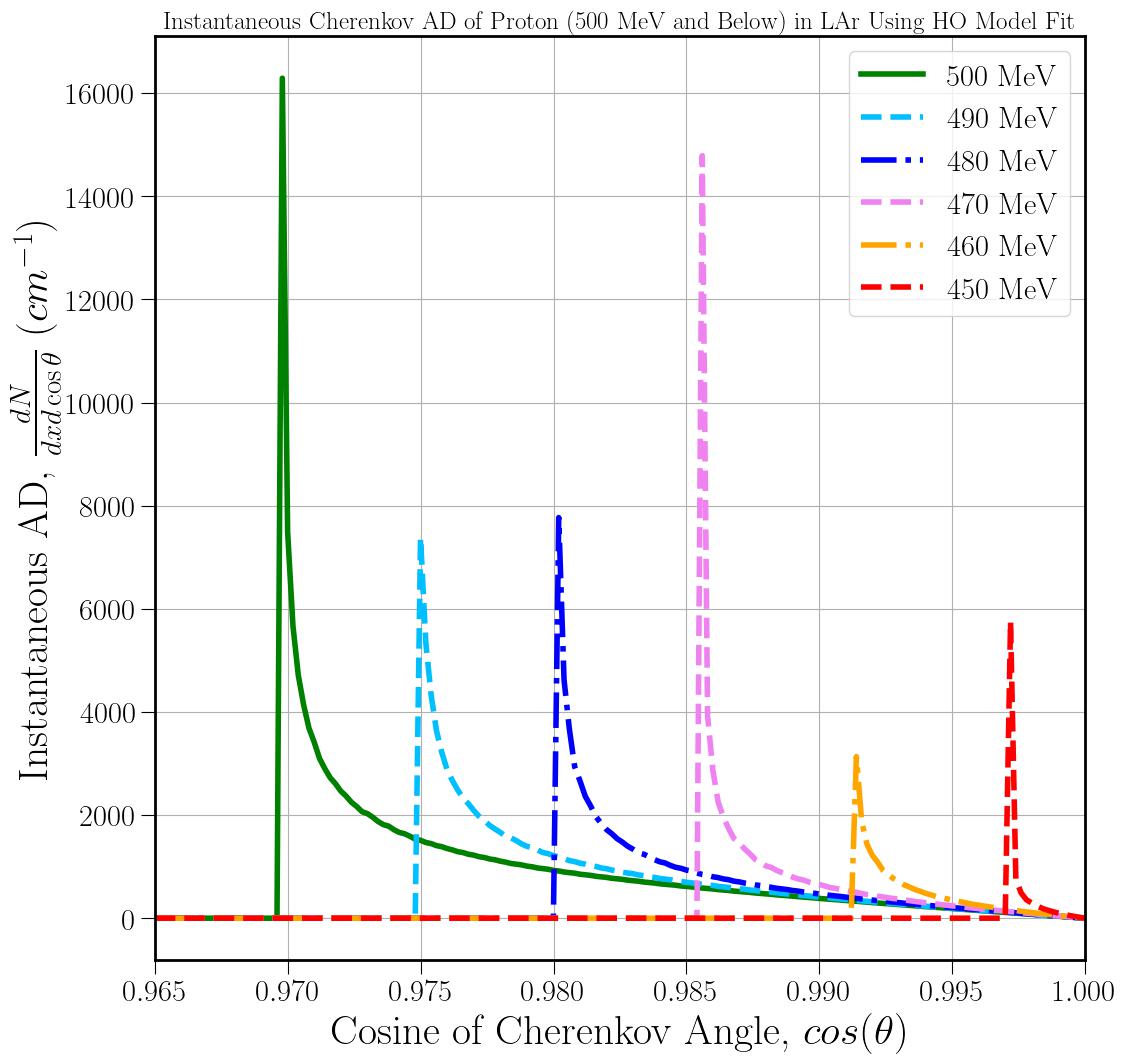}
\caption{IAD for $T < 500$ MeV}
\label{f:ADinstHO500nblw}
\end{subfigure}%
\begin{subfigure}{.5\textwidth}
\centering
\includegraphics[width=1\linewidth]{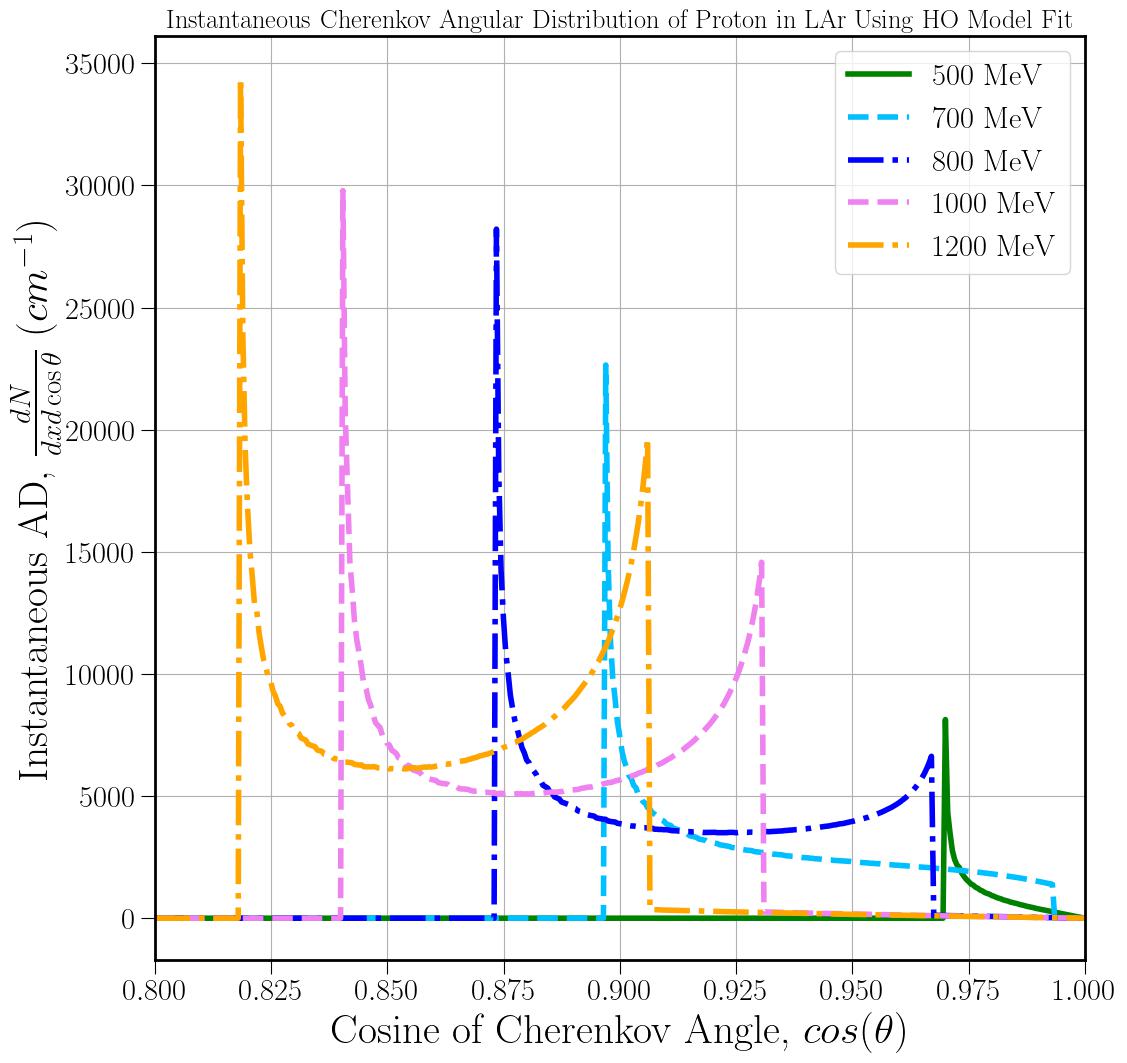}
\caption{IAD for $1200 \: \mathrm{MeV} > T > 500 \: \mathrm{MeV}$}
\label{f:ADinstHO500nabv}
\end{subfigure}

\medskip

\begin{subfigure}{.5\textwidth}
\centering
\includegraphics[width=1\linewidth]{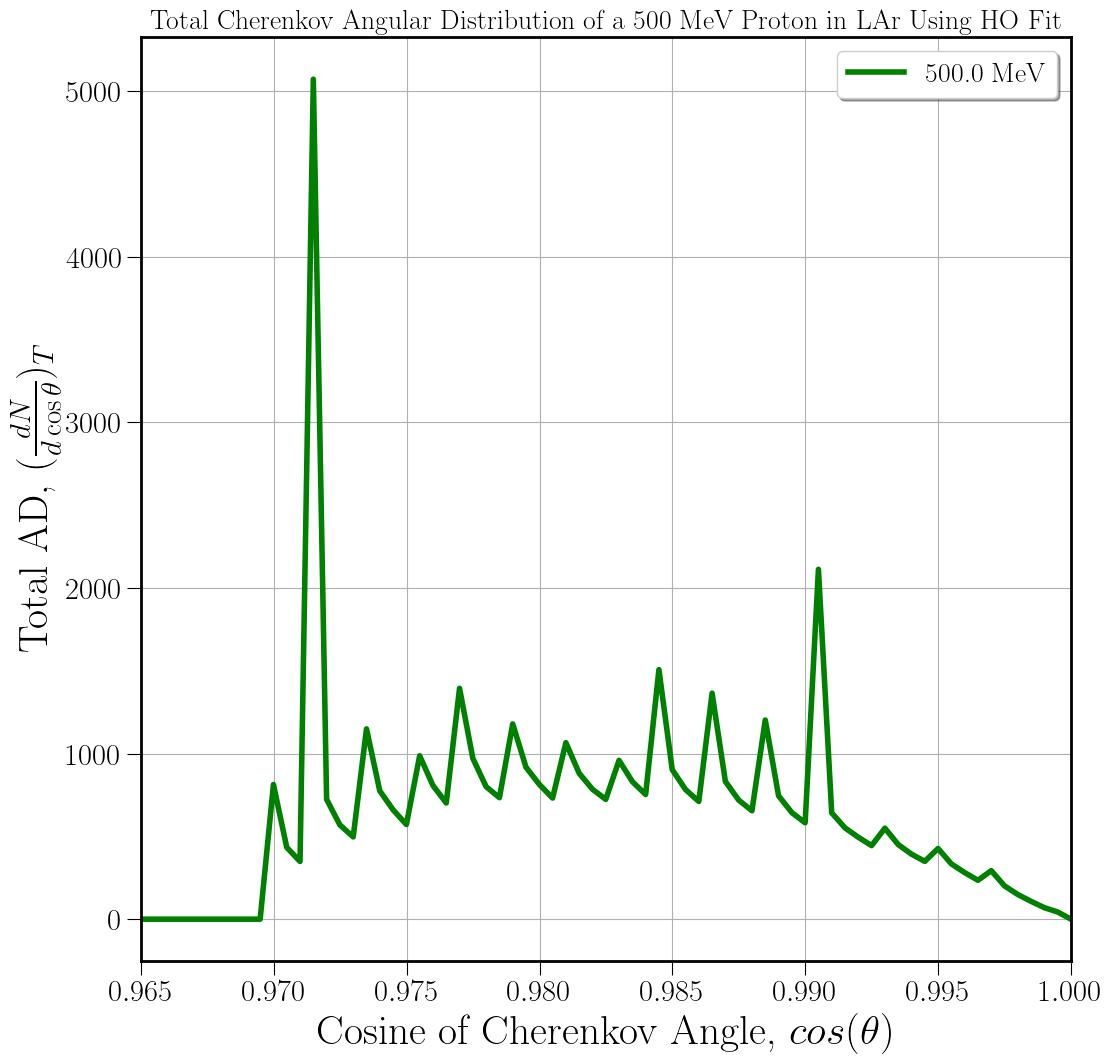}
\caption{Total AD of 500 MeV (low res)}
\label{f:adtot500MeVnblw}
\end{subfigure}%
\begin{subfigure}{.5\textwidth}
\centering
\includegraphics[width=1\linewidth]{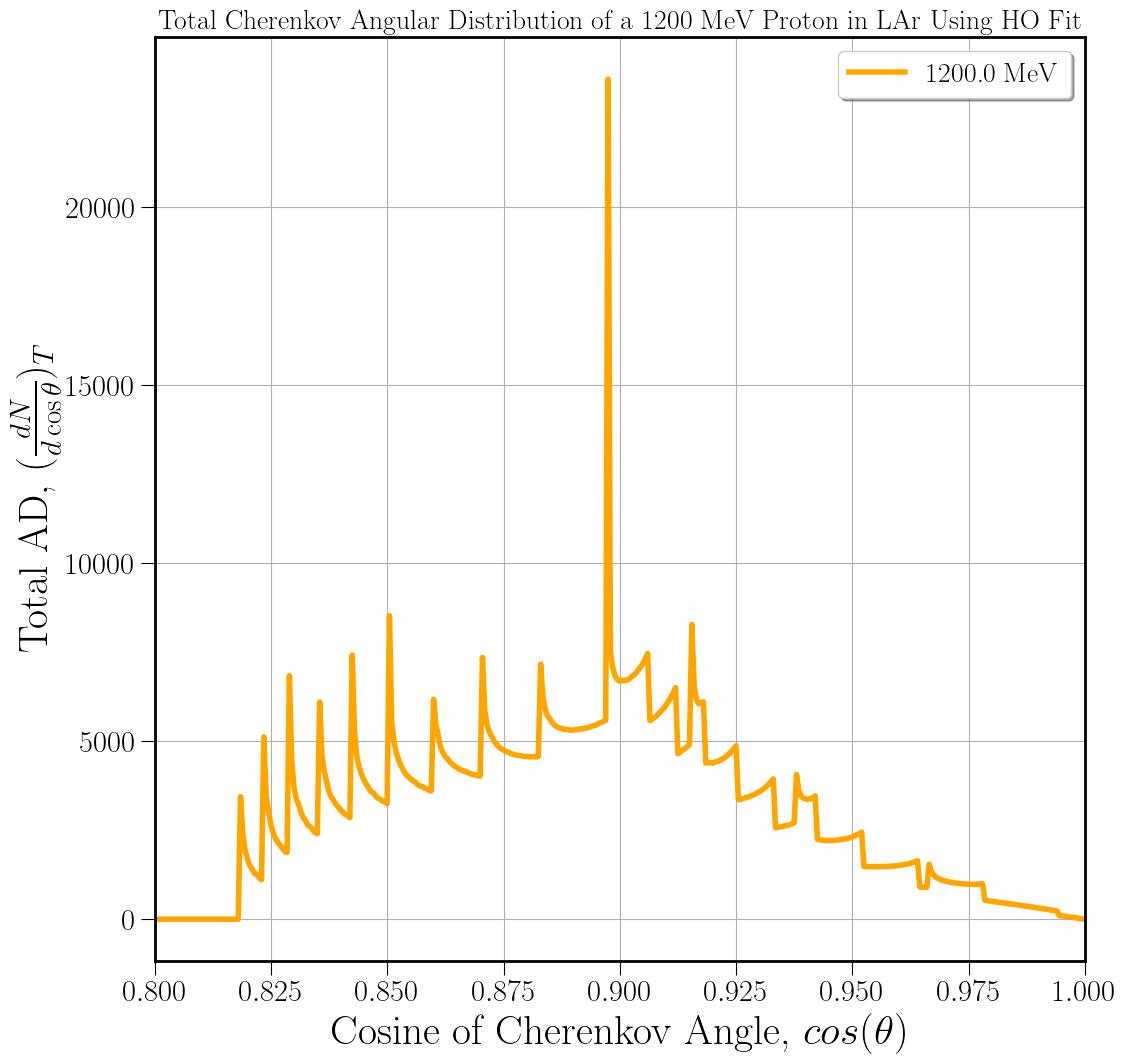}
\caption{Total AD of 1200 MeV (low res)}
\label{f:adtot1200MeVnblw}
\end{subfigure}
\caption{Instantaneous angular distributions (IAD) for a given proton initial energy (top row) and a coarse-grained calculation of the total angular distribution (AD) (bottom row) with a step size $dx = 1 \: \mathrm{cm}$.}
\label{f:ADinsttointeg} 
\end{figure}

The presence of spiky features in the IAD makes the calculation of the integrated angular distribution $\frac{dN}{d\cos\theta}$ numerically challenging, as illustrated in Fig.~\ref{f:ADinstHO500nabv}.  Here show explicitly the instantaneous angular distributions (IAD) for a few specific energies to illustrate how they build up to the total integrated angular distribution (AD).  For instance, in Fig.~\ref{f:ADinstHO500nblw} we show the IAD for kinetic energies $T < 500$ MeV in steps of $10$ MeV which contribute to the total AD.  The IAD for each energy below $500$ MeV is similar in shape: sharply peaked at the outer edge of the Cherenkov cone (minimum value of $\cos\theta$) and skew right, filling the whole interior of the cone with photons.  Adding up many such skewed distributions (Fig.~\ref{f:adtot500MeVnblw}) produces an integrated AD which smooths out the peaks, but remains skewed with most photons concentrated toward the outer edge.  

\begin{figure} [p]
\centering
\begin{subfigure}{.48\textwidth}
  \centering
  \includegraphics[width=1\linewidth]{images/IAD_Cherenkov_HOmodel_500MeV.jpg}
  \caption{Instantaneous AD (500 MeV) 
  \label{f:adho500mevinst}
  }
  \end{subfigure}%
\begin{subfigure}{.48\textwidth}
    \centering
    \includegraphics[width=1\textwidth]{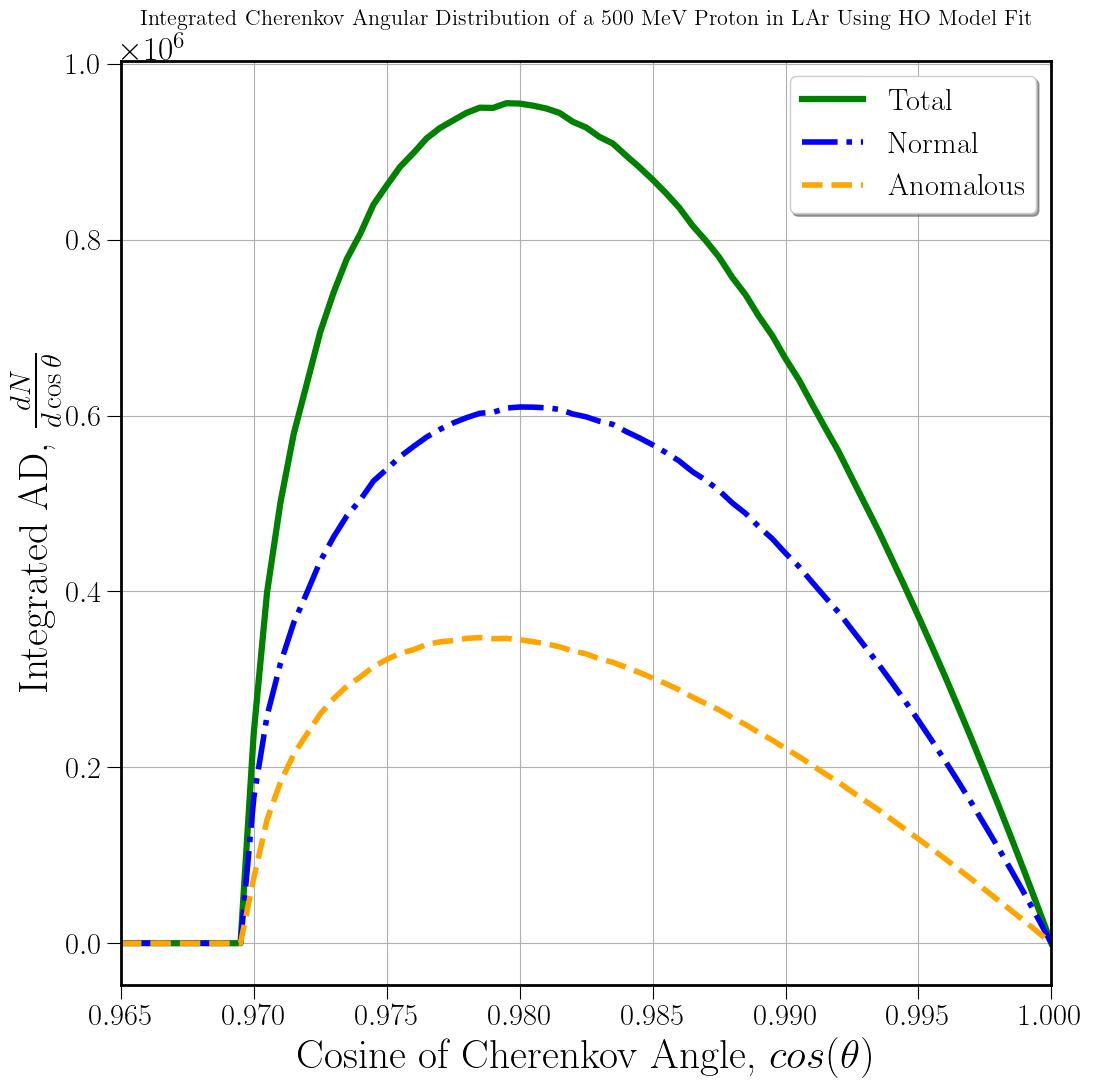}
    \caption{Integrated AD (500 MeV)
    \label{f:adho500mevinteg}
    }
    \end{subfigure}
%

%
\medskip

\begin{subfigure}{.49\textwidth}
  \centering
  \includegraphics[width=1\linewidth]{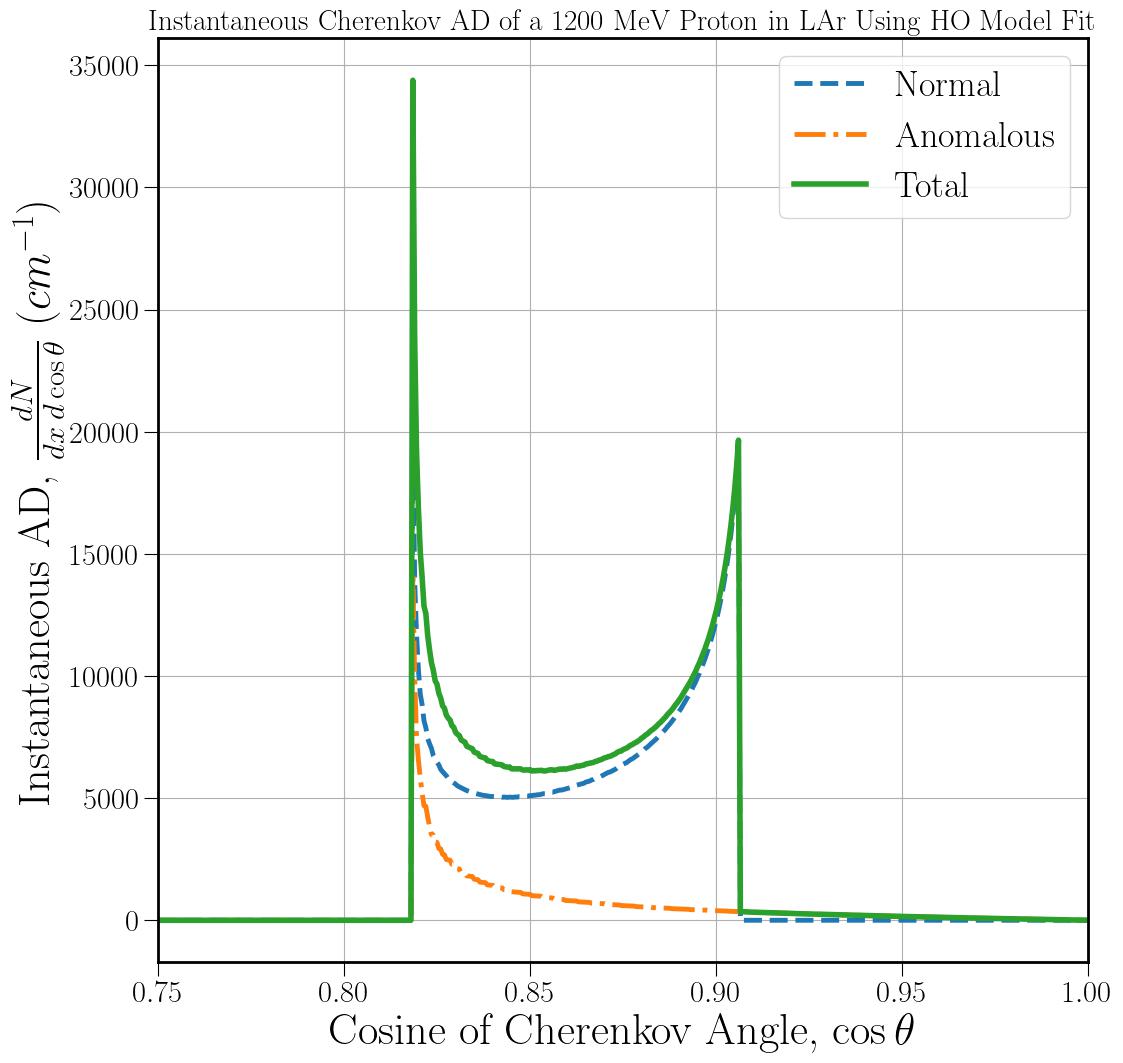}
  \caption{Instantaneous AD (1200 MeV) 
  \label{f:adho1200mevinst}
  }
  \end{subfigure}%
\begin{subfigure}{.48\textwidth}
    \centering
    \includegraphics[width=1\textwidth]{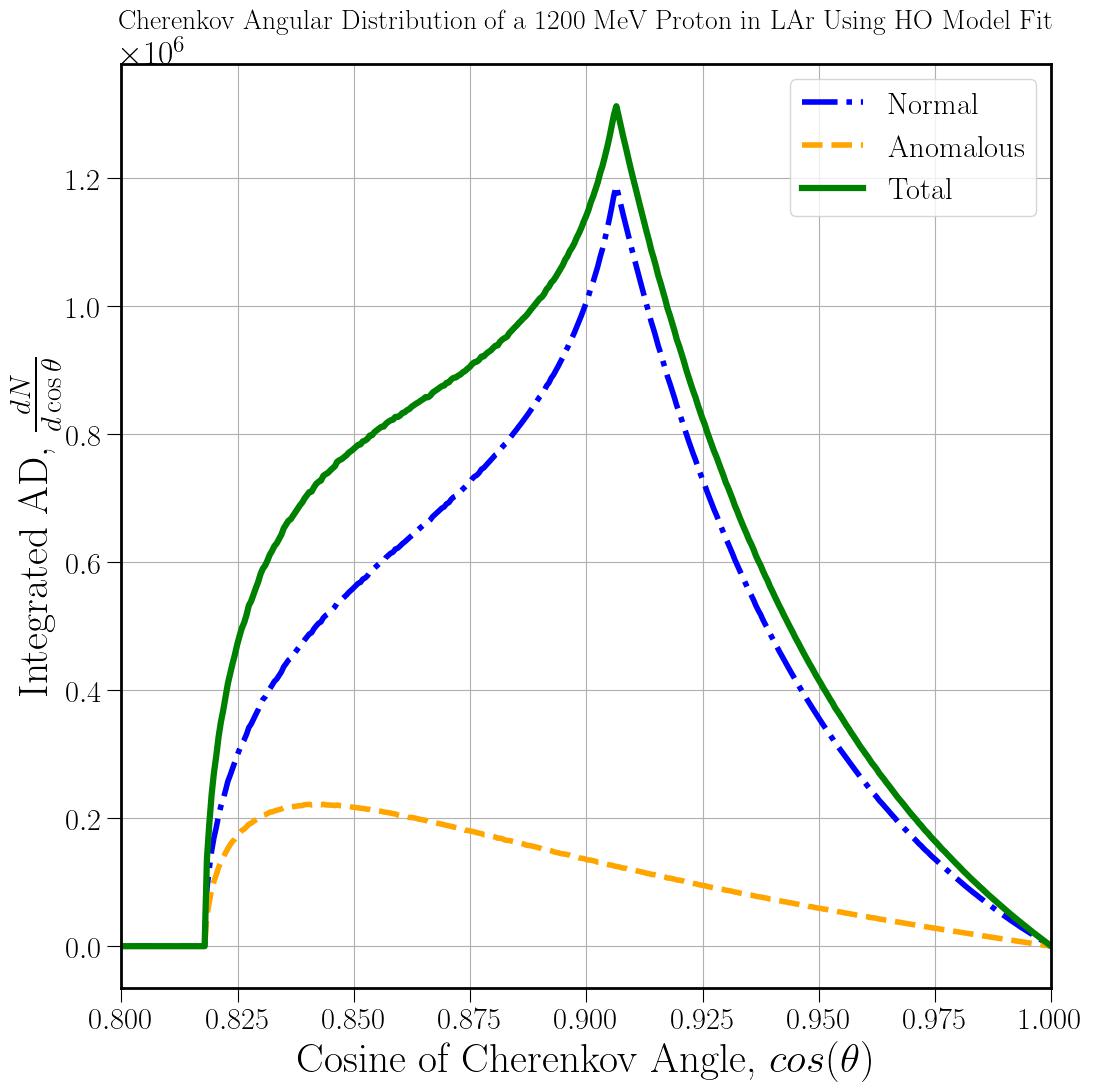}
    \caption{Integrated AD (1200 MeV)
    \label{f:adho1200mevinteg}
    }
    \end{subfigure}
\caption{Normal and Anomalous contributions to the instantaneous (left) and total (right) angular distributions for a 500 MeV ($\beta= 0.7576$) (top) and 1200 MeV ($\beta= 0.899$) (bottom) Proton in LAr calculated using harmonic oscillator (HO) model fit of refractive index.
\label{f:adho1200mev} }
\end{figure}

In contrast, for $T > 500 $ MeV, the shape of the IAD begins to change substantially, as shown in Fig.~\ref{f:ADinstHO500nabv}.  With the entire wavelength range $\lambda_{UV} < \lambda < \lambda_{max}$ saturated, the IAD moves to larger angles with increasing energy, but no longer fills in the entire interior of the cone.  Instead, a second peak develops on the inner edge of the cone (maximum $\cos\theta$), and the new double-peak structure moves as a function of energy.  Consequently, for protons with initial kinetic energies far above this threshold, the total AD reflects a distribution that is not only skewed towards larger angles but changes slope to exhibit a sharper cutoff on the cone interior (Fig.~\ref{f:adtot1200MeVnblw}).  This is a direct consequence of the IR cutoff $\lambda_{max}$ and the kinematic boundary $\cos\theta_{max}$ of the IAD.

When the above procedure is carried out with fine step sizes ($dx = 0.01 \: cm$, $d(\cos\theta) = 5 \times 10^{-4}$), the spiky features of the individual IADs are washed out, resulting in the smooth integrated angular distributions $dN/d\cos \theta$ shown in Fig.~\ref{f:adho1200mev}.  For the integrated AD of a $500$ MeV proton (Fig.~\ref{f:adho500mevinteg}), the spikes seen in the IAD for individual energies in Fig.~\ref{f:adtot500MeVnblw} are smoothed out, resulting in a continuous distribution which fills the whole interior of the cone $0 < \cos\theta < 1 / \beta n_{peak}$ and is skewed toward the cone's exterior (smaller $\cos\theta$).

Similarly, the integrated AD of a 1200 MeV proton (Fig.~\ref{f:adho1200mevinteg}) smooths most of the individual spikes seen in Fig.~\ref{f:adtot1200MeVnblw}, except for a sharp kink at $\cos \theta \approx 0.907$.  This kink coincides exactly with the innermost edge of the IAD of the initial 1200 MeV proton (Fig.~\ref{f:adho1200mevinst}).  As the proton energy decreases from 1200 MeV, the inner edge of the Cherenkov cone moves inward, always tracking the boundary condition $\cos\theta = 1 / \beta n(\lambda_{max})$.  Thus, the sharp decrease of the integrated AD between $0.907 \leq \cos\theta \leq 1$ exactly tracks the IR boundary of the proton's Cherenkov radiation across its entire trajectory.  Moreover, this IR contribution occurs entirely from the ``normal dispersion'' part of the refractive index, dominating the total AD in the interior of the cone.


\begin{figure} [p]
\centering
\begin{subfigure}{.5\textwidth}
  \centering
  \includegraphics[width=1\linewidth]{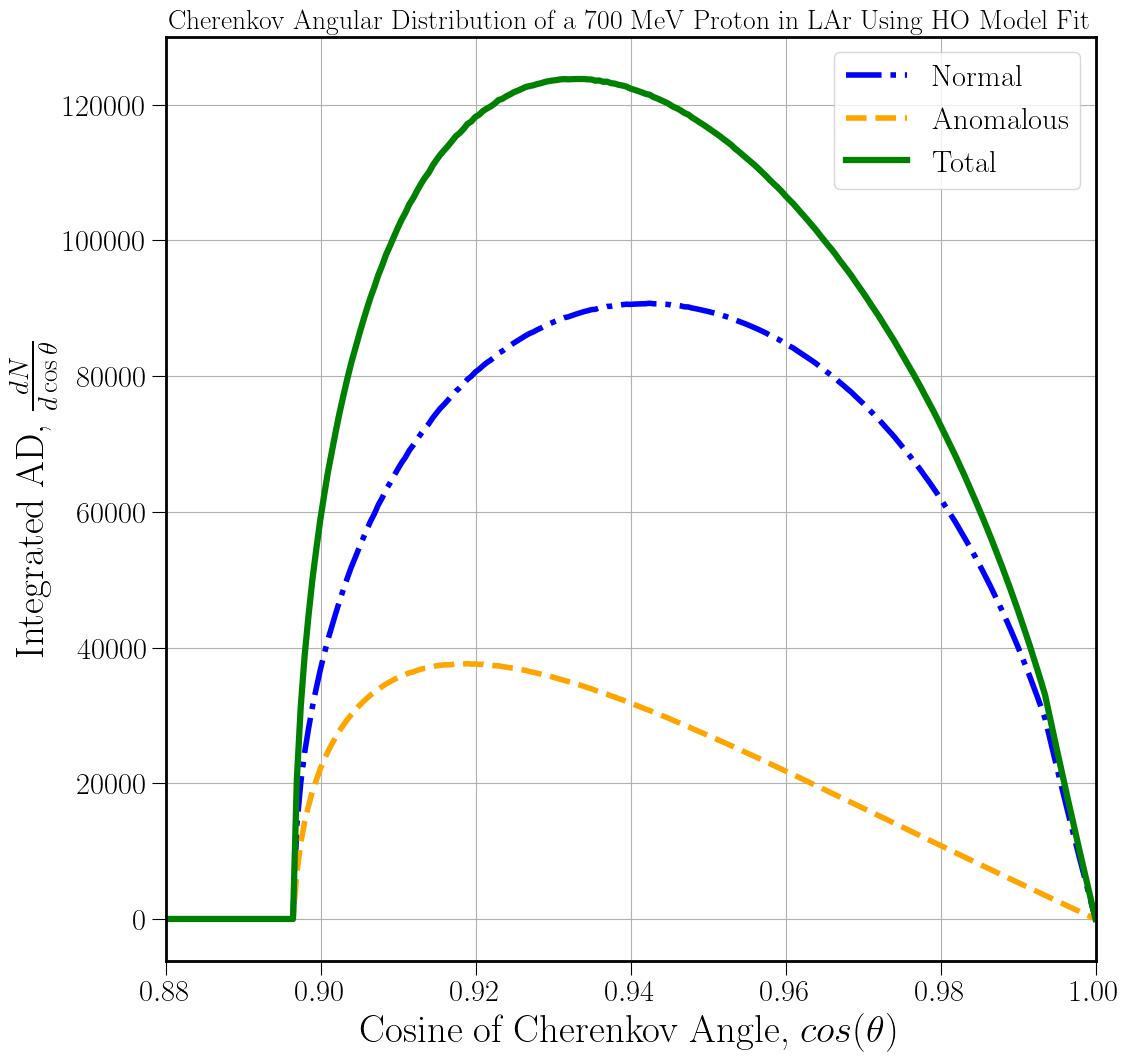}
  \caption{700 MeV}
  \label{f:ADtotHO700MeV}
\end{subfigure}%
\begin{subfigure}{.5\textwidth}
  \centering
  \includegraphics[width=1\linewidth]{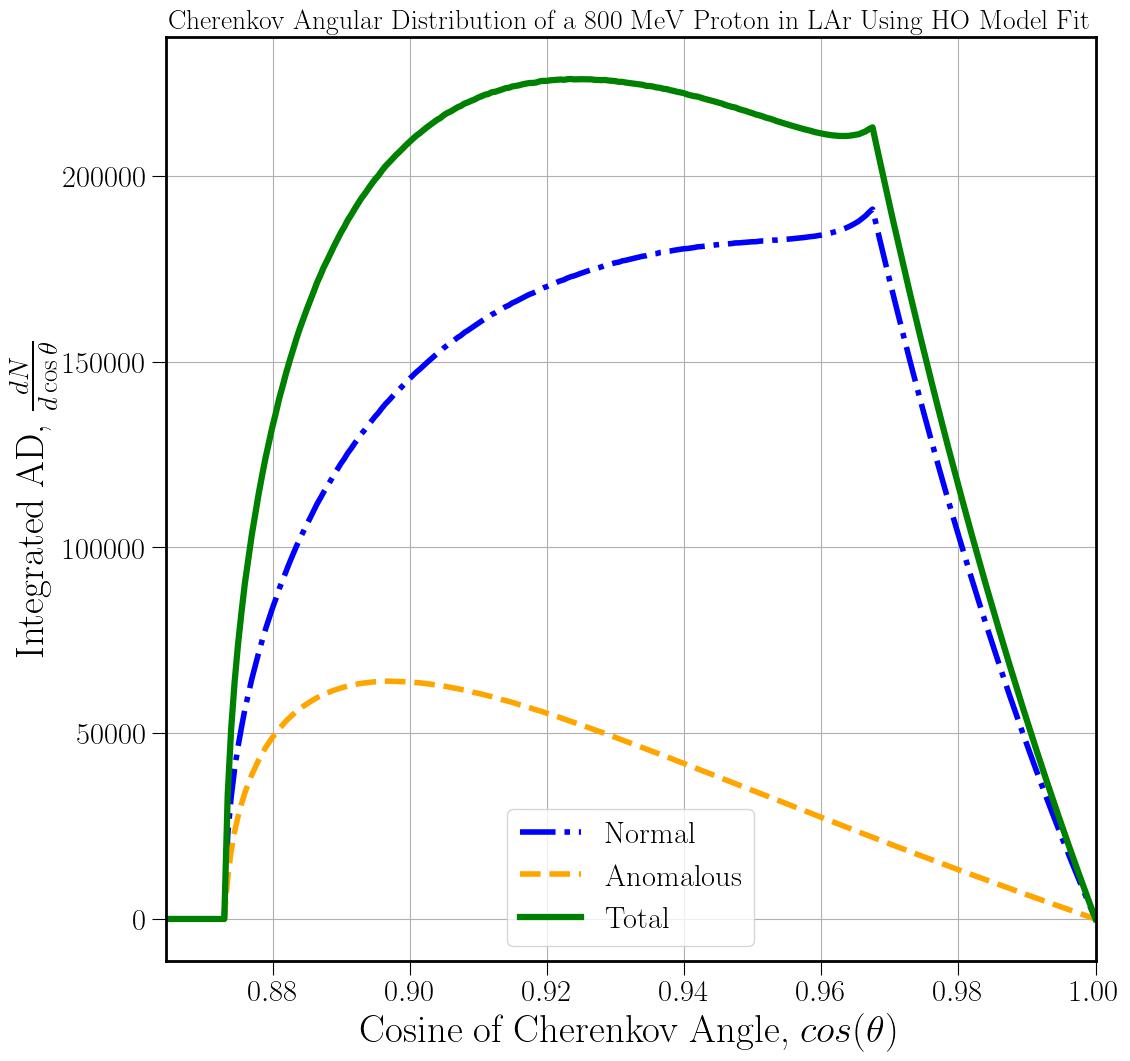}
  \caption{800 MeV}
  \label{f:ADtotHO800MeV}
  
\end{subfigure}

\medskip

\begin{subfigure}{.5\textwidth}
  \centering
  \includegraphics[width=1\linewidth]{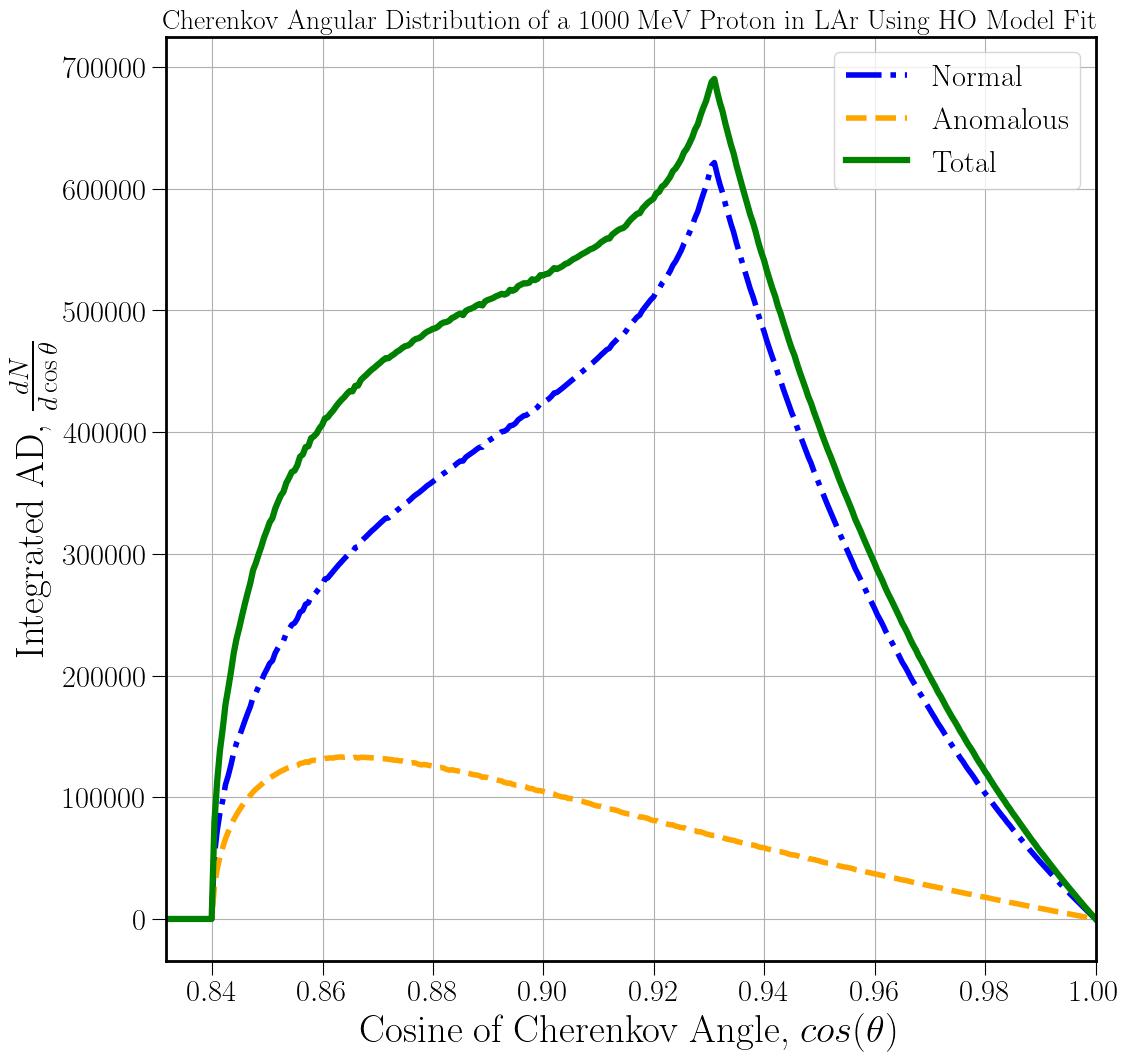}
  \caption{1000 MeV}
  \label{f:ADtotHO1000MeV}
\end{subfigure}%
\begin{subfigure}{.5\textwidth}
  \centering
  \includegraphics[width=1\linewidth]{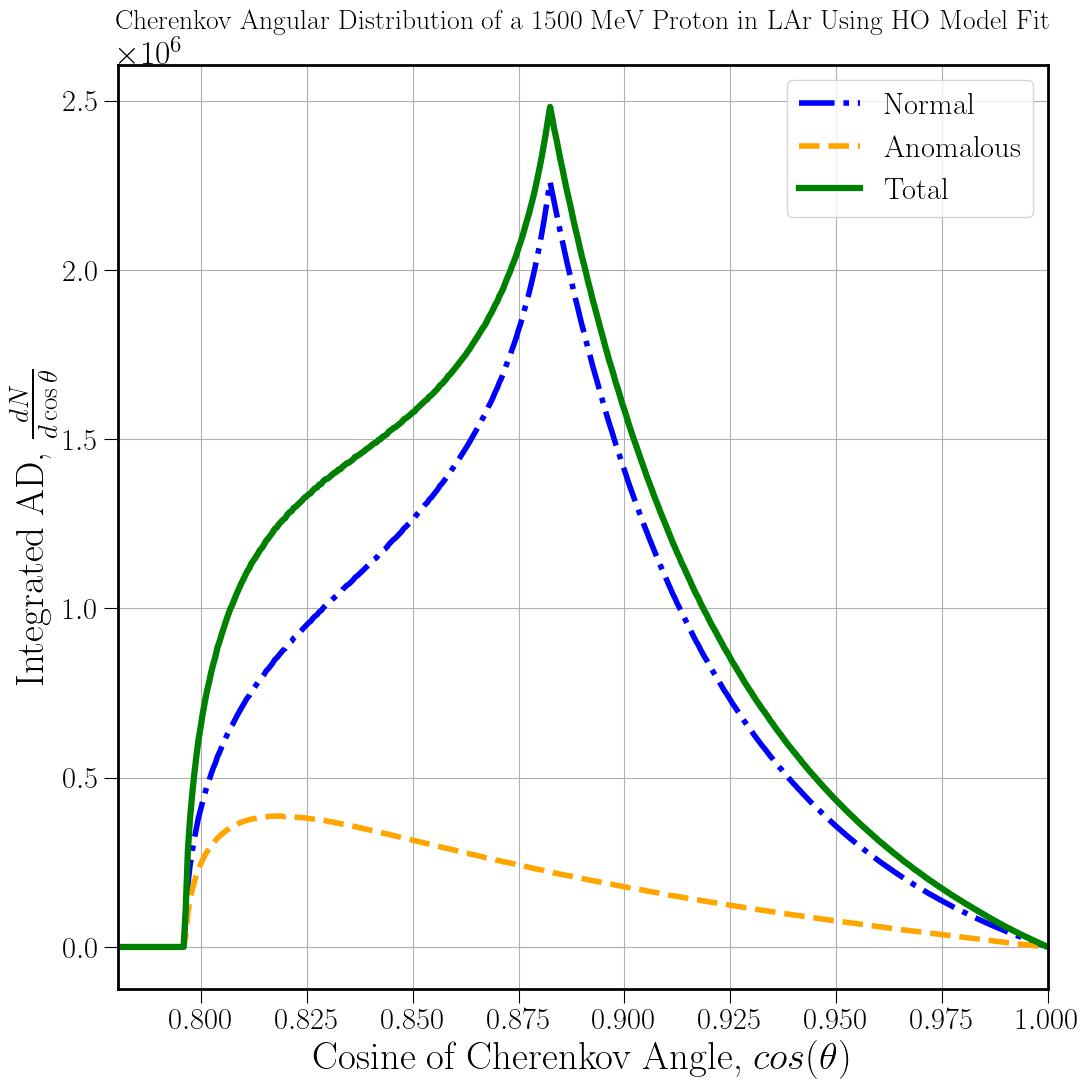}
  \caption{1500 MeV}
  \label{f:ADtotHO1500MeV}
  
\end{subfigure}
\caption{Integrated Cherenkov AD (total, normal, anomalous) calculated using absorptive (HO model) Fit}
\label{f:ADtotHO} 
\end{figure}

We summarize the results for the integrated AD in Fig.~\ref{f:ADtotHO} for various initial energies between 700 and 1500 MeV, along with the separate contributions from the normal (blue dash-dotted curve) and anomalous (orange dashed curve) components of dispersion. In Fig.~\ref{f:ADtotHO700MeV}, the integrated AD of a 700 MeV proton has similar shape as that for a 500 MeV proton shown in Fig.~\ref{f:adho500mevinteg} previously, just extending the Cherenkov cone to larger angles.  However, at these energies the cone is fully filled: $\theta_{min} = 0$ for all $T< 700 \: \mathrm{MeV}$.  The normal component dominates the total AD over the full angular region and is concentrated toward the middle of the cone, while the anomalous component is skewed towards larger angles (lower $\cos \theta$). 

For $T \geq 750$ MeV, the normal component of the AD changes its shape as in Fig.~\ref{f:ADtotHO800MeV}.  Due to the saturation of the IR wavelength range, the normal component develops a peak at larger values of $\cos \theta$ (smaller $\theta$); the anomalous AD, however, is insensitive to the IR wavelength region and maintains its shape over all proton energies.  These qualitative features persist for higher proton energies (1000 MeV: Fig.~\ref{f:ADtotHO1000MeV}, 1500 MeV: Fig.~\ref{f:ADtotHO1500MeV}), with the overall distribution being stretched to larger and larger angles (smaller $\cos\theta$).  The relative height of the ``IR-onset peak'' increases with increasing energy, such that a single bright line at $\cos\theta \approx 0.88$ dominates the distribution by $1500$ MeV.








Finally, we note that we have conducted the same numerical consistency check as in Chapter \ref{mathform}, comparing the total integrated yield $N_{tot}$ explicitly from the Frank-Tamm formula with the result of integrating the angular distribution $\frac{dN}{d\cos\theta}$.  As with Tables~\ref{tab:NGrace} and \ref{tab:NBabicz}, we compare the two methods for obtaining the total yield for the absorptive harmonic oscillator model in Table~\ref{tab:Nhofit}.  We again see consistency of the two results within a numerical error that decreases with increasing kinetic energy.  Since the two results converge to each other, this gives confidence that the numerical uncertainty is under control.  

\begin{figure}[h]
\begin{centering}
\begin{subfigure} {.49\textwidth}
\centering
\includegraphics[width=1\textwidth]{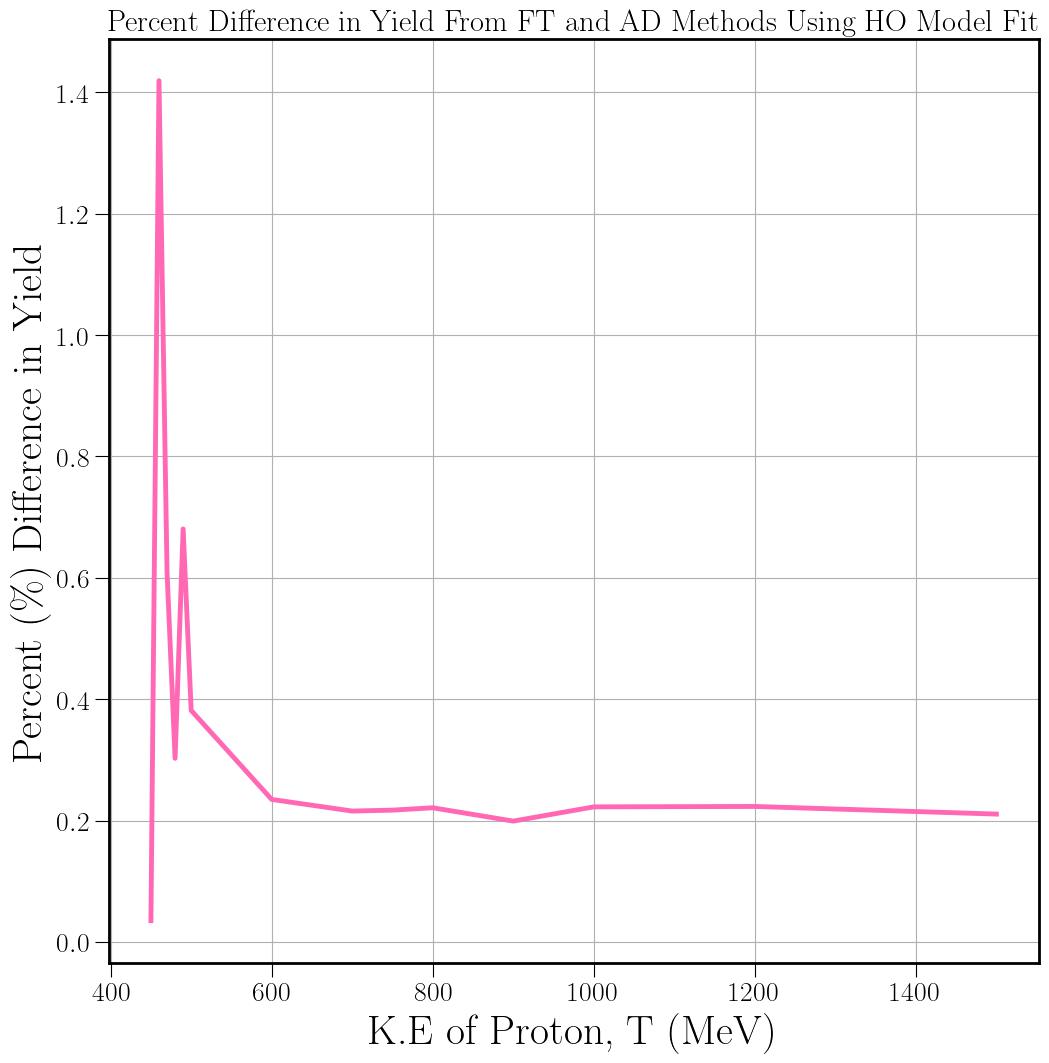}
\caption{
\label{f:perdiffho}
}
\end{subfigure}
\begin{subfigure} {.49\textwidth} 
\centering
\includegraphics[width=1\textwidth]{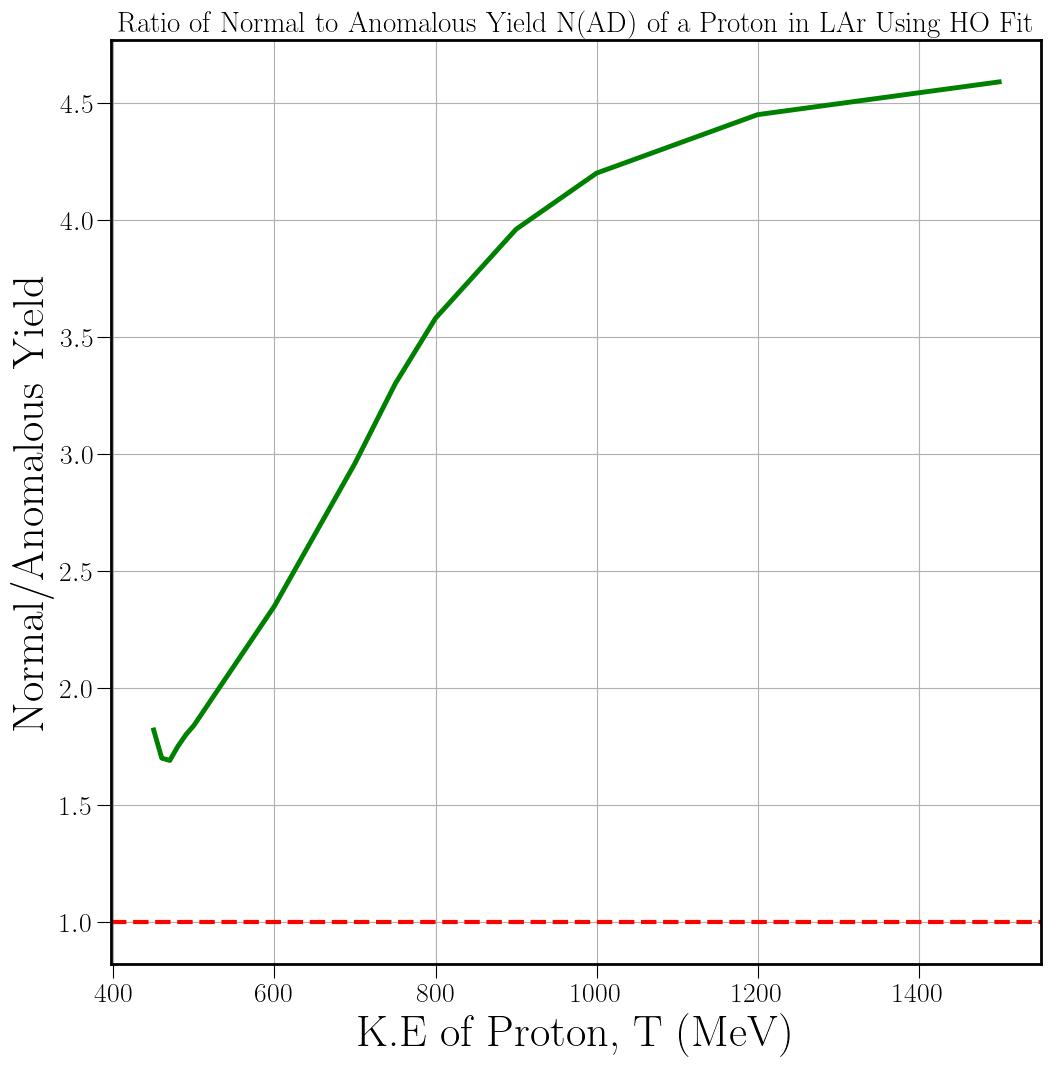}
\caption{
\label{f:nbyaho}
}
\end{subfigure}
\caption{(a) Comparison (\% difference) of Total Cherenkov yield (N) from Two Different Methods: Frank-Tamm Integral (FT) and Angular Distribution (AD).
(b) Ratio of normal and anomalous components of the total Cherenkov yield (N) 
of Protons with different kinetic energies in LAr using Our Absorptive Fit (HO Model).
\label{f:diffnbya}
}
\end{centering}
\end{figure}

We also record in Table~\ref{tab:Nhofit} the total yield of normal and anomalous photons at different energies, along with their ratios.  The ratio of normal to anomalous photons is an interesting measure of how sensitive the Cherenkov spectrum at a given energy is to the details of the index of refraction, like anomalous dispersion.  This result, as well as the convergence of the two numerical methods, is shown in Fig.~\ref{f:diffnbya}.  We see that the normal-to-anomalous ratio is a monotonically-increasing function of the energy.  On one hand, this implies that the energy range most sensitive to anomalous dispersion is the \textit{low}-energy range of protons near to the Cherenkov threshold.  On the other hand, while the normal-to-anomalous yield ratio increases with increasing energy, it again exhibits a noticeable slowdown at high kinetic energies.  This can again be traced to a change in curvature of the yield around $700$ MeV due to the IR saturation of the emitting wavelength range.  Interestingly, this phenomenon only affects the normal part of the dispersion.  As a result, the normal-dispersion yield slows down significantly at high energies, such that the normal-to-anomalous yield ratio flattens out to approximately $5:1$ at high energies, rather than diverging to infinity.  This leads to the anomalous part of the dispersion contributing a nontrivial amount of the total Cherenkov yield, even at high energies.

\begin{table}[h!]
  \begin{center}
    \begin{tabular}{|c|c|c|c|c|c|c|} 
      \hline
      \textbf{$T$}(MeV) & \textbf{$N(FT)_{HO}$ } & \textbf{$N(AD)_{HO}$} & \textbf{\% diff} & \textbf{normal} & \textbf{anomalous} & \textbf{norm/anom} \\
      \hline
      400	& 0.00	& 0.00	& n/a & 0.00 & 0.00	& n/a  \\
      \hline
      450 & 0.42 & 0.42 & 0.03 & 0.27 & 0.15 & 1.82 \\
      \hline
      500 & 199.40 & 200.16 & 0.38 & 129.75 & 70.41 & 1.84 \\
      \hline
      600	& 2675.34 & 2669.06 & 0.23 & 1871.20 & 797.86 & 2.35 \\
      \hline
      700 & 9547.35	& 9526.77	& 0.79 & 7121.57 & 2405.19 & 2.96 \\
      \hline
      800 & 22245.09 & 22195.93 & 0.22 & 17346.14 & 4849.79	& 3.58 \\
      \hline
      1000 & 62168.71	& 62030.37 & 0.22 & 50111.50 & 11918.88 & 4.20 \\
      \hline
      1200 & 116528.29 & 116268.20 & 0.22	& 94945.57 & 21322.63 & 4.45 \\
      \hline
    \end{tabular}
    \caption{Comparison (\% difference) of Total Cherenkov yield (N) using Our HO Model Fit from Two Different Methods: Frank-Tamm Integral (FT) and Angular Distribution (AD) and contributions of normal and anomalous components to the total N(AD).}
    \label{tab:Nhofit}
  \end{center}
\end{table}

%


\subsubsection{Comparison: Number Distribution vs. Energy Distribution} \label{ADvsED}




\begin{figure}[p]
\centering
\begin{subfigure}{.49\textwidth}
\includegraphics[width=1\textwidth]{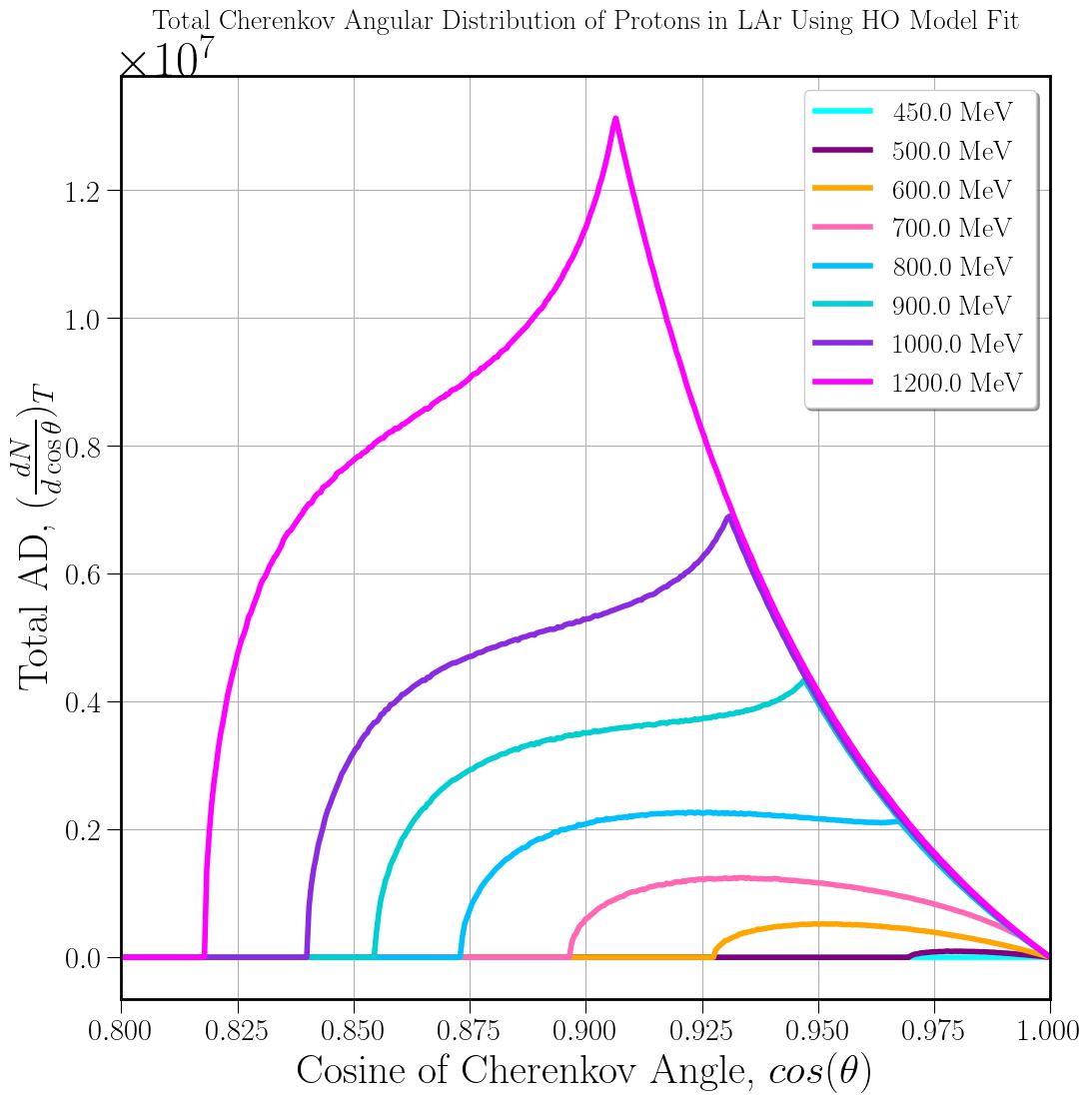
}
\caption{Total AD 
\label{f:adtotalHO450to1200MeV}
}
\end{subfigure}
\begin{subfigure}{.49\textwidth}
\centering
\includegraphics[width=1\textwidth]{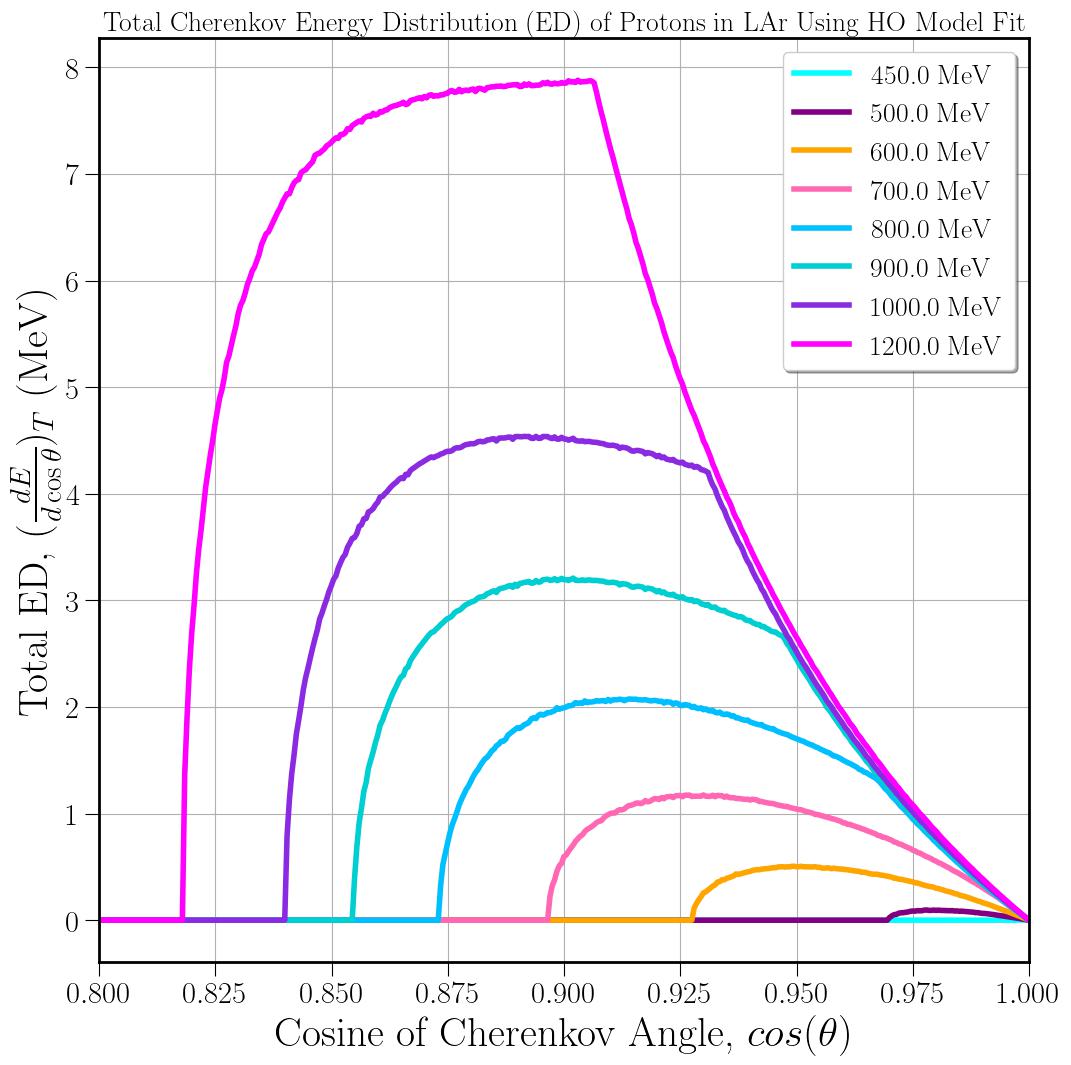}
\caption{Total ED
\label{f:edtotalHO450to1200MeV}
}
\end{subfigure}
\medskip

\begin{subfigure}{.49\textwidth}
\includegraphics[width=1\textwidth]{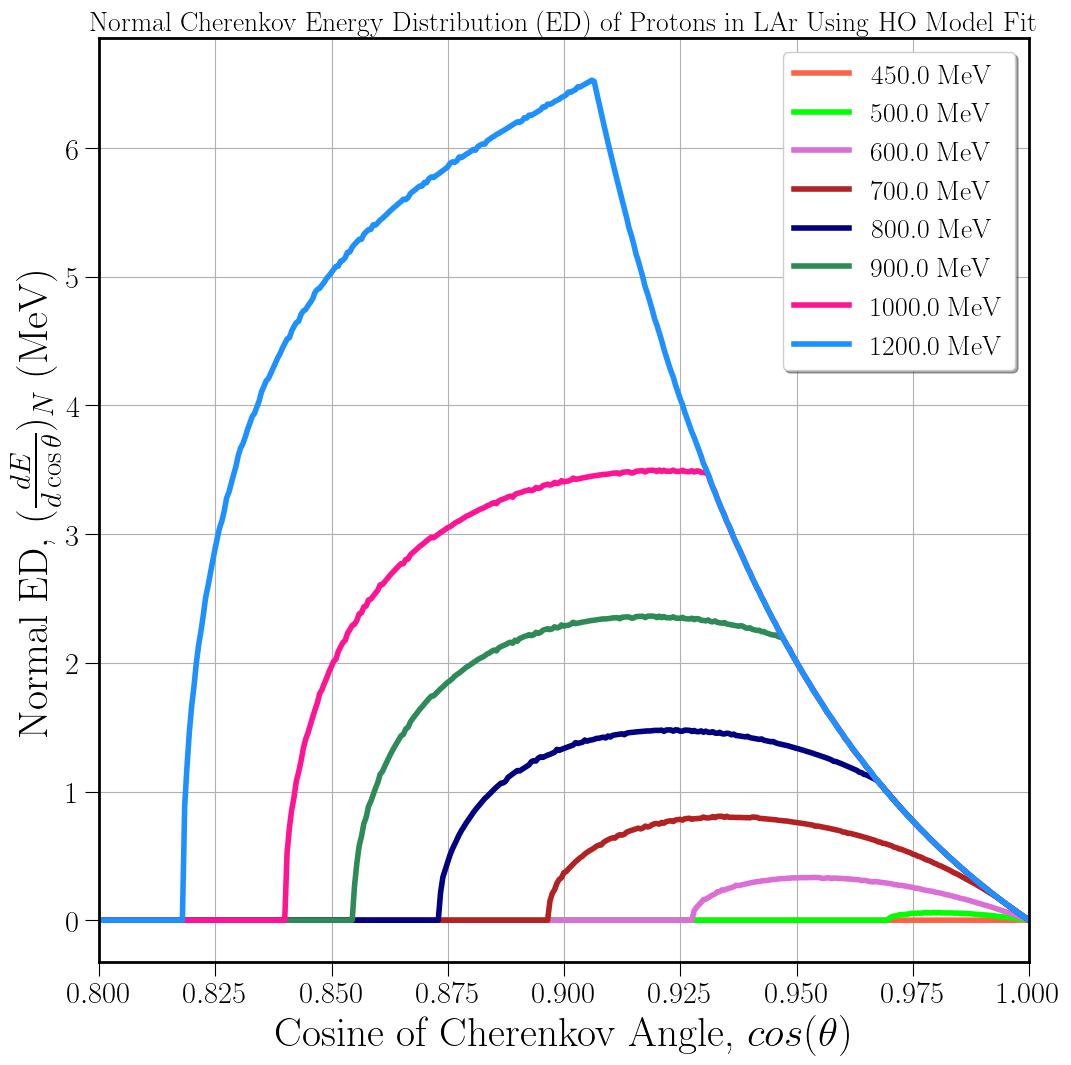
}
\caption{Normal ED 
\label{f:ednormalHO450to1200MeVdxp01}
}
\end{subfigure}
\begin{subfigure}{.49\textwidth}
\centering
\includegraphics[width=1\textwidth]{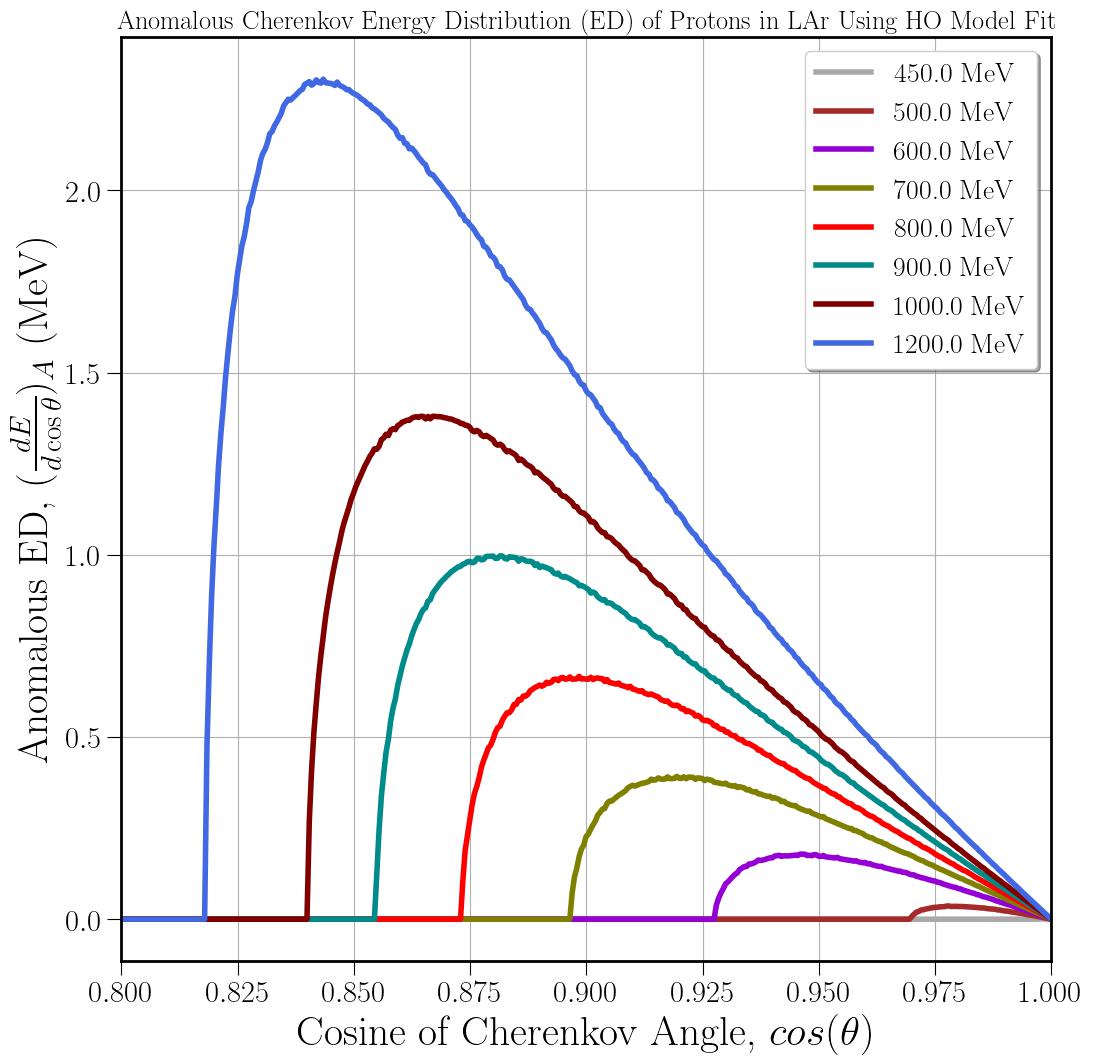}
\caption{Anomalous ED
\label{f:edanomlsHO450to1200dxp01}
}
\end{subfigure}
\caption{Total Cherenkov Angular Distribution (AD) and Energy Distribution (ED) of Protons with Different K.Es Travelling in LAr using HO model fit  (dx = 0.01 cm) including Normal and Anomalous component of the total ED. 
\label{f:advsedHO450to1200}
}
\end{figure}

We can highlight certain features of the angular distribution by examining not just the \textit{number distribution} $\frac{dN}{d\cos\theta}$ discussed previously, but also the \textit{energy distribution} (ED) of Cherenkov radiation.  Since we already know the angular distribution (AD) of Cherenkov photon number, it is a simple matter to multiply the number of photons of a given energy and wavelength by the photon energy ($hc/\lambda$):
\begin{align}
    \frac{dE}{dx \: d\lambda \: d\cos\theta} =
    \left(\frac{hc}{\lambda}\right) \:
    \frac{dN}{dx \: d\lambda \: d\cos\theta}    \: .
\end{align}
Then, integrating over the wavelength and enforcing the Cherenkov condition by delta function as before, we obtain the integrated energy distribution,
\begin{align}   \label{e:energydistformulagenho}
    \frac{dE}{d\cos\theta } &= 
    \int d\lambda \: 
    \left(\frac{hc}{\lambda}\right) \:
    \frac{dN}{dx \: d\lambda \: d\cos\theta}
    \notag \\ &=
    2\pi \: \alpha_{EM} \, \int_{\lambda_{min}}^{\lambda_{max}} \frac{d\lambda}{\lambda^2}
    \left(\frac{hc}{\lambda}\right) \left( 1 - \frac{1}{\beta^2 n^2(\lambda)} \right)
     \delta\left( \cos\theta - \frac{1}{\beta n(\lambda)} \right)   \:,
    \notag \\
    &= \frac{4\pi  h c \alpha_{EM}}{\beta^2} \frac{1}{\lambda_\theta^3} \left(\frac{\sin^2 \theta}{\cos^3 \theta}\right)
    \frac{1}{\left| \frac{dn^2}{d\lambda} \right|_{\lambda= \lambda_{\theta}}}
    \notag \\ &=
    \left(\frac{hc}{\lambda_\theta}\right) \:
    \frac{dN}{d\cos\theta}
    \: .
\end{align} 
One can further integrate Eq.~\ref{e:energydistformulagenho} over all angular regions to get the total energy $E_{Cherenkov}$ radiated in Cherenkov photons.

In Fig.~\ref{f:advsedHO450to1200}, we compare the angular distribution (AD) of photon number versus the energy distribution (ED) to understand the differences between these two observables.  One immediately notices by eye that the shape of the ED is (Fig.~\ref{f:edtotalHO450to1200MeV}) is different from the AD (Fig.~\ref{f:adtotalHO450to1200MeV}). 
While both distributions cover the same angular region, the ED is significantly flattened toward the outer edge of the Cherenkov cone, $\cos \theta < \cos \theta_{max}$, whereas the AD sharply falls beyond its peak at $\cos \theta_{max}$.  This flattening in the ED curve arises from the photon energy factor $h c / \lambda_\theta$ which multiplies the AD
Eq.~\ref{e:energydistformulagenho}.  Because $n(\lambda)$ is a decreasing function of $\lambda$ (normal dispersion) over most of its range, the enhancement at small $\lambda$ of the ED occurs at larger angles $\cos\theta = 1 / \beta n$ as dictated by the Cherenkov condition.  The biggest effect is a smearing of the peak from the normal dispersion, while the shape of the anomalous contribution is mostly unchanged.  Thus, the energy distribution can be used as a distinct measure of the Cherenkov spectrum and can highlight different features of the radiation.

%
%
%
%
\begin{figure}[t]
\begin{centering}
\includegraphics[width=0.5\textwidth]{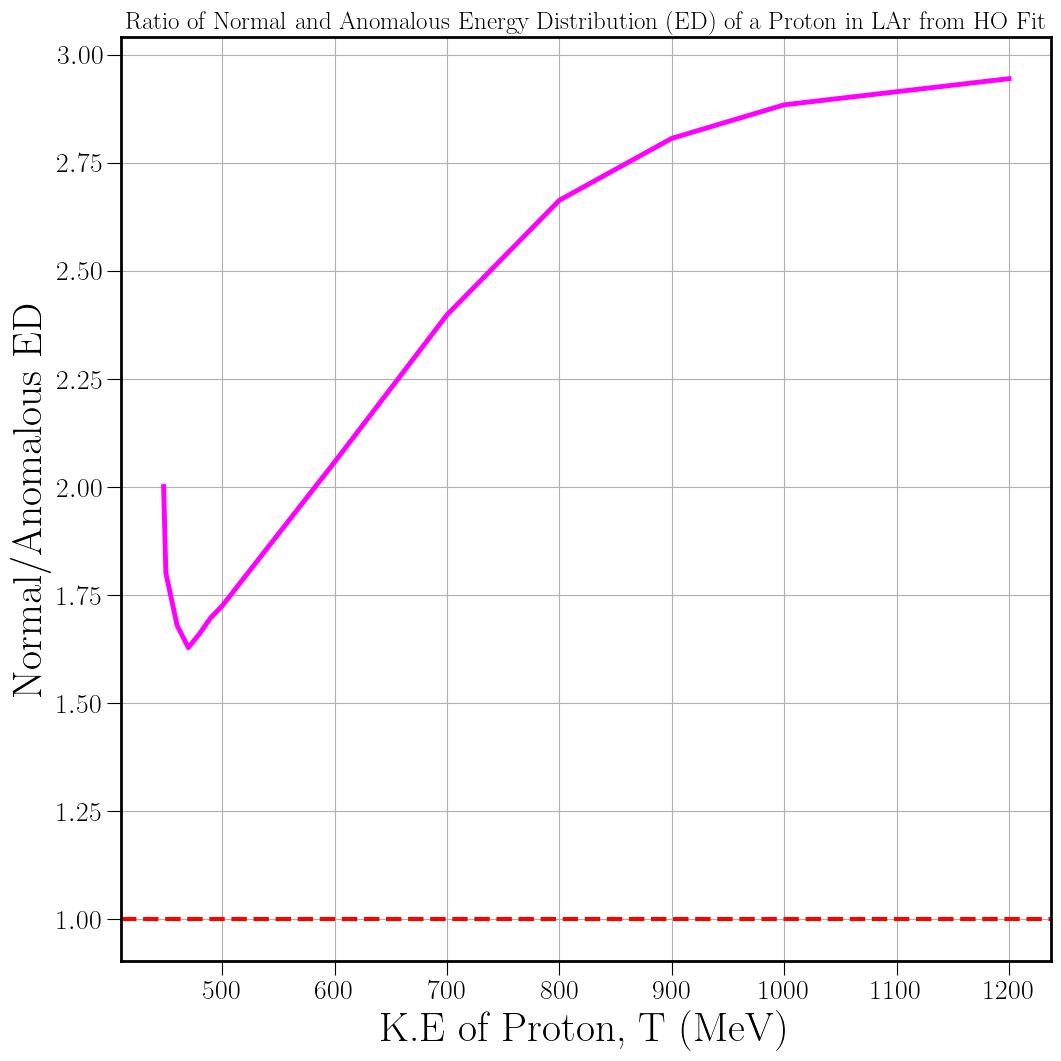}
\caption{Ratio of normal and anomalous components of the total energy distribution (ED) of Protons with different kinetic energies in LAr using Our Absorptive Fit (HO Model). 
\label{f:enbyahofit}
}
\end{centering}
\end{figure}
In Fig.~\ref{f:enbyahofit}, we show the ratio of the normal-to-anomalous energy distributions (ED). Similar to the normal-to-anomalous ratio of the angular distribution (AD) shown previously in Fig.~\ref{f:nbyaho}, the ratio computed from the ED also increases with the proton energy, but then flattens out to
below 3 : 1 around 1200 MeV.  This flattening of the ratio, rather than diverging to infinity, is due to the wavelength saturation discussed previously in Sec.~\ref{Sec:AbsorptiveAD}.  As a result, the anomalous part of the dispersion contributes a nontrivial amount of the total energy contained in the Cherenkov
distribution.  Comparing the ED ratio from Fig.~\ref{f:enbyahofit} with the AD ratio from  Fig.~\ref{f:nbyaho}, we note that the normal dispersion component contributes more to the AD than to the ED, leading to normal-to-anomalous ratios which are larger for the AD than for the ED.  In other words, the ED is more sensitive to the contributions of anomalous dispersion, because those photons are produced near the resonance, with shorter wavelengths than the normal dispersion.

Another interesting feature one can notice readily from Fig.~\ref{f:enbyahofit} is the sudden onset of non-monotonic behavior below 500 MeV.  While the normal-to-anomalous ratio is always greater than 1, it exhibits a minimum around 470 MeV, both for the  ED ratio and AD ratio.  In this regime very close to threshold, the range of radiating wavelengths is approximately symmetric between the normal and anomalous components of the dispersion ( see the intersection method shown in Fig.~\ref{f:ho450MeV} ).  Thus, very close to threshold, the normal-to-anomalous ratio is not controlled by differences in the phase space to emit normal versus anomalous photons.  Rather, it is dominated by differences in the instantaneous yields at a given energy, arising from the different wavelengths $\lambda_{N,A}$ and different dispersion parameters $|\frac{dn^2}{d\lambda}|_{N, A}$ (see Eqs.~\eqref{e:angdistformulhonormanom}).  The greatly increased phase space for normal photon emission, however, quickly takes over with increasing energy, such that by 500 MeV the normal-to-anomalous ratio is a monotonically increasing function of energy.

%

%


\subsubsection{Comparison: Absorptive vs. Non-absorptive Fits}

\begin{figure}[h!]
\centering
\begin{subfigure}{.48\textwidth}
    \centering
    \includegraphics[width=1\textwidth]{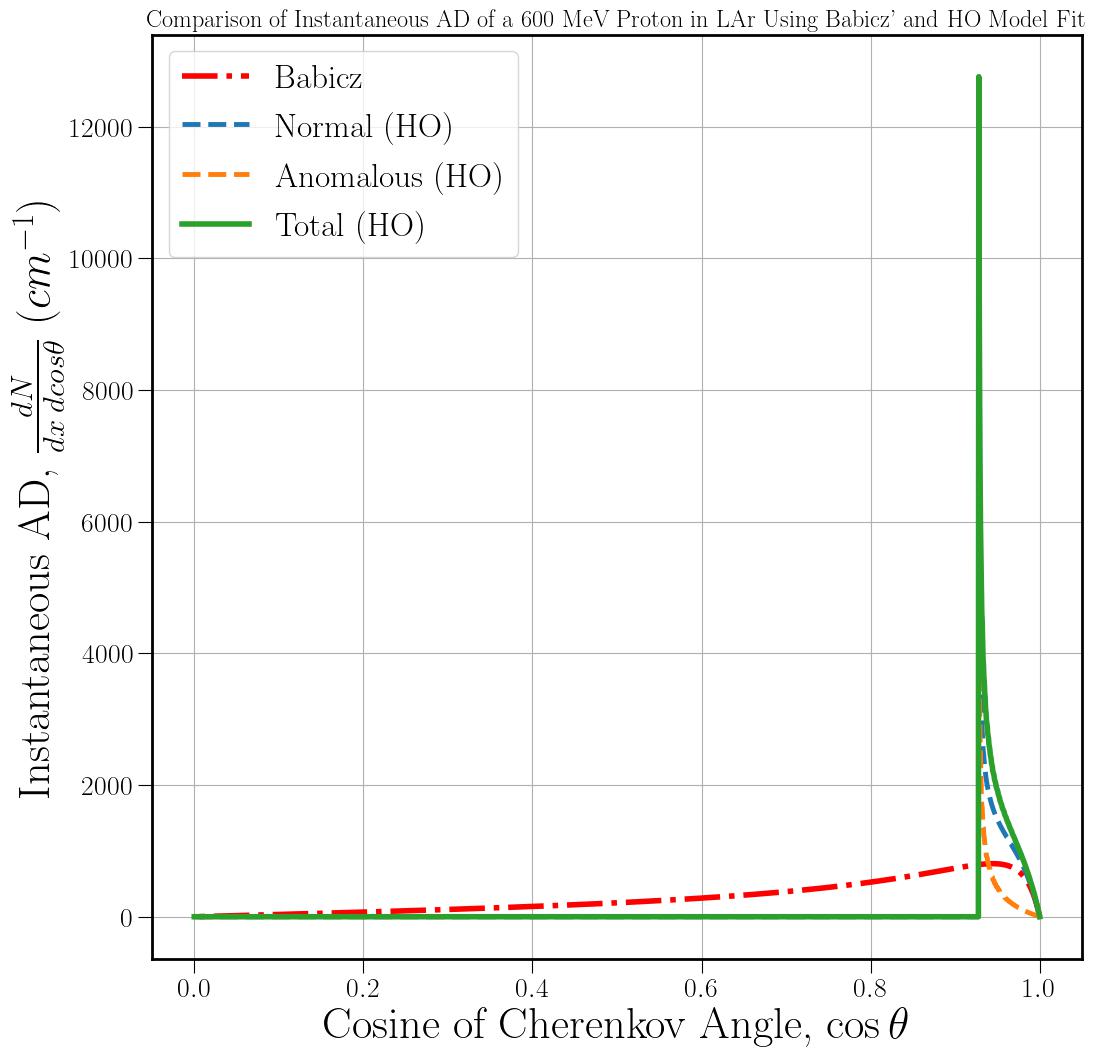}
    \caption{Instantaneous AD
    \label{f:IAD_Babiczvsho_bp792}
    }
    \end{subfigure}
\begin{subfigure}{.48\textwidth}
    \centering
    \includegraphics[width=1\textwidth]{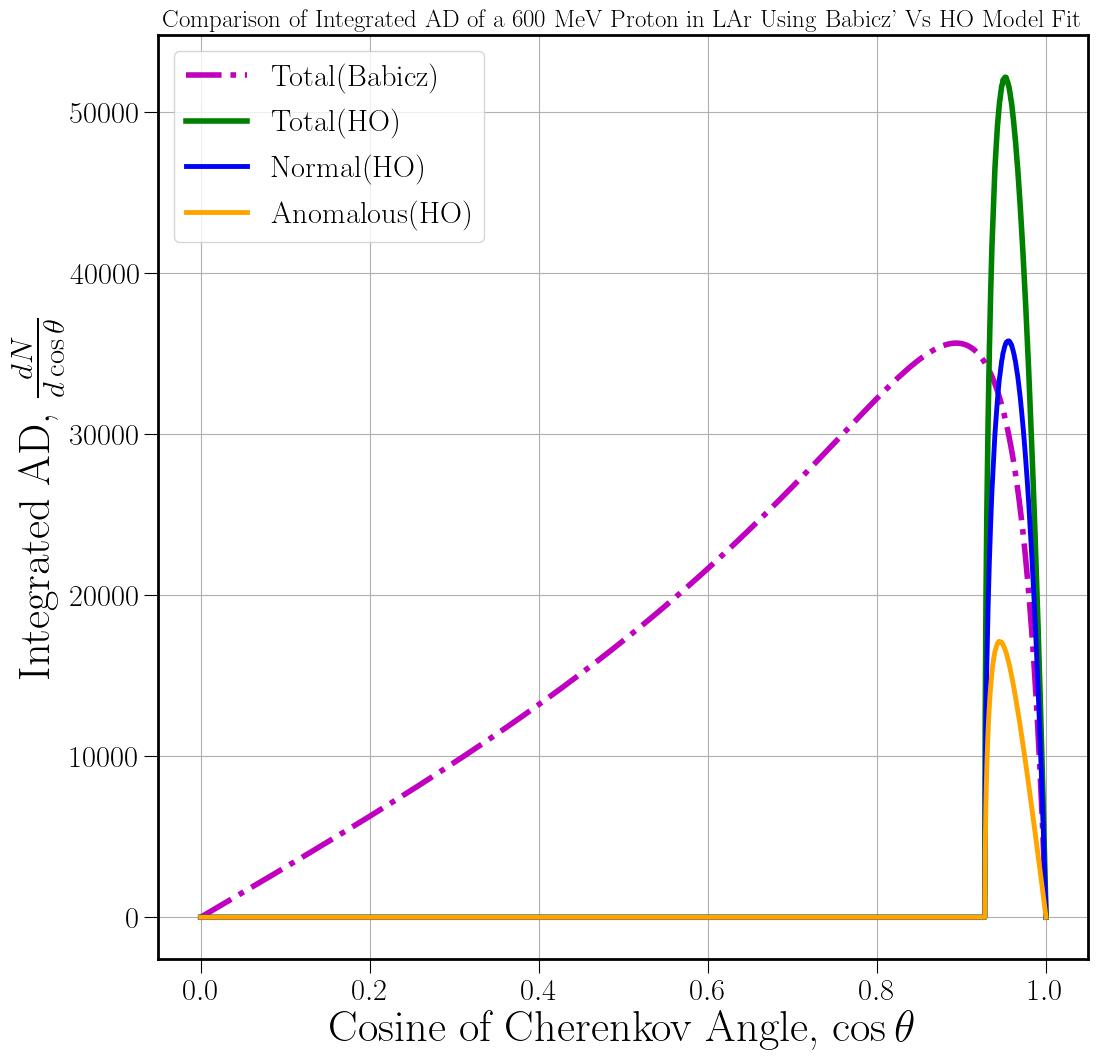}
    \caption{Integrated AD
    \label{f:AD_Babiczvsho_600MeV}
    }
    \end{subfigure}%
\caption{Comparison between Instantaneous and Integrated Cherenkov Angular Distribution of a 600 MeV Proton traveling in LAr derived Using Babicz's (Non-absorptive) refractive index fit Vs this work (Absorptive: HO model). 
\label{f:ADcompBabiczvsHO600MeV}
}
\end{figure}

Having computed the angular distributions for both the existing resonant fits (Chapter \ref{mathform}) and for our own absorptive fit based on the first-principles harmonic oscillator model (Sec.~\ref{Sec:AbsorptiveAD}), we can now compare the two models directly.  We plot the instantaneous and integrated angular distributions for the two models side by side in Fig.~\ref{f:ADcompBabiczvsHO600MeV} for a 600 MeV proton. 

As previously noted, the difference in shape between the resonant fits such as Babicz' discussed in Chap.~\ref{mathform} and the absorptive fits is striking.  While the angular distribution of Babicz' fit spans the entire angular region $0 \leq \theta \leq \pi/2$, the absorptive harmonic oscillator fit is cut off sharply at a maximum emission angle $\theta_{max}$.  This enormous difference is directly due to the divergence of the refractive index $n \rightarrow \infty$ in the resonant fit versus its finiteness $n \rightarrow n_{peak} < \infty$ in the absorptive fit.  Despite the small emission region for the absorptive fit, when integrated over the proton range, many such photons are emitted in the same angular direction, resulting in substantially large densities of photons in the narrow Cherenkov cone.  For the instantaneous AD (Fig.~\ref{f:IAD_Babiczvsho_bp792}), the result is striking: the peak density of Cherenkov photons in the HO model is so large that even the overly-generous Babicz model is hardly visible on the same scale.  For the integrated AD at 600 MeV (Fig.~\ref{f:AD_Babiczvsho_600MeV}), these differences are washed out somewhat, with the peak Cherenkov density being comparable between the HO and Babicz fits.  The HO peak continues to grow compared with the Babicz peak with increasing energy, up until the saturation of the wavelength range (see also Fig.~\ref{f:ADcompabsvsnonabs}).



\subsubsection{Comparison: Different Absorptive Fits}

While we focus primarily on the first-principles harmonic oscillator model \eqref{e:nlambdauvhofinal} for the refractive index, it is also instructive to compare the results with those for a similar absorptive fit: the \textit{Approximate} fit given by Eq.~\eqref{e:nmodifiedlambda}.  The approximate form, first introduced in Sec.~\ref{compabs}, differs from the HO model in its precise functional form, but is qualitatively similar and can fit the available experimental refractive index ($n$) data of LAr equally well.  Clearly, while working with a formula like the HO model obtained from first principles is desirable, the approximate formula has the advantage of being analytically tractable, and one can derive the Cherenkov angular distribution (AD) formula algebraically using this fit. This analyticity can make the yield and AD calculations faster and more precise, while not deviating too much quantitatively from the predictions of the first-principles HO model.  Comparing two qualitatively similar but quantitatively different models offers a unique opportunity to study the model dependence of Cherenkov radiation with multiple realistic models.


%
\begin{figure}[h!]
\centering
\begin{subfigure}   {.48\textwidth}
    \centering
    \includegraphics[width=1\textwidth]{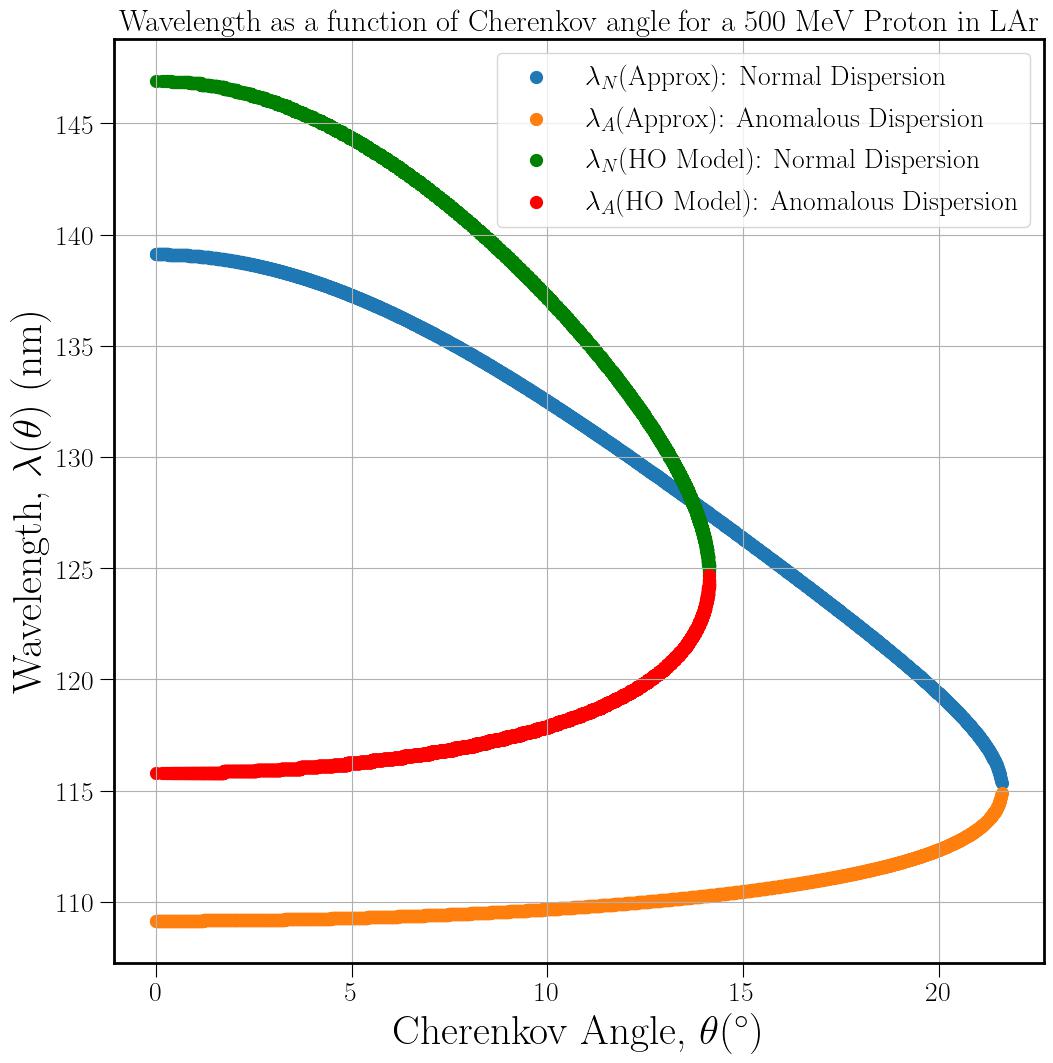}
    \caption{Wavelength solutions
    \label{f:lambdana500MeVHOfit}
    }
\end{subfigure}
\begin{subfigure}   {.50\textwidth}
    \centering
    \includegraphics[width=1\textwidth]{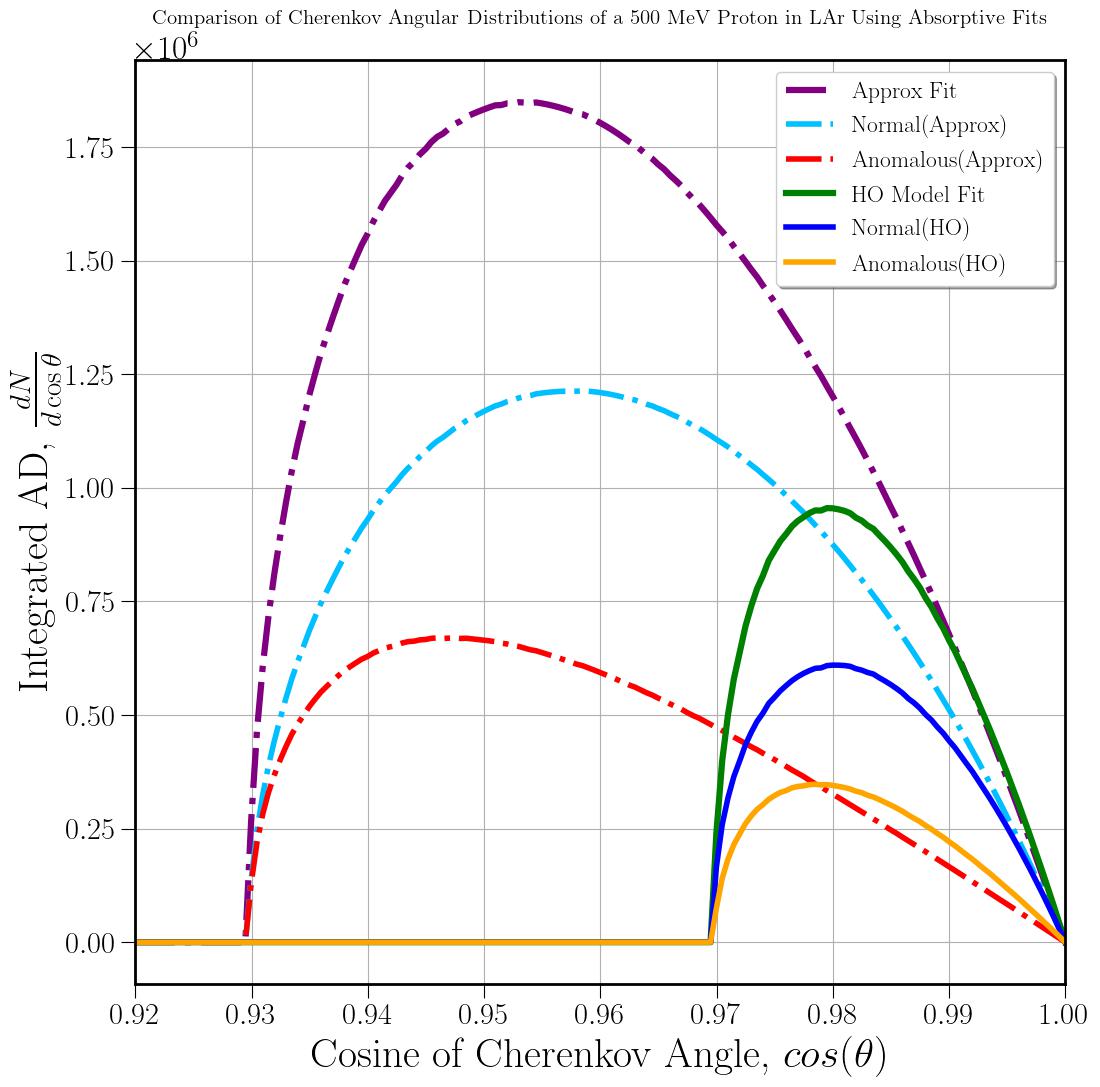}
    \caption{AD comparison
    \label{f:ADcomps500MeVHOvsPVS}
    }
\end{subfigure}
\caption{Normal and anomalous wavelength solutions and AD comparison of a 500 MeV proton ($\beta = 0.758$) travelling in LAr calculated using absorptive Fits (HO Model VS Approx fit).
\label{f:WLAD500MeVHOvsPVS}
}
\end{figure}

For the approximate fit, the Cherenkov condition is
\begin{align}   \label{e:nourfit2}
    n_{approx} (\lambda) = a_0 + a_{approx} \left( \frac{\lambda_{UV}^{-1} - \lambda^{-1}}{(\lambda_{UV}^{-1} - \lambda^{-1})^2 + \Gamma^2} 
    \right) = \frac{1}{\beta \cos\theta} \: ,
\end{align}
which we solve to obtain the wavelength solutions
\begin{align}   \label{e:lambdatheta}
    \lambda_{\theta} = \frac{2 (\beta a_{0} \cos\theta - 1)}{2\lambda_{UV}^{-1} (\beta a_{0} \cos\theta - 1) + \beta a_{approx} \cos\theta \mp \sqrt{\beta^2 a_{approx}^2 \cos^2(\theta) - 4 \Gamma^2 (\beta a_{0} \cos\theta - 1)^2 }} \: .
\end{align}
The two solutions \eqref{e:lambdatheta} to the quadratic equation \eqref{e:nourfit2} correspond to the normal and anomalous solutions, which respectively are given by
\begin{subequations}
\begin{align}   \label{e:lambdathetanormal}
    \lambda_{\theta}^{normal} = \frac{2 (\beta a_{0} \cos\theta - 1)}{2\lambda_{UV}^{-1} (\beta a_{0} \cos\theta - 1) + \beta a_{approx} \cos\theta - \sqrt{\beta^2 a_{approx}^2 \cos^2\theta - 4 \Gamma^2 (\beta a_{0} \cos\theta - 1)^2 }}    \: ,
    \\
    \label{e:lambdathetaanomalous}
    \lambda_{\theta}^{anomalous} = \frac{2 (\beta a_{0} \cos\theta - 1)}{2\lambda_{UV}^{-1} (\beta a_{0} \cos\theta - 1) + \beta a_{approx} \cos\theta + \sqrt{\beta^2 a_{approx}^2 \cos^2\theta - 4 \Gamma^2 (\beta a_{0} \cos\theta - 1)^2 }} \: .
\end{align}
\end{subequations}
Here we solved the Cherenkov condition $n(\lambda) = 1 / \beta \cos\theta$ which is relevant for emission at a particular angle $\theta$, but by simply putting $\cos \theta = 1$, we can obtain the absolute maximum or minimum wavelengths for a given velocity $\beta$:
\begin{align}   \label{e:lambdasolnsourfit}
  \lambda_{max/min} = \frac{2\lambda_{UV}(a_0 \beta - 1)}{2(a_0 \beta - 1) + \lambda_{UV} (a_{approx} \beta \mp \sqrt{a_{approx}^2 \beta^2 - 4(a_0 \beta - 1)^2 \Gamma^2})}  \: ,  
\end{align}
which is used to integrate the Frank-Tamm formula \eqref{e:ftintegrand}.

In Fig.~\ref{f:lambdana500MeVHOfit}, we show the normal and anomalous wavelength solutions of a 500 MeV proton ($\beta = 0.758$) travelling in LAr using the refractive index calculated from the HO model versus the approximate fit.  We can see that the HO fit gives wavelength solutions $\lambda_\theta$ which are slightly higher than the solutions $\lambda_\theta$ for the approximate fit. 
On the other hand, the approximate fit covers a wider angular region than the HO model fit, with the Cherenkov cone extending an extra $\sim 8^\circ$ beyond that of the HO fit.  This difference in wavelength channels for emitting Cherenkov photons and angular domain of the wavelength solutions can be directly attributed to the difference in the refractive indices itself, particularly the peak height $n_{peak}$.  The shift to lower wavelengths in the approximate fit, together with a Frank-Tamm intensity which scales like $dN/dx d\cos\theta \propto 1 / \lambda_\theta^2$, leads to an enhancement in the number of Cherenkov photons emitted in the approximate fit, as shown in Fig.~\ref{f:ADcomps500MeVHOvsPVS}.

\begin{figure}[p]
\centering
\begin{subfigure}{.49\textwidth}
\centering
\includegraphics[width=1\textwidth]{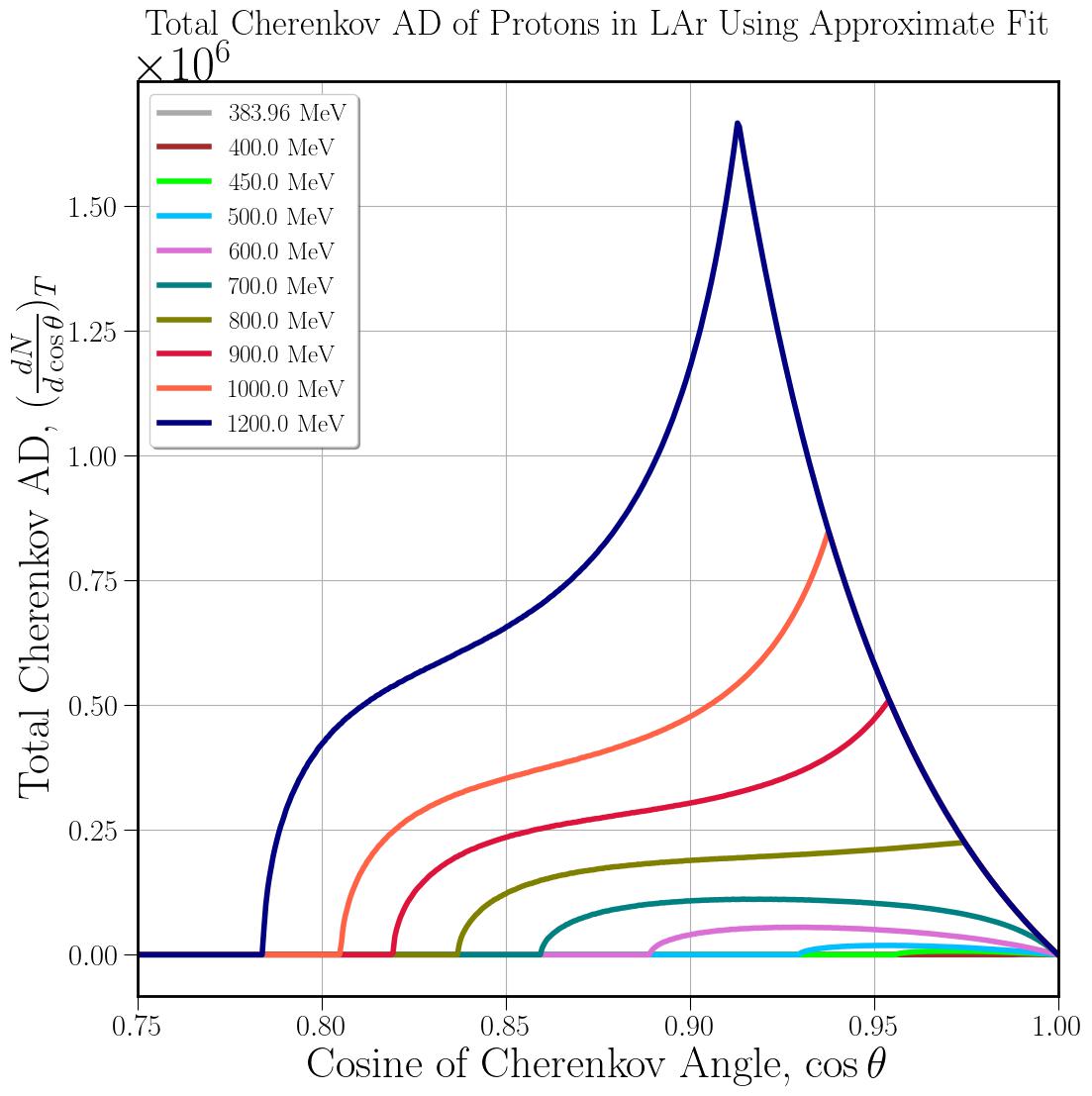
}
\caption{Total AD (Normal Scale) 
\label{f:adapprox383to1200MeV}
}
\end{subfigure}
\begin{subfigure}{.49\textwidth}
\centering
\includegraphics[width=1\textwidth]{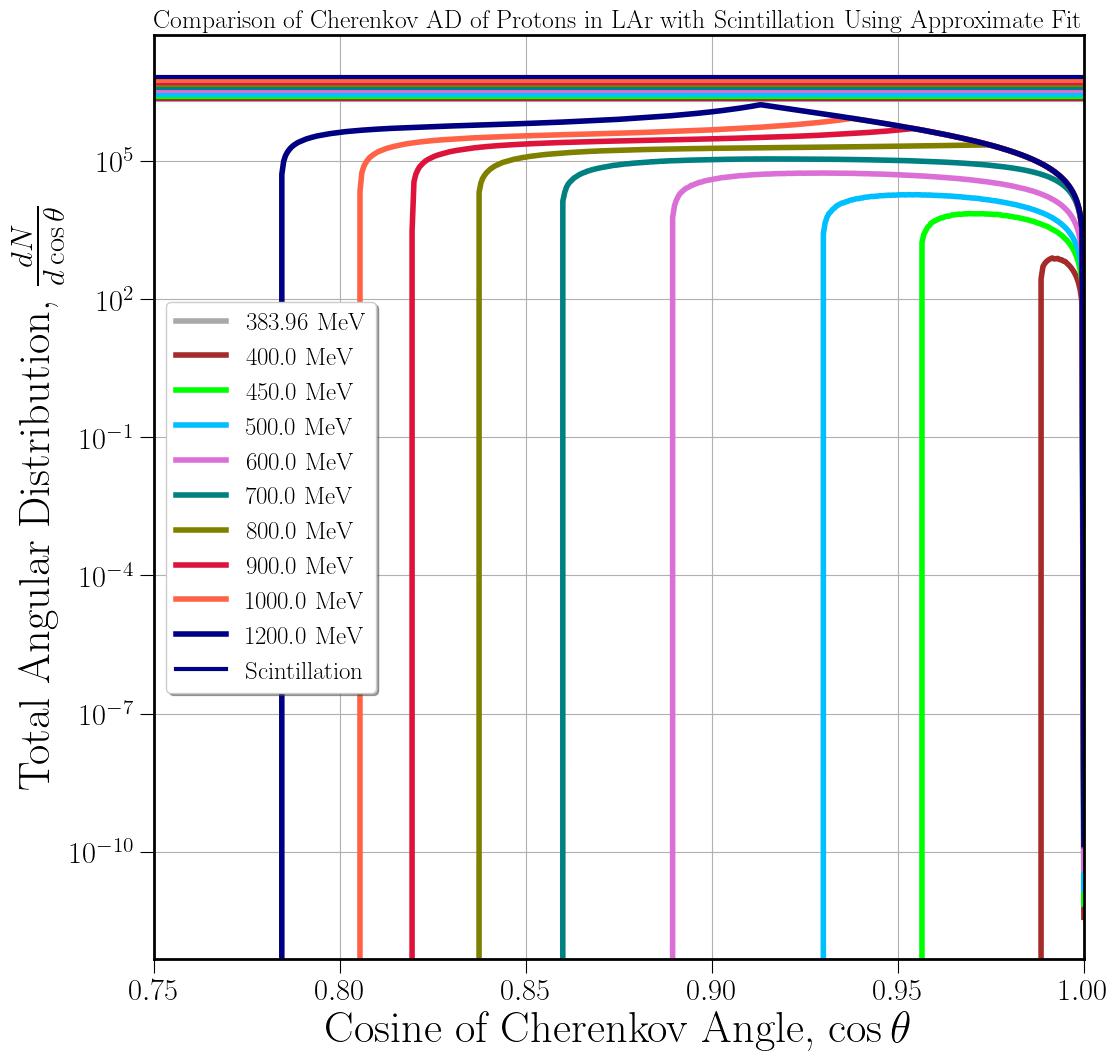
}
\caption{Total AD (Log Scale)
\label{f:adapproxlog383to1200MeV}
}
\end{subfigure}
\medskip
\begin{subfigure}{.49\textwidth}
\centering
\includegraphics[width=1\textwidth]{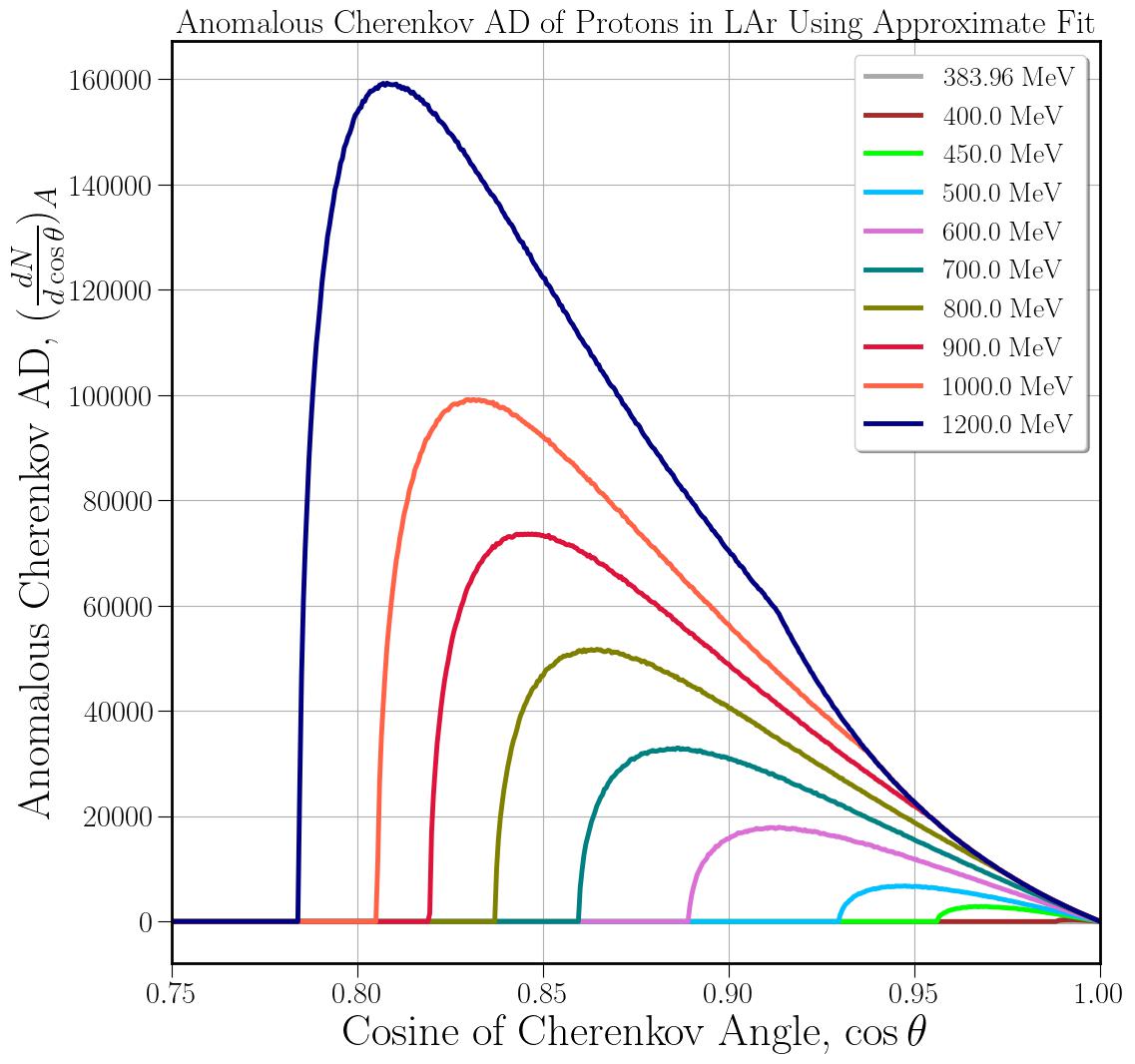
}
\caption{Anomalous AD  
\label{f:adapproxanomalous383to1200MeV}
}
\end{subfigure}
\begin{subfigure}{.49\textwidth}
\centering
\includegraphics[width=1\textwidth]{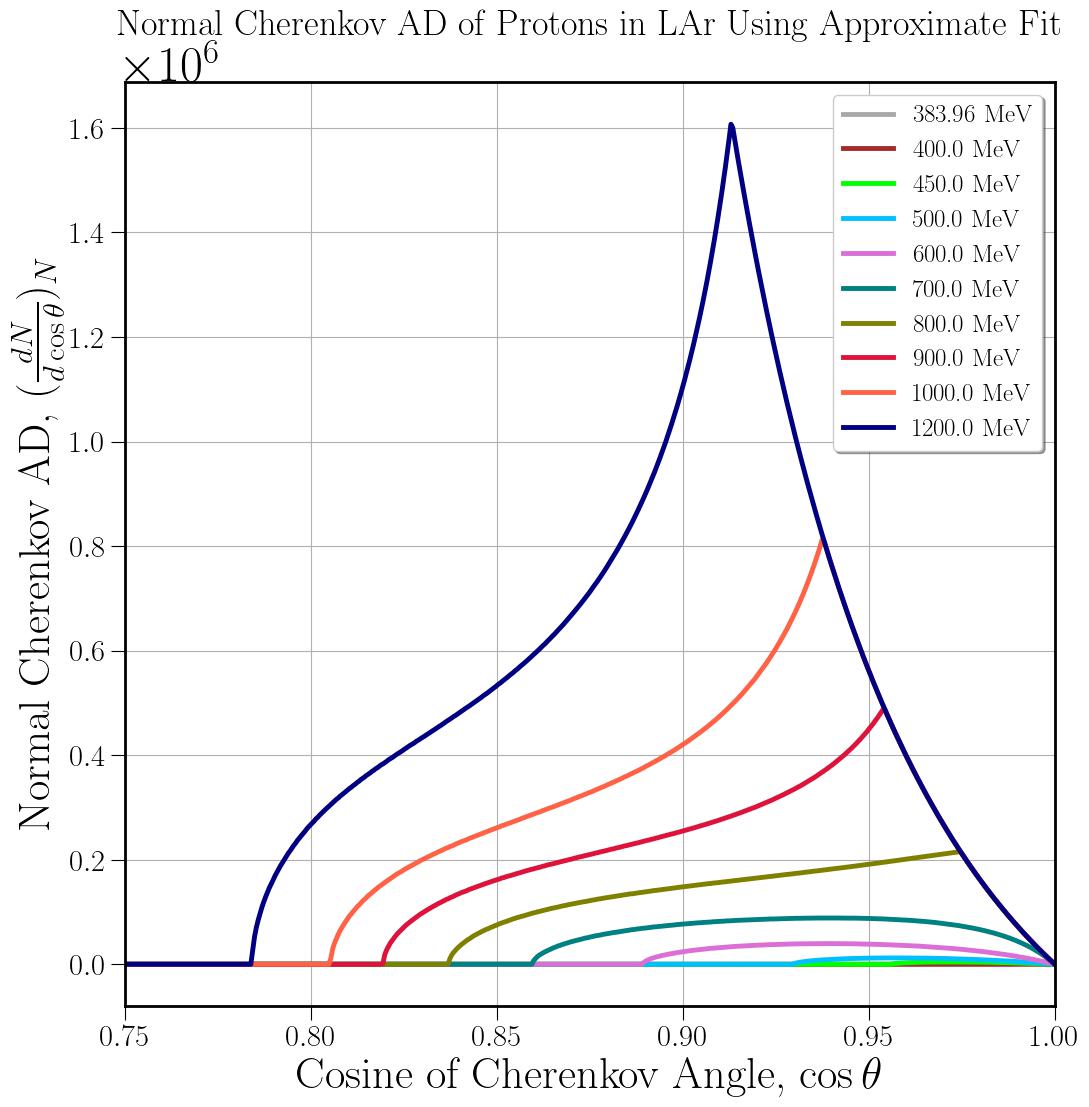}
\caption{Normal AD 
\label{f:adnormalapprox383to1200MeV}
}
\end{subfigure}
\caption{Total Cherenkov Angular Distribution of protons with different kinetic energies (T) travelling in LAr using approximate absorptive fit. Bottom plots shows the respective normal and anomalous components of the total AD in same color scheme for a given T. 
\label{f:ad_Cherenkov_383to1200MeV}
}
\end{figure}

Now we use the general AD formula given by Eq.~\eqref{e:angdistdergen2n} deduced in \ref{compabs} which is valid for any fit $n(\lambda)$ to the index of refraction:
\begin{align}   \label{e:angdistdergen2n}
    \frac{dN}{d\cos\theta} &=
    2\pi \: \alpha_{EM} \: \tan^2 \theta \:
    \int\limits_0^R dx \:
    \frac{1}{\beta \: \lambda_\theta^2 \:  \left|\frac{dn}{d\lambda} \right|_{\lambda= \lambda_{\theta}}}
    \: .
\end{align}
We now simply take the derivative of $n_{approx} (\lambda)$ given in Eq.~\eqref{e:nourfit2} (or its square) with respect to $\lambda$ and plug it back in Eq.~\ref{e:angdistdergen2n} to get the IAD analytically using this fit.  Doing so gives 
\begin{align}   \label{e:nderivativeapprox}
    \frac{dn}{d\lambda} &= a_{approx} \frac{\lambda^{-2} [((\lambda_{UV}^{-1} - \lambda^{-1})^2 + \Gamma^2) - 2 (\lambda_{UV}^{-1} - \lambda^{-1})^2]}{((\lambda_{UV}^{-1} - \lambda^{-1})^2 + \Gamma^2)^2}
    \: .
\end{align}
Plugging this derivative (Eq.~\eqref{e:nderivativeapprox}) and the wavelength solutions $\lambda_\theta$ (Eq.~\eqref{e:lambdatheta}) back into Eq.~\eqref{e:angdistdergen2n}, we obtain the integrated angular distributions shown in Fig.~\ref{f:ad_Cherenkov_383to1200MeV}.

In Fig.~\ref{f:ad_Cherenkov_383to1200MeV}, the integrated angular distribution $\frac{dN}{d\cos\theta}$ is plotted against $\cos\theta$ for different kinetic energies. Fig.~\ref{f:adapprox383to1200MeV} we show the total AD of Cherenkov radiation emitted from protons with energies ranging from 384 MeV, barely above the Cherenkov threshold, up to 1200 MeV where the full wavelength range has been saturated.  One can clearly notice that the shape of the integrated AD changes with increasing kinetic energy, particularly 
above 750 MeV, where it develops a strong peak near the inner edge of the Cherenkov cone (i.e. $\theta \rightarrow 0$ or $\cos \theta \rightarrow 1$). 

This effect can be better understood in terms of the normal and anomalous components shown in Figs.~\ref{f:adnormalapprox383to1200MeV} and \ref{f:adapproxanomalous383to1200MeV}, respectively.  The total yield of normal and anomalous photons are comparable, however, as seen in Fig.~\ref{f:diffnbya}, with increasing energy, the relative contribution of normal versus anomalous photons increases.  Thus, with increasing kinetic energy, the \textit{normal} AD component comes to dominate the overall shape of the total AD.   As a result, the total AD starts off significantly skewed to the left, but becomes more symmetric for intermediate proton energies and ultimately becomes skewed towards the right.  For higher energies, the shape of the AD significantly mimics the shape of the normal dispersion component. 
While the anomalous component loses the battle with the normal component at higher energies, it still leaves an imprint on the total AD, enhancing the number of Cherenkov photons at the outer edge (small $\cos\theta$) of the Cherenkov cone.

Fig.~\ref{f:adapproxlog383to1200MeV} also shows the integrated AD for the same energies plotted in Fig.~\ref{f:ad_Cherenkov_383to1200MeV} but on a logarithmic scale to show the absolute comparison with the scintillation yield, shown as a colorful band. The scintillation yield increases linearly with T, whereas the Cherenkov yield increases non-linearly with increasing T.  Clearly, the scintillation yield still dominates over Cherenkov photons over all of the energies shown.  We will elaborate on this comparison in the next Sec.~\ref{scintcomp}.



%
\begin{figure}[t]
\begin{centering}
\begin{subfigure} {.50\textwidth}
\centering
\includegraphics[width=1\textwidth]{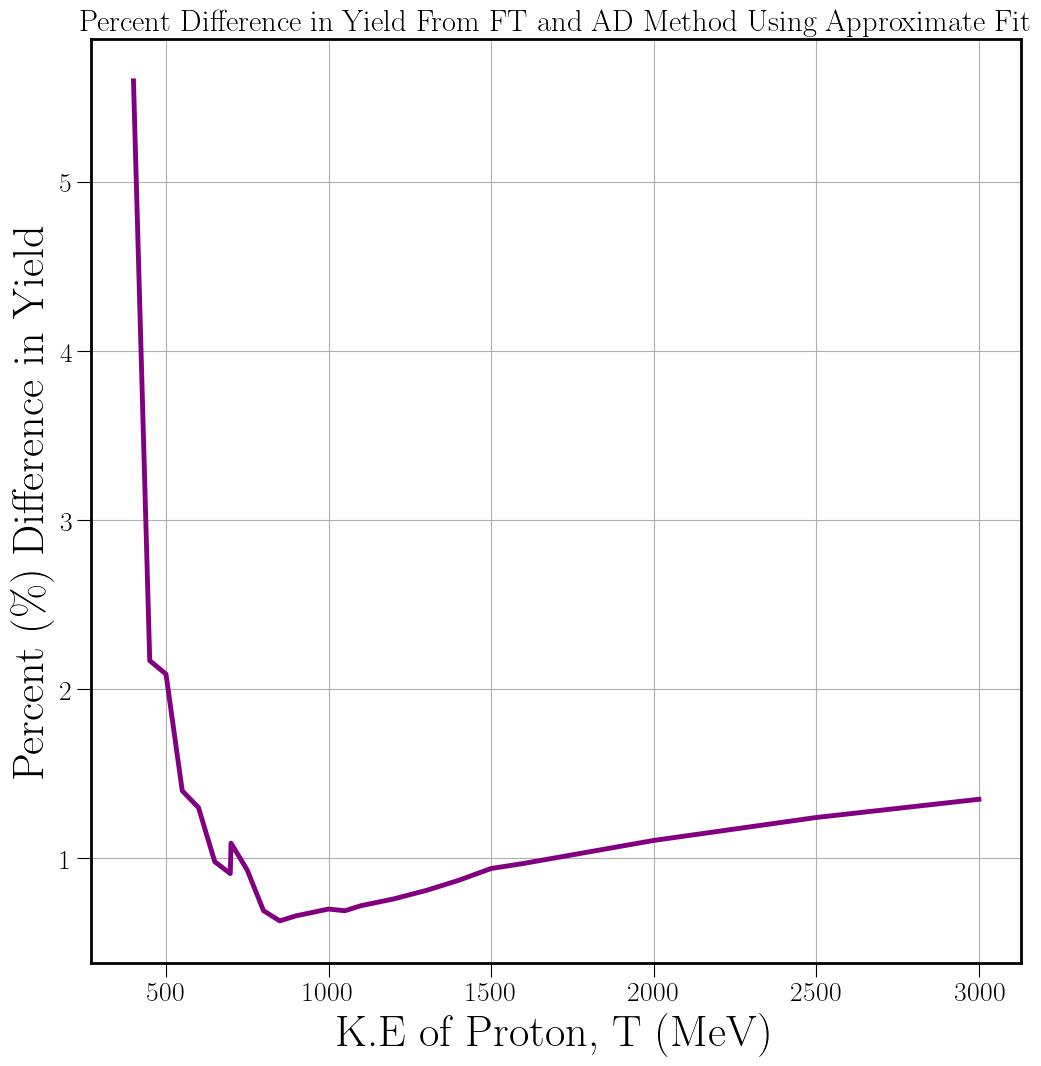}
\caption{
\label{f:perdiffapprox}
}
\end{subfigure}
\begin{subfigure} {.48\textwidth} 
\centering
\includegraphics[width=1\textwidth]{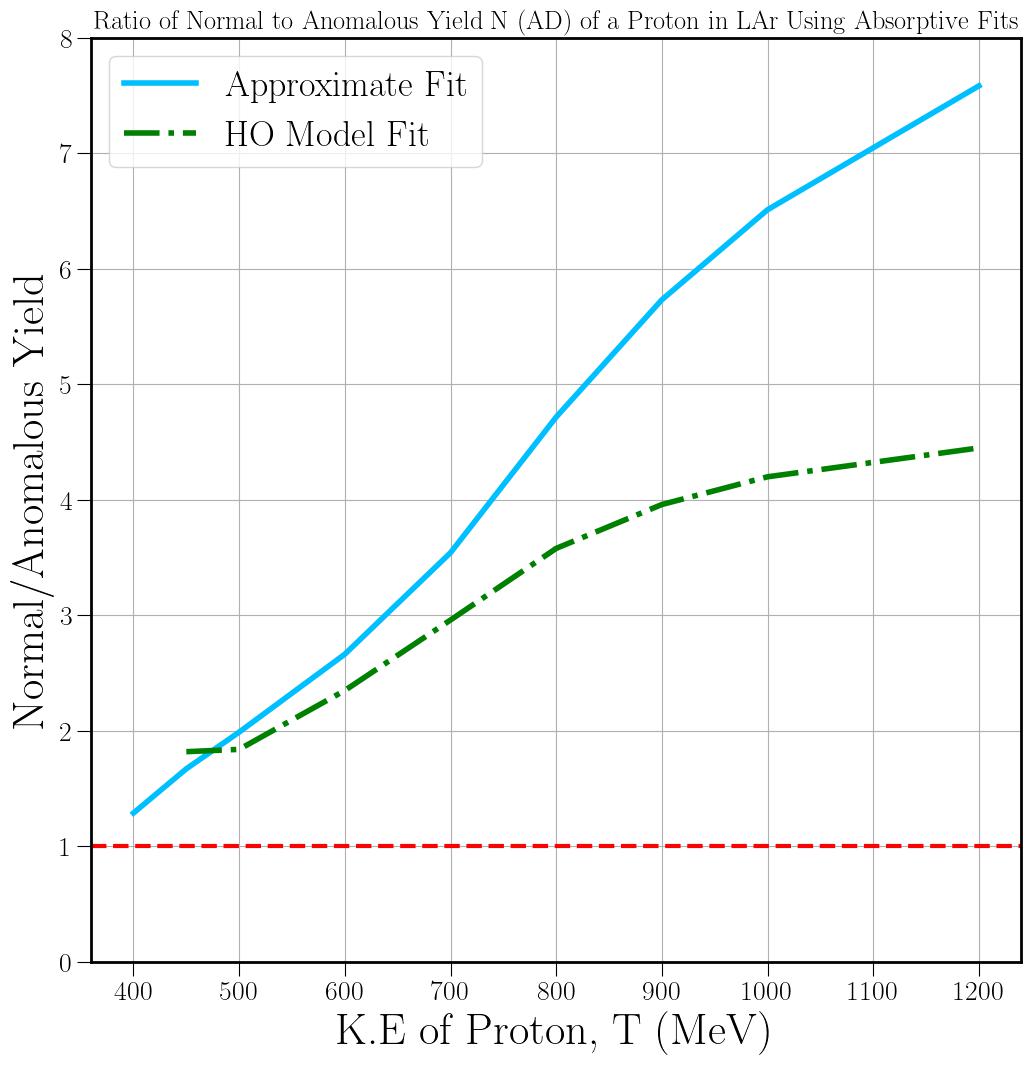}
\caption{
\label{f:nbyahovsapprox}
}
\end{subfigure}
\caption{(a) Comparison (\% difference) of total Cherenkov yield (N) from two different methods: Frank-Tamm Integral (FT) and Angular Distribution (AD) using the approximate fit.
(b) Ratio of normal to anomalous components of the total Cherenkov yield (N) of Protons with different kinetic energies in LAr using the approximate (absorptive) fit.
\label{f:diffnbyaapprox}
}
\end{centering}
\end{figure}

Now the crosschecks we performed with the previous fits can also be done for this approximate fit. Again, we verify the numerical convergence of the two approaches; the percent diﬀerence between the total Cherenkov yield calculated previously from integrating Frank-Tamm formula N (FT) and Angular Distribution method N(AD) for Proton with a
range of K.E.s is shown in Fig.~\ref{f:perdiffapprox}.  Both methods are in good agreement ($< 2 \%$) for the energy range we are interested in. 

In Fig.~\ref{f:nbyahovsapprox}, we plot the  normal-to-anomalous yield ratio for the approximate fit and compare with the HO model fit.  The two curves are qualitatively similar, growing with increasing energy quickly at first, and then slowing, with a change in curvature around $\sim 750$ MeV due to the IR saturation of the emitting wavelength ranges.  The major source of differences between the approximate at HO model fits is the difference in thresholds:  $\sim 384$ MeV versus $\sim 445$ MeV, respectively.  This means that, for the same initial proton energy $T$, the approximate fit is deeper into the phase-space-dominated regime than the HO mode.  This, along with other small differences in the shape of the fit, result in a normal-to-anomalous ratio which gets closer to unity at low energies and is almost twice as large at high energies energies $\sim 1200$ MeV.

\subsubsection{AD Comparison: Cherenkov  Vs. Scintillation} \label{scintcomp}

\hspace{\parindent}

    
    

To be detected, the Cherenkov photons discussed in this work must be significant in comparison to the dominant source of photons produced during the passage of protons through liquid argon: scintillation photons.  We recall that, unlike Cherenkov radiation, scintillation light is isotropically emitted by the argon atoms after being excited by the passage of the charged proton.  The ratio of integrated yields $\frac{N_{Cherenkov}}{N_{Scintillation}}$ as shown in Fig.~\ref{f:allFTvsscintback_log} gives a first measure of the signal-to-background ratio achievable for Cherenkov emission.



%
\begin{figure}[t]
\begin{centering}
\begin{subfigure}{.48\textwidth}
    \centering
    \includegraphics[width=1\textwidth]    {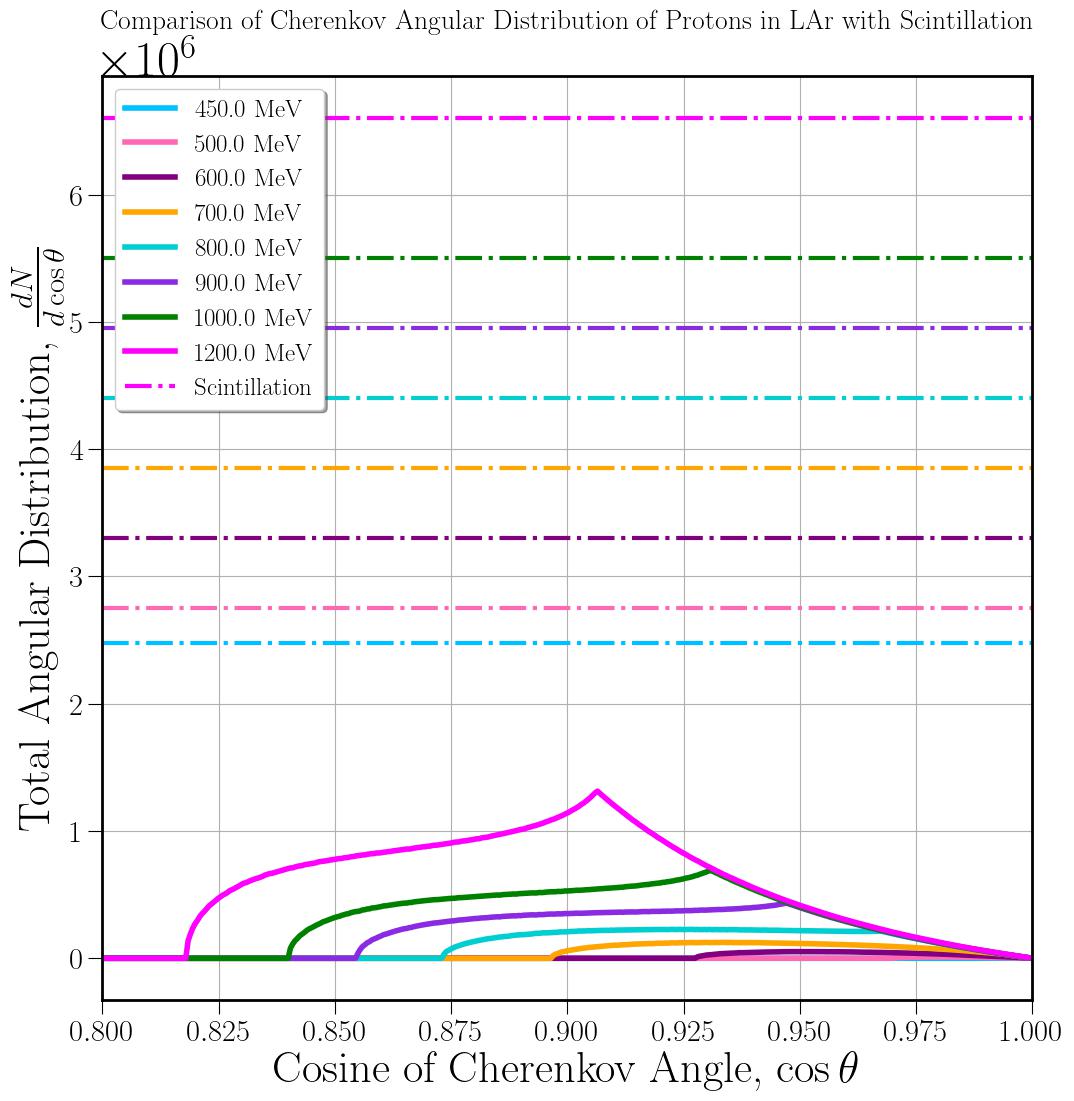}
    \caption{Normal Plot 
    \label{f:compadhoscintnorm}
    }
    \end{subfigure}
\begin{subfigure}{.49\textwidth}
    \centering
    \includegraphics[width=1\textwidth]{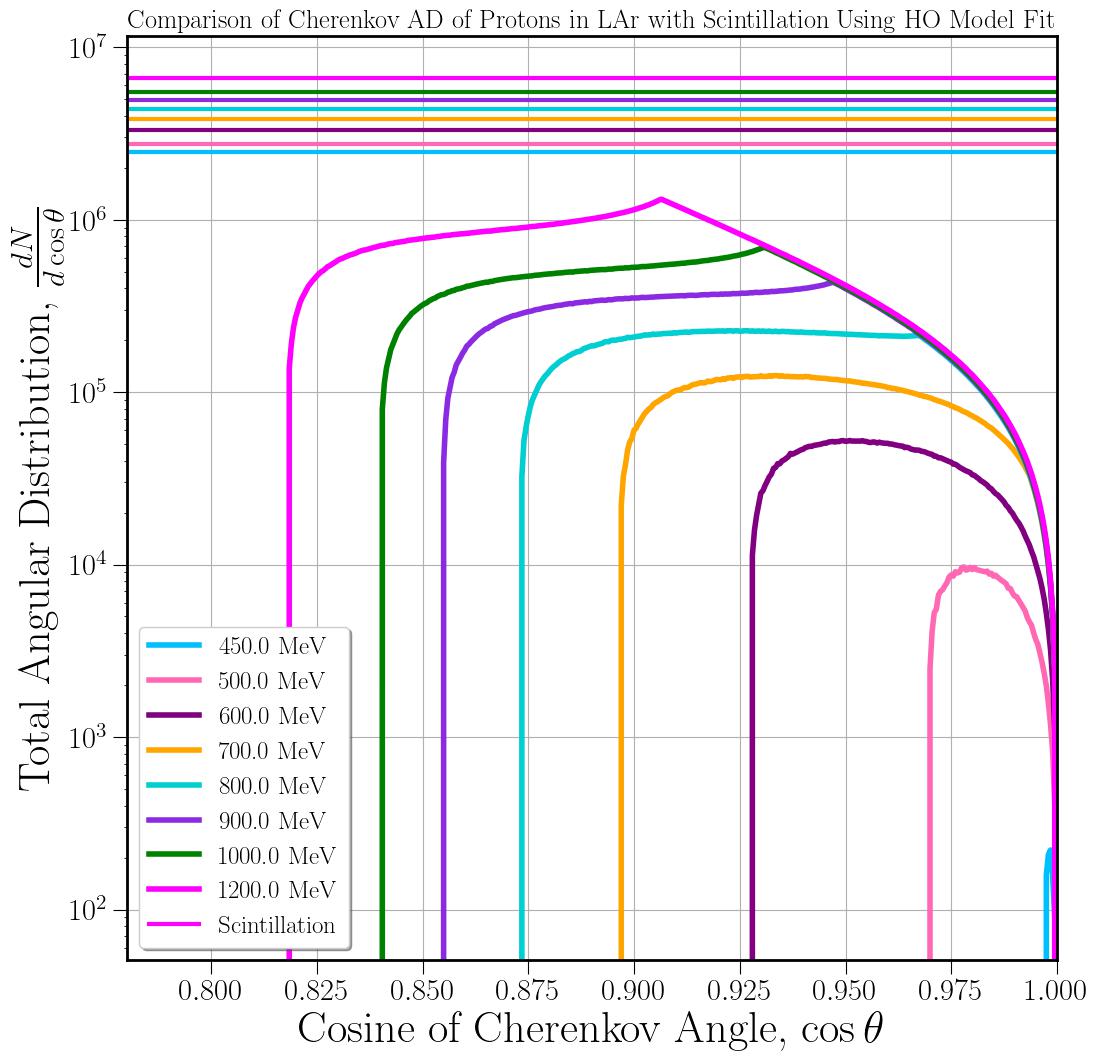}
    \caption{Log Plot
    \label{f:compadhoscintlog}
    }
    \end{subfigure}
\caption{Comparison of Cherenkov and Scintillation Angular Distribution of Protons (with Different K.Es) Travelling in LAr.   
\label{f:compadhoscint1}
}
\end{centering}
\end{figure}
%

\begin{figure}[t!]
\centering
\begin{subfigure}{.5\textwidth}
  \centering
  \includegraphics[width=1\linewidth]{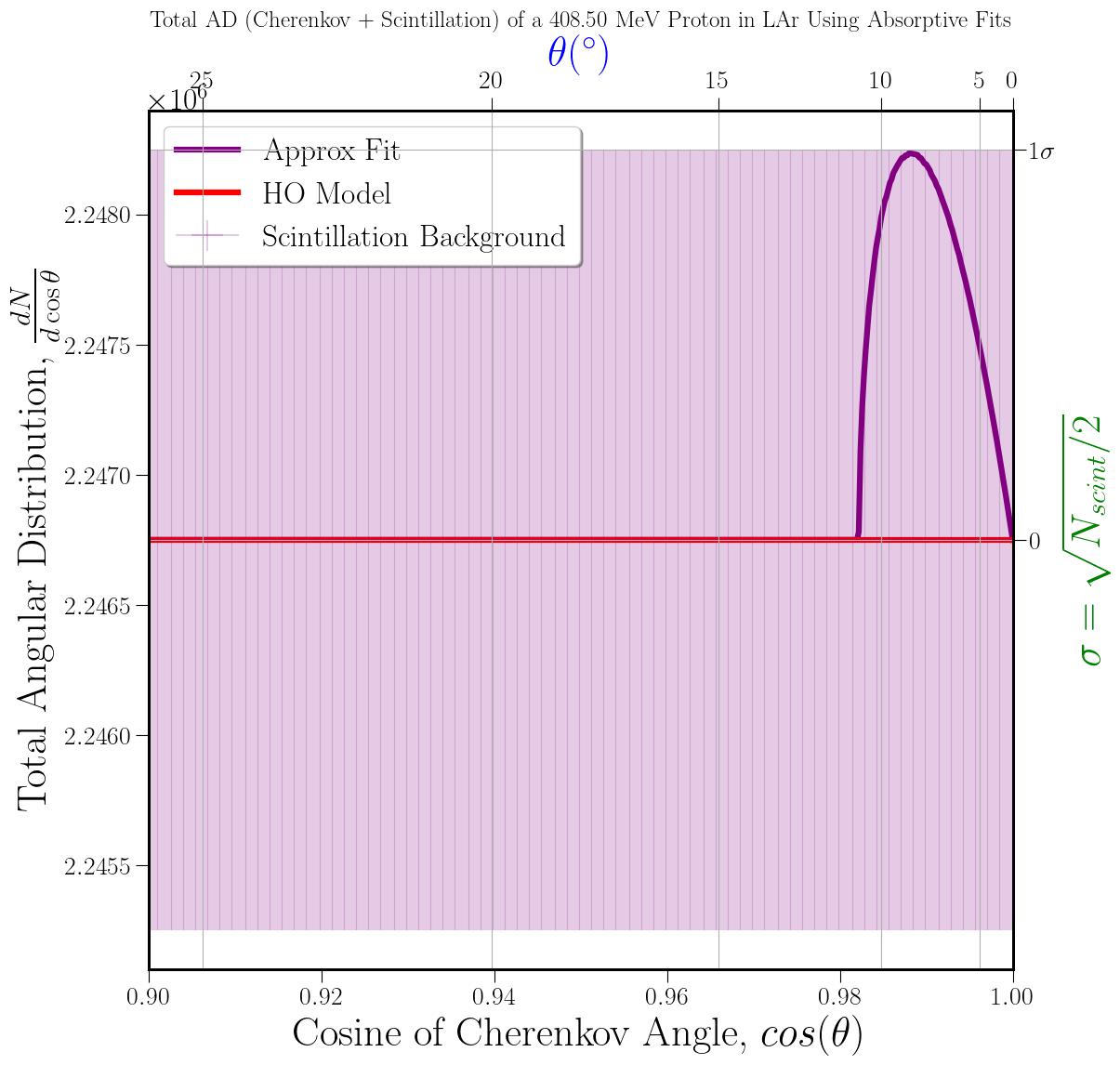}
  \caption{408.50 MeV}
  \label{f:sigoverbackabs408.5MeV}
\end{subfigure}%
\begin{subfigure}{.5\textwidth}
  \centering
  \includegraphics[width=1\linewidth]{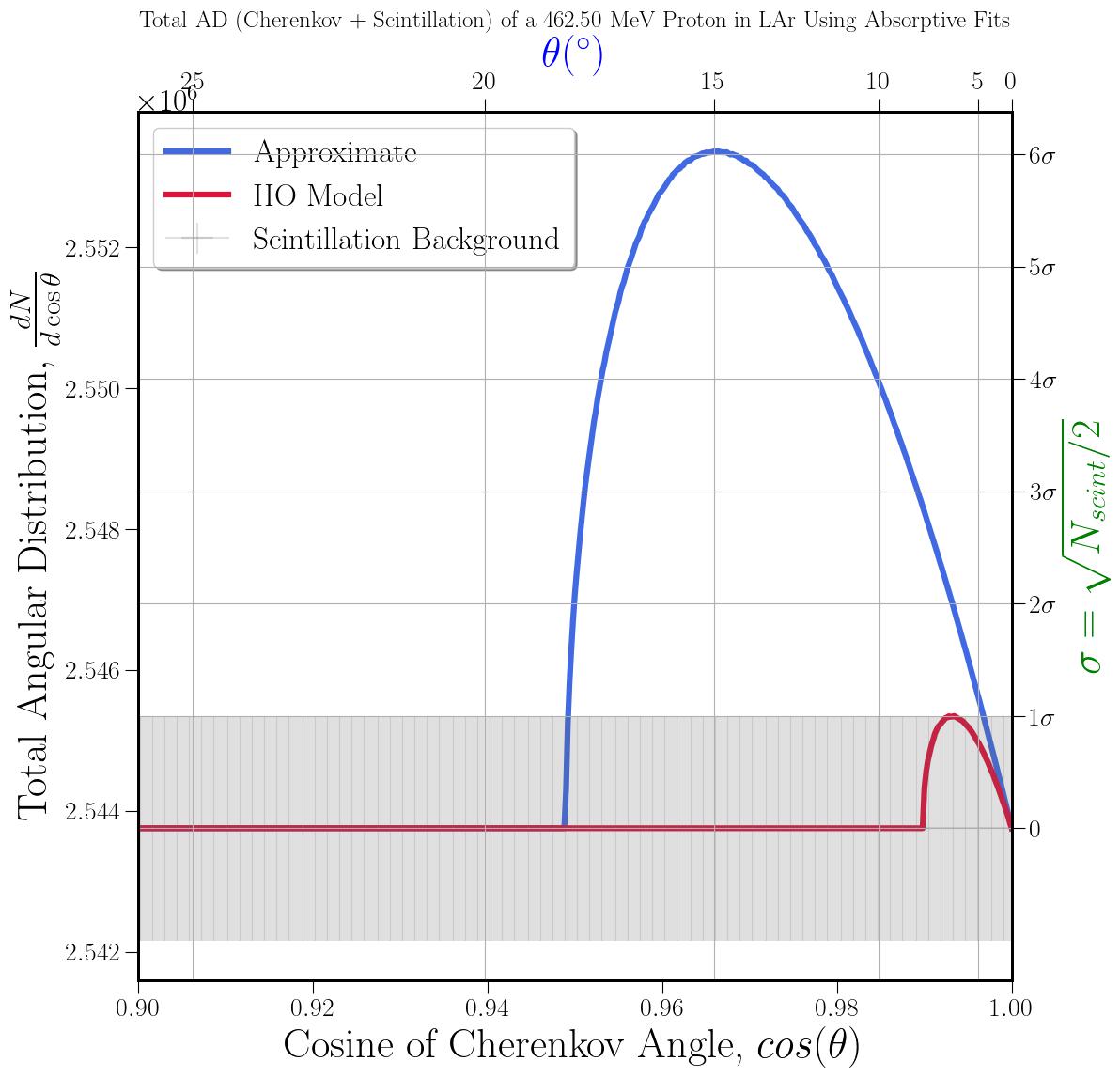}
  \caption{462.50 MeV}
  \label{f:sigoverbackabs462.5MeV}
  \end{subfigure}
\caption{Comparison of Cherenkov AD derived from the absorptive refractive fits with Scintillation light}
\label{f:sigoverbackabs} 
\end{figure}

A far more differential picture of the Cherenkov and scintillation yields is provided by the integrated angular distributions shown in Fig.~\ref{f:compadhoscint1}.  There, we plot $dN/d\cos\theta$ for the HO model fit for protons with initial kinetic energies ranging from 450 MeV to 1200 MeV.  The scintillation photons, being isotropic, are distributed uniformly in $\cos\theta$:
\begin{align}
    \frac{dN_{scint}}{d\Omega} &= \frac{dN_{scint}}{d\cos\theta \, d\phi} = \frac{N_{scint}}{4\pi}
    \notag \\ \notag \\ \rightarrow \qquad
    \frac{dN_{scint}}{d\cos\theta} &= 2\pi \, \frac{dN_{scint}}{d\Omega} = \frac{N_{scint}}{2}  \: ,
\end{align}
with the total integrated yield $N_{scint}$ being given by 
\begin{align} \label{scintcountrep}
    N_{scint} &\approx T \times \left(\frac{40,000 \, \gamma}{\mathrm{MeV}}\right) \times \Big( 27.5\% \, \mathrm{prompt} \Big) 
    \notag \\ &=
    \left(\frac{T}{\mathrm{MeV}}\right) \times 11,000   \: .
\end{align}
We plot this (energy-dependent) constant in the same color as the corresponding Cherenkov AD both on a linear (Fig.~\ref{f:compadhoscintnorm}) and logarithmic scale (Fig.~\ref{f:compadhoscintlog}).  The energy dependence of the yields is easier to see on the linear scale, while the logarithmic plot shows more clearly that the scintillation AD is roughly an order of magnitude higher than the Cherenkov AD shown in Fig.~\ref{f:compadhoscintlog}.

The ADs can be difficult to compare due to the scale difference between scintillation and Cherenkov emission.  However, to be detected, the Cherenkov photons do not need to exceed the total number of scintillation photons in a given angular bin; they simply have to produce an excess which is measurable above the statistical fluctuations of the scintillation AD.  Since these fluctuations are Poissonian, the scintillation background is just the \textit{square root} of the scintillation AD:   $\sigma = \sqrt{\frac{d N_{scint}}{d\cos\theta}} = \sqrt{\frac{N_{scint}}{2}}$, which is a substantially reduced detection threshold compared to the scintillation AD itself.  
\begin{align}   \label{e:ScintError2}
    \sigma \equiv \sqrt{\frac{N_{scint}}{2}} = \sqrt{\left(\frac{T}{\mathrm{MeV}}\right) \times 5,500} \: .
\end{align}
This results in a significantly more optimistic picture of the Cherenkov ``signal'' to be detected against the ``scintillation background''.






%
\begin{figure}[t]
\begin{centering}
\includegraphics[width=0.7\textwidth]{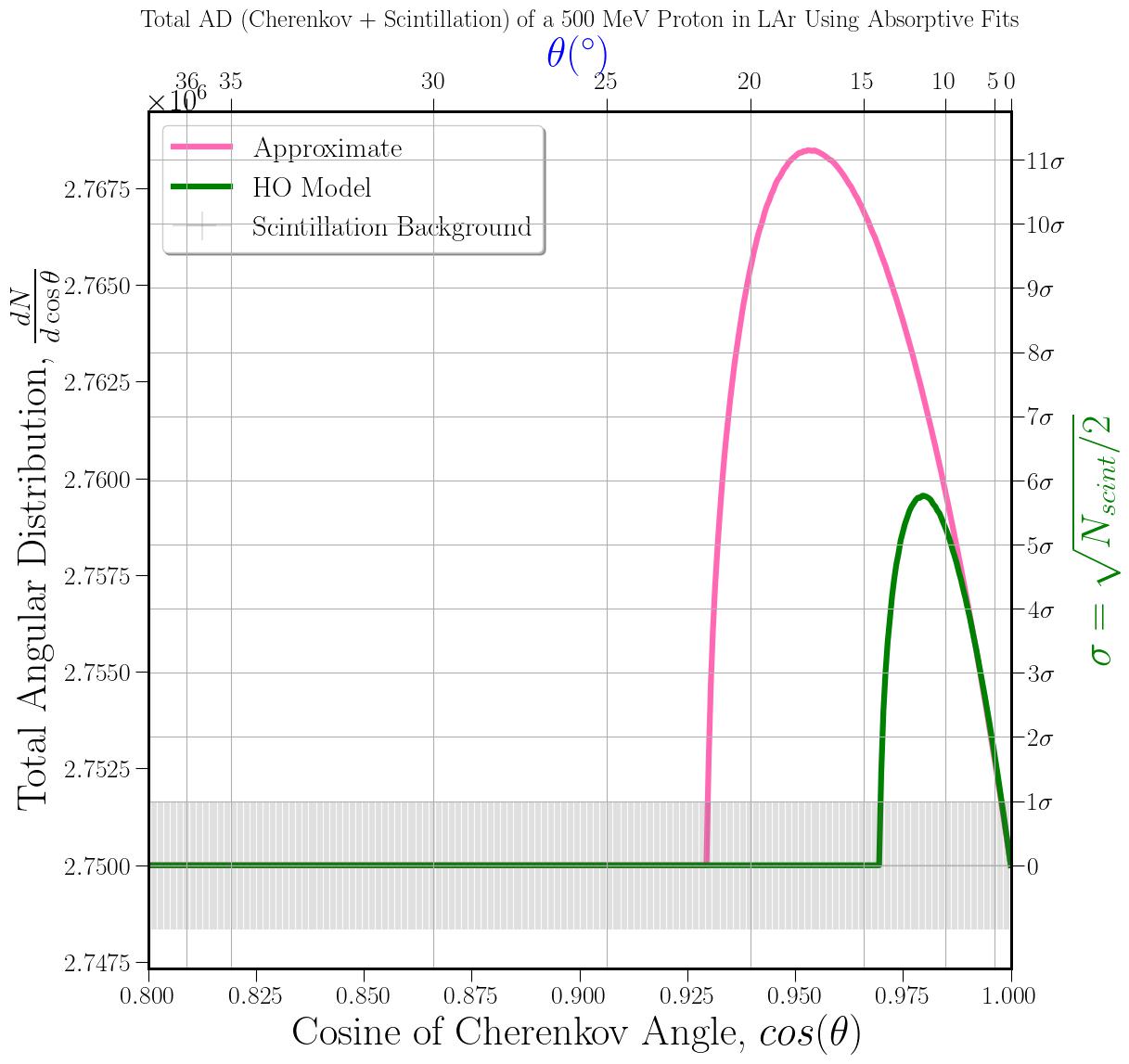
}
\caption{Comparison of Cherenkov Radiation with Scintillation Light for a 500 MeV proton.
\label{f:absfitssigoverback500MeV}
}
\end{centering}
\end{figure}

In Fig.~\ref{f:sigoverbackabs}, such comparison between the Total (Cherenkov + Scintillation) AD of photons and the uncertainty $\sigma$  in the scintillation background is presented for two different initial proton energies $T$.  We focus first on the threshold cases of protons at the lowest kinetic energies capable of emitting Cherenkov photons for a given fit of the refractive index.  Fig.~\ref{f:sigoverbackabs408.5MeV} shows the Cherenkov AD of a $408.50$ MeV proton computed within the approximate model, which just reaches the $1\sigma$ level of significance at a characteristic angle of about $8 ^{\circ}$.  Above this point, the Cherenkov signal will exceed the typical $1\sigma$ statistical fluctuations of the scintillation background; we consider this to be the ``detection threshold'' of the Cherenkov photons above the scintillation background.  Note that, at $T < 445$ MeV, we are still below the minimum threshold to emit Cherenkov photons for the HO model, which therefore predicts no Cherenkov radiation.  

The detection threshold in the HO model is at $T = 462.50$ MeV, as shown in Fig.~\ref{f:sigoverbackabs462.5MeV}.  At this energy, the HO model (crimson curve) peaks at a characteristic angle of approximately $6 ^{\circ}$.  Meanwhile, the Cherenkov signal predicted by the approximate fit to the refractive index (royal blue curve) has already reached $6 \sigma$ of significance above the scintillation background.  At this energy, the AD of the approximate model peaks at $15 ^{\circ}$.

A similar comparison of Cherenkov signal versus scintillation background is shown for a 500 MeV Proton in Fig.~\ref{f:absfitssigoverback500MeV}.  At this energy, not too far above threshold, the Cherenkov AD computed from either model is substantially above the scintillation background.  The HO model (green curve) peaks at a significance of nearly $6 \sigma$ at about $12 ^{\circ}$, whereas the approximate fit (pink curve) crosses a striking $11 \sigma$ at about $17 ^{\circ}$.  

The key difference between the HO and approximate fits is the peak of the refractive index (1.36 for HO model and 1.42 for the approximate model; see Fig.~\ref{f:n_vs_lambda_absfits}).  This determines the threshold kinetic energy (velocity) such that $\beta_{min} = 1 / n_{peak}$.  Consequently, since the threshold energy for Cherenkov emission is lower ($T = 384$ MeV) for the approximate fit compared to HO model fit ($T = 445$ MeV), the approximate fit produces almost twice as strong a Cherenkov signal compared to the HO model fit for a 500 MeV Proton.

Not only does $n_{peak}$ control the height of the Cherenkov AD; it also controls the width through the kinematic boundary $\cos\theta > 1 / \beta  n_{peak}$.  As a result, the fits with different $n_{peak}$ values generate angular distributions with different widths: the approximate fit AD is spread over a wider angular range up to $22^{\circ}$, whereas the HO model's AD only extends out to $14^{\circ}$.  This can simply be visualized as much smaller Cherenkov cone size for the HO model of the refractive index compared to the approximate fit.  Despite this difference in angular spread, for a 500 MeV proton, both models of the refractive index predict a similar AD in the innermost region of the Cherenkov cone, below $8^{\circ}$ above which the distribution gets seemingly different pertaining similar functional shape.  This may be due to the fact that, in the infrared, both models for the refractive index have similar functional forms.



\begin{figure}[p]
\centering
\begin{subfigure}{0.5\columnwidth}
\centering
\includegraphics[width=\textwidth]{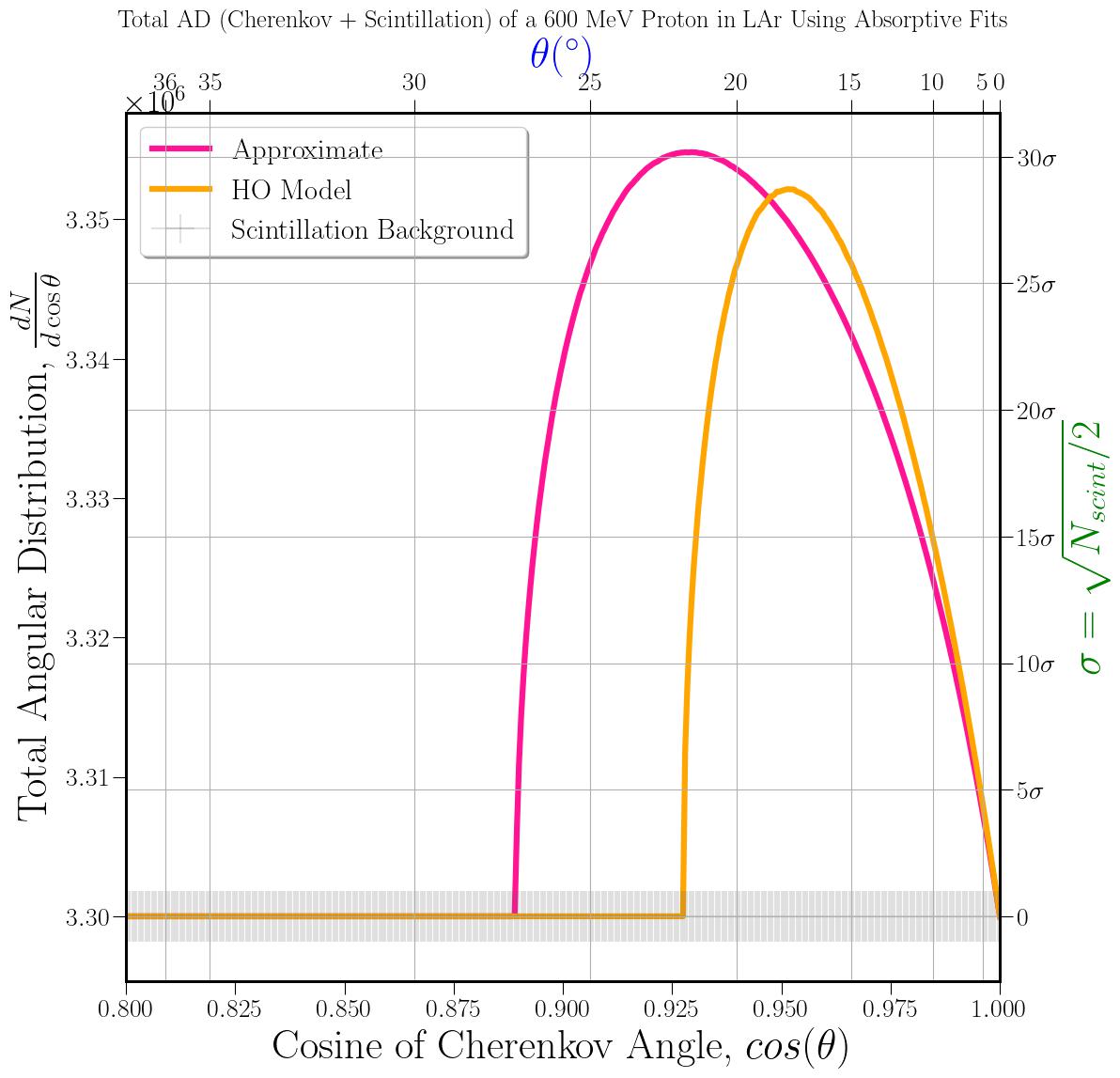}
\caption{600 MeV Proton}
\label{f:absfitssigoverback600MeV}
\end{subfigure}\hfill
\begin{subfigure}{0.5\columnwidth}
\centering
\includegraphics[width=\textwidth]{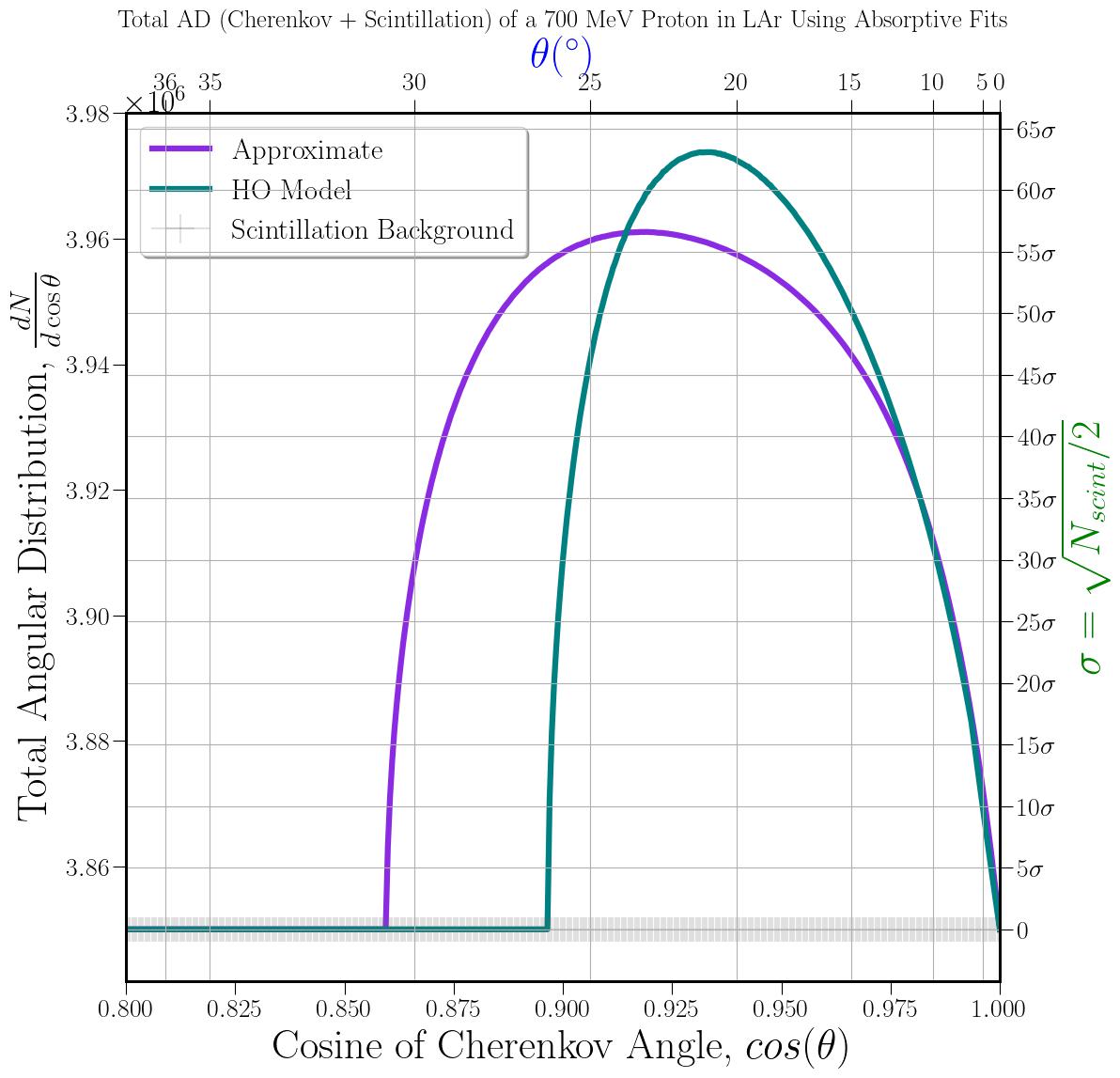}
\caption{700 MeV Proton}
\label{f:absfitssigoverback700MeV}
\end{subfigure} 

\medskip

\begin{subfigure}{0.5\columnwidth}
\centering
\includegraphics[width=\textwidth]{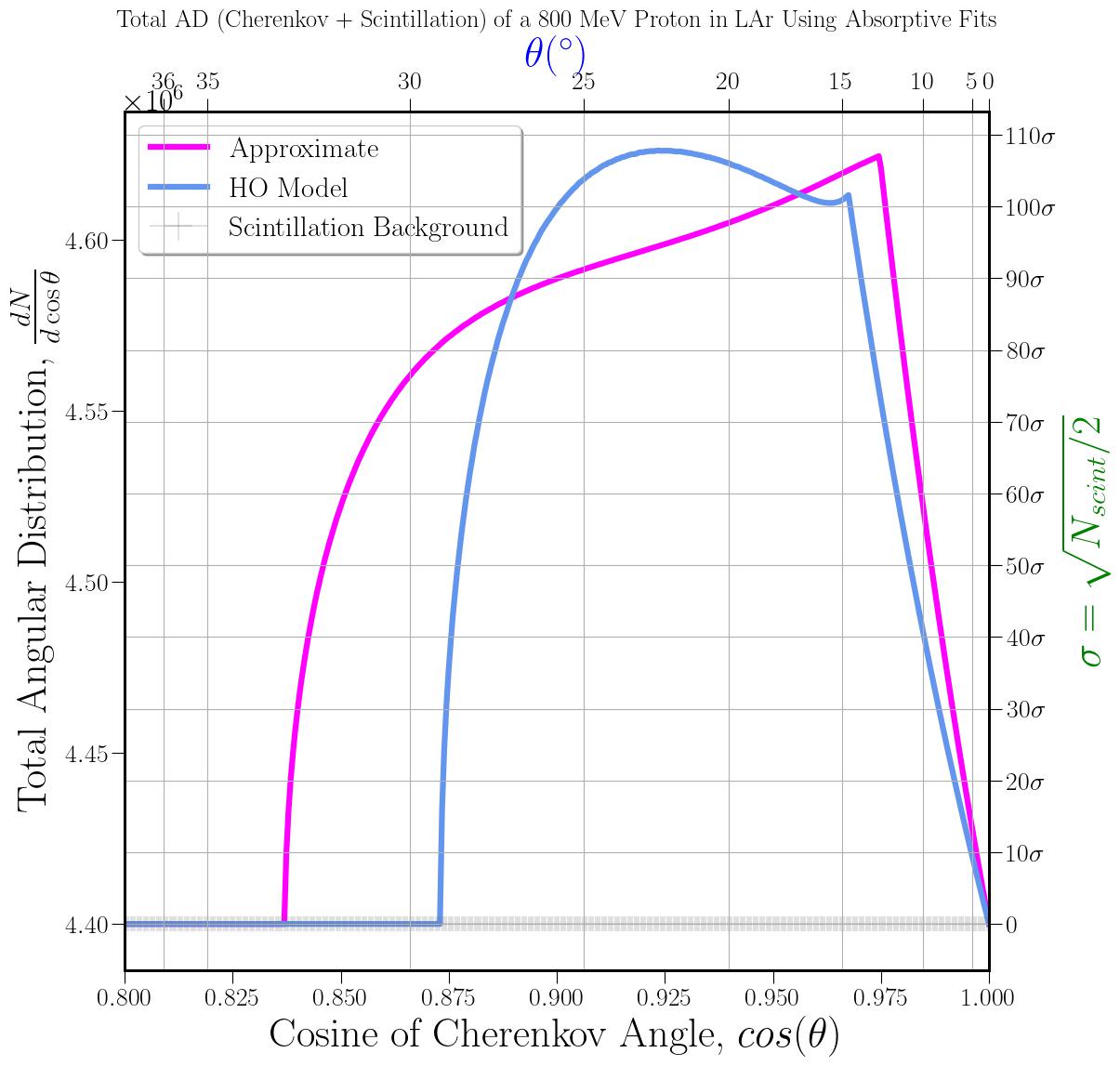}
\caption{800 MeV Proton}
\label{fig:absfitssigoverback800MeV}
\end{subfigure}\hfill
\begin{subfigure}{0.5\columnwidth}
\centering
\includegraphics[width=\textwidth]{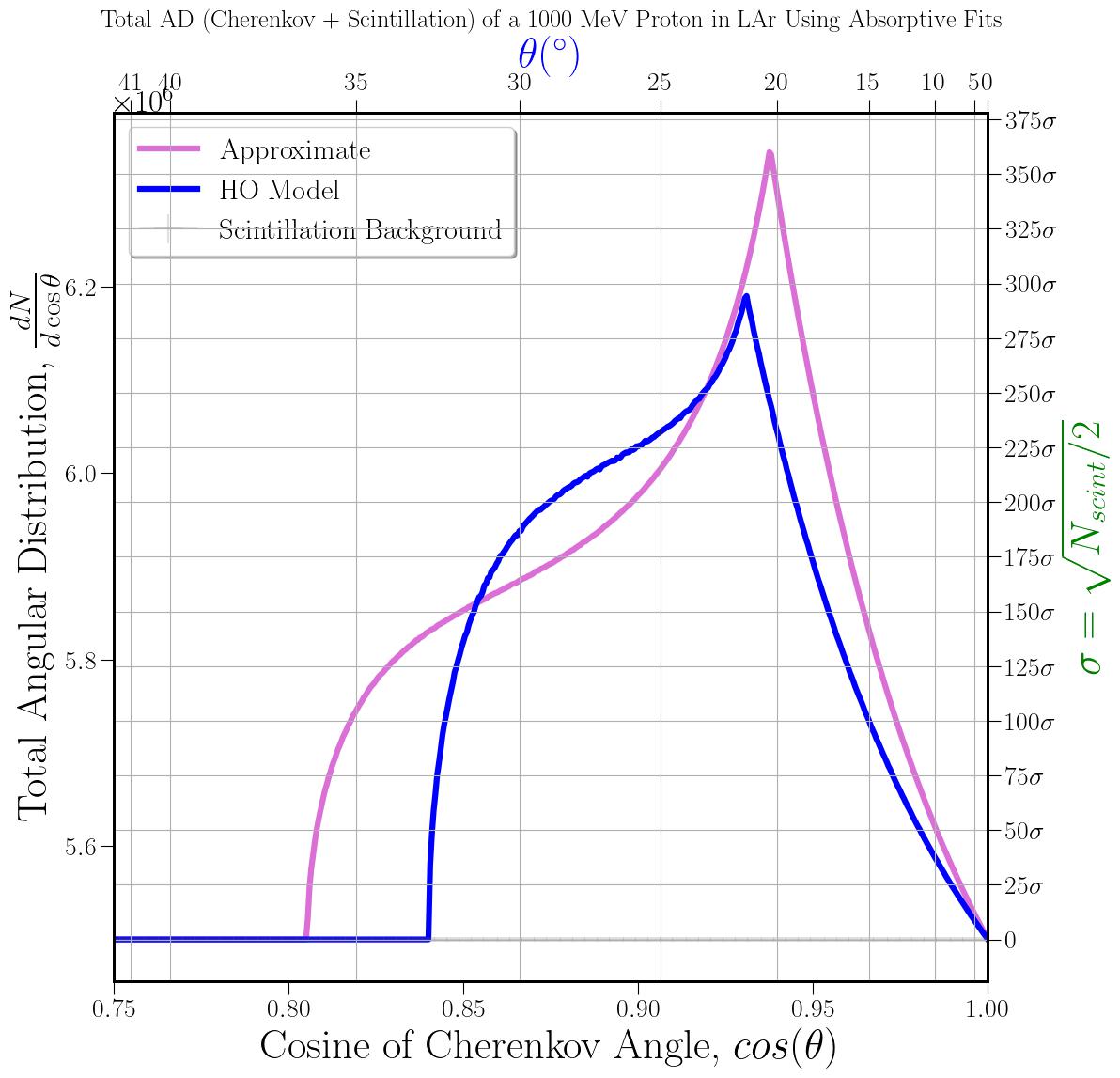}
\caption{1000 MeV Proton}
\label{f:absfitssigoverback1000MeV}
\end{subfigure}
\caption{Comparison of Cherenkov AD Using HO model fit with Scintillation Background}
\label{f:absfitssigoverback2}
\end{figure}


In Fig.~\ref{f:absfitssigoverback2}, we similarly compare the Cherenkov signal and scintillation background for both models over a range of kinetic energies (600 - 1000 MeV).  For $T = 600$ MeV, (Fig.~\ref{f:absfitssigoverback600MeV}), the situation is similar to the 500 MeV proton depicted in Fig.~\ref{f:absfitssigoverback500MeV}.  By 600 MeV, the peak of the HO model at $\sim 18^{\circ} $ is comparable to the peak of the approximate model around $22^\circ$, reaching a significance of $27 \sigma$ above the scintillation background.  By 700 MeV (Fig.~\ref{f:absfitssigoverback700MeV}),  the HO model now exceeds the approximate fit, reaching a peak of $\sim 63\sigma$ at an angle of $\sim 22^\circ$, with the latter reaching a broad peak of $\sim 56 \sigma$ at an angle of $\sim 24 ^{\circ}$.  The two angular distributions coincide for $\theta \leq 12^\circ$, with the HO model signal exceeding that of the approximate fit between $12 ^{\circ}< \theta < 20 ^{\circ}$.  Thereafter, the angular distribution for the HO model reaches its outer kinematic boundary, while the approximate fit still produces sizeable Cherenkov radiation for an additional $\sim 5^\circ$.  At this energy, the HO model catches up with the yield from the approximate fit, but since the HO model's Cherenkov emission is concentrated into a narrower angular region than the approximate model, the intensity of the Cherenkov signal is enhanced.

With further increase in the energy of the proton, the shape of the Cherenkov angular distribution from both models get significantly distorted, as shown in Fig.~\ref{fig:absfitssigoverback800MeV} for $T = 800$ MeV.  This change in the overall shape of AD at this intermediate T using absorptive fits is due to the saturation of the emitting wavelength range $\lambda_{UV} \leq \lambda \leq 500 \: \mathrm{MeV}$, as previously discussed in Sec.~\ref{Sec:AbsorptiveAD}.  This important transition occurs around 678 MeV for the HO model and 697 MeV for the approximate model, so that Fig.~\ref{f:absfitssigoverback700MeV} is just barely above this threshold, while the significant shape modifications appear at higher energies.

At 800 MeV, both fits produce very distinct shapes which are in transition between the smooth, low-energy angular distributions and the peaked high-energy angular distributions.  The shape change leads to a double crossing of the two angular distributions, once at an intensity of $\sim 100 \sigma$ at $\sim 17 ^{\circ}$ and again at an intensity of $\sim 87 \sigma$ at $\sim 27 ^{\circ}$.  Both fits produce maximum signal $> 100 \sigma$ level, but at notably different angles.  The shape of the Cherenkov signal continues to evolve with  increasing energy, producing qualitatively similar distributions for both models at $T = 1000$ MeV (Fig.~\ref{f:absfitssigoverback1000MeV}).  The new shape of the Cherenkov AD possesses a sharp, pointy peak toward the interior of the Cherenkov cone, along with a broad ``shoulder'' of intensity extending out to higher angles.  The sharp peak and subsequent decrease of the Cherenkov intensity in the innermost core of the Cherenkov cone is due to the kinematic bound $\cos\theta < 1 / \beta n_{min}$ and is sensitive to the infrared regime $n(\lambda)$ for large wavelengths $\lambda$.  At this energy, the peak Cherenkov signals of the HO and approximate models exceed a staggering  $\sim 300 \sigma$ and $\sim 360 \sigma$, respectively, at corresponding angles of $\sim 21 ^{\circ}$ and $\sim 20 ^{\circ}$. 

\begin{figure}[t]
\centering
\begin{subfigure}{0.5\columnwidth}
\centering
\includegraphics[width=\textwidth]{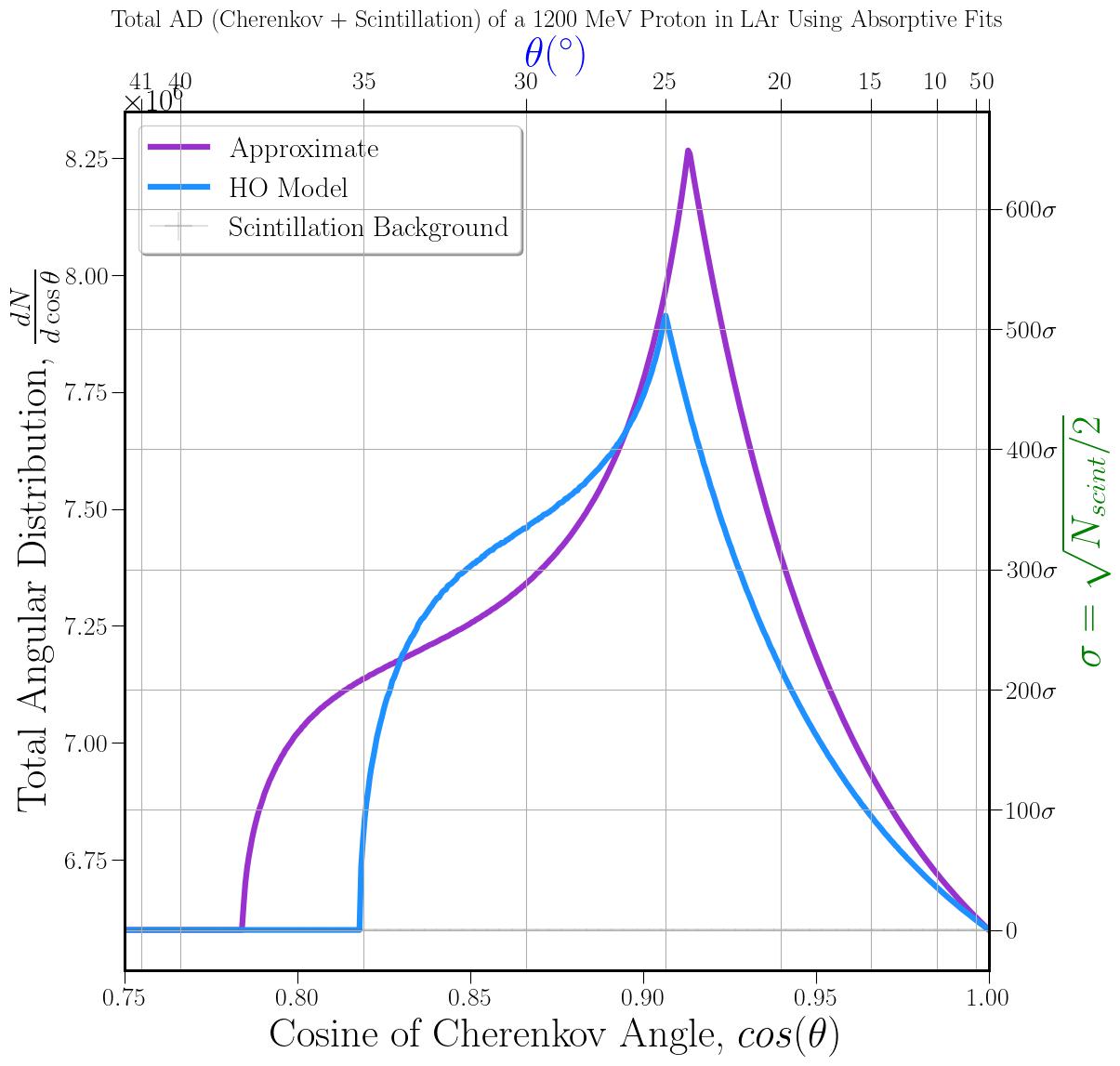}
\caption{1200 MeV Proton}
\label{f:absfitssigoverback1200MeV}
\end{subfigure}\hfill
\begin{subfigure}{0.5\columnwidth}
\centering
\includegraphics[width=\textwidth]{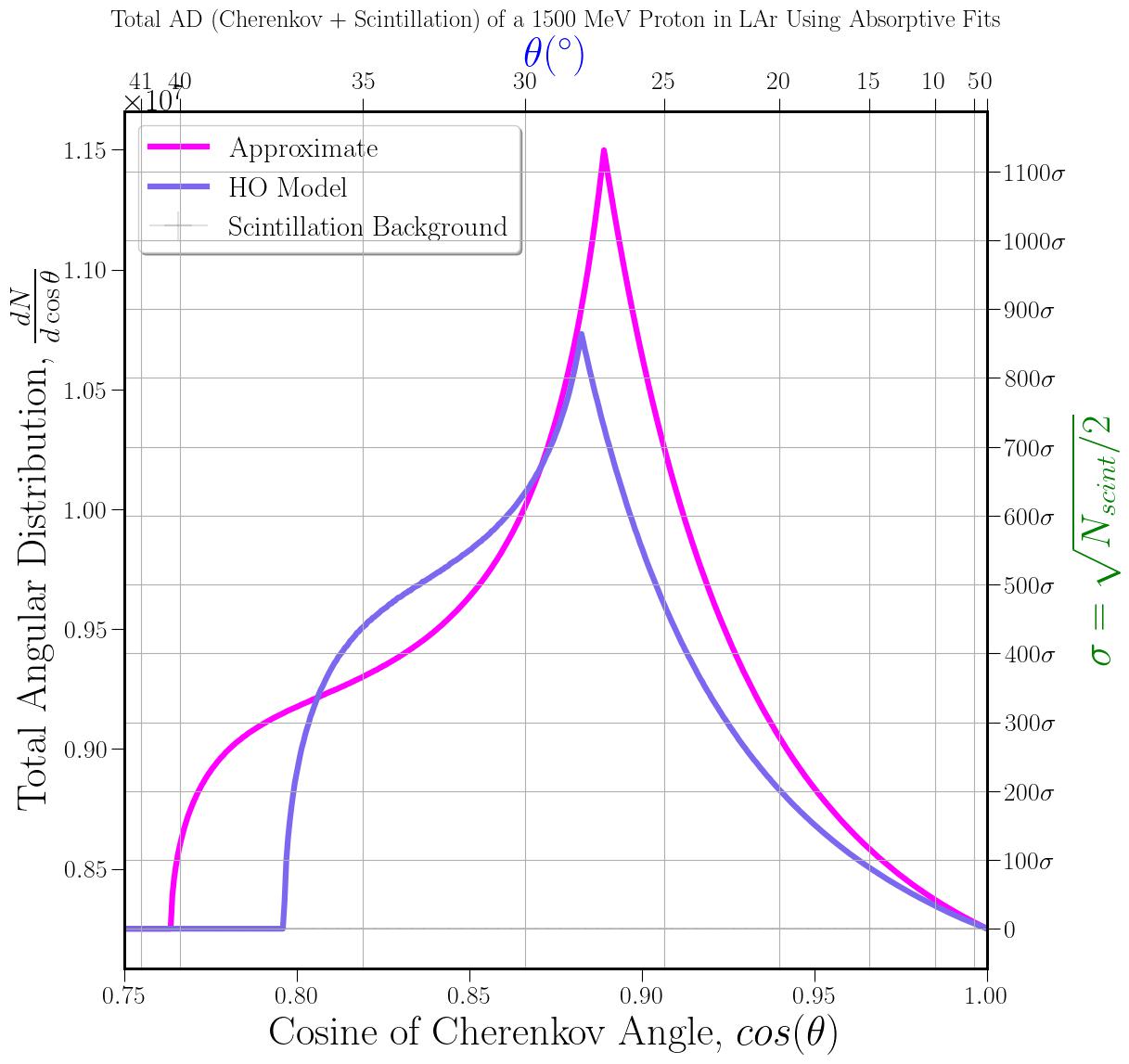}
\caption{1500 MeV Proton}
\label{f:absfitssigoverback1500MeV}
\end{subfigure}
\caption{Comparison of Cherenkov AD Using HO model fit with Scintillation Background}
\label{f:absfitssigoverbackhighT}
\end{figure}
Finally, in Fig.~\ref{f:absfitssigoverbackhighT}, we compare the Cherenkov signal and scintillation background for even higher energies: $T = 1200$ MeV in Fig.~\ref{f:absfitssigoverback1200MeV} and $T = 1500$ MeV in Fig.~\ref{f:absfitssigoverback1500MeV}. The sharp triangular pattern seen at 1000 MeV still prevails for higher T.  After converging to this shape, the angular distribution only continues to increase in height (intensity) and width (angular extent).  Both of the absorptive fits produce huge signal compared to the background at such ultra high $T$.  Both absorptive fits peak at comparable angles: $\sim 24-25 ^{\circ}$ for 1200 MeV and  $\sim 27-28 ^{\circ}$ for 1500 MeV. The height of these peaks and also their angular positions continue to increase with increasing $T$ and are quite large (tens of degrees) for $T \geq 500$ MeV.  Given the extreme levels of statistical significance predicted by even conservative fits to the refractive index, and the robustness of the physical picture that emerges to our modeling assumptions, it seems highly likely that Cherenkov radiation from low-energy protons in LAr is eminently measurable and carries detailed physical signatures of the refractive index.  This carries great promise for potential applications in existing scintillation-based particle detectors worldwide at different neutrino facilities.



\newpage
\newpage


\hspace{\parindent}
\section{CONCLUSION}    \label{sum}



\subsection{Summary}


%
\begin{figure}[h!]
\begin{centering}
\includegraphics[width=0.65\textwidth]{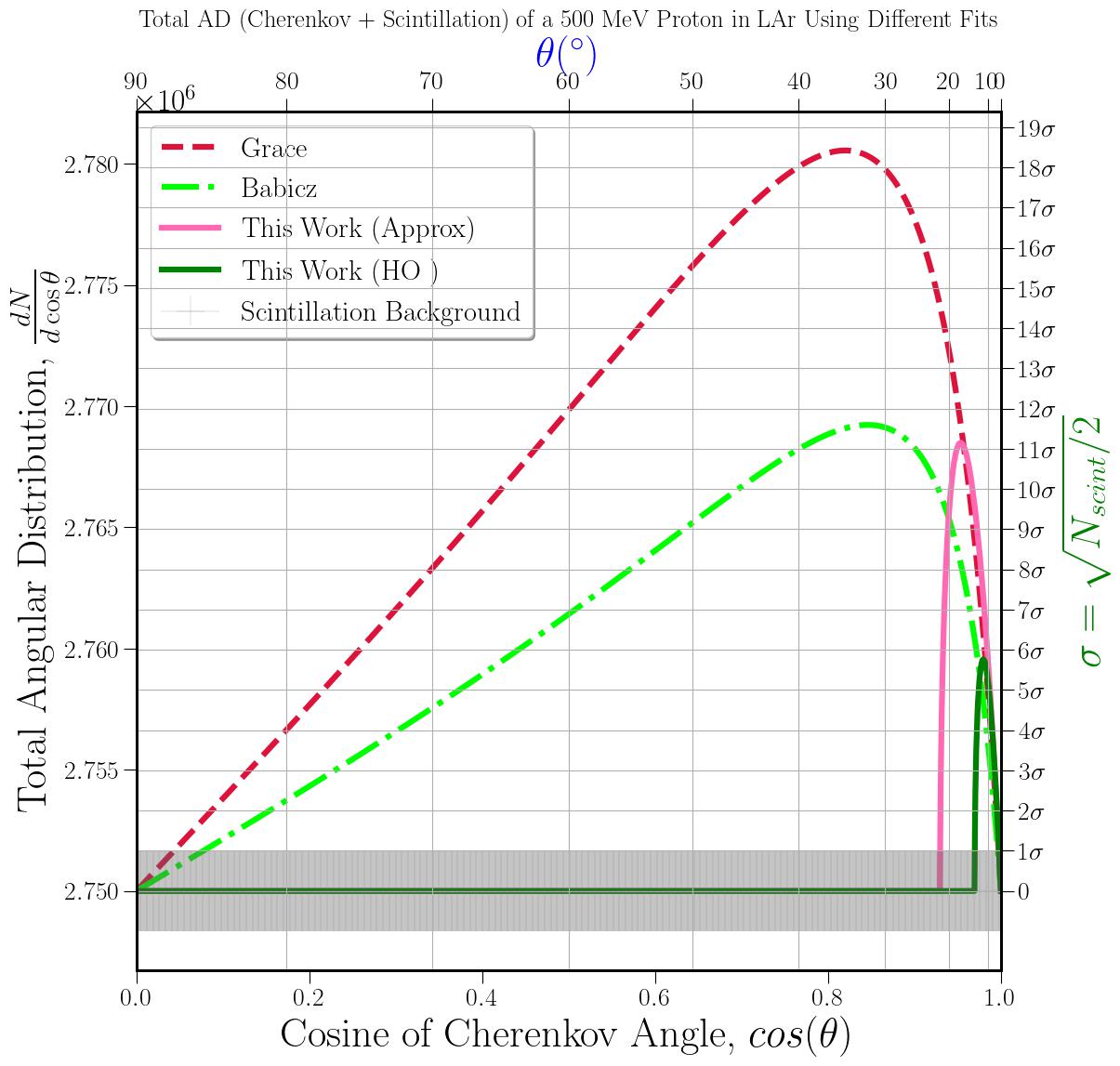
}
\caption{Comparison of Cherenkov Radiation with Scintillation Light using different refractive index fits. 
\label{f:ADcomp500MeV}
}
\end{centering}
\end{figure}

In Chapter \ref{mathform}, we have discussed how the Cherenkov signal compares with the scintillation background when a resonant fit (Babicz, Grace) to the refractive index is employed.  We have seen that this sort of resonant fit overestimates the yield to a great extent due to the fact that they do not contain the correct physics near the resonance.  In contrast, in Chapter \ref{cerenkovwabs}, we used our own absorptive fits (HO model, approximate fit) to calculate the Cherenkov radiation, but this time correctly encoding the physics of absorption and anomalous dispersion near the resonance.  This change in the form of the fit made a substantial cut to the yield, which resulted in a much smaller integrated yield compared to the ones from resonant fits.  



\begin{figure}[p!]
\centering
\begin{subfigure}{.5\textwidth}
  \centering
  \includegraphics[width=1\linewidth]{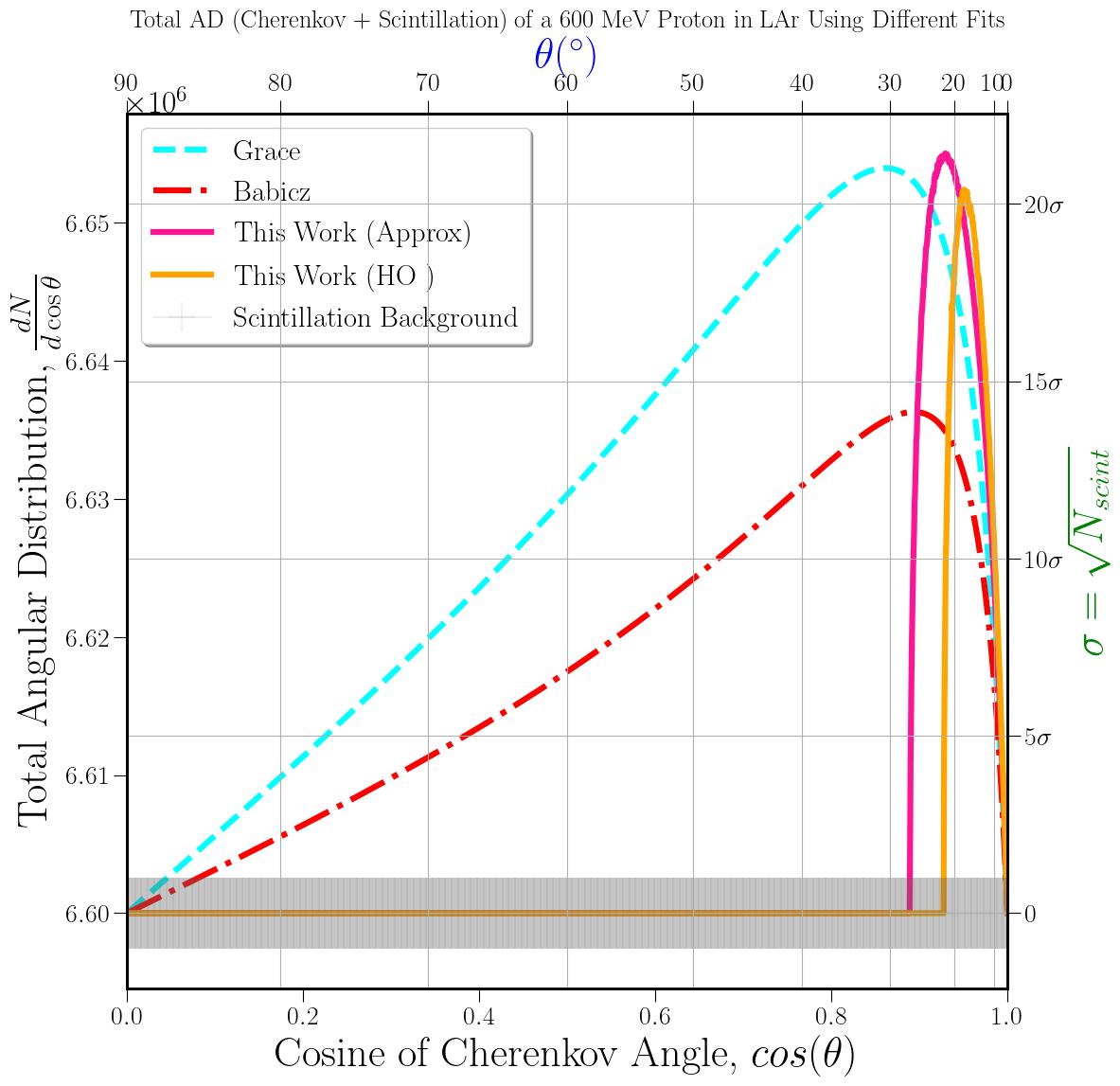}
  \caption{600 MeV}
  \label{f:ADcomp600MeV}
\end{subfigure}%
\begin{subfigure}{.5\textwidth}
  \centering
  \includegraphics[width=1\linewidth]{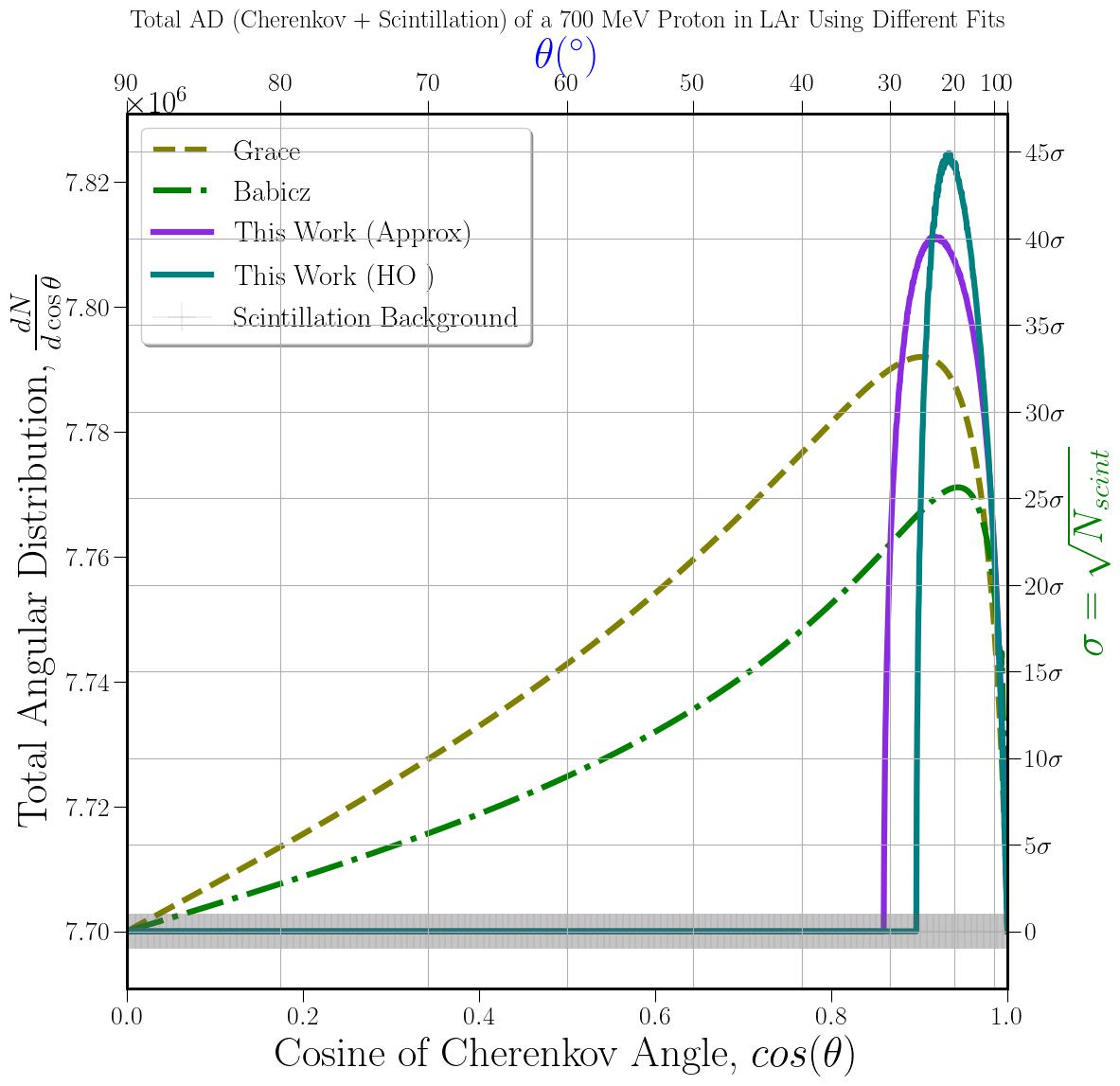}
  \caption{700 MeV}
  \label{f:ADcomp700MeV}
  
\end{subfigure}

\medskip

\begin{subfigure}{.5\textwidth}
  \centering
  \includegraphics[width=1\linewidth]{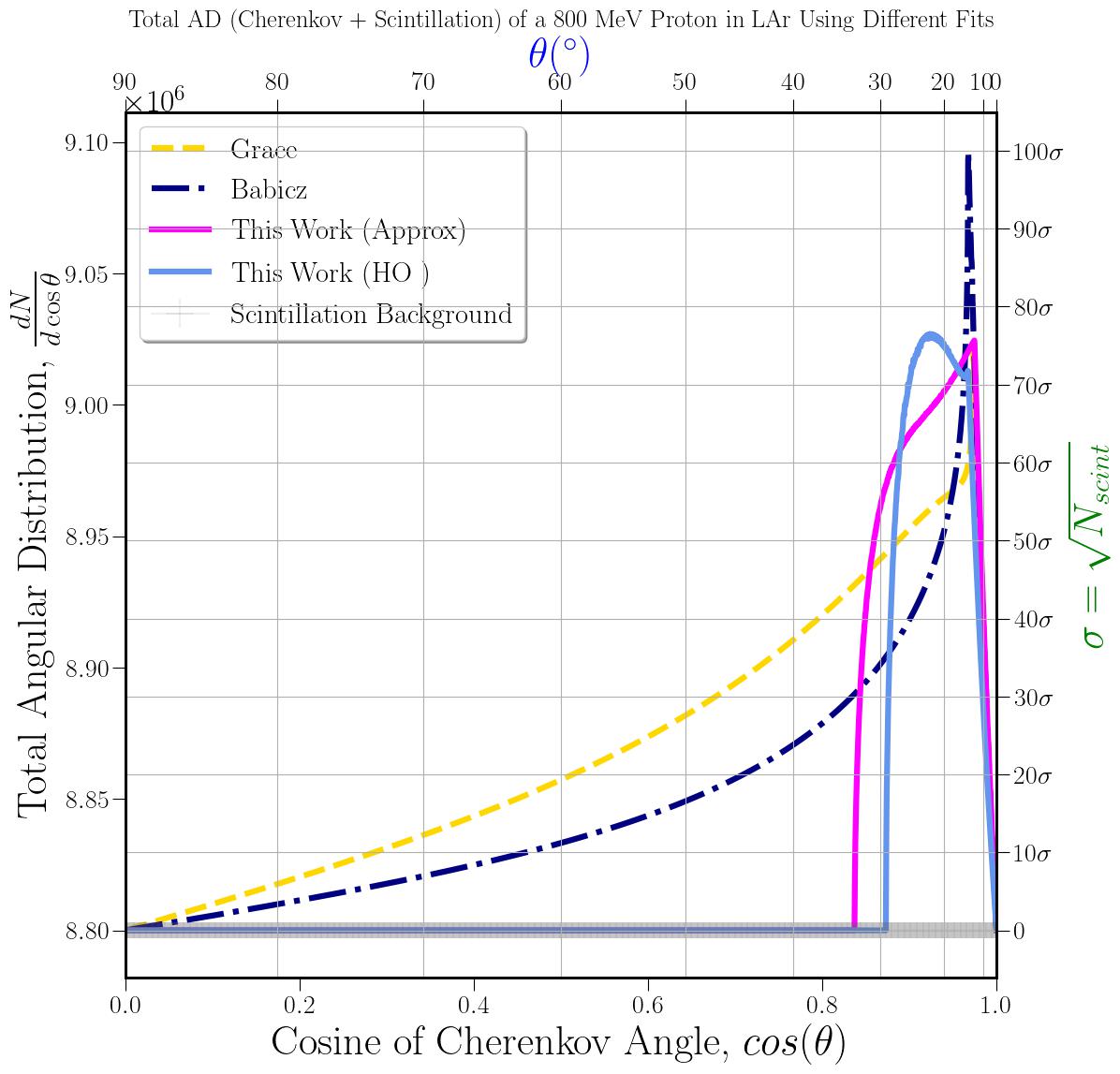}
  \caption{800 MeV}
  \label{f:ADcomp800MeV}
\end{subfigure}%
\begin{subfigure}{.5\textwidth}
  \centering
  \includegraphics[width=1\linewidth]{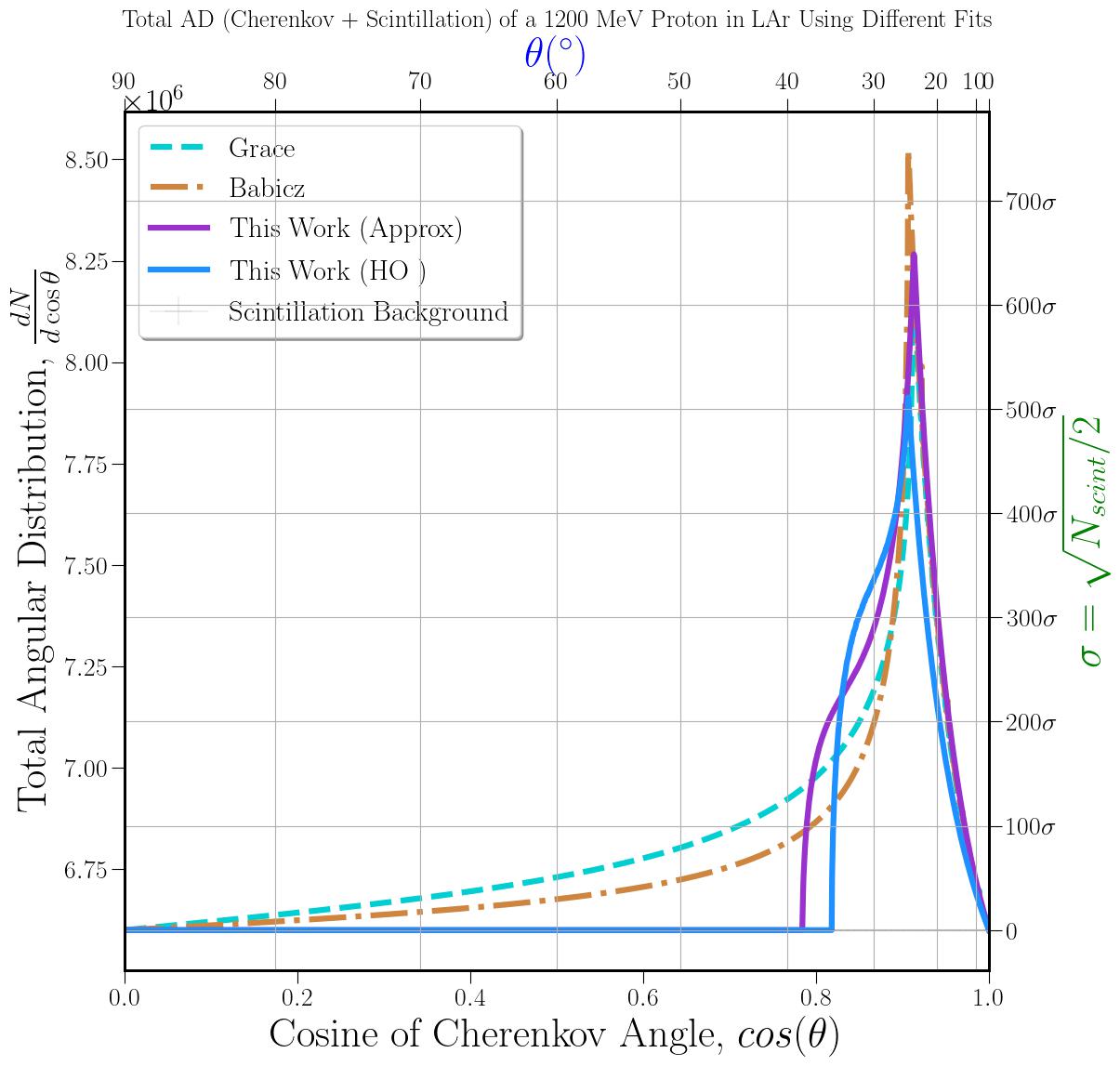}
  \caption{1200 MeV}
  \label{f:ADcomp1200MeV}
  
\end{subfigure}
\caption{Comparison of Cherenkov AD calculated using different fits against scintillation background.}
\label{f:ADcompabsvsnonabs} 
\end{figure}

Despite the simplistic picture of the reduction in yield, the angular distribution of Cherenkov photons reveals significant structure, as summarized in Fig.~\ref{f:ADcomp500MeV}.  There, we compare the total AD (Cherenkov + scintillation) of a 500 MeV proton against the gray uncertainty band of the background due to scintillation alone.  We also mark the vertical axis in units of this uncertainty $\sigma$. A key difference between absorptive (HO model and approximate) fits and resonant (Babicz, Grace) fits is the width of the angular distribution.  Both the HO model and the approximate fit produce narrow spikes with moderate height compared to the non-absorptive resonant fits.  Both the Grace and Babicz resonant fits predict Cherenkov radiation spanning the entire angular range $0 \leq \theta \leq \tfrac{\pi}{2}$, and they produce a higher (unrealistic) signal than either absorptive fit. For a 500 MeV proton, our HO model and approximate peak at $~ 6 \sigma$ and $~ 11 \sigma$ significance, respectively whereas the Grace and Babicz fits peak at $~ 12 \sigma$ and $> 18 \sigma$ respectively. The angular position of the peak intensity also varies for the different fits, with the HO model lying closest to the proton trajectory ($\sim 12 ^{\circ}$ opening angle), followed by the approximate fit ($18 ^{\circ}$) and  the two resonant fits (both $> 30 ^{\circ}$).

In Fig.~\ref{f:ADcompabsvsnonabs}, we depict the energy dependence of the total photon distribution (Cherenkov + scintillation) for proton kinetic energies $T$ ranging from 600 MeV to 1000 MeV.  Already by 600 MeV (Fig.~\ref{f:ADcomp600MeV}), the peaks of the absorptive fits exceed $20 \sigma$ of significance over the scintillation background -- which remarkably is even comparable or greater than the peak intensity of the absorptive fits.  At 700 MeV (Fig.~\ref{f:ADcomp700MeV}), the peak signals from the absorptive fits fully exceed those from resonant fits.  Similarly, despite producing less integrated yield than the more generous approximate model (see Fig.~\ref{f:comparisoncherenkovreabs}), in the differential angular distribution the peak of the HO model ($45 \sigma$ signal) overcomes this handicap and is greater than the approximate fit's peak signal ($\sim 40 \sigma$).  The general conclusion is that the smaller the peak of the refractive index ($n_{peak}^{(HO)} < n_{peak}^{(Approx)}$), the smaller the angular region ($ \frac{1}{\beta n_{peak}^{(Approx)}} < \frac{1}{\beta n_{peak}^{(HO)}} < \cos\theta < 1$) into which the Cherenkov photons are concentrated.  While this results in a \textit{smaller integrated yield} for smaller $n_{peak}$, it inversely results in a \textit{greater angular density} of photons in the narrower window into which they are emitted.

A qualitative change in the shape of the instantaneous angular distribution occurs around 700 MeV for all fits, resulting in a substantial distortion of the integrated angular distributions seen at 800 MeV (Fig.~\ref{f:ADcomp800MeV}) and 1000 MeV (Fig.~\ref{f:ADcomp1200MeV}).  This occurs because the wavelength range capable of emitting Cherenkov radiation eventually becomes saturated with increasing energy, $\lambda_\theta \rightarrow \lambda_{max} 
$, so that higher-energy protons emit in finite angular regions with $(\cos\theta)_{max} < 1$.  At 800 MeV, both absorptive fits produce comparable maximum signals ($> 70 \sigma$) but peak at different angles, which has been elaborated in Chap.~\ref{cerenkovwabs}.  For the resonant fits, the shape is different, but a qualitative change in the integrated AD is again seen around 700 MeV.  Interestingly, near the center of the Cherenkov cone (around $15 ^{\circ}$), there is an inversion of the hierarchy $dN_{Grace}/d\cos\theta > dN_{Babicz}/d\cos\theta$ which results in the Babicz AD having a greater peak signal ($\sim 100 \sigma$) compared to the Grace AD.  With increasing energy (Fig.~\ref{f:ADcomp1200MeV}), the distortion in the shape of the AD continues to grow, resulting in a more triangular profile of the integrated AD.  For $T = 1200 \mathrm{MeV}$, all four fits of the refractive index produce a maximum signal above $500 \sigma$, peaking at similar angles ($\sim 20 - 30 ^{\circ}$).  The location of this high-energy peak, and the shape of the interior edge of the Cherenkov cone, are determined by the $\cos \theta_{max}$ effect.

\subsection{Visualization:  Projection onto a Screen}

%
\begin{figure}[t]
\begin{centering}
\includegraphics[width=0.65\textwidth]{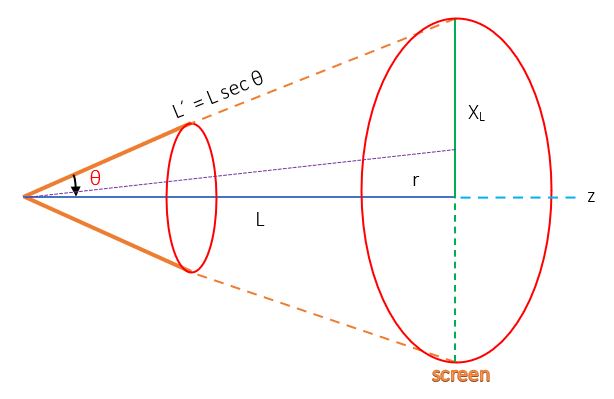
}
\caption{Projection of a Cherenkov cone onto a screen at L distance away from the incoming proton that causes Cherenkov radiation. The radius of the projected ring $x_L = L \tan \theta$ is expressed in terms of the distance L and cosine of the Cherenkov angle $\theta$.
\label{f:ringonthescreengraphic}
}
\end{centering}
\end{figure}

Another way to visualize our results for the Cherenkov angular distribution is to examine its projection onto a screen a distance $L$ away from the incoming proton (Fig.~\ref{f:ringonthescreengraphic}). Let the radius of the ring projected on the screen be $x_L$, which can be expressed in terms of the distance $L$ and Cherenkov angle $\theta$ as $L \tan \theta$ such that
\begin{align}   \label{e:xL}
    x_L \equiv L \tan \theta = L \frac{\sin \theta}{\cos \theta} = L \frac{\sqrt{1 - \cos^2\theta}}{\cos\theta}  \: .
\end{align}
Then we can re-express the angular distribution in terms of $dN/dx_L$ by performing the change of variables from $d\cos\theta$ to $dx_L$,
\begin{align}   \label{e:dndxL1}
    \frac{dN}{dx_L} = \frac{dN}{d (\cos \theta)} . \left| \frac{d \cos \theta}{dx_L}\right|
    \: ,
\end{align}
where the Jacobian can be expressed as
\begin{align}   \label{e:dxLdcostheta}
    \left| \frac{d\cos\theta}{dx_L} \right| &=
    \left| \frac{dx_L}{d\cos\theta} \right|^{-1} 
    %
    \notag \\ &=
    %
    \frac{\cos^2\theta \, \sin\theta}{L} 
    \notag \\ &=
    %
    %
    \frac{1}{L} \: \frac{(x_L / L)}{ \left[1 + \left(\frac{x_L}{L} \right)^2 \right]^{3/2}}
    \: .
\end{align}

\begin{figure}[t!]
\centering
\begin{subfigure}{.4\textwidth}
  \centering
  \includegraphics[width=1\textwidth]{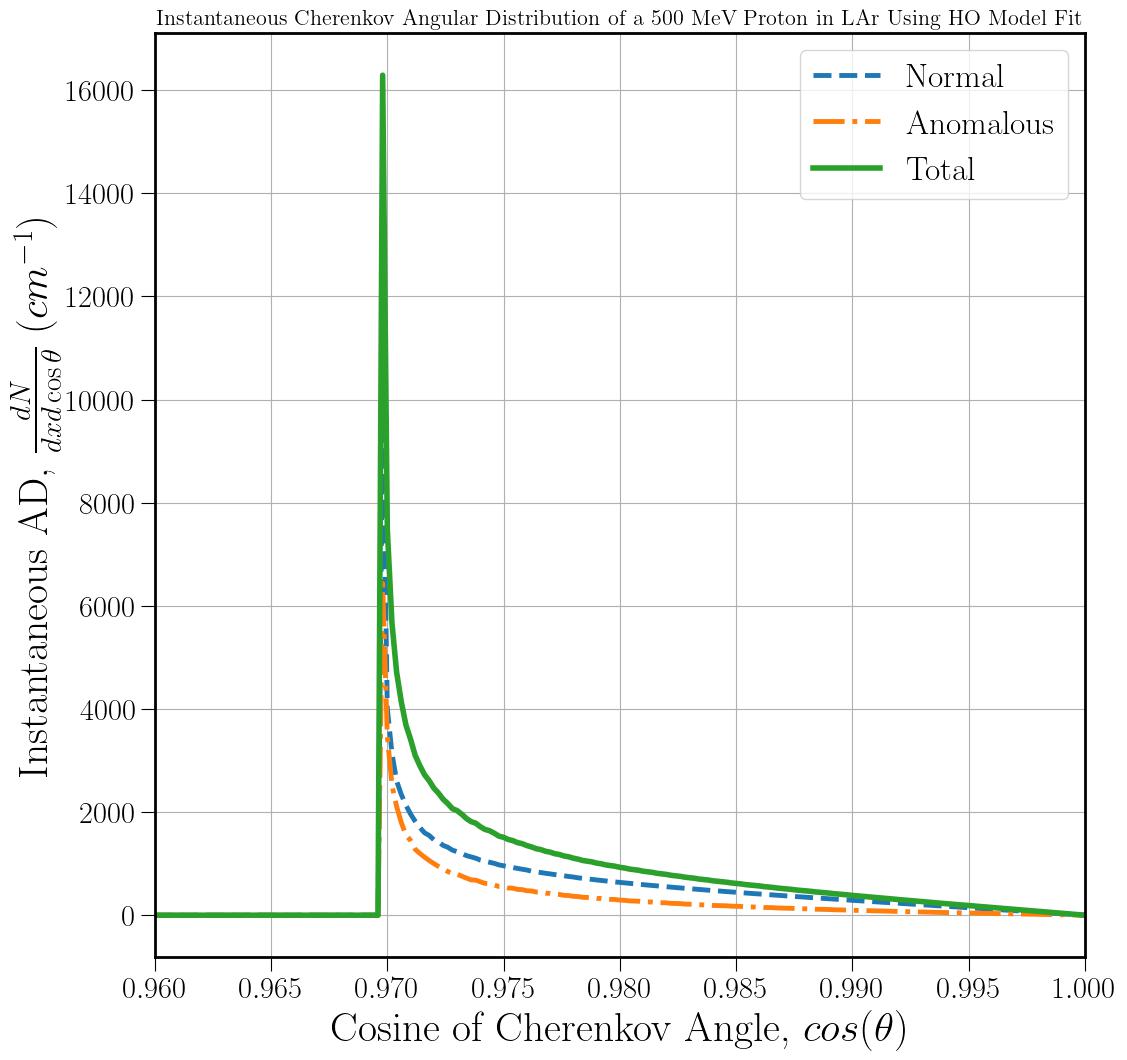
} 
\caption{Cherenkov IAD (500 MeV)
\label{f:IADho500MeVoriginal}
}
\end{subfigure}%
\begin{subfigure}{.6\textwidth}
  \centering
  \includegraphics[width=1\linewidth]{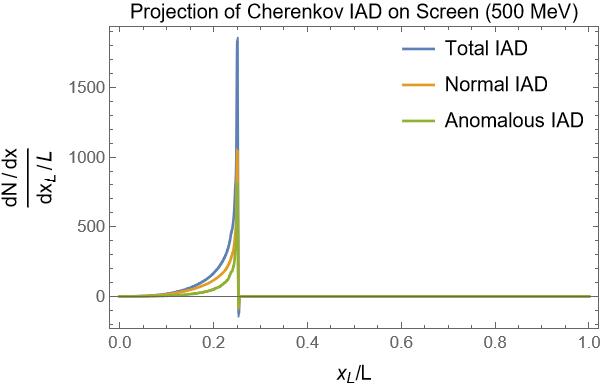}
  \caption{Projected IAD (500 MeV)}
  \label{f:IADho500MeVonscreen}
\end{subfigure}

\medskip

\begin{subfigure}{.4\textwidth}
  \centering
  \includegraphics[width=1\textwidth]{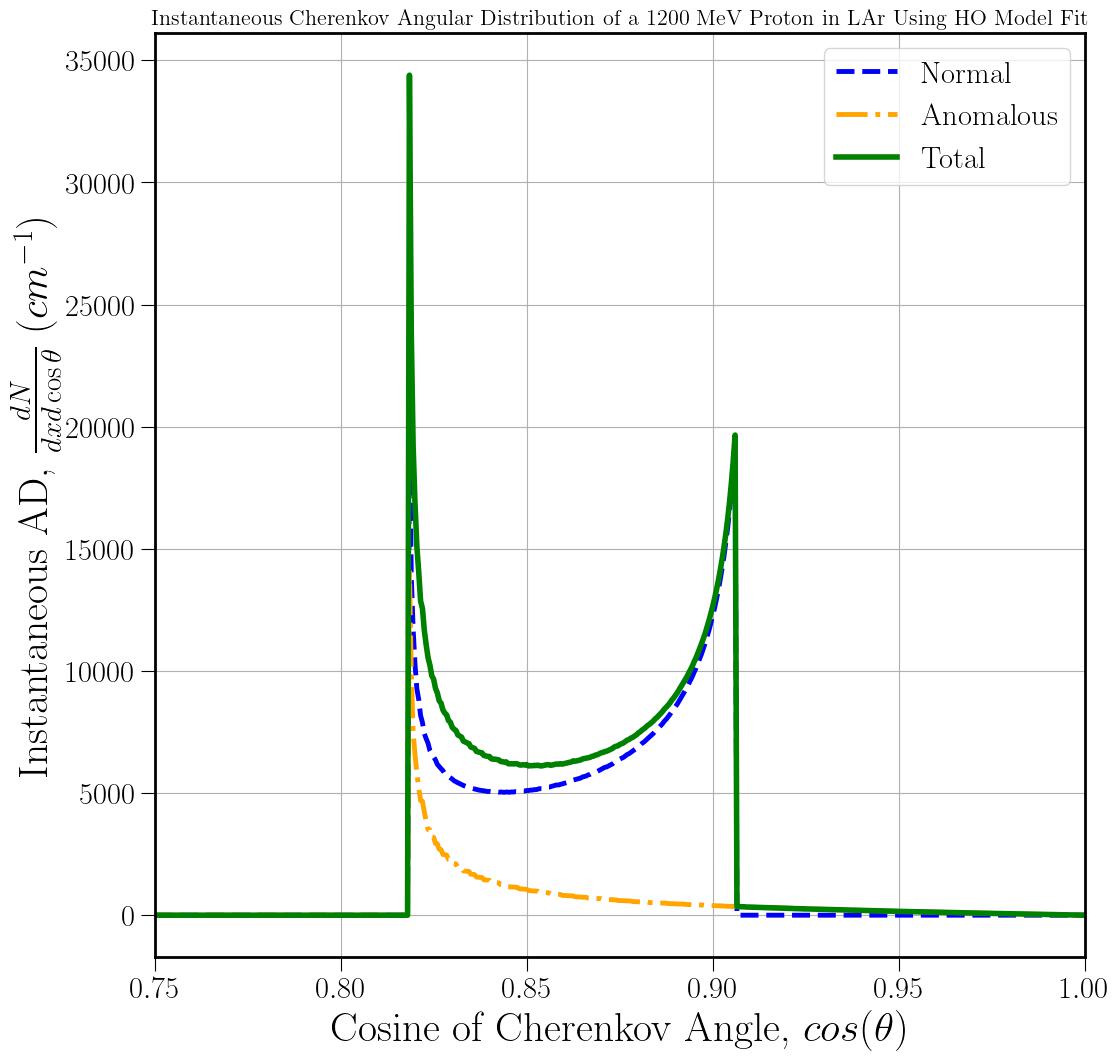}
\caption{Cherenkov IAD (1200 MeV)
\label{f:IADho1200MeVoriginal}
}
\end{subfigure}%
\begin{subfigure}{.6\textwidth}
  \centering
  \includegraphics[width=1\linewidth]{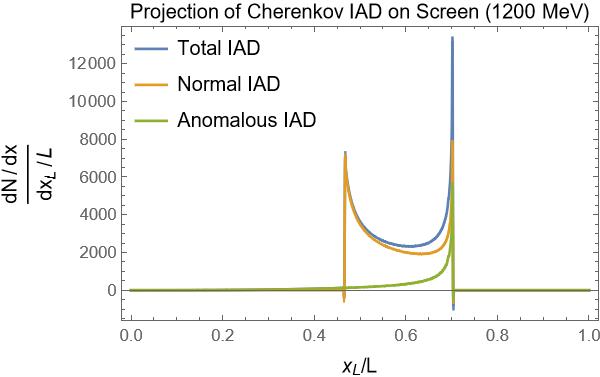}
  \caption{Projected IAD (1200 MeV)}
  \label{f:IADho1200MeVonscreen}
\end{subfigure}



\caption{Instantaneous Angular Distribution (IAD) of Cherenkov Radiation of proton in LAr and its corresponding projection on a screen for total, normal, and anomalous photons.}
\label{f:IADandonscreenIAD} 
\end{figure}
Here we have used the identities
\begin{subequations}
\begin{align}   \label{e:cos^2theta}
    \cos\theta &= \frac{1}{\left[ 1 + \left(\frac{x_L}{L} \right)^2 \right]^{1/2}}    \: ,
    \\  \label{e:sintheta}
    \sin \theta &= 
    \frac{(x_L / L)}{ \left[1 + \left(\frac{x_L}{L} \right)^2 \right]^{1/2}}   \: ,
\end{align}
\end{subequations}
which follow from \eqref{e:xL}.  This allows us to express the projected distribution as
\begin{align}   \label{e:dndxLfinal2}
    \frac{dN}{d (x_L/L)} &= \left(\frac{x_L}{L}\right) \left[1 + \left(\frac{x_L}{L} \right)^2 \right]^{-3/2}  \left. \frac{dN}{d\cos\theta}\right|_{\cos \theta = \left[1 + \left(\tfrac{x_L}{L} \right)^2 \right]^{-1/2}}    \: ,
\end{align}
which is a simple conversion from the angular distributions shown previously.  As we will see, the nontrivial Jacobian \eqref{e:dxLdcostheta} significantly modifies the shape of the Cherenkov profile when projecting on the screen, obscuring some features but emphasizing others.  Notably, the prefactor $(x_L / L)$ ensures that the projected distribution vanishes at $x_L = 0$, even if the angular distribution is nonzero at $\cos\theta = 1$.


\subsubsection{Instantaneous Projected Distributions}
\label{sec:projIAD}





In Fig.~\ref{f:IADandonscreenIAD}, we show a side-by-side comparison of the instantaneous angular distribution (IAD) $\frac{dN}{dx d\cos\theta}$ with the instantaneous projected distribution $dN/dx d(x_L / L)$ for 500 MeV and 1200 MeV protons.  Due to the mapping \eqref{e:xL}, the projected distribution is reversed along the horizontal axis, with the center of the Cherenkov cone occurring at $x_L = 0$ as opposed to $\cos\theta = 1$.  The shape is also modulated non-trivially due to the Jacobian \eqref{e:dxLdcostheta}.


\begin{figure}[p!]
\centering
\begin{subfigure}{.5\textwidth}
  \centering
  \includegraphics[width=1\textwidth]{images/iadprojectiononscreen_500MeV.jpg} 
\caption{Projected IAD (500 MeV)
\label{f:iadho500mevonscreen}
}
\end{subfigure}%
\begin{subfigure}{.5\textwidth}
  \centering
  \includegraphics[width=1\linewidth]{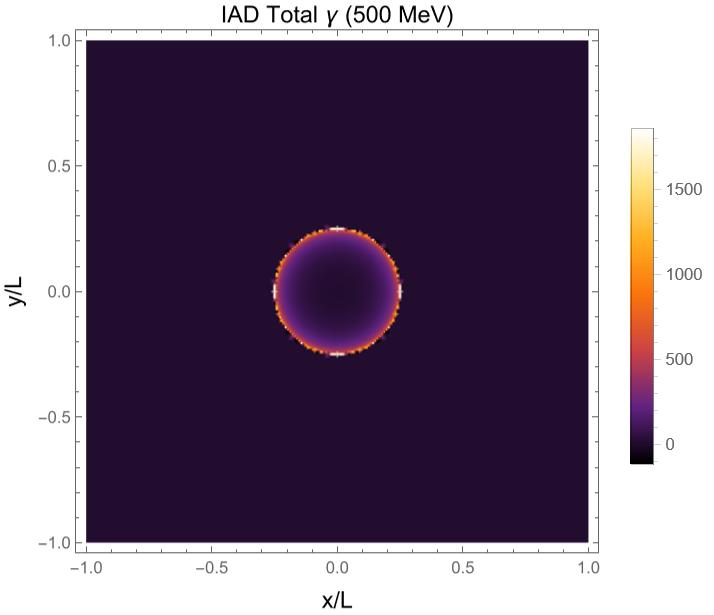}
  \caption{Total IAD ring (500 MeV)}
  \label{f:iadtotalringondascreen500MeV}
  
\end{subfigure}

\medskip

\begin{subfigure}{.5\textwidth}
  \centering
  \includegraphics[width=1\linewidth]{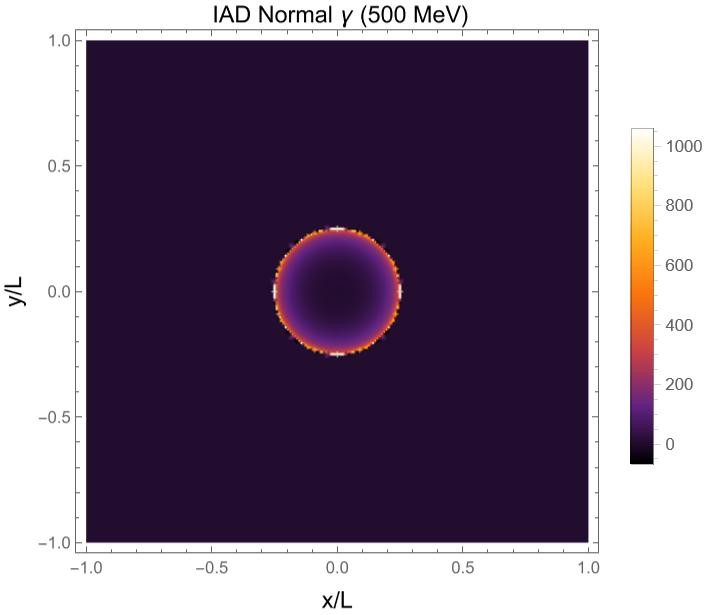}
  \caption{Normal IAD ring (500 MeV)}
  \label{f:iadnormalringondascreen500MeV}
\end{subfigure}%
\begin{subfigure}{.5\textwidth}
  \centering
  \includegraphics[width=1\linewidth]{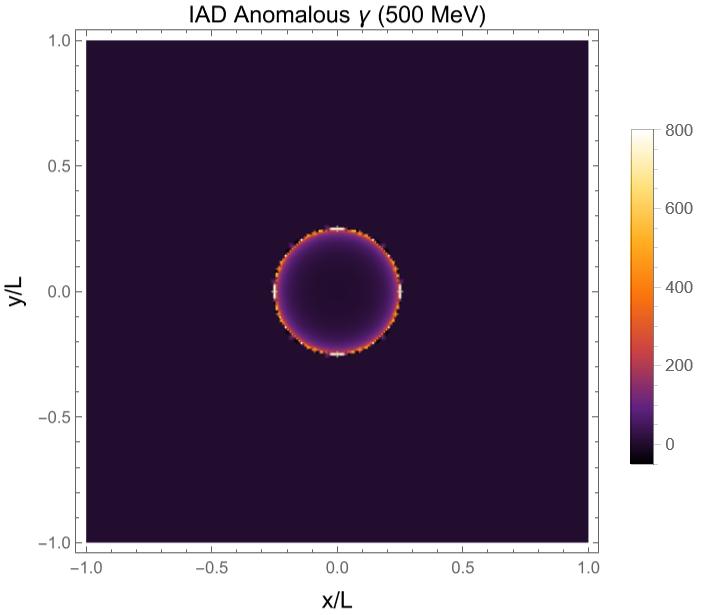}
  \caption{Anomalous IAD ring (500 MeV)}
  \label{f:iadanomringondascreen500MeV}
  
\end{subfigure}
\caption{Instantaneous Angular Distribution (IAD) of Cherenkov Radiation of a 500 MeV proton in LAr using HO model fit and its corresponding projection on a screen for total, normal, and anomalous photons.}
\label{f:iadringondascreen500MeV} 
\end{figure}


\begin{figure}[p!]
\centering
\begin{subfigure}{.5\textwidth}
  \centering
  \includegraphics[width=1\textwidth]{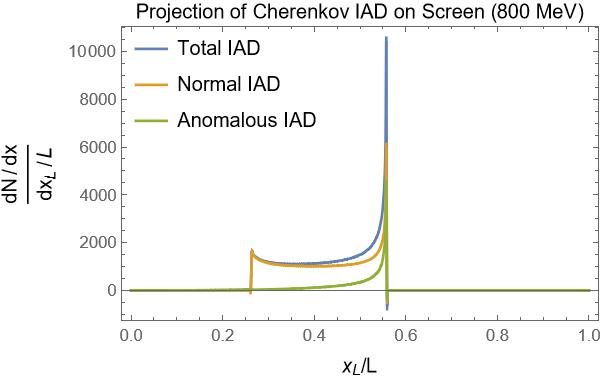} 
\caption{Projected IAD (800 MeV)
\label{f:iadho800mevonscreen}
}
\end{subfigure}%
\begin{subfigure}{.5\textwidth}
  \centering
  \includegraphics[width=1\linewidth]{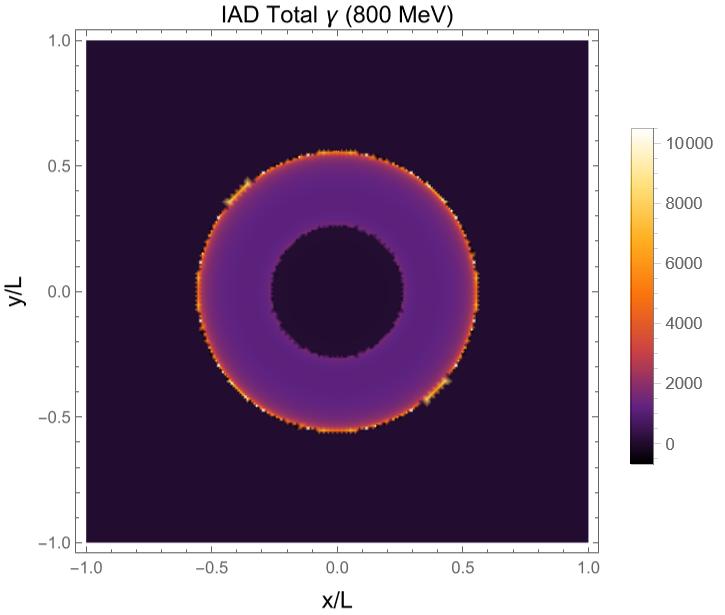}
  \caption{Total IAD ring (800 MeV)}
  \label{f:iadtotalringondascreen800MeV}
  
\end{subfigure}

\medskip

\begin{subfigure}{.5\textwidth}
  \centering
  \includegraphics[width=1\linewidth]{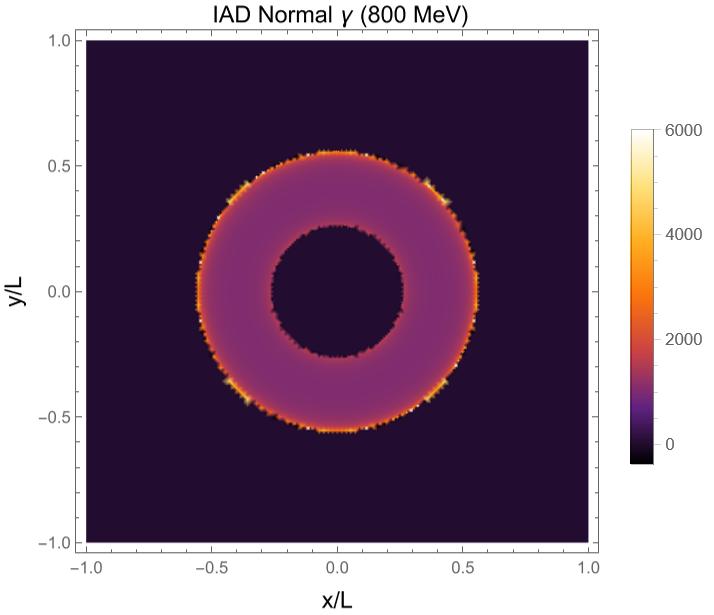}
  \caption{Normal IAD ring (800 MeV)}
  \label{f:iadnormalringondascreen800MeV}
\end{subfigure}%
\begin{subfigure}{.5\textwidth}
  \centering
  \includegraphics[width=1\linewidth]{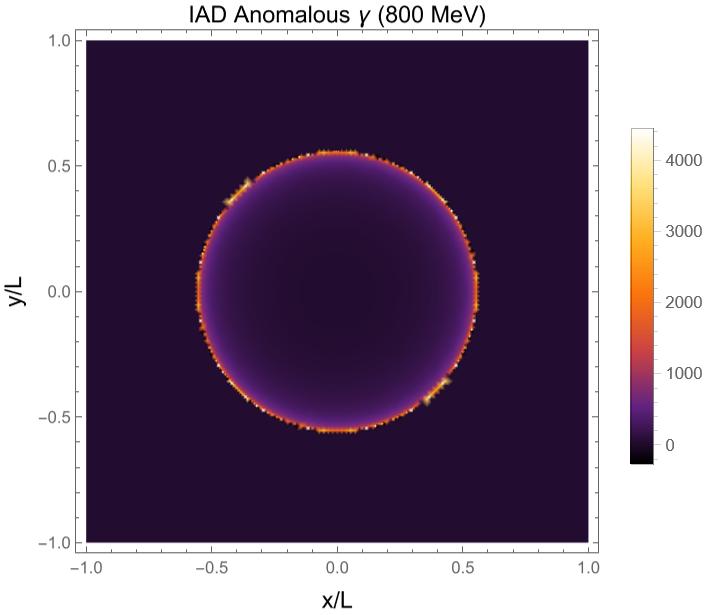}
  \caption{Anomalous IAD ring (800 MeV)}
  \label{f:iadanomringondascreen800MeV}
  
\end{subfigure}
\caption{Instantaneous Angular Distribution (IAD) of Cherenkov Radiation of an 800 MeV proton in LAr using HO model fit and its corresponding projection on a screen for total, normal, and anomalous photons.}
\label{f:iadringondascreen800MeV} 
\end{figure}


\begin{figure}[p!]
\centering
\begin{subfigure}{.5\textwidth}
  \centering
  \includegraphics[width=1\textwidth]{images/iadprojectiononscreen_1200MeV.jpg}
\caption{Projected IAD (1200 MeV)
\label{f:iadho1200mevonscreen}
}
\end{subfigure}%
\begin{subfigure}{.5\textwidth}
  \centering
  \includegraphics[width=1\linewidth]{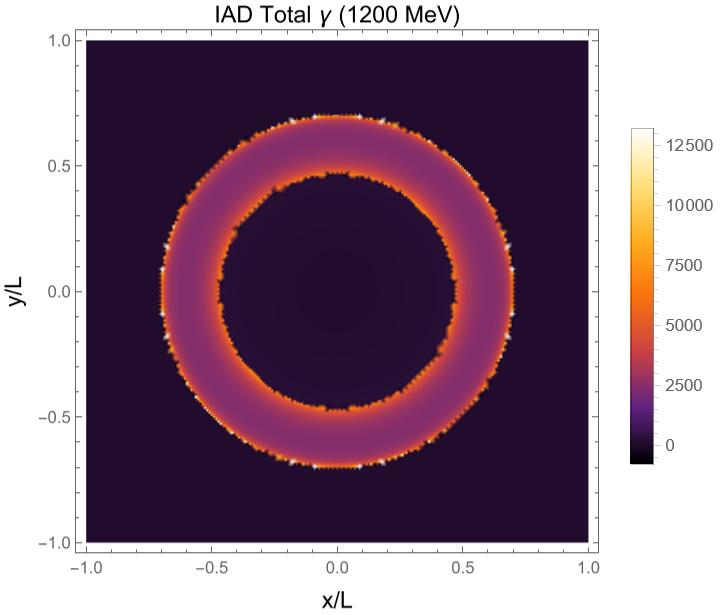}
  \caption{Total IAD ring (1200 MeV)}
  \label{f:iadtotalringondascreen1200MeV}
  
\end{subfigure}

\medskip

\begin{subfigure}{.5\textwidth}
  \centering
  \includegraphics[width=1\linewidth]{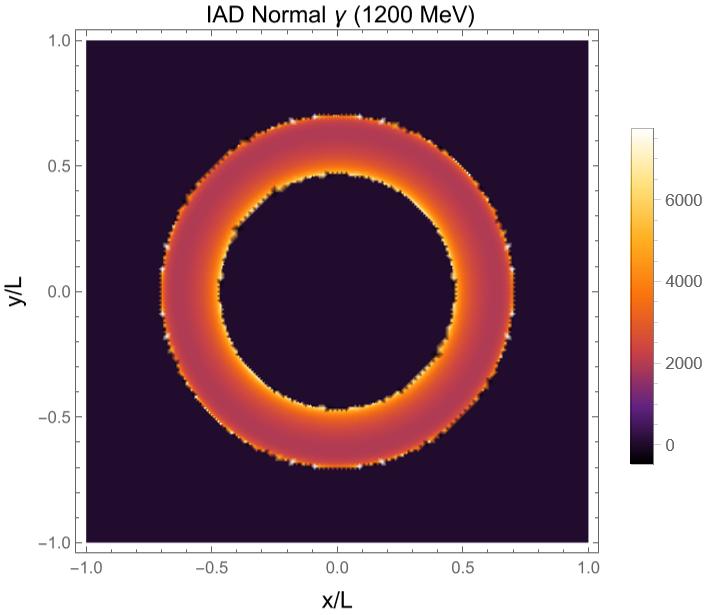}
  \caption{Normal IAD ring (1200 MeV)}
  \label{f:iadnormalringondascreen1200MeV}
\end{subfigure}%
\begin{subfigure}{.5\textwidth}
  \centering
  \includegraphics[width=1\linewidth]{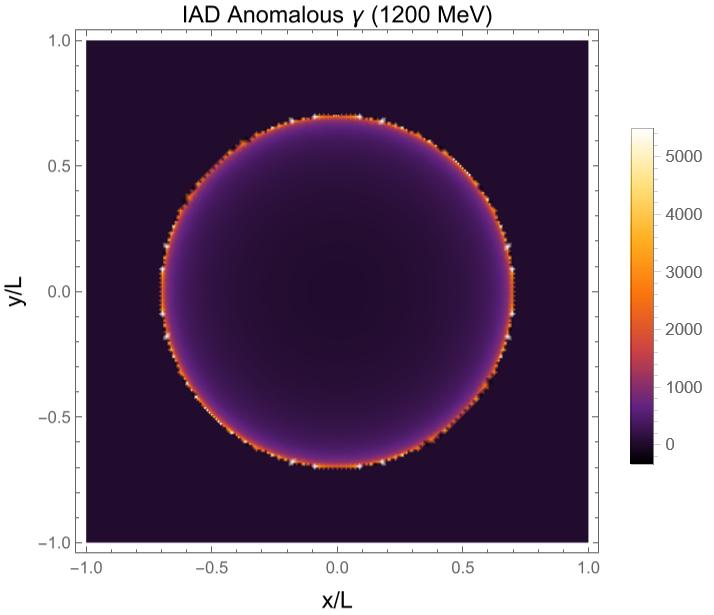}
  \caption{Anomalous IAD ring (1200 MeV)}
  \label{f:iadanomringondascreen1200MeV}  
\end{subfigure}
\caption{Instantaneous Angular Distribution of Cherenkov Radiation of a 1200 MeV proton in LAr and its corresponding projection on a screen for total, normal, and anomalous photons.}
\label{f:iadringondascreen1200MeV} 
\end{figure}


\begin{figure}[p!]
\centering
\begin{subfigure}{.5\textwidth}
  \centering
  \includegraphics[width=1\textwidth]{images/ringiad_500MeV_total.jpg}
\caption{Total IAD ring (500 MeV)
\label{f:iadringondascreen500MeV2}
}
\end{subfigure}%
\begin{subfigure}{.5\textwidth}
  \centering
  \includegraphics[width=1\linewidth]{images/ringiad_800MeV_total.jpg}
  \caption{Total IAD ring (800 MeV)}
  \label{f:iadringondascreen800MeV2}
  
\end{subfigure}

\medskip

\begin{subfigure}{.5\textwidth}
  \centering
  \includegraphics[width=1\linewidth]{images/ringiad_1200MeV_total.jpg}
  \caption{Total IAD ring (1200 MeV)}
  \label{f:iadringondascreen1200MeV2}
\end{subfigure}%
\begin{subfigure}{.5\textwidth}
  \centering
  \includegraphics[width=1\linewidth]{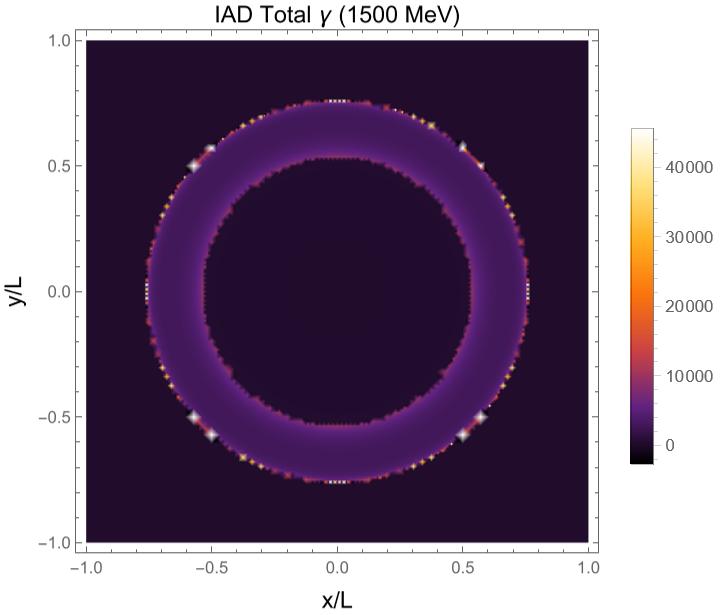}
  \caption{Total IAD ring (1500 MeV)}
  \label{f:iadringondascreen1500MeV2}
\end{subfigure}
\caption{Projection of total Cherenkov instantaneous angular distribution (IAD) of protons with different kinetic energies (T) in LAr on a detector screen.  Note the changing color scale from panel to panel.}
\label{f:iadringondascreenTMeV} 
\end{figure}


Fig.~\ref{f:iadringondascreen500MeV} expresses the projected distribution as a 2D intensity plot, comparing the shapes of the normal, anomalous, and instantaneous projected distributions for a 500 MeV proton. At such low energies close to the Cherenkov threshold, the normal and anomalous distributions have very similar shapes and contribute almost equally to the total distribution.  We note that the Cherenkov intensity in the projected distributions always goes to zero intensity at the center due to the Jacobian phase-space factor $(x_L / L)$.  At 500 MeV, the Cherenkov intensity gradually increases from zero at the center to a peak at the outer edge at around $x_L/L \sim 0.25$.

The shape of the instantaneous projected distributions shown in Fig.~\ref{f:iadringondascreen800MeV} for an 800 MeV proton is substantially different than that of the 500 MeV seen in Fig.~\ref{f:iadringondascreen500MeV}.  The finite kinematic bounds on $\cos\theta$ in the angular distribution result in a Cherenkov ring of finite thickness, with all three distributions vanishing for $0 \geq x_L/L < 0.25$  (c.f. Fig.~\ref{f:ADinstHO800MeV}). The normal component dominates the Cherenkov ring on the interior, so the total instantaneous projected distribution reflects the double peak structure of the normal component; a lower peak occurrs at the inner boundary $x_L/L = 0.25$ and smoothly increases to a higher peak at the outer edge at $x_L/L = 0.55$.  The anomalous projected distribution, on the other hand, only has the single peak at the exterior $x_L/L = 0.55$ of the Cherenkov ring.

With increasing energy (Fig.~\ref{f:iadringondascreen1200MeV}), the size of the instantaneous Cherenkov ring increases, but the kinematically excluded region (central dark circle) grows faster, making the Cherenkov ring comparatively narrower than at lower energies.  The intensity of the Cherenkov ring also increases with increasing energy, especially at the inner and outer peaks.  The increase in the inner peak is particularly noticeable, resulting in a brighter inner edge of the Cherenkov cone compared to an 800 MeV proton.  On the other hand, the shape of the anomalous distribution (Fig.~\ref{f:iadanomringondascreen1200MeV}) is unchanged from lower energies, but with a substantial increase in its intensity, which is concentrated on the outer edge.  This pattern continues for even higher energies of the proton.  We summarize the energy dependence of the total instantaneous projected distributions (normal + anomalous) in Fig.~\ref{f:iadringondascreenTMeV}.

\subsubsection{Integrated Projected Distributions}
\label{sec:projAD}







\begin{figure}[p!]
\centering
\begin{subfigure}{.5\textwidth}
  \centering
  \includegraphics[width=1\textwidth]{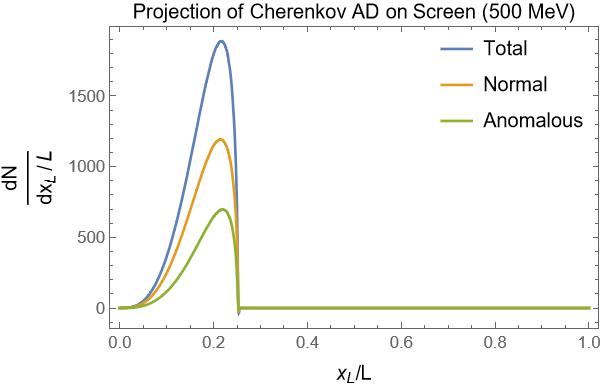} 
\caption{Projected AD (500 MeV)
\label{f:adho500mevonscreen}
}
\end{subfigure}%
\begin{subfigure}{.5\textwidth}
  \centering
  \includegraphics[width=1\linewidth]{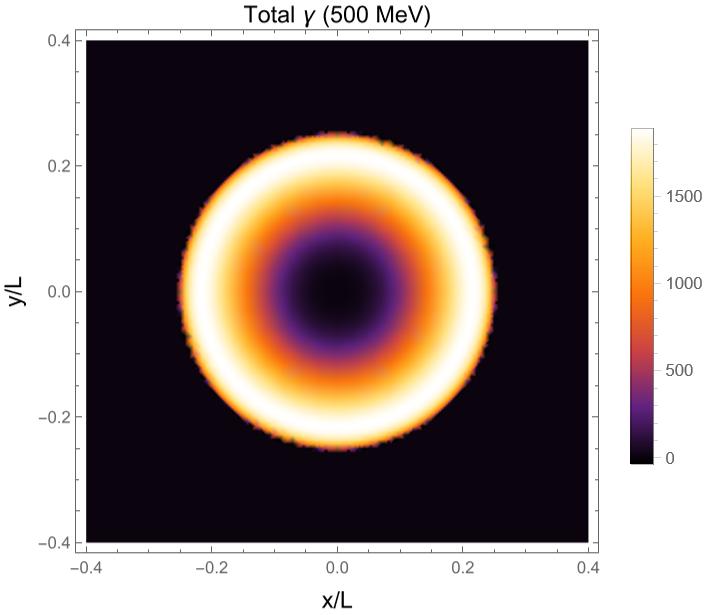}
  \caption{Total Cherenkov ring (500 MeV)}
  \label{f:totalringondascreen500MeV}
  
\end{subfigure}

\medskip

\begin{subfigure}{.5\textwidth}
  \centering
  \includegraphics[width=1\linewidth]{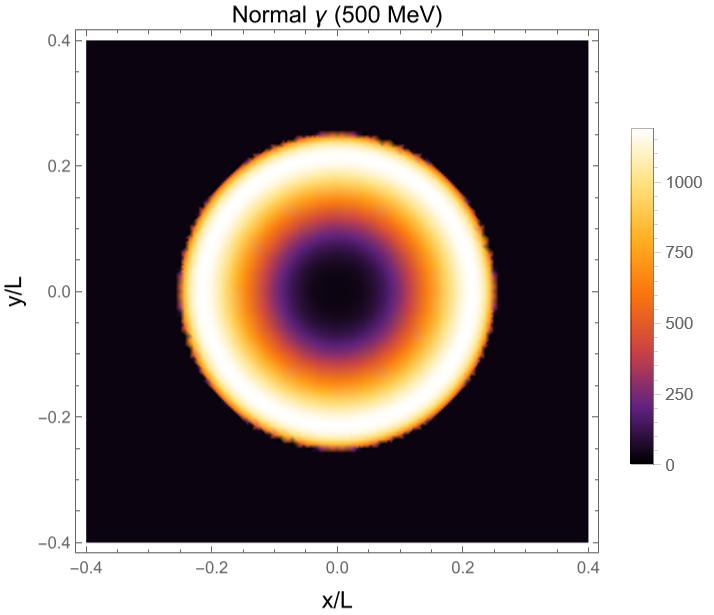}
  \caption{Normal Cherenkov ring (500 MeV)}
  \label{f:normalringondascreen500MeV}
\end{subfigure}%
\begin{subfigure}{.5\textwidth}
  \centering
  \includegraphics[width=1\linewidth]{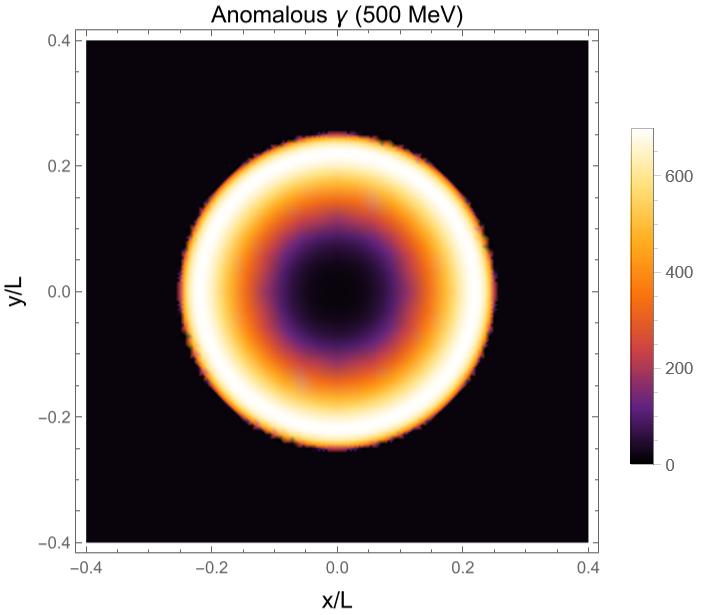}
  \caption{Anomalous Cherenkov ring (500 MeV)}
  \label{f:anomringondascreen500MeV}
  
\end{subfigure}
\caption{Angular Distribution of Cherenkov Radiation of a 500 MeV proton in LAr using HO model fit and its corresponding projection on a screen for total, normal, and anomalous photons. }
\label{f:ringondascreen500MeV} 
\end{figure}


  


  


\begin{figure}[p!]
\centering
\begin{subfigure}{.5\textwidth}
  \centering
  \includegraphics[width=1\textwidth]{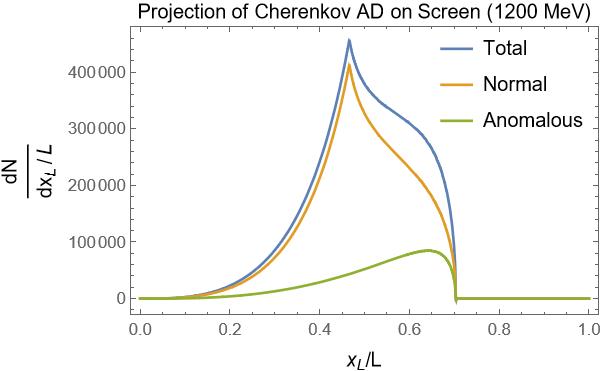}
\caption{Projected AD (1200 MeV)
\label{f:adho1200mevonscreen}
}
\end{subfigure}%
\begin{subfigure}{.5\textwidth}
  \centering
  \includegraphics[width=1\linewidth]{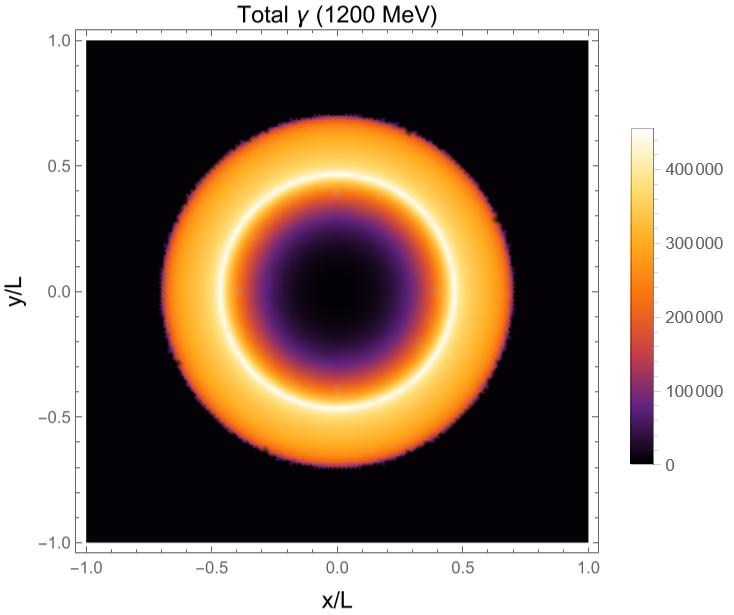}
  \caption{Total Cherenkov ring (1200 MeV)}
  \label{f:totalringondascreen1200MeV}
  
\end{subfigure}

\medskip

\begin{subfigure}{.5\textwidth}
  \centering
  \includegraphics[width=1\linewidth]{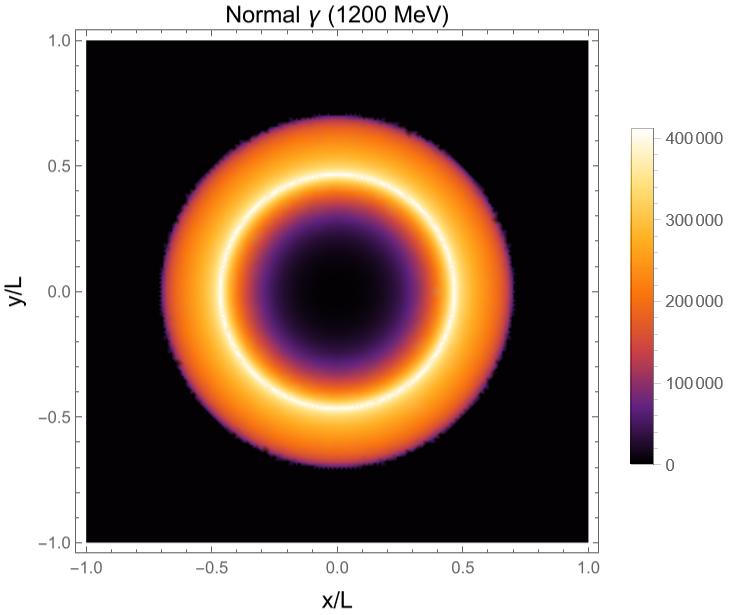}
  \caption{Normal Cherenkov ring (1200 MeV)}
  \label{f:normalringondascreen1200MeV}
\end{subfigure}%
\begin{subfigure}{.5\textwidth}
  \centering
  \includegraphics[width=1\linewidth]{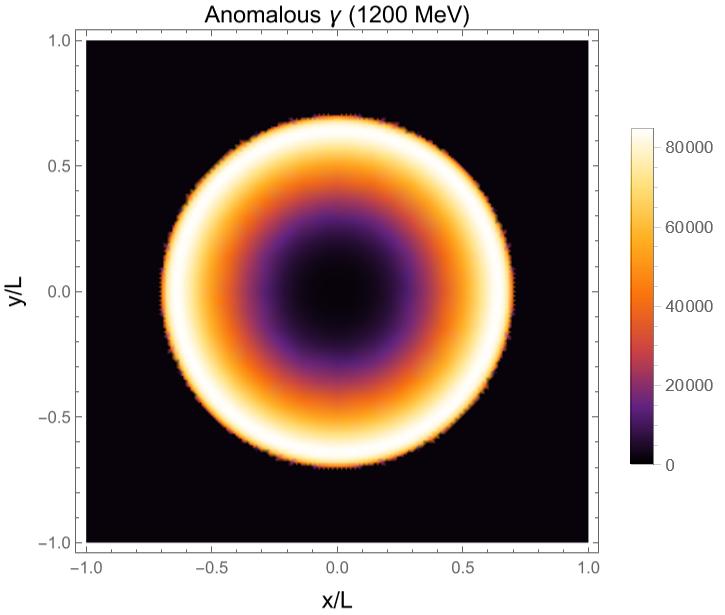}
  \caption{Anomalous Cherenkov ring (1200 MeV)}
  \label{f:anomringondascreen1200MeV}
  
\end{subfigure}
\caption{Angular Distribution of Cherenkov Radiation of a 1200 MeV proton in LAr and its corresponding projection on a screen for total, normal, and anomalous photons.}
\label{f:ringondascreen1200MeV} 
\end{figure}


\begin{figure}[p!]
\centering
\begin{subfigure}{.5\textwidth}
  \centering
  \includegraphics[width=1\textwidth]{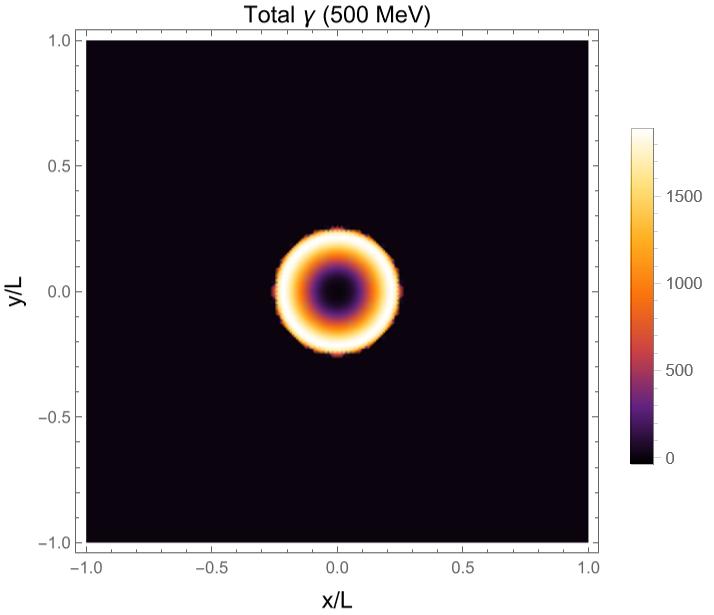}
\caption{Total Cherenkov ring (500 MeV)
\label{f:ringondascreen500MeV2}
}
\end{subfigure}%
\begin{subfigure}{.5\textwidth}
  \centering
  \includegraphics[width=1\linewidth]{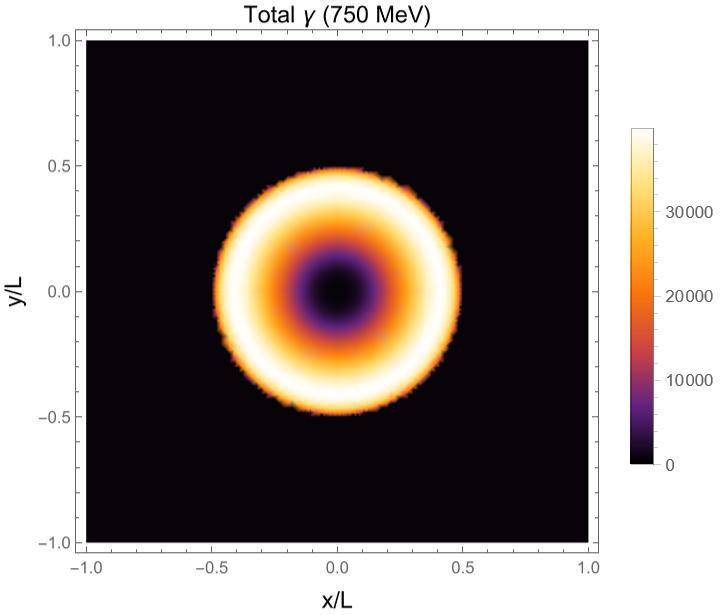}
  \caption{Total Cherenkov ring (750 MeV)}
  \label{f:ringondascreen750MeV2}
  
\end{subfigure}

\medskip

\begin{subfigure}{.5\textwidth}
  \centering
  \includegraphics[width=1\linewidth]{images/ring_1200MeV_totaln.jpg}
  \caption{Total Cherenkov ring (1200 MeV)}
  \label{f:ringondascreen1200MeV2}
\end{subfigure}%
\begin{subfigure}{.5\textwidth}
  \centering
  \includegraphics[width=1\linewidth]{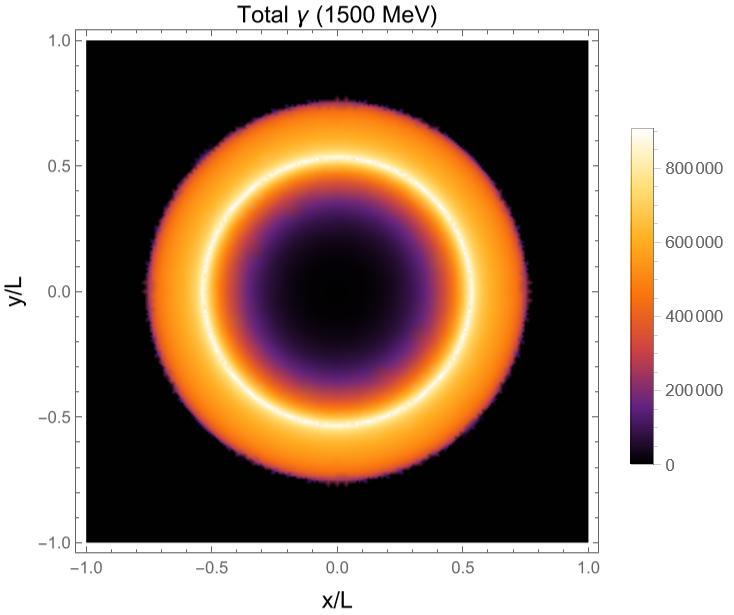}
  \caption{Total Cherenkov ring (1500 MeV)}
  \label{f:ringondascreen1500MeV}
  
\end{subfigure}
\caption{Projection of total Cherenkov angular distribution of protons with different kinetic energies (T) in LAr on a detector screen.  Note the changing color scale from panel to panel.}
\label{f:ringondascreenTMeV} 
\end{figure}


\begin{figure}[p!]
\centering
\begin{subfigure}{.5\textwidth}
  \centering
  \includegraphics[width=1\textwidth]{images/ring_500MeV_totalnzoom.jpg}
\caption{HO model fit (500 MeV)
\label{f:ringondascreen500MeVho}
}
\end{subfigure}%
\begin{subfigure}{.5\textwidth}
  \centering
  \includegraphics[width=1\linewidth]{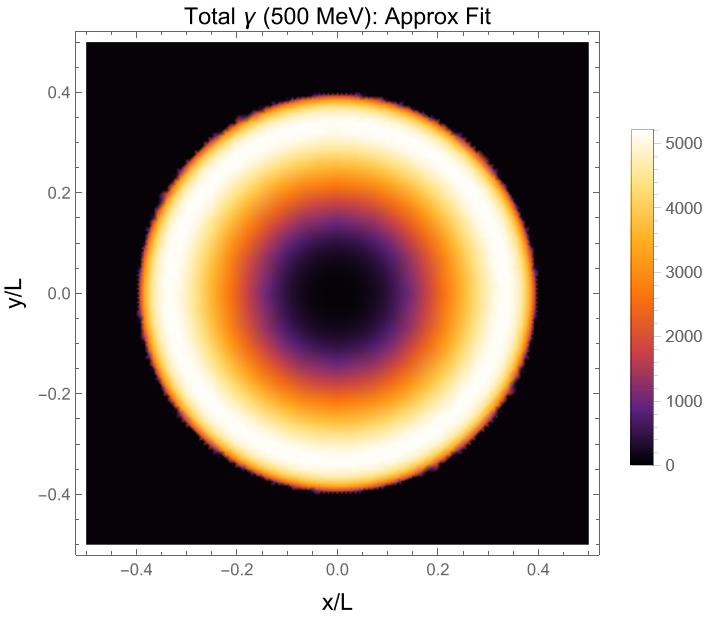}
  \caption{Approx fit (500 MeV)}
  \label{f:ringondascreen500MeVapprox}
  
\end{subfigure}

\medskip

\begin{subfigure}{.5\textwidth}
  \centering
  \includegraphics[width=1\linewidth]{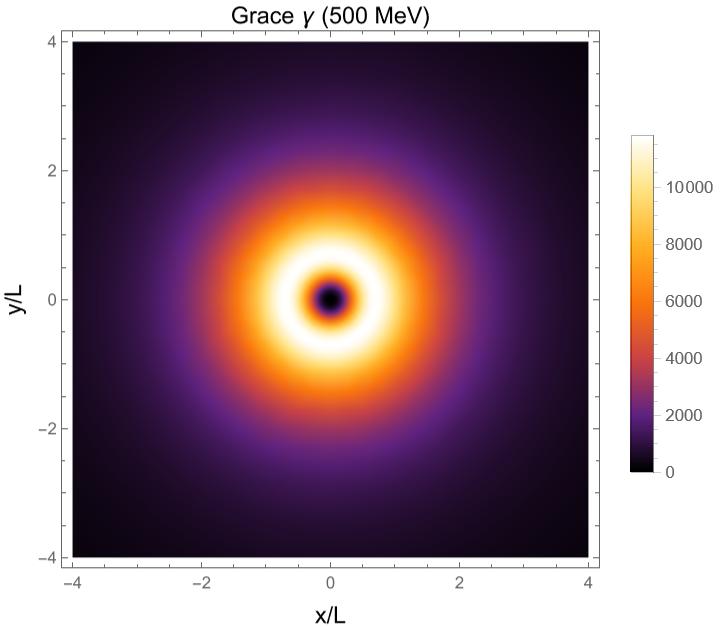}
  \caption{Grace (500 MeV)}
  \label{f:ringondascreen500MeVGrace}
\end{subfigure}%
\begin{subfigure}{.5\textwidth}
  \centering
  \includegraphics[width=1\linewidth]{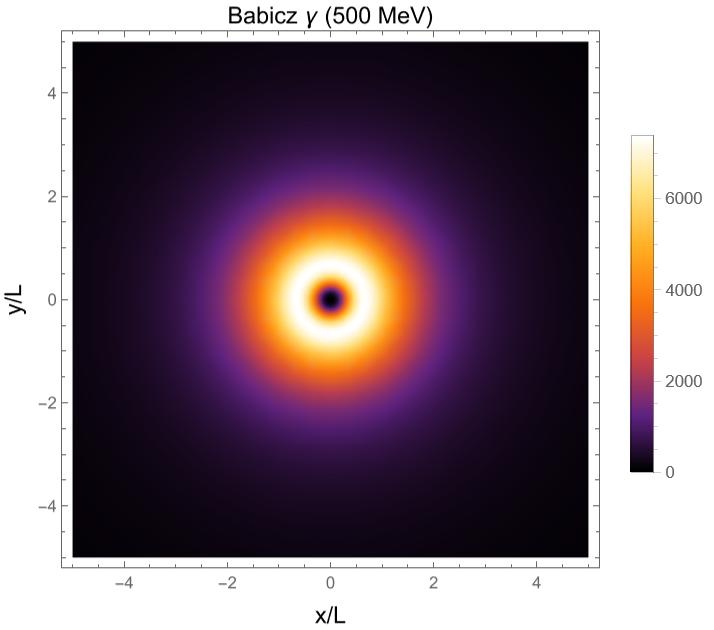}
  \caption{Babicz (500 MeV)}
  \label{f:ringondascreen500MeVBabicz}
  
\end{subfigure}
\caption{Projection of total Cherenkov angular distribution of a 500 MeV proton in LAr on a detector screen using various refractive index fits.  Note the large differences in both spatial scale and color scale from panel to panel.}
\label{f:ringondascreenallfits500MeV} 
\end{figure}


\begin{figure}[p!]
\centering
\begin{subfigure}{.5\textwidth}
    \centering
    \includegraphics[width=1\textwidth]{images/ring_1200MeV_totaln.jpg}
    \caption{HO model fit (1200 MeV)
    \label{f:ringondascreen1200MeVho}
    }
    \end{subfigure}%
\begin{subfigure}{.5\textwidth}
  \centering
  \includegraphics[width=1\linewidth]{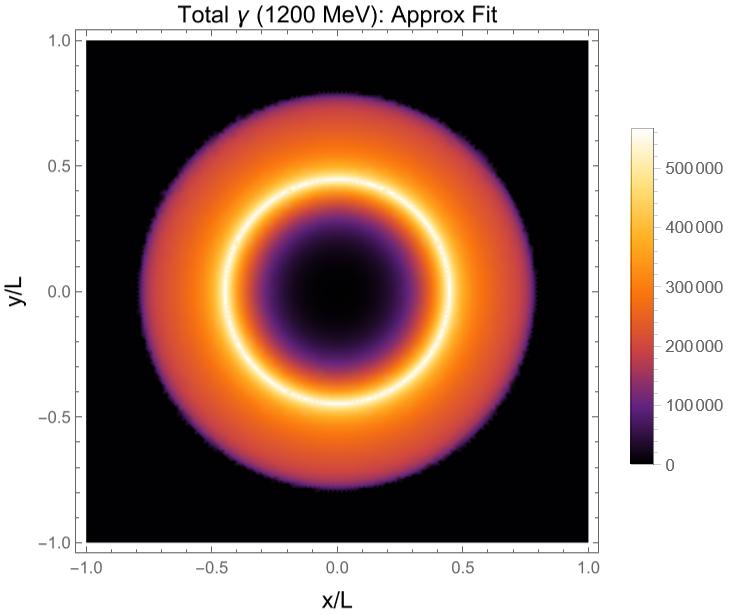}
  \caption{Approx fit (1200 MeV)}
  \label{f:ringondascreen1200MeVapprox}
  \end{subfigure}
\medskip
\begin{subfigure}{.5\textwidth}
    \centering
    \includegraphics[width=1\linewidth]{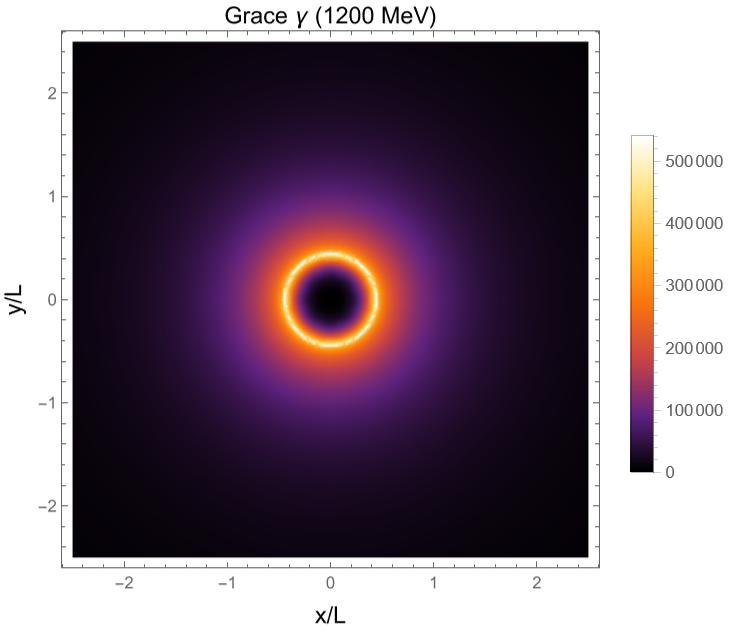}
    \caption{Grace (1200 MeV)}
    \label{f:ringondascreen1200MeVGrace}
    \end{subfigure}%
\begin{subfigure}{.5\textwidth}
    \centering
    \includegraphics[width=1\linewidth]{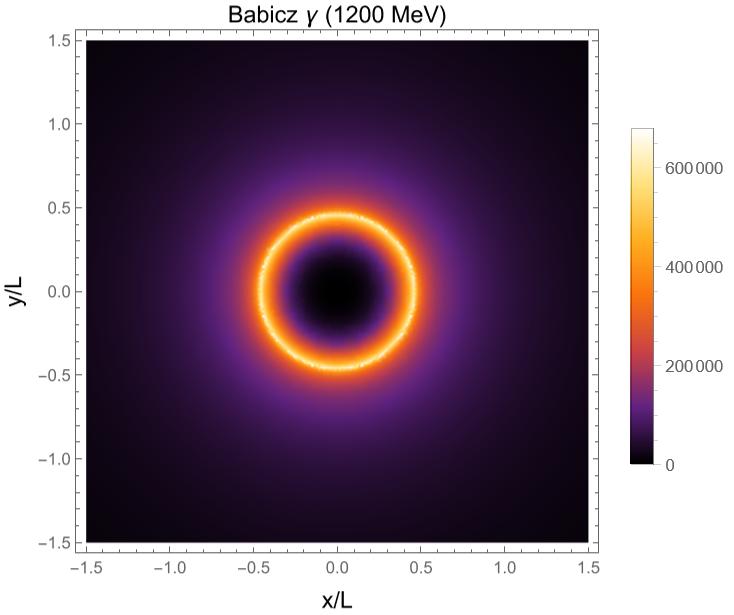}
    \caption{Babicz (1200 MeV)}
    \label{f:ringondascreen1200MeVBabicz}
    \end{subfigure}
\caption{Projection of total Cherenkov angular distribution of a 1200 MeV proton in LAr on a detector screen using various refractive index fits.  Note the changing spatial and color scales from panel to panel.}
\label{f:ringondascreenallfits1200MeV} 
\end{figure}


\begin{figure}[p!]
\centering
\begin{subfigure}{.5\textwidth}
  \centering
  \includegraphics[width=1\textwidth]{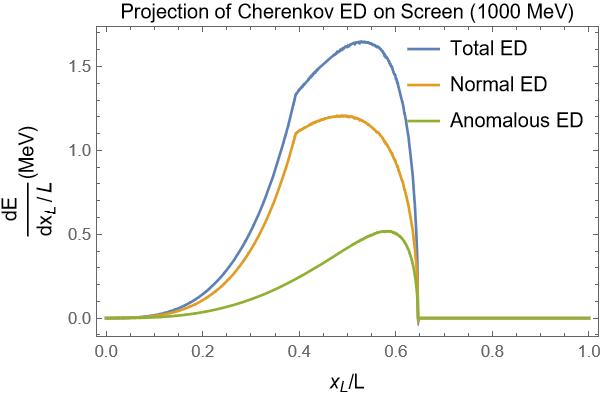}
\caption{Projected ED (1000 MeV)
\label{f:edho1000mevonscreen}
}
\end{subfigure}%
\begin{subfigure}{.5\textwidth}
  \centering
  \includegraphics[width=1\linewidth]{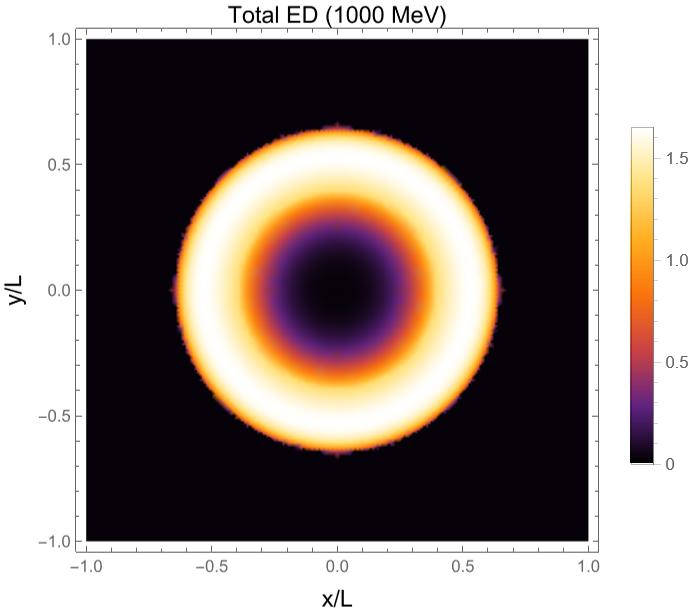}
  \caption{Total ED ring (1000 MeV)}
  \label{f:etotalringondascreen1000MeV}
  
\end{subfigure}

\medskip

\begin{subfigure}{.5\textwidth}
  \centering
  \includegraphics[width=1\linewidth]{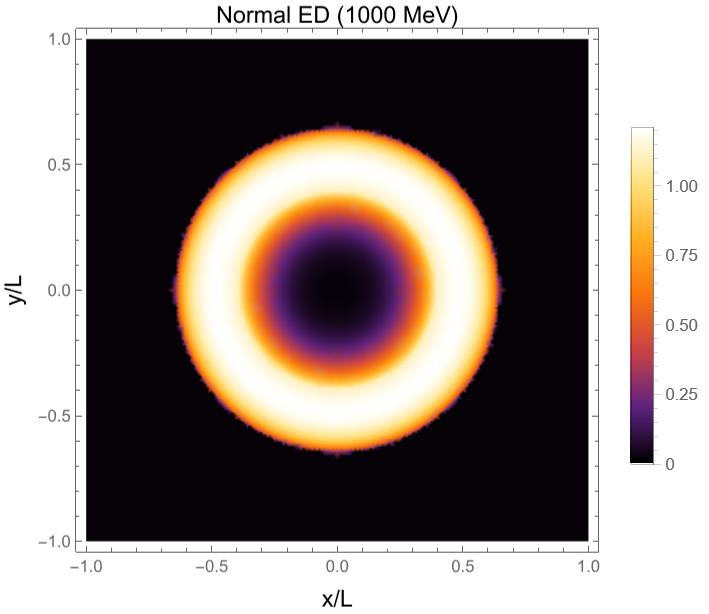}
  \caption{Normal ED ring (1000 MeV)}
  \label{f:enormalringondascreen1000MeV}
\end{subfigure}%
\begin{subfigure}{.5\textwidth}
  \centering
  \includegraphics[width=1\linewidth]{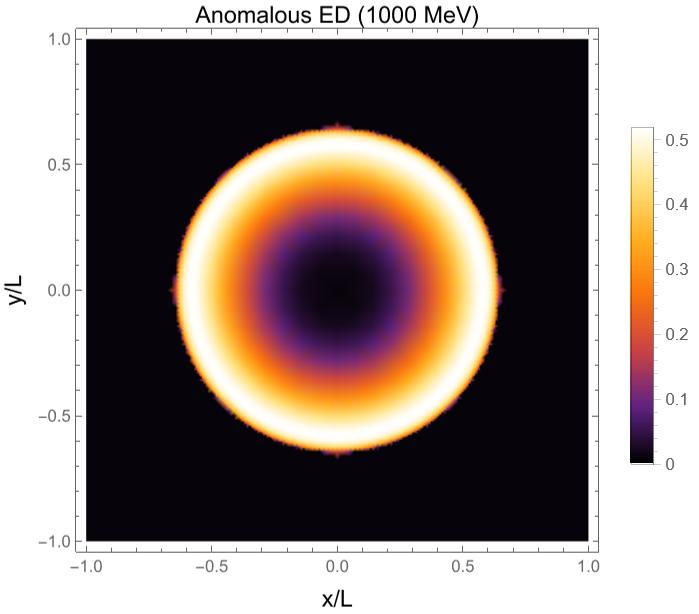}
  \caption{Anomalous ED ring (1000 MeV)}
  \label{f:eanomringondascreen1000MeV}  
\end{subfigure}
\caption{Energy Distribution (ED) of Cherenkov Radiation of a 1000 MeV proton in LAr and its corresponding projection on a screen for total, normal, and anomalous photons.}
\label{f:eringondascreen1000MeV} 
\end{figure}


\begin{figure}[h!]
\centering
\begin{subfigure}{.49\textwidth}
  \centering
  \includegraphics[width=1\linewidth]{images/ering_1000MeV_total.jpg}
  \caption{Total ED ring (1000 MeV)}
  \label{f:etotalringondascreen1000MeV3}  
\end{subfigure}
\begin{subfigure}{.49\textwidth}
  \centering
  \includegraphics[width=1\linewidth]{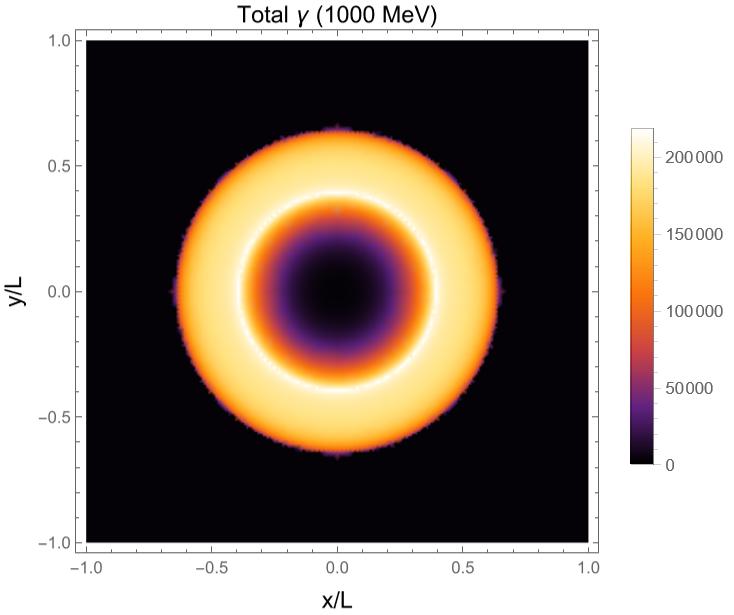}
  \caption{Total AD ring (1000 MeV)}
  \label{f:totalringondascreen1000MeV3}  
\end{subfigure}

\caption{Comparison between Energy Distribution (ED) and Number Distribution (AD) projection of Cherenkov radiation of a 1000 MeV proton in LAr on a screen.}
\label{f:EDADringondascreen1000MeV} 
\end{figure}


Now we similarly consider the \textit{integrated} projected Cherenkov distributions, which are formed by the superposition of the instantaneous projected distributions shown in Sec.~\ref{sec:projIAD} over the trajectory of the proton.

In Fig.~\ref{f:ringondascreen500MeV}, we show the integrated projected distributions for a 500 MeV proton.  At this low energy, the integrated Cherenkov ring spans the range $0 \leq x_L/L < 0.25$.  All three distributions (normal, anomalous, and total) have an intensity which goes to zero in the center of the Cherenkov ring and increases smoothly in intensity with increasing radius.  As a result, the peak Cherenkov intensity occurs very close to the outer edge, before sharply falling to zero at the kinematic boundary.  This integrated shape clearly arises from the superposition of many low-energy spectra of the form seen in Fig.~\ref{f:iadringondascreen500MeV}, all of which are concentrated toward the outer edge.

In contrast, for a 1200 MeV proton, the projected distributions change drastically, as shown in Fig.~\ref{f:ringondascreen1200MeV}.  At this energy, the Cherenkov distribution spans a larger range from $0 \leq x_L/L < 0.7$.  The anomalous distribution, which always preserves the shape seen at low energies, continues to be concentrated toward the outer edge of the Cherenkov ring, just as in Fig.~\ref{f:ringondascreen500MeV}.  But the normal component undergoes major qualitative changes in shape over the trajectory of the jet, evolving from the double-peak instantaneous structure at the initial energy (Fig.~\ref{f:iadringondascreen1200MeV}) down to the single-peak shape close to the Cherenkov threshold (Fig.~\ref{f:iadringondascreen500MeV}).  As a consequence, the normal distribution acquires a pronounced, triangular shape, with the peak intensity occurring in the \textit{middle} of the Cherenkov ring, at about $x_L / L = 0.46$.  The total Cherenkov ring reflects both of these features, with a pronounced peak intensity in middle of the ring and a ``shoulder'' which enhances the Cherenkov intensity toward the outer edge.

We summarize the energy dependence of the integrated projected distributions in Fig.~\ref{f:ringondascreenTMeV}, for 500 MeV $< T <$ 1500 MeV.  The transition from the low-energy profile (which is monotonically concentrated toward the outer edge of the ring) to the high-energy profile (characterized by a bright ring of peak intensity in the middle of the Cherenkov ring) occurs around 800 MeV.

We illustrate the sensitivity of the Cherenkov light to differences in the fit to the refractive index in Fig.~\ref{f:ringondascreenallfits500MeV} for a 500 MeV proton and in Fig.~\ref{f:ringondascreenallfits1200MeV} for a 1200 MeV proton.  In both cases, we compare all four different fits considered in this work: the absorptive fits (HO, Approx) as well as the resonant fits (Grace, Babicz).  The striking differences between the absorptive and resonant fits are apparent to the naked eye.  In contrast to the features of the absorptive fits discussed previously, the resonant fits have a wide ring of peak intensity which gradually decreases radially over a long distance, with no sharp boundary.  Due to the different thresholds of the fits (and the lack of any threshold in the resonant fits), at low energy (Fig.~\ref{f:ringondascreenallfits500MeV}), the fits differ tremendously both in size and intensity.

At higher energy (Fig.~\ref{f:ringondascreenallfits1200MeV}), the shape differences between the absorptive and resonant fits remain, but now that all fits are far from their thresholds, the size and intensity of the rings begins to converge.  These pronounced differences in the Cherenkov profile depending on the behavior of the refractive index, especially near threshold, illustrate the potential of Cherenkov light to impose new constraints on the refractive index.




Finally, in Fig.~\ref{f:eringondascreen1000MeV}, we illustrate the different features of the Cherenkov distribution which are emphasized by the projected energy distribution (ED) compared to the number distribution.  As discussed in Sec.~\ref{ADvsED}, higher-energy Cherenkov photons are concentrated toward the outer edge of the ring, substantially broadening the radius of peak energy compared to the radius of peak intensity seen in Fig.~\ref{f:ringondascreen1200MeV}.



\hspace{\parindent}

\subsection{Implications and Outlook}

The motivation for this study was to theoretically investigate the amount of Cherenkov radiation that emerges from medium-energy protons traveling in LAr. In neutrino experiments often the detected neutrinos are measured from a correlation of the secondary particle shower created in the medium.  These secondary charged particles, such as protons, deposit energy in the LAr medium as they travel which appears primarily as scintillation light \cite{Segreto_2021}. As scintillation photons are currently the only signal utilized even in state-of-the-art neutrino detectors like the CCM and DUNE experiments, we were interested to explore the physics potential of using Cherenkov light at these energies and compare it to the scintillation light.  We found that, remarkably, the Cherenkov contributes a substantial excess of detected photons above the scintillation background and carries significant information in its angular structure.  For the intermediate energy range of interest, $T \geq 500$ MeV, we found a $\sim 6\sigma$ photon excess, peaking at an angle of about $10^\circ$ -- even using the most conservative possible fit to the index of refraction. 

Previous existing fits in the literature lead to completely unreasonable predictions for the Cherenkov yield if taken at face value.  Our new fits incorporate the fundamental physical principles of absorption and anomalous dispersion and give much more realistic (and conservative) predictions.  Unlike older resonant models, the use of an absorptive fit to the refractive index leads to qualitatively new structures in the angular distribution, arising from anomalous dispersion.  While one might expect the region of anomalous dispersion, living close to the resonance, would only be important at low energies; however, we find that the anomalous contribution to the total yield is sizeable, remaining of order $\sim 20\%$ even at higher energies.  The differences in angular structure between the normal and anomalous contributions, however, is striking: the anomalous contribution is largest on the outer edge of the Cherenkov ring, while the normal contribution is concentrated toward the center and is more symmetric.  The total angular distribution is a nontrivial superposition of the two.



This study has already opened new interest in Cherenkov photons as a tool for experimental vertex reconstruction, not only in neutrino searches, but more broadly in other particle physics experiments. The current analytical and computational framework that we have developed for the study of protons can be applied as well to any other secondary particles of interest, such as electrons and muons.  By simply replacing the particle mass and charge, we can straightforwardly generalize our results to study showers of other charged particles. Therefore, this study can open up multiple follow-up projects and could be developed into a comprehensive shower simulation for applications in experimental particle physics. 


This research finding demands further event-by-event Monte Carlo (MC) study of this phenomenon for further integration into experimental analyses. In particular, Monte Carlo implementation for specific experimental conditions will be crucial to incorporate the effects of multiple scattering (Rayleigh scattering), which is expected to further smear the angular distribution \cite{Rayleigh_1871}. Including the Cherenkov radiation in the analyzed signal, as well as the scintillation light, can provide increased precision as well as new directional information for vertex reconstruction.  The major advantage of experimentally searching for the Cherenkov effect described in this study is that it can be implemented in the current experimental schemes of scintillation-based neutrino detectors like CCM and DUNE. One does not need to build a new set of detectors to be able to take advantage of the Cherenkov photons studied in this work, so its addition to the physics program can be quick and inexpensive.

The physics of absorption, and its impact on the Cherenkov radiation, was a major focus of this study.  In the future, one could also do a more complete, dynamical calculation of reabsorption.  In this study, we considered only the extreme cases of no reabsorption and a ``black disk'' model of reabsorption which could be easily implemented with cuts to the wavelength range.  This more complete calculation will require evolving the full angular distribution, also differential in $\lambda$, with normal and anomalous photons being absorbed differently (even though they come out at the same angle).  However, even without performing such a dedicated analysis, our study estimated the potential losses due to reabsorption using a conservative black disk model and found that the Cherenkov yield is still a significant excess over the background.

This study also found that the yield and AD are very sensitive to details of the refractive index -- especially the peak which controls the Cherenkov threshold.  Small differences in $n(\lambda)$ can lead to massive differences, especially in the AD.  This may even provide a pathway to perform new, more precise measurements of the refractive index using the Cherenkov emission itself.  While performing new measurements of the refractive index near a resonance is extremely difficult to achieve with traditional methods, our theoretical framework can be used to constrain the refractive index fit/model in precisely this region using Cherenkov light, for which the detectors already exist.  Because of its sensitivity to the refractive index and its ease of implementation, we believe that Cherenkov light can serve as a powerful alternative to the expensive experimental procedure needed previously to constrain the refractive index.  If successful, this method could greatly extend the coverage of the refractive index, which is currently limited to wavelengths above the scintillation wavelength of 128 nm.

For all of the reasons discussed above, we believe that the study presented here has multiple significant implications for experimental high-energy physics, especially concerning vertex reconstruction and particle identification.  The Cherenkov yield is substantial, especially in certain angular bins, and it contains important directional information about the momentum of the proton.  Our absorptive fit to the refractive index significantly improves the theoretical description compared to previous fits, which omitted the physics of anomalous dispersion.  And the Cherenkov yield is a sensitive probe of the refractive index -- so much so that it can plausibly be used to perform novel measurements of the refractive index in unexplored wavelength regimes.  These results are timely and can be immediately applied to benefit international collaborations like CCM and DUNE.  If our calculation of Cherenkov radiation in LAr is successfully integrated into the worldwide experimental neutrino physics program, it will constitute a groundbreaking advance that has added a whole new instrument to the analysis toolkit.

\hspace{\parindent}

\newpage



\newpage

\appendix
\newpage

\hspace{\parindent}

\section{Harmonic Oscillator Model: The Frequency Dependence of Refractive Index}
\label{sec:HarmonicOscillator}

\hspace{\parindent}

\begin{figure}[h!] 
\begin{centering}
\includegraphics[width=0.5\textwidth]{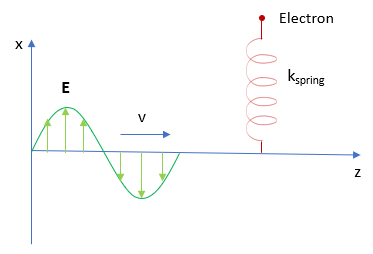}
\caption{Simplified Model of Electrons in Dielectrics  
\label{f:electronsindielectrics}
}
\end{centering}
\end{figure}

In this Appendix, we summarize the derivation of the particular (absorptive) form of the refractive index used in this work, following the canonical textbook treatment \cite{Jackson:1998nia, griffiths_2017}.

Three parameters govern the propagation of electromagnetic waves through matter: the permittivity $\epsilon$, the permeability $\mu$, and the conductivity $\sigma$.  In general, all three quantities can be frequency dependent: $\epsilon (\omega), \mu(\omega), \sigma(\omega)$.  The frequency dependence of the refractive index $n(\omega)$ inherited from these three parameters leads to the phenomenon of dispersion, such as the different angles taken by different colors of light in a prism.



Consider an elementary, classical model of in a dielectric material.  For small displacements from equilibrium, the forces which bind electrons to their atomic nuclei can be approximated as a simple harmonic oscillator (HO) (~Fig \ref{f:electronsindielectrics}): 
\begin{align}   \label{e:fbinding}
    F_{binding} = - k_{spring} \, x = - m\omega_{0}^2 \, x,
\end{align}
where $x$ is the displacement from equilibrium, $m$ is the electron's mass, and $\omega_{0} = \sqrt{k_{spring}/m}$ is the natural oscillation frequency.  (Geometrically, the statement is that any function can be approximated by a suitable parabola near a local minimum.)  In the presence of an electromagnetic wave of frequency $\omega$ propagating through the medium (Fig. \ref{f:electronsindielectrics}), the electron is also subject to a driving force due to the oscillating electric field, 
\begin{align}   \label{e:fdriving}
    F_{driving} = qE = qE_0 \cos(\omega t) \: ,
\end{align}
where q is the charge of the electron and $E_0$ is the amplitude of the electric field.  Meanwhile, dissipative effects can be approximated through a linear damping term of the form
\begin{align}   \label{e:fdamping}
    F_{damping} = -m\gamma \frac{dx}{dt} \: ,
\end{align}
with $\gamma$ the damping constant.  Altogether, this gives for the equation of motion
\begin{align}   \label{e:ftot}
    \qquad m \frac{d^2x}{dt^2} + m\gamma \frac{dx}{dt} +  m\omega_{0}^2 x = qE_0 \cos(\omega t)  \: .
\end{align}
We emphasize that the ``harmonic oscillator model'' (HO) encapsulated in the equation of motion \eqref{e:ftot} is derived from first principles as approximations to the exact physics; if desired, Eq.~\eqref{e:ftot} can be systematically improved to higher accuracy.

Eq.~\eqref{e:ftot} may be regarded as the real part of a complex equation $x(t) = \mathrm{Re} \: \tilde{x} (t)$ given by
\begin{align}   \label{e:ftotcomplex}
    \frac{d^2\tilde{x}}{dt^2} + \gamma \frac{d\tilde{x}}{dt} +  \omega_{0}^2 \tilde{x} = \frac{q}{m} E_0 \: e^{-i\omega t} \: .
\end{align}
In the steady state, the system will oscillate at the driving frequency $\omega$ with some amplitude $\tilde{x}_0$:
\begin{align}   \label{e:xcomplex}
    \tilde{x} (t) = \tilde{x}_0 \: e^{-i\omega t}, 
\end{align}
which gives the solution for the amplitude as
\begin{align}   \label{e:x0complex}
    \tilde{x}_0 = \frac{q/m}{\omega_{0}^2 - \omega^2 - i\gamma\omega} E_{0}. 
\end{align}
Then the induced (complex) dipole moment $\tilde{p} (t)$ is just proportional to the displacement:
\begin{align}   \label{e:dipolemoment}
    \tilde{p} (t) = q \: \tilde{x} (t) = \frac{q^2/m}{\omega_{0}^2 - \omega^2 - i\gamma\omega} 
    \: E_{0} \: e^{-i\omega t}. 
\end{align}
We note that the imaginary term in the denominator produces a phase shift of the induced dipole moment with respect to the external field.  The electron oscillations being out of phase with the driving force is associated with the dissipation of energy as the external field now does some amount of negative work on the oscillating electron.

To generalize Eq.~\eqref{e:dipolemoment} to a material, we consider $N$ molecules per unit volume with $f_j$ of electrons per molecule\footnote{We use the uppercase $N$ for the number of molecules per unit volume because we reserve the lowercase $n$ to refer to the refractive index.}.  Rather than assuming that each electron can oscillate about a single local minimum, we consider oscillations about a range $j$ of possible minima.  Describing each of these possible oscillation modes by a   frequency $\omega_j$ and damping factor $\gamma_j$, the total polarization density $\textbf{P}$ generated by the electromagnetic wave is
\begin{align}   \label{e:dipolemomentN}
    \tilde{\textbf{P}} = \frac{Nq^2}{m} {\left( \sum_{j}\frac{f_j}{\omega_{j}^2 - \omega^2 - i\gamma_j \omega}\right)} \tilde{\textbf{E}}. 
\end{align}
Eq.~\eqref{e:dipolemomentN} treats the many electrons as independent; this is valid for the case of a dilute gas and may receive corrections at high density.

The complex susceptibility $\tilde{\chi}_e$ is defined as the constant of proportionality between the electric field and polarization:
\begin{align}   \label{e:dipolemomentcomplexN}
    \tilde{\textbf{P}} = \epsilon_0 \tilde{\chi_e} \tilde{\textbf{E}},
\end{align}
so we may read off from Eq.~\eqref{e:dipolemomentN}
\begin{align}   \label{e:complxsusc}
    \tilde{\chi}_e = \frac{Nq^2}{m \epsilon_0} {\left( \sum_{j}\frac{f_j}{\omega_{j}^2 - \omega^2 - i\gamma_j \omega}\right)} \: . 
\end{align}
Then the susceptibility enters the permittivity through
\begin{align}   \label{e:idielectricconst}
    \frac{\tilde{\epsilon}}{\epsilon_0} = 
    1 + \tilde{\chi}_e =
    1 + \frac{Nq^2}{m\epsilon_0} {\left( \sum_{j}\frac{f_j}{\omega_{j}^2 - \omega^2 - i\gamma_j \omega}\right)}. 
\end{align}
Ordinarily, the imaginary term is negligible; however, when $\omega$ is very close to one of the resonant frequencies ($\omega_j$) it describes absorption, as we shall see. In a dispersive medium, the plane wave solution of the wave equation ($\nabla^2\tilde{\textbf{E}} = \tilde{\epsilon} \mu_0 \frac{\partial^2\tilde{\textbf{E}}}{\partial t^2}$) for a given frequency reads, 



%
\begin{align}   \label{e:planewavesoln}
    \tilde{\textbf{E}} (z, t) = \tilde{\textbf{E}}_0 \: e^{i(\tilde{k} z - \omega t)}, 
\end{align}

with the complex wave number given by, 

\begin{align}   \label{e:iwaveno}
    \tilde{k} = k + i \kappa \equiv \sqrt{\tilde{\epsilon} \mu_0} \: \omega. 
\end{align}
%


Thus, Eq.~\eqref{e:planewavesoln} becomes, 
\begin{align}   \label{e:planewavesolnfinal}
    \tilde{\textbf{E}} (z, t) = \tilde{\textbf{E}}_0 \: e^{-\kappa z} \: e^{i(\tilde{k} z - \omega t)}, 
\end{align}

The wave is attenuated due to the absorption of energy from damping. Because the intensity is proportional to $E^2$ (and hence to $e^{-2\kappa z})$, the quantity $\alpha \equiv 2 \kappa$, is called the \textit{absorption coefficient}.  

Moreover, the wave velocity is $\omega/k$, and the refractive index, $n = \frac{ck}{\omega}$; where k and $\kappa$ are determined by the parameters of our damped harmonic oscillator. The second term in Eq.~\eqref{e:idielectricconst} is small for gases, and we can approximate the square root (Eq.~\eqref{e:iwaveno}) by the first term in the binomial expansion. $ \sqrt{1+\epsilon} \cong 1 + \frac{1}{2}\epsilon$. Then,


%
\begin{align}   \label{e:iwavenofinal}
    \tilde{k} = \frac{\omega}{c} \sqrt{\tilde{\epsilon}_r} \cong \frac{\omega}{c} {\left[1 + \frac{Nq^2}{2m\epsilon_0} \sum_{j}\frac{f_j}{\omega_{j}^2 - \omega^2 - i\gamma_j \omega} \right]}, 
\end{align}

Finally, we derive general expressions for $n$ and $\alpha$ given by,   

\begin{align}   \label{e:nfinal}
     n = \frac{ck}{\omega} \cong 1 + \frac{Nq^2}{2m\epsilon_0} \sum_{j}\frac{f_j (\omega_{j}^2 - \omega^2)}{(\omega_{j}^2 - \omega^2)^2 + \gamma_j^2 \omega^2}, 
\end{align}

and 

\begin{align}   \label{e:alphafinal}
    \alpha \equiv 2 \kappa \cong \frac{Nq^2 \omega^2}{m \epsilon_0 c} \sum_{j}\frac{f_j \gamma_j}{(\omega_{j}^2 - \omega^2)^2 + \gamma_j^2 \omega^2}. 
\end{align}

If we stay far from any resonances and ignore damping, the formula for refractive index further simplifies to, 

\begin{align}   \label{e:nfinalsimplified}
     n \cong 1 + \frac{Nq^2}{2m\epsilon_0} \sum_{j}\frac{f_j }{\omega_{j}^2 - \omega^2}.
\end{align}

For most substances, the natural frequencies $\omega_j$ are scattered all over the spectrum in a somewhat chaotic fashion. But the nearest significant resonances typically lie in the ultraviolet for transparent materials, so that $\omega < \omega_j$. 


Now we will derive the wavelength equivalent expression of \eqref{e:nfinal} and \eqref{e:alphafinal} by simply noting that the angular frequency ($\omega$) is related to the wavelength ($\lambda$) as, $\omega = 2\pi c/\lambda$
and $\omega_j = 2\pi c/\lambda_j$ so by substituting this in \eqref{e:nfinal} and \eqref{e:alphafinal} we have, 

\begin{align}   \label{e:nlambda}
     n = 1 + \frac{Nq^2}{2m\epsilon_0} \sum_{j}\frac{f_j \lambda^2 \lambda_{j}^2 (\lambda^2 - \lambda_{j}^2)}{(2 \pi c)^2 (\lambda^2 - \lambda_{j}^2)^2 + \gamma_j^2 \lambda^2 \lambda_{j}^2}, 
\end{align}

and 

\begin{align}   \label{e:alphalambda}
    \alpha \cong \frac{Nq^2}{m \epsilon_0 c} \sum_{j}\frac{f_j \gamma_j \lambda^2 \lambda_{j}^4}{(2 \pi c)^2 (\lambda^2 - \lambda_{j}^2)^2 + \gamma_j^2 \lambda^2 \lambda_{j}^4}. 
\end{align}

Now, if we consider only UV and IR resonances, we have the corresponding UV and IR terms from the summation from \eqref{e:nfinal} and \eqref{e:alphafinal} as follows,

\begin{align}   \label{e:nlambdauvir}
     n = 1 + \frac{Nq^2}{2m\epsilon_0} \left( \frac{f_{UV} \lambda^2 \lambda_{UV}^2 (\lambda^2 - \lambda_{UV}^2)}{(2 \pi c)^2 (\lambda^2 - \lambda_{UV}^2)^2 + \gamma_{UV}^2 \lambda^2 \lambda_{UV}^2} + \frac{f_{IR} \lambda^2 \lambda_{IR}^2 (\lambda^2 - \lambda_{IR}^2)}{(2 \pi c)^2 (\lambda^2 - \lambda_{IR}^2)^2 + \gamma_{IR}^2 \lambda^2 \lambda_{IR}^2} \right), 
\end{align}

and 

\begin{align}   \label{e:alphalambdauvir}
    \alpha \cong \frac{Nq^2}{m \epsilon_0 c} \left( \frac{f_{UV} \gamma_{UV} \lambda^2 \lambda_{UV}^4}{(2 \pi c)^2 (\lambda^2 - \lambda_{UV}^2)^2 + \gamma_{UV}^2 \lambda^2 \lambda_{UV}^4} + \frac{f_{IR} \gamma_{IR} \lambda^2 \lambda_{IR}^4}{(2 \pi c)^2 (\lambda^2 - \lambda_{IR}^2)^2 + \gamma_{IR}^2 \lambda^2 \lambda_{IR}^4} \right). 
\end{align}

\begin{figure}[h!] 
\begin{centering}
\includegraphics[width=0.5\textwidth]{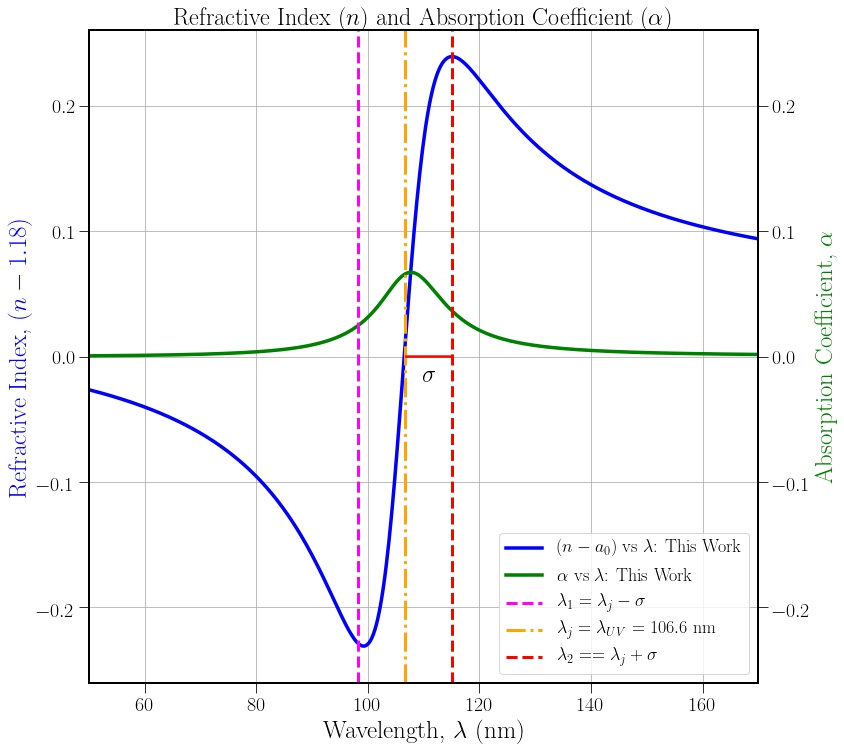}
\caption{Refractive Index and Absorption Coefficients in the vicinity of UV resonance ($\lambda_{UV} = $ 106.6 nm)   
\label{f:absorption_coeff_theoryplot2}
}
\end{centering}
\end{figure}

The index of refraction and the absorption coefficient given by Eq.~\eqref{e:nlambdauvir} and \eqref{e:alphalambdauvir} in the vicinity of one of the resonances ($\lambda_{UV} = $ 106.6 nm) is plotted here in Fig. \ref{f:absorption_coeff_theoryplot2}. 


%

Generally, the index of refraction rises gradually with increasing frequency or decreasing wavelength, however, near a resonance the index of refraction drops sharply. This is called \textit{anomalous dispersion} because of this unusual behavior. Notice that the region of anomalous dispersion ($\lambda_1 < \lambda_{UV} < \lambda_2$, in the figure) coincides with the region of maximum absorption. In this wavelength range, the material can be nearly impenetrable. Because we are now driving the electrons at their ``preferred'' frequency/wavelength, their oscillations have a comparatively significant amplitude, which causes the damping process to dissipate a substantial amount of energy.


\vspace{1cm}

\noindent

\section{Numerical Convergence Test of Step-size and Angular Resolution }

\label{App:ConvergenceTest}

\hspace{\parindent}

\begin{figure}[tp]
\centering

\begin{subfigure}{0.32\columnwidth}
\centering
\includegraphics[width=\textwidth]{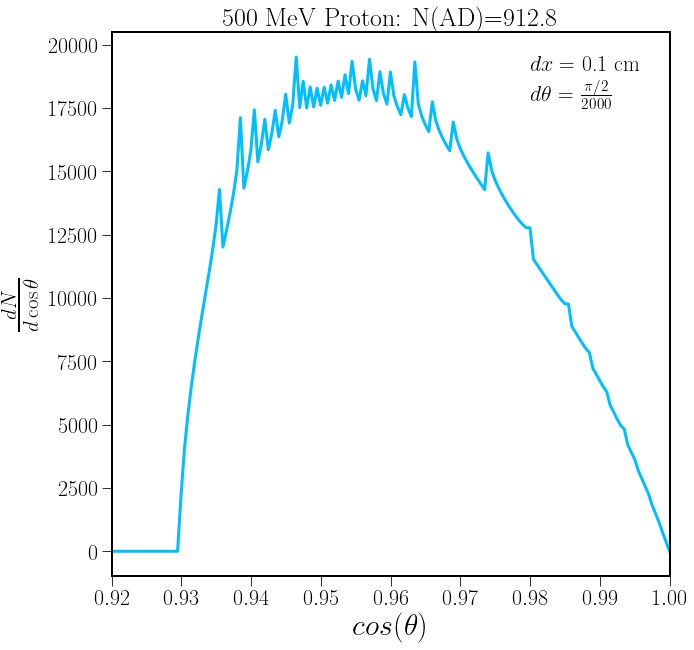}
\caption{$dx=0.1, d\theta=\frac{\pi/2}{2000}$}
\label{fig:PVS500MeVdxp1dtheta2000}
\end{subfigure} 
\begin{subfigure}{0.32\columnwidth}
\centering
\includegraphics[width=\textwidth]{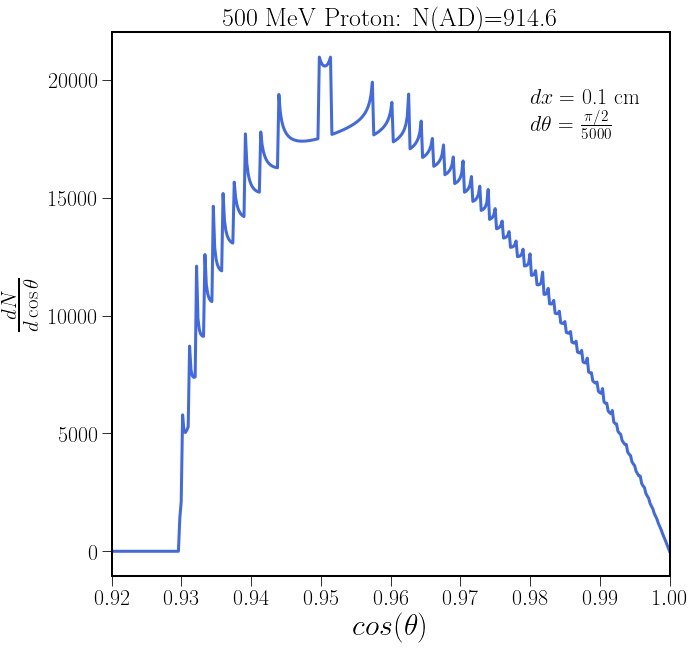}
\caption{$dx=0.1, d\theta=\frac{\pi/2}{5000}$}
\label{fig:PVS500MeVdxp1dtheta5000}
\end{subfigure}
\begin{subfigure}{0.32\columnwidth}
\centering
\includegraphics[width=\textwidth]{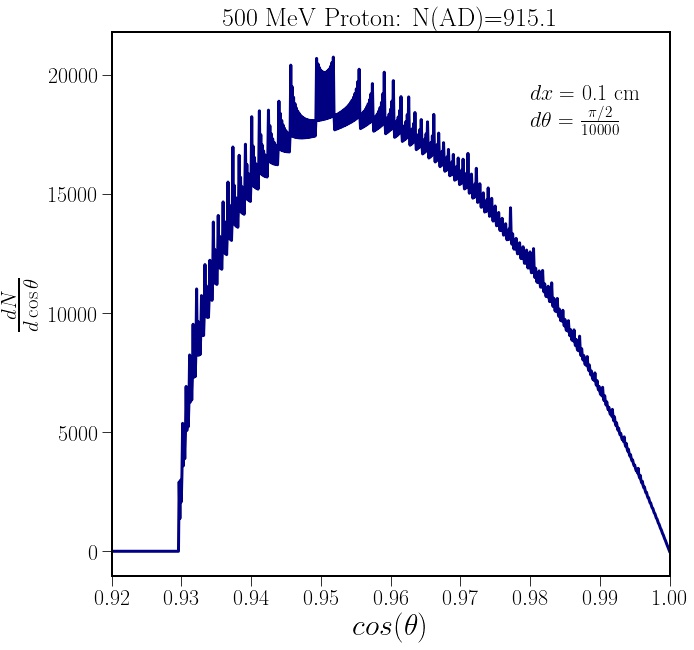}
\caption{$dx=0.1, d\theta=\frac{\pi/2}{10000}$}
\label{fig:PVS500MeVdxp1dtheta10000}
\end{subfigure}

\medskip

\begin{subfigure}{0.32\columnwidth}
\centering
\includegraphics[width=\textwidth]{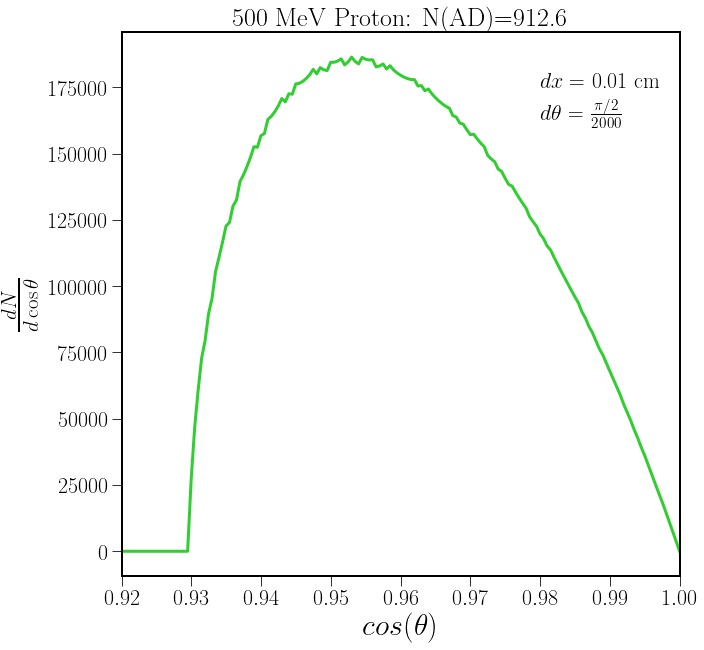}
\caption{$dx=0.01, d\theta=\frac{\pi/2}{2000}$}
\label{fig:PVS500MeVdxp01dtheta2000}
\end{subfigure} 
\begin{subfigure}{0.32\columnwidth}
\centering
\includegraphics[width=\textwidth]{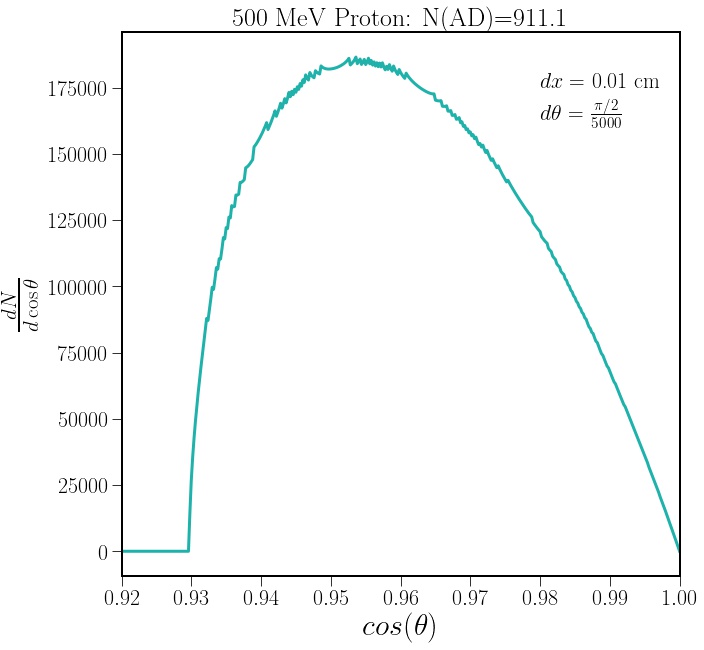}
\caption{$dx=0.01, d\theta=\frac{\pi/2}{5000}$}
\label{fig:PVS500MeVdxp01dtheta5000}
\end{subfigure}
\begin{subfigure}{0.32\columnwidth}
\centering
\includegraphics[width=\textwidth]{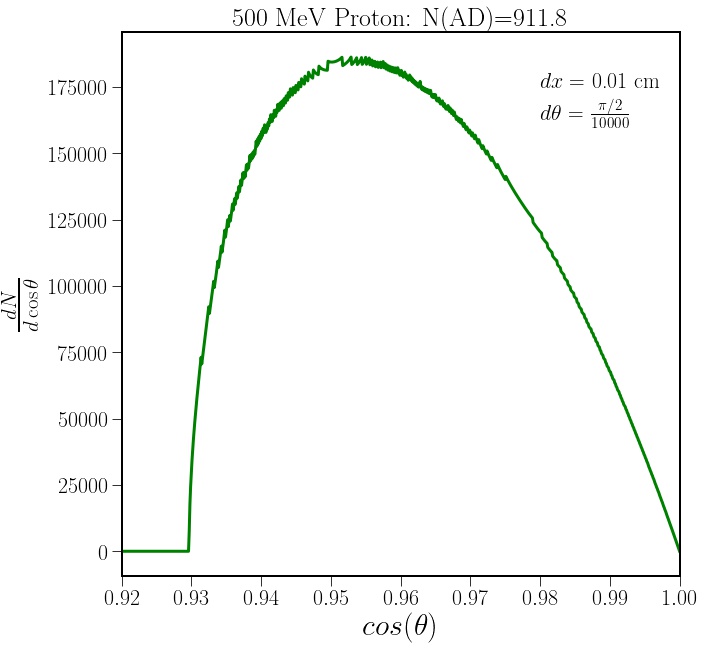}
\caption{$dx=0.01, d\theta=\frac{\pi/2}{10000}$}
\label{fig:PVS500MeVdxp01dtheta10000}
\end{subfigure}

\medskip

\begin{subfigure}{0.32\columnwidth}
\centering
\includegraphics[width=\textwidth]{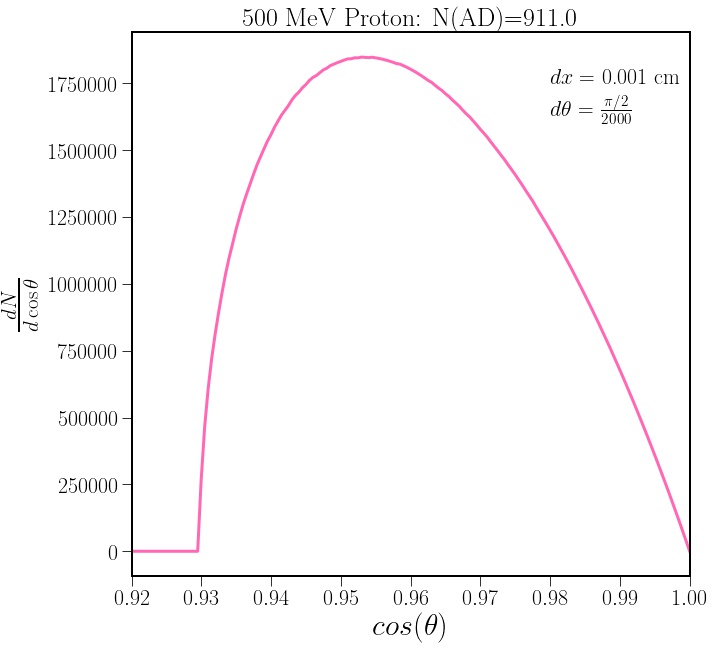}
\caption{$dx=0.001, d\theta=\frac{\pi/2}{2000}$}
\label{fig:PVS500MeVdxp001dtheta2000}
\end{subfigure} 
\begin{subfigure}{0.32\columnwidth}
\centering
\includegraphics[width=\textwidth]{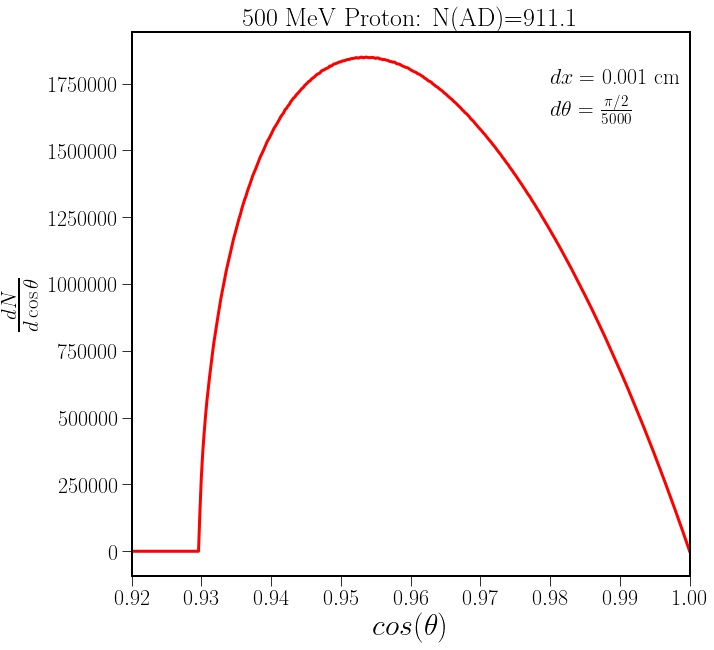}
\caption{$dx=0.001, d\theta=\frac{\pi/2}{5000}$}
\label{fig:PVS500MeVdxp001dtheta5000}
\end{subfigure}
\begin{subfigure}{0.32\columnwidth}
\centering
\includegraphics[width=\textwidth]{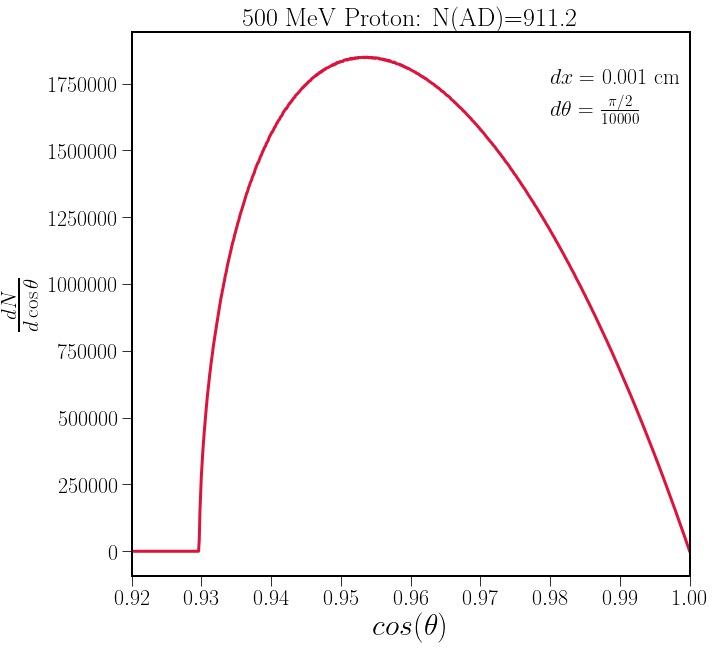}
\caption{$dx=0.001, d\theta=\frac{\pi/2}{10000}$}
\label{fig:PVS500MeVdxp001dtheta10000}
\end{subfigure}

\caption{Effect of stepsize ($dx$) and angular resolution ($d\theta$) on total Cherenkov yield N(AD) calculated for a 500 MeV proton in LAr from AD method using our (approx) refractive fit.}
\label{fig:cherenkresolution_pvs}
\end{figure}

We noted in Chapter 2 that the total yield N(AD) calculated from the area under the Angular Distribution (AD) curve depends on our choice of step size $dx$ used to integrate over the proton trajectory and also on the angular resolution $d\theta$ used to compute the angular distribution. Smaller step sizes $dx$ and $d\theta$ will yield more precise results, but increase the cost in terms of computational time.  It is thus important to optimize this resolution to achieve the desired precision most efficiently. For most of our calculations, we have used a bigger step size of $dx=$ 0.1 cm and $d\theta= \frac{\pi/2}{2000}$, but we increased the resolution by decreasing dx and $d\theta$ wherever any calculation demanded better precision.  The convergence of the angular distribution is shown in Fig.~\ref{fig:cherenkresolution_pvs}, and the improvement of the quantitative agreement in the yield between the angular distribution and direct Frank-Tamm calculation with decreasing step sizes is summarized in Table~\ref{tab:appstepsize}.




\begin{table}[h!]
  \begin{center}
    \begin{tabular}{|c|c|c|c|c|} 
      \hline
      \textbf{$dx$}(cm) & \textbf{$d\theta$} &  \textbf{$N(AD)$} & \textbf{N(FT)} & \textbf{\% Error} \\
      \hline
      0.1	& $\frac{\pi/2}{2000}$ & 912.81	& 932.31 & 2.09 \\
      \hline
      0.1	& $\frac{\pi/2}{5000}$ & 914.58	& 932.31 & 1.90 \\
      \hline
      0.1	& $\frac{\pi/2}{10000}$ & 915.11 & 932.31 & 1.84 \\
      \hline
      \hline
      0.01 & $\frac{\pi/2}{2000}$ & 912.57 & 929.03 & 1.77 \\
      \hline
      0.01 & $\frac{\pi/2}{5000}$ & 911.07 & 929.03 & 1.93 \\
      \hline
      0.01 & $\frac{\pi/2}{10000}$ & 911.80 & 929.03 & 1.85 \\
      \hline
      \hline
      0.001 & $\frac{\pi/2}{2000}$ & 910.98 & 928.70 & 1.91 \\
      \hline
      0.001 & $\frac{\pi/2}{5000}$ & 911.14 & 928.70 & 1.89 \\
      \hline
      0.001 & $\frac{\pi/2}{10000}$ & 911.24 & 928.70 & 1.88 \\
      \hline
    \end{tabular}
  \end{center}
  \caption{Quantitative cross-check of the yield and angular distribution for a 500 MeV Proton in LAr using our (approx) fit for different choices of stepsize ($dx$) and angular resolution ($d\theta$). The total Cherenkov yield $N(FT)$ computed directly from the Frank-Tamm formula is compared with the total yield $N(AD)$ obtained from numerically integrating the angular distribution. The agreement between the two methods is good and gets better with decreasing step size.}
  \label{tab:appstepsize}
\end{table}

%

%

\newpage

\hspace{\parindent}


\newpage



\addcontentsline{toc}{section}{REFERENCES}
\setlength{\baselineskip}{\singlespace}

%

\end{document}